%% file: thesis_revised.tex
\label{key}\documentclass [12pt,a4,twosided]{book}
\usepackage{fancyhdr}
\usepackage{multirow}\label{key}
\usepackage{wrapfig}
\usepackage[toc,page]{appendix}
\usepackage{ragged2e}
\usepackage{graphicx}
\usepackage{float}
\usepackage{anysize}
\usepackage{subfigure}
\usepackage{setspace}
\usepackage{cancel}
\usepackage{rotating}
\usepackage{lscape}
\usepackage{amssymb}
\usepackage{amsmath}

\usepackage{wasysym}
\usepackage{latexsym}
\usepackage[margin=28pt,font={small,stretch=1.25},labelfont=sc]{caption}
\usepackage{afterpage}
\usepackage{rotating}
\usepackage[usenames,dvipsnames]{color}
\usepackage[table]{xcolor}
\usepackage[left=3.5cm, right=2.5cm, bottom=3.6cm, top=2.70cm]{geometry}
\usepackage{float}
\usepackage{multirow}
\usepackage{adjustbox}

\usepackage{url}
\usepackage{imakeidx}
\makeindex
\usepackage{setspace}
\usepackage{mathpazo}
\usepackage{multirow}
\usepackage{enumitem}
\usepackage{cite}
\usepackage{hyperref}
\usepackage{gensymb,bm}
\usepackage{widetext}
\usepackage{array}
\newcolumntype{P}[1]{>{\centering\arraybackslash}p{#1}}
\newcolumntype{M}[1]{>{\centering\arraybackslash}m{#1}}

\hypersetup{
	linktocpage,
	linktoc=all,			
	pagebackref,
	pdfpagemode={UseOutlines},	
	colorlinks=true,
	linkcolor=Blue,
	citecolor=blue,			
	urlcolor=blue,			
	filecolor=blue,
	anchorcolor = green,
	pdfstartview={FitV},
	unicode,
	breaklinks=true,
	bookmarks=true,
	bookmarksopen=true,         
	bookmarksopenlevel=0,
	bookmarksnumbered=true,     
	hypertexnames=false,
	hyperindex,
	pdftitle={Pushpendra~Yadav~Ph.D. Thesis~IITK},
	pdfauthor={Pushpendra~Yadav},
	pdfcreator={pdflatex},
}

\usepackage{titlesec}
\titleformat{\chapter}[display]
  {\normalfont\bfseries\Huge}
  {\chaptertitlename~\thechapter}{3pc}
  {
	{\color{brown}\titlerule[2pt]}
  }
  [{\color{brown}\titlerule[2pt]}]
\titleformat{name=\chapter,numberless}[display]
  {\normalfont\bfseries\Huge}{}{2pc}
  {}

\makeatletter
\renewcommand\cleardoublepage{%
  \clearpage
  \if@twoside
    \ifodd\c@page
    \else
      \hbox{}\thispagestyle{numberleft}\newpage 
      \if@twocolumn
        \hbox{}\newpage
      \fi
    \fi
  \fi}
\makeatother
\fancypagestyle{numberleft}{%
  \fancyhf{} 
  \renewcommand{\headrulewidth}{0pt}
  }

\renewcommand{\Re}{\mathrm{Re}\,}
\renewcommand{\Im}{\mathrm{Im}\,}
\renewcommand{\i}{\mathrm{i}}

\renewcommand{\Re}{\mathrm{Re}\,}
\renewcommand{\Im}{\mathrm{Im}\,}

\DeclareMathAlphabet{\bi}{OML}{cmm}{b}{it}
\fancyheadoffset{0.005\textwidth}

\def\be{\begin{equation}}
\def\ee{\end{equation}}

\def\bea{\begin{eqnarray}}
\def\eea{\end{eqnarray}}

\def \formula{$\mathrm{MoSi}_{2}{\mathrm{Z}}_{4}$}
\def \formulaAs{$\mathrm{MoSi}_{2}{\mathrm{As}}_{4}$}
\def \formulaN{$\mathrm{MoSi}_{2}{\mathrm{N}}_{4}$}
\def \formulaP{$\mathrm{MoSi}_{2}{\mathrm{P}}_{4}$}
\def \hsp{\hspace{5cm}}

\def\bearr{\begin{eqnarray}}
\def\eearr{\end{eqnarray}}

\makeindex
\begin{document}

\begin{center} \thispagestyle{empty}
{\LARGE {\bf Exciton dynamics in equilibrium and nonequilibrium regimes}}

\vskip 2.0cm

{\large A Thesis Submitted\\
in Partial Fulfilment of the Requirements\\
for the Degree of\\
{\bf Doctor of Philosophy}}

\vskip 2.0cm

\large {by}\\
\Large{\bf{Pushpendra Kumar Yadav}} \\
\large{Roll No: 18209268}

\end{center}

\vspace{1.0cm}
\begin{figure}[h]
\begin{center}
\includegraphics[scale=.5]{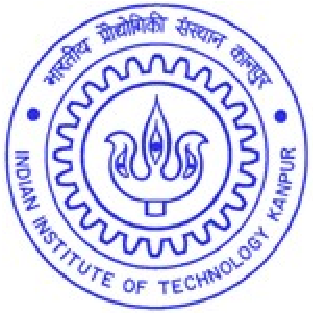}
\end{center}
\end{figure}  

\vskip 0.5cm

\begin{center}
\large{to the}\\
\large{\bf{DEPARTMENT OF PHYSICS}}\\
\large{{INDIAN INSTITUTE OF TECHNOLOGY KANPUR}}\\
\large{\bf{February, 2024}}\\
\end{center} \newpage \thispagestyle{empty} \mbox{} 
\newpage

\pagenumbering{roman}
\setcounter{page}{2}
\renewcommand{\headrulewidth}{0pt}
\begin{figure*}[t]
	\centering
	\hspace*{-1.2in}
	\includegraphics[scale=1.0]{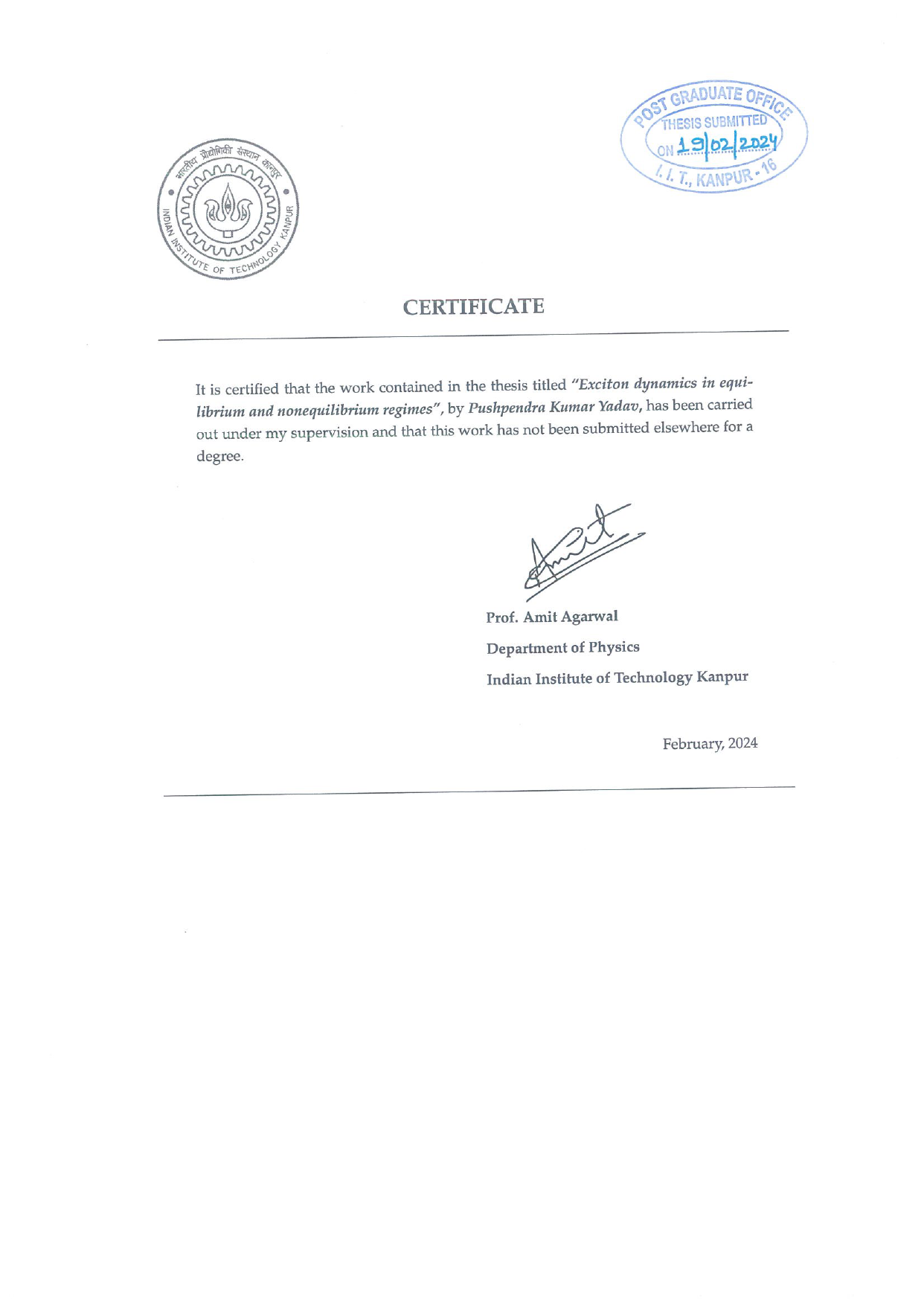}
\end{figure*} 

\cleardoublepage
\newpage\leavevmode\thispagestyle{empty}
\newpage
\renewcommand{\headrulewidth}{0pt}
\begin{figure*}[t]
	\centering
	\hspace*{-1.2in}
	\includegraphics[scale=1.0]{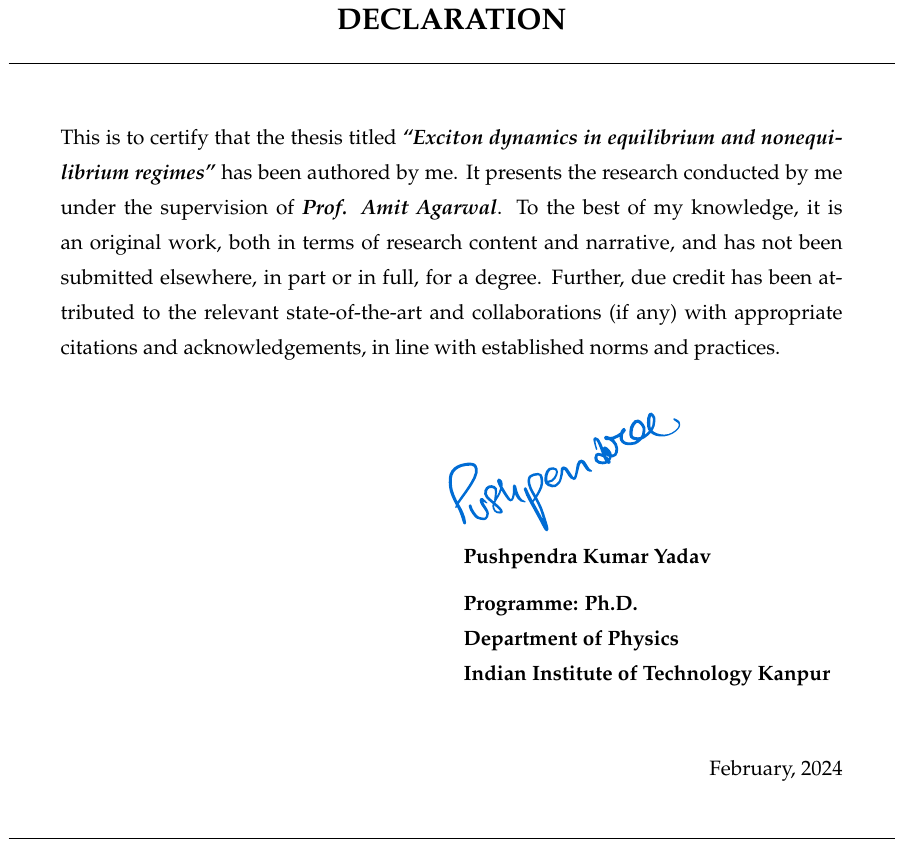}
\end{figure*} 

\cleardoublepage
\clearpage
\newpage\leavevmode\thispagestyle{empty}\newpage

\addcontentsline{toc}{chapter}{Thesis Synopsis}
\renewcommand{\headrulewidth}{0.7pt}

\input{synopsis.tex}

\cleardoublepage

\phantomsection
\begin{spacing}{1.4}
	\addcontentsline{toc}{chapter}{Acknowledgements}
	\input{acknowledge.tex}
\end{spacing}

\input{ded.tex}
\cleardoublepage
\phantomsection
\input{publication.tex}

\cleardoublepage
\phantomsection

\addtolength{\parskip}{-0.5\baselineskip}
\begin{spacing}{1.5}
	\renewcommand{\contentsname}{Contents}
	\tableofcontents
	
	\cleardoublepage
	\phantomsection
	\addtolength{\parskip}{0.4\baselineskip}
	\addcontentsline{toc}{chapter}{\listfigurename}
	\listoffigures
	
	\cleardoublepage
	\phantomsection
	\addcontentsline{toc}{chapter}{\listtablename}
	\listoftables
	\cleardoublepage
	\phantomsection
	
\end{spacing}

\pagenumbering{arabic}%

\addtolength{\parskip}{0.5\baselineskip}

\cleardoublepage
\phantomsection
\input{chap1.tex}

\input{chap2.tex}
\input{chap3.tex}

\input{chap4.tex}
\input{chap5.tex}
\input{chap6.tex}


\cleardoublepage
\phantomsection
\begin{spacing}{1.25}
\addcontentsline{toc}{chapter}{Bibliography}%

\bibliographystyle{abbrv_ashu}

\bibliography{ref_thesis}



\end{spacing}

\end{document}

%% file: acknowledge.tex

\thispagestyle{plain}
\begin{center}
	{\Huge \bf Acknowledgements}
\end{center}

\vskip 50pt
"Life is a journey" feels more realistic now. When I came to this beautiful campus of IIT Kanpur in January 2019 for my Ph.D., it felt like a home here. Just a couple of my undergraduate friends were like a trigger for a chain reaction of making friends for life on campus. Their company and cheerfulness made me fall in love with this vibrant and soulful campus, and I spent the five most beautiful years of my life. It has been an honor to pursue physics within this intellectually rich research environment of IITK, which has helped me grow academically and learn life lessons. Now, as this beautiful journey is about to end, I would like to express my heartfelt gratitude to everyone who has been a part of it.

First and foremost, I sincerely thank my thesis supervisor, Prof. Amit Agarwal. His support has been integral; without him, I could not have achieved what I have so far. I joined Prof. Amit for my PhD in January 2019. Throughout these years, I have learned a lot of lessons, academic and non-academic, which I will remember for my life. Engaging in physics discussions with him has consistently been a valuable learning experience for me. His guidance in breaking complex problems and calculations in the easiest possible way and his positive perspective has helped me transform into a different personality. His open-mindedness for new projects and learning new tools has always encouraged me to learn new things that immensely helped my Ph.D. training. He has played a crucial role in improving my academic writing and project management skills with his constant advice. He is a great mentor, a thoughtful advisor, and an excellent motivational speaker. He always reminded me of my privileges as a researcher whenever I faced ups and downs during these years. I would also like to thank Prof. Amit and his family for the cheerful moments shared during Diwali and several other dinner occasions.

I want to express my sincere gratitude to Prof. K V Adarsh and Prof. Subham Majumdar. Working with these outstanding experimental physicists during our collaborative projects provided me with valuable insights into the experimental aspects of condensed matter physics. I extend my heartfelt thanks to Prof. Sitangshu Bhattacharya and Prof. Bramhachari Khamari for their guidance in learning about the theoretical understanding of exciton physics.

I am grateful to the Late Prof. Amit Dutta, Prof. Manoj K. Harbola, Prof. Soumik Mukhopadhyay, Prof. Arijit Kundu, Prof. Adhip Agarwala, Prof. Sudeep Ghosh, and Prof. Koushik Pal for invaluable feedback during seminars and peer reviews, which has significantly contributed in refining my research.

I express gratitude to my M.Sc. thesis supervisor, Prof. Ranjit Nanda, and Prof. Pinaki Majumdar for their influential guidance in computational condensed matter physics.

I extend my sincere gratitude to Dr. Gyan Prakash, Dr. Sougata Mardanya, Dr. Barun Ghosh, Dr. Shailendra Rathor, Dr. Manohar Sharma, Dr. Ashish, and Dr. Atasi Chakraborty for their constant support, encouragement, and insightful technical and conceptual discussions. They have served as mentors, guiding me through the challenges in academics and life. I also want to express my appreciation to other collaborators, Dr. Imran Ahamed, Dr. Suman Mondal, Ajay Poonia, Riyanka Karmakar, and Pravrati Taank, from whom I have gained valuable insights into various aspects of condensed matter physics.

I am grateful to my friends and colleagues in the QTT group Raihan da, Kamal, Debottam, Debasis, Maneesh, Nirmalya, Sunit, Harsh, Sanjay, Dibyanandan, Koushik, and Sayan. Our daily tea break has always been a delightful time when we step outside to research, share jokes, discuss life, and lighten the mood, an experience I will always cherish. Having Kamal, Debasis, and Debottam as friends in the group has been a blessing for me. Our discussions extended beyond physics, involving random dinner outings and supporting each other whenever needed. I am grateful to everyone in the group for their help in my PhD seminar and thesis.

I am thankful to my friends on the campus: Rohit, Tahir, Kishori, Ashish, Poonam, Vinayak, Hari, Pooja, Soumyadeep, Ayesha, Swati, Santu, Ekta, Suraina, and Suman, for creating the most memorable moments during midnight birthday celebrations, especially in the chilling months of December and January. I am also grateful to my friends from the 103D Lab, Samrat, Vikash, Bheem, Arunava, and Joy, and others, for the philosophical discussions and humorous moments in and outside the lab. I sincerely thank my friends at Hall-XI, Kamal, Yogesh, Rohit, Himanshu, and Bashab, for late-night parties and for the best adventuras memory of playing volleyball games on the coldest nights of Kanpur. I am also very grateful to Ashok bhaiya, Gaganjot bhaiya,  Shuryakant bhaiya, Samiullah, Amitabh, Sanatan and others for the dinner table discussions, which have always helped me feeling livelihood in Hall XI. I am also very grateful to my college friends Saurabh, Sunil, Dheeraj, Nauratan, Anjanay, Krishna, Uma, Shambhavi, Deepika, Prem, Satya, Shashank, and Anand.  

I am grateful to the Physics office staff, especially Pooja ma'am, for their academic assistance. Additionally, I thank the Hall XI staff members; without them, I could not imagine such a comfortable life on campus.

I am grateful to all my teachers, especially Mr. Mohan Lal Yadav, who always has been a guardian, a mentor, and a source of inspiration for me. Throughout this journey, I am also grateful to my childhood friend Sunil and all my classmates.

I give special thanks to Vijaya, my friend and companion, for her unconditional love and support. She has always been there for me, listening, encouraging, and cheering me up whenever I worried about anything. With her, this journey has been more beautiful and full of joy.

Above all, I express my gratitude to my family members. My father has had a remarkable impact on my life by providing me with his invaluable support and encouragement, especially for his efforts in waking us up at 4 AM in the morning and making the best tea ever for us so that we could study. He still makes the best tea that I can find anywhere. Special thanks to my mother for her unwavering love and support. I'm very thankful to her for helping me learn cooking. I am very grateful to my elder brothers for their support, encouragement, and belief in me. Last but not least, I'm thankful to my younger brother, who has always been there for any help and support I needed, especially for taking care of our parents. Without them, it would not have been easy for me to accomplish my goals. I dedicate my achievements to them.

\begin{flushleft}
	February, 2024 \hfill {Pushpendra Yadav}
\end{flushleft}

%% file: ded.tex
\chapter*{}\thispagestyle{empty}
\vskip 2cm

\begin{center}
\begin{tabular}{ll}
&{\Huge {To my family}}\\
\end{tabular}
\end{center} \newpage \thispagestyle{empty}\mbox{}\newpage

%% file: publication.tex


\thispagestyle{plain}
\begin{center}
{\Huge \bf List of  Publications}
\end{center}
{\bf Included in the thesis}
\begin{enumerate}

\item {{\bf Pushpendra Yadav}, Bramhachari Khamari, Bahadur Singh, K. V. Adarsh, and Amit Agarwal,
\newblock {\it Fluence dependent dynamics of excitons in monolayer MoSi$_2$Z$_4$ (Z = pnictogen)},
\newblock {\href{https://iopscience.iop.org/article/10.1088/1361-648X/acc43f/meta}{{J. Phys.: Condens. Matter} {\bf 35}, 235701 (2023)}.}}

\item {{\bf Pushpendra Yadav}, K. V. Adarsh, and Amit Agarwal,
	\newblock {\it Room temperature electron-hole liquid phase
		in monolayer MoSi$_2$Z$_4$ (Z = Pinctogen)},
	\newblock {\href{https://iopscience.iop.org/article/10.1088/2053-1583/ace83b}{{2D Mater} {\bf 10}, 045007 (2023)}.}}

\item {{\bf Pushpendra Yadav}, Amit Agarwal, and Sitangshu Bhattacharya,
	\newblock {\it Phonon-assisted photoluminescence and exciton recombination in monolayer aluminum nitride}
	\newblock {\href{https://iopscience.iop.org/article/10.1088/2053-1583/adb8be}{{2D Mater} {\bf 12}, 025022 (2025)}.}}

\end{enumerate}

\noindent{\bf Not included in the thesis}

\begin{enumerate}

\item {{Suman Mondal, {\bf Pushpendra Yadav}, Anan Bari Sarkar, Prabir Dutta, Saurav Giri, Amit Agarwal, and Subham Majumdar}
\newblock {\it Competing magnetic interactions and magnetocaloric effect in Ho$_5$Sn$_3$},
\newblock {\href{https://iopscience.iop.org/article/10.1088/1361-648X/ac2cf1/meta}{{J. Phys.: Condens. Matter} {\bf 34}, 0258011 (2022)}.}}

\item {{Ajay Poonia, {\bf Pushpendra Yadav}, Barnali Mondal, Dipendranath; Pravrati Taank, Megha Shrivastava, Angshuman Nag, A. Agarwal, and K. V. Adarsh,}
	\newblock {\it Room-Temperature Electron-Hole Condensation
		in Direct-Band-Gap Semiconductor Nanocrystals},
	\newblock {\href{https://journals.aps.org/prapplied/abstract/10.1103/PhysRevApplied.20.L021002}{{Phys. Rev. Applied} {\bf 20}, L021002 (2023)}.}}

\item {Imran Ahamed, Atasi Chakraborty, {\bf Pushpendra Yadav}, Rik Dey, Yogesh Singh Chauhan, Somnath Bhowmick, and Amit Agarwal,
	\newblock {\it Room Temperature Ferroelectricity, Ferromagnetism, and Anomalous
		Hall Effect in Half-metallic Monolayer CrTe},
	\newblock {\href{https://doi.org/10.1021/acsaelm.4c00205}{{ACS Applied Electronic Materials, 6, 5, 3810-3819} (2024)}.}}

\item {Riyanka Karmakar, Pravrati Taank, Debjit Ghoshal, {\bf Pushpendra Yadav}, Dipendranath Mandal, Megha Shrivastava, Amit Agarwal, Mathew C. Beard, Elisa M. Miller, and K. V. Adarsh,
	\newblock {\it Multiple Carrier Generation at an Exceptionally Low Energy Threshold}
	\newblock {\href{https://journals.aps.org/prl/abstract/10.1103/PhysRevLett.134.026903}{{Phys. Rev. Lett.} {\bf 134}, 026903 (2025)}.}}

\end{enumerate}

%% file: chap1.tex
\chapter[Introduction]{Introduction}
\label{chap1}
\pagestyle{fancy}
A bound state of an electron-hole pair is called an exciton. Excitons are charge-neutral bosonic quasiparticles, and they play an essential role in evaluating the optical properties of insulating materials~\cite{Mueller2018,walker87,Hanke1979-PRL}. Fundamentally, an exciton is similar to an atom in a crystal lattice. Like a hydrogen atom, excitons hold discrete excited energy states and possess inherent internal fine structures. This makes them promising for applications in excitonic-carrier devices for quantum computation and circuits. Given their bosonic nature, excitons can assemble into Bose-Einstein condensates, showing transport characteristics like a superfluid~\cite{Kasprzak2006, Eisenstein2004}. Excitons are essential in understanding all optical properties of various condensed matter systems, and they play a crucial role in the fundamental understanding of light-matter interaction and optoelectronic device applications ~\cite{Knox1983, Man2021, Quintela2022,Diana2013,Bernardi2013-graphene-light}. The large exciton oscillator strength and strong light-matter interaction are the key ingredients for efficient absorption and emission of light~\cite{Feldmann1987}. Therefore, understanding and tuning the excitons are essential for predicting the potential of quantum materials for future optoelectronic and photonic devices. 

Solid-state devices, like electronics, use particles and their quantum properties to work. There is a constant search for new ways to make devices and circuits more energy-efficient. Although spintronics and photonics are promising for this purpose, interestingly, excitons hold more promise. The initial excitonic devices, comprising GaAs double quantum wells~\cite{BUTOV20172}, marked a pivotal connection between photonics and electronics. These devices can directly process optical data streams without converting them into charge currents and light~\cite{High07,Alex2008,Grosso2009}. Excitonic transistors, akin to electronic field-effect transistors, possess a compact architecture, leveraging the nanoscale exciton size (Bohr radius)~\cite{Stier2016}. Furthermore, excitonic devices show promise for integration with photonic~\cite{GonzalezMarin2019} or plasmonic waveguides~\cite{Lee2016}, offering the potential for better-optimized structures and overcoming issues related to optical crosstalk and bulky input–output gratings~\cite{Ciarrocchi2022}. Therefore, understanding exciton characterstics in materials is crucial. Below, we discuss some interesting phenomena that arise in semiconducting solids upon photoexcitations and the challenges involved in investigating the electron-hole system at a higher electron-hole density at room temperature.

In semiconductors, photo-excited electrons and holes can form bound electron–hole pairs (excitons), trions, and biexcitons~\cite{lampert1958mobile,kheng1993observation,mak2013tightly}. Some recent optical experiments have reported that by increasing the exciton density with the help of pump fluence, the electron-hole system can be brought into a nonequilibrium regime of exciton density~\cite{Sei,MoS2-EHL-EXP-ACS-Nano-2019-Yu}. This leads to density-dependent renormalization of optical bandgap, exciton binding energy, and exciton radius with a change in the pump fluence or photo-excited carrier density. It was found that varying exciton density can renormalize the Coulomb interaction among the electron-hole pairs, and hence, the optical properties of the materials can significantly change~\cite{Sei,amit-exciton-2}. The fluence control also brings the photo-excited quasiparticles into exotic phases of electron-hole systems such as electron-hole plasma (EHP) and electron-hole liquid (EHL)~\cite{Arp2019,MoS2-EHL-EXP-ACS-Nano-2019-Yu,Dey-MoS2-2023}. This thesis addresses the theoretical understanding of such density-dependent optical properties in semiconductors within the excitons. 

Compared to bulk conventional semiconductors, the exciton binding energies of two-dimensional (2D) materials, particularly transition metal dichalcogenides (TMDs) and hexagonal boron nitride (hBN), is one order of magnitude larger as per theoretical~\cite{optical-MoS2-PRL,Davide,Molina2013,Galvani2016} and experimental~\cite{Hanbickia2015,Hill2015,Robert2016} observations. When the dielectric screening between the electron-hole pair is lower in a reduced dimensionality system, excitons are held together with exceptionally large binding energy, making excitons stable even at room temperature~\cite{Thygesen2017}. Therefore, exciton physics has attracted an enormous interest in low-dimensional materials. In this thesis, we keep two-dimensional semiconductors in the center to explore different excitonic phenomena. 

Given the potential implications of excitonic devices, the current interest lies towards its operations i) at higher temperatures~\cite{Marini2008,Molina2015}, ii) tunability of the optical response by varying exciton density~\cite{Sei,Pflug2021}, and iii) investigating different electron-hole phases at room-temperature for optoelectronic devices~\cite{Arp2019}. But, before further discussing the specific problems, we will review some basic concepts of the exciton formations, theoretical models, and exciton classifications. 
\section{Introduction to excitons}
\begin{figure}[t]
	\centering
	\includegraphics[width =0.95\linewidth]{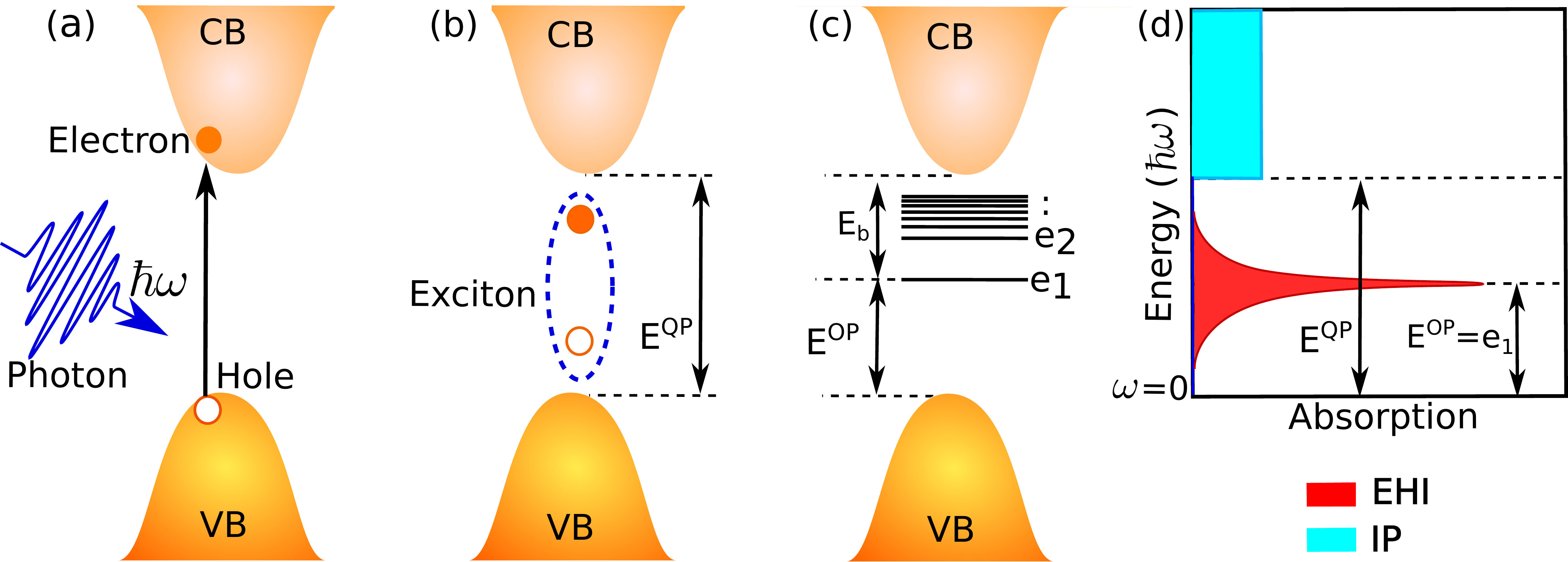}
	\caption {Illustration of exciton formation in an insulating material. (a) Optical excitation of an electron from the valence band to the conduction band via an external pump with photon energy ($\hbar\omega$), and (b) exciton formation with binding energy $E_b$. (c) The energy levels $e_1$, $e_2$, and so on, of the bound excitons below the quasiparticle band gap ($E^{QP}$). (d) Optical absorption spectrum in an insulating material with excitonic effects (red) and without electron-hole interaction (EHI) in blue color.     
		\label{ch1.fig1}}
\end{figure}

When a photon excites an electron from the valence band to the conduction band in an insulating solid, it creates a hole in the valence band. The negatively charged electron attracts the positively charged hole by their electrostatic potential [see Figure~\ref{ch1.fig1} (a)]. 
The interaction between the electron-hole pair is taken into account as a bound state leads to the formation of a bound state called an exciton, as presented in Figure~\ref{ch1.fig1} (b). Excitons are charged neutral bosonic quasiparticles, which play an essential role in calculating the optical properties of insulating materials. The significance of excitons lies in their binding energy, an attractive force that positions them in a lower energy state (\textit{$e_1, e_2, ...$}) compared to the unoccupied energy (conduction) states as shown schematically in Figure~\ref{ch1.fig1} (c). 

To understand the exciton formation from a quantum mechanical model, we review the \textit{effective mass approximation}~\cite{DRESSELHAUS195614}, which qualitatively defines excitons in an insulating material. \\

In examining the electronic properties of various solids, a model with translational symmetry is commonly employed, eliminating the need for specific structural details. The optical and transport properties of materials due to electrons and holes moving freely in regions of macroscopic size potential can be studied by their effective masses. The foundational concept of excitons, initially introduced by Slater and Shockley~\cite{Slater_exciton}, Wannier~\cite{Wannier_exciton}, and Mott~\cite{Mott1938} in the 1930s, revolves around the effective mass (EM) theory. The EM theory simplifies the description of electrons and holes in a region with constant average potential energy. For electrons, the Hamiltonian is represented as $-\hbar^2\nabla^2_e/2m^*_e$, while for holes, it is $-\hbar^2\nabla^2_h/2m^*_h$. This drastic simplification ignores the influence of the rapidly varying crystal potential on these model particles. The understanding of the EM approximation evolved with the fact that carriers, in maintaining orthogonal wave functions to the core electrons, employ an effective potential energy. This effective potential reduces a significant portion of the actual core potential, leading to a minimal net pseudopotential~\cite{Knox1983}.
At a moderate carrier density, the Coulomb repulsion effectively maintains a spatial separation between electrons (holes). Additionally, collisions between electrons and holes lead to recombination. Before the recombination process for a specific electron-hole pair, we can write the Hamiltonian governing this pair in line with the EM theory. The Schr\"odinger equation governing the two-body electron-hole pair ( i.e., exciton) wavefunction ($\Psi$) can be expressed as~\cite{DRESSELHAUS195614}
\begin{equation}
	\left[ -\frac{\hbar^2}{2m^*_e}\nabla^2_e - \frac{\hbar^2}{2m^*_h}\nabla^2_h - \frac{e^2}{\epsilon|\vec{r}_e - \vec{r}_h|} \right] \Psi = E \Psi~.
	\label{ch1.eq1}
\end{equation}
To simplify, we assume the dielectric constant ($\epsilon$) to be independent of $\vec{r}_e$ and $\vec{r}_h$, indicating a resonably large radius of the exciton in the semiconductor. Further, we introduce new coordinates for the electron-hole seperation as $\vec{r} = \vec{r}_e - \vec{r}_h$, and a cordinate for the center of mass is defined as:
\begin{equation}
	\vec{\lambda} = \frac{m^*_e \vec{r}_e + m^*_h \vec{r}_h}{m^*_e + m^*_h}~.
	\label{ch1.eq2}
\end{equation}
We proceed by breaking down the Schr\"odinger equation (Equation~\ref{ch1.eq1}) into two parts: one describing relative motion the of exciton wave packet, denoted as $S(\vec{r})$, and another for the motion of the center of mass, represented by $T(\vec{\lambda})$. This separation is expressed as:
\begin{equation}
	\Psi(\vec{r}_e, \vec{r}_h) = S(\vec{r})T(\vec{\lambda}) \label{ch1.eq3}~.
\end{equation}
Subsequently, Equation~\ref{ch1.eq1} transforms into the following equation:
\begin{equation}
	\left[ \frac{\hbar^2}{2(m^*_e + m^*_h)}\nabla^2_{\lambda} + \frac{\hbar^2}{2m^*_r}\nabla^2_{r} - \frac{e^2}{\epsilon r} \right] S(\vec{r})T(\vec{\lambda}) = E S(\vec{r})T(\vec{\lambda})~. \label{ch1.eq4}
\end{equation}
Here, the reduced effective mass $m_{r}^*$,
\begin{equation}
	\frac{1}{m^*_r} = \frac{1}{m^*_e} + \frac{1}{m^*_h}~. 
	\label{ch1.eq5}
\end{equation}
This leads to an eigenvalue problem for $T(\vec{\lambda})$:
\begin{equation}
	\frac{\hbar^2}{2(m^*_e + m^*_h)}\nabla^2_{\lambda} T(\vec{\lambda}) = \Lambda T(\vec{\lambda})~. \label{ch1.eq6}
\end{equation}
Having the form of a free particle, Equation~\ref{ch1.eq6} yields eigenvalues:
\begin{equation}
	\Lambda(K) = \frac{\hbar^2 K^2}{2(m^*_e + m^*_h)} \label{ch1.eq7}~,
\end{equation}
where $K$ is exciton center of mass momentum. Solutions to the center of mass problem suggest that the exciton can traverse the crystal freely as a unified entity. During this movement, excitons can carry energy without carrying a charge, as they are neutral quasiparticles.
To obtain the complete exciton wavefunction and energies of the electron-hole system as defined in Equation~\ref{ch1.eq1}, we now proceed to review the solution for the relative motion coordinate system $S(\vec{r})$ as defined in Equation~\ref{ch1.eq3},
\begin{equation}
	\left[ \frac{\hbar^2}{2m^*_r}\nabla^2_{r} - \frac{e^2}{\epsilon r} \right] S(\vec{r}) = E_n S(\vec{r}) ~.
	\label{ch1.eq8}
\end{equation}
Equation~\ref{ch1.eq8} appears to similar to the Schrödinger equation for a hydrogen atom, having eigenvalues $E_n$ corresponding to quantum numbers $n$ (where $n = 1, 2, \ldots$) represented as~\cite{DRESSELHAUS195614}:
\begin{equation}
	E_n = -\frac{m^*_r e^4}{2\hbar^2 \epsilon^2 n^2}~.
	\label{ch1.eq9}
\end{equation}
The total energy for the exciton can be expressed as $E = \Lambda(K) + E_n$.\\
The $E_n$ can also be written as,
\begin{equation}
	E_n = -13.6 \frac{m^*_r}{\epsilon^2_r}\frac{1} {n^2}~.
	\label{ch1.eq10}
\end{equation}
for three-dimensional systems. Here, $\epsilon_r = \epsilon/\epsilon_0$, with $\epsilon$ being the dielectric constant of the material. Similarly, the exciton energies for 2D materials follow, 
\begin{equation}
	E_n = -13.6 \frac{m^*_r}{\epsilon^2_r}\frac{1}{ (n-\frac{1}{2})^2}~.
	\label{ch1.eq11}
\end{equation}
Once we have the exciton energies, we can obtain the optical spectra in the following way. Optical absorption and conductivity are proportional to the imaginary part of the macroscopic dielectric function~\cite{DFT_GWA_Exciton-GW3},
\begin{equation}
	\epsilon_2(\omega) = \frac{8\pi^2e^2}{\omega^2}\sum_{n}\left |\sum_{vc\textbf{k}}A_{vc\textbf{k}}^{n}\hat{\kappa}\cdot\langle v\textbf{k}|\textbf{v}|c\textbf{k}\rangle \right|^2\delta(\hbar\omega - E_n)~.
	\label{ch1.eq12}
\end{equation}
Here, $A_{vc\textbf{k}}$ is the dipole oscillator strength of the $n^{th}$ exciton made up between the valence band ($v$) and conduction band ($c$), upon electromagnetic radiation having velocity operator $\textbf{v}$ with polarization vector $\hat{\kappa}$, at the lattice momentum $\textbf{k}$, at the lattice momentum $\textbf{k}$. A schematic representation of optical absorption as a function of photon energy ($\hbar\omega$) is shown in Figure~\ref{ch1.fig1} (d). The peak at energy $e_1$, below the single particle bandgap ($E^{QP}$), in the absorption spectrum in the red color represents the impact of electron-hole interaction (EHI). It corresponds to the lowest energy exciton, known as the optical bandgap $(E^{OP} =e_1)$. The difference between the single particle bandgap and optical bandgap represents the exciton binding energy ($E_b = E^{QP}-E^{OP}$). Once the EHI is switched off then Equation~\ref{ch1.eq12} takes the following form, 
\begin{equation}
	\epsilon_2(\omega) = \frac{8\pi^2e^2}{\omega^2}\sum_{n}\left|\sum_{vc\textbf{k}}A_{vc\textbf{k}}^{n}\hat{\kappa}\cdot\langle v\textbf{k}|\textbf{v}|c\textbf{k}\rangle \right|^2\delta(\hbar\omega - E^{QP})~,
	\label{ch1.eq13}
\end{equation}
with zero exciton binding energy. A schematic representation of the single particle optical spectrum is shown in Figure~\ref{ch1.fig1} (d), in the cyan color. We highlight that the experimentally observed optical spectrum can only be explained if the excitonic effects are incorporated while calculating the dielectric function.
Next, we will discuss the types of excitons in an insulating material based on their exciton radius, dipole oscillator strength, and optical selection rules. 

\section{Classification of excitons}
Depending on the exciton radius, the electron-hole pair can either be localized or spread over several unit cells \textit{i.e.} the exciton can be categorized based on their localized or delocalized characterstics~\cite{SHIAU2023169431}. Below, we discuss them briefly.  
\subsection{Frenkel and Wannier-Mott excitons}
Frenkel excitons are named after Yakov Frenkel~\cite{Frenkel1931} and occur in materials where there is a strong attraction between electron-hole pairs, like in ionic crystals~\cite{Ng2020,Combescot2015}. Frenkel excitons are made from highly localized excitations due to the large effective masses of well-localized carriers. 
However, Frankel excitons still obey translational invariance and not localized at any particular lattice site. 
When the material has a low dielectric constant, the interaction between electrons and holes is strong, resulting in small radius excitons, approximately the size of the unit cell. They typically have a binding energy ranging from 0.1 to 1 eV, highlighting the strong pairing of electrons and holes in these systems~\cite{AGRANOVICH2003317}.
\begin{figure}[t]
	\centering
	\includegraphics[width =.95\linewidth]{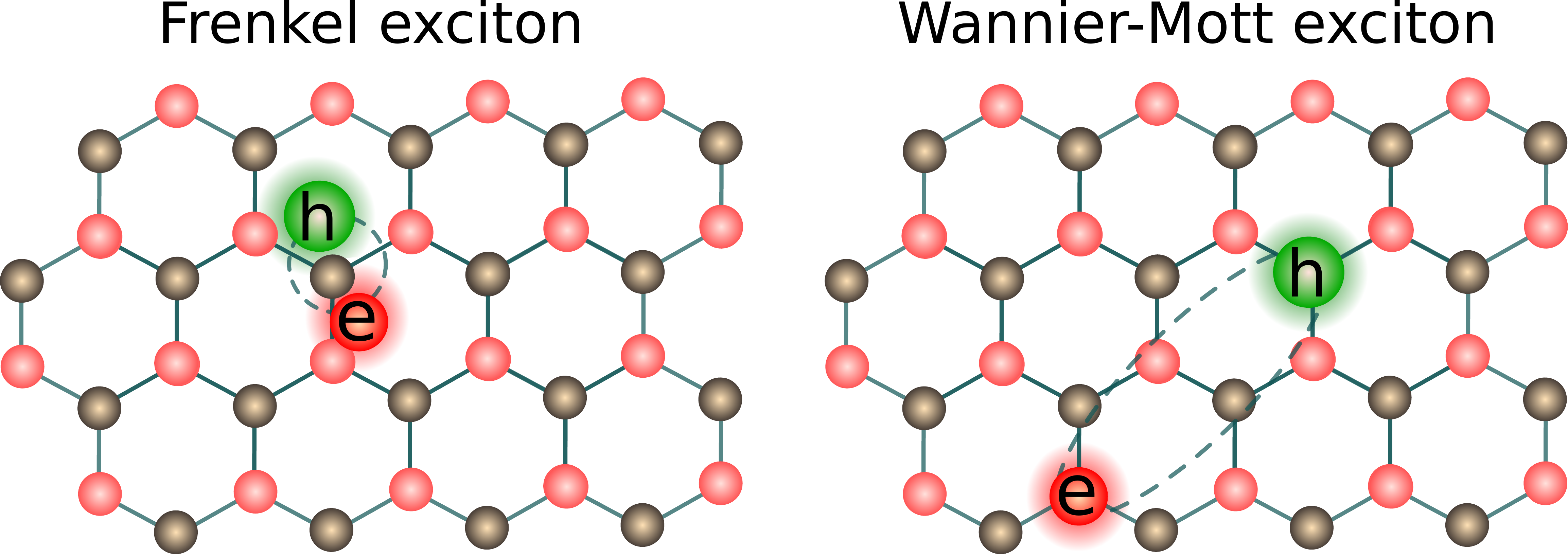}
	\caption {Schematic representation of a (left) tightly bound Frenkel exciton that has a  radius of around the size of the unit cell; (right) Wannier-Mott exciton has a large radius that exceeds the unit cell size.}
	\label{ch1.fig2}
\end{figure}
On the other hand, Wannier-Mott excitons are made from delocalized conduction electrons and delocalized valence holes. They are characterized by their internal structure resembling hydrogen-like wave functions, arising from the Coulombic interaction between electrons and holes in a crystalline periodic potential~\cite{Wannier_exciton, Mott1938}. 

Unlike Frenkel excitons, Wannier-Mott excitons exhibit a relatively large mean electron-hole distance, notably greater than the lattice constant~\cite{LAROCCA200397}. In most semiconductors, where the Coulomb interaction is screened by valence electrons due to a large dielectric constant, electrons and holes are weakly bound. These large excitons are commonly found in semiconductor crystals with small energy gaps and high dielectric constants, such as liquid Xenon. The dielectric constant in semiconductors tends to diminish the Coulomb interaction, especially since the radius of these excitons surpasses the lattice spacing. The effective mass of electrons in semiconductors, typically small, further contributes to the formation of large exciton radii. Calculations of Wannier-Mott excitons and their properties often employ the effective mass approximation, treating electrons and holes as particles with the effective masses of the conduction and valence bands, respectively. In this thesis, we will deal with the excitons in low-dimensional crystalline solids. Hence, excitons in these systems are likely to be strongly bound and localized within a few unit cells. 
Along with the energy conservation requirement in an exciton formation, the momentum conservation requirement is also crucial in an absorption process. Below, we discuss the role of momentum in optical transitions and its impact on optical properties.  

\subsection{Direct and indirect excitons}
\begin{figure}[t]
	\centering
	\includegraphics[width =.95\linewidth]{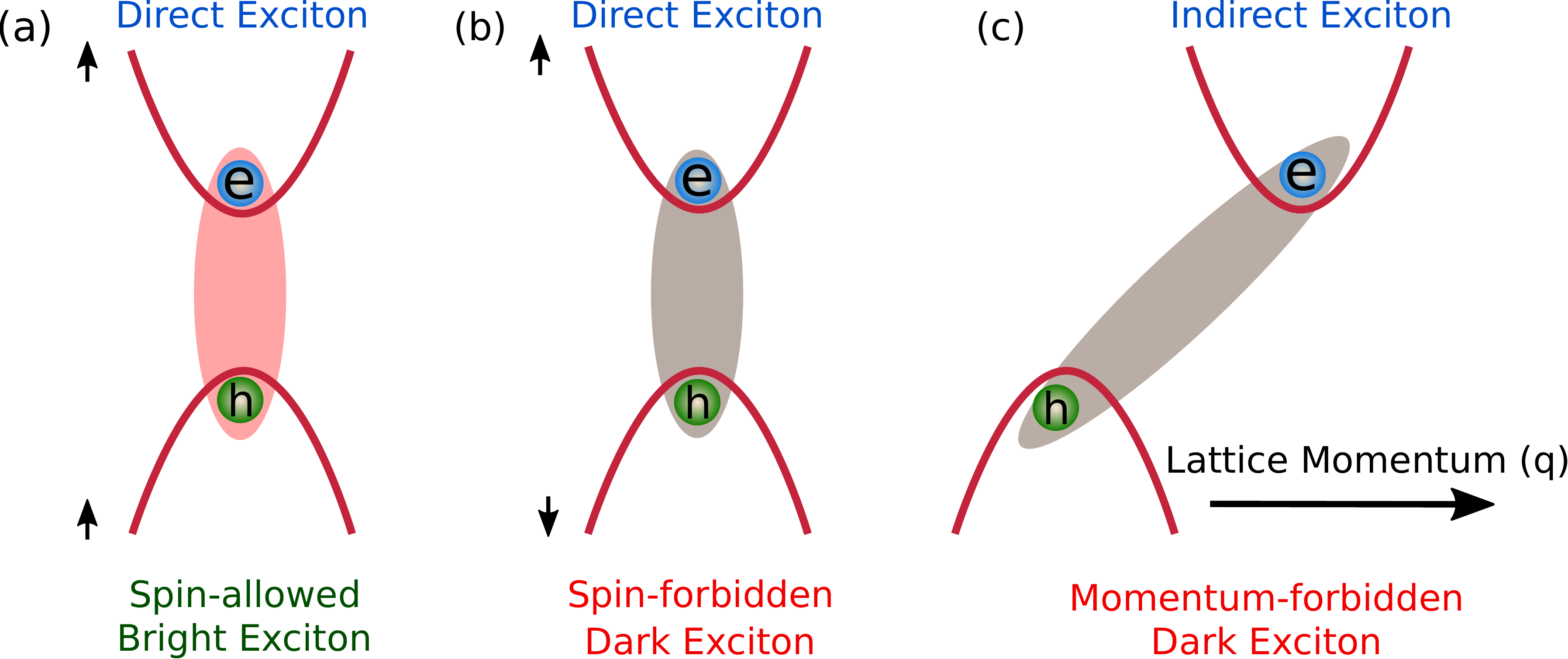}
	\caption {Illustration of exciton types and optical transitions. (a) and (b) depict direct excitons with vertical optical transitions with conservation of momentum for excited electron-hole pairs. (c) Represents an indirect exciton, where an electron (or it could be a hole) has a lattice momentum $\hbar q$ with respect to its counterpart. The optical selection rules classify excitons into bright and dark types based on the spin configuration of the conduction electron and valence hole. Bright excitons, shown in (a), have the same spin, allowing them to absorb photons with zero momentum linear polarized light. Dark excitons, illustrated in (b), have opposite spins, leading to spin-forbidden optical transitions for linearly polarized light.}
	\label{ch1.fig3}
\end{figure}

Since the incident photons carry almost no momentum, therefore the optical transitions are usually considered vertical [see Figure~\ref{ch1.fig3} (a) and (b)], and they form direct excitons~\cite{Mueller2018}. In other words, when the excited electron-hole pairs conserve their momentum individually in their bound state, they are called a direct exciton. In contrast, for indirect exciton, an electron (or a hole) has a lattice momentum $\hbar q$ with respect to a hole (or an electron), as depicted in Figure ~\ref{ch1.fig3} (c)~\cite{Mueller2018}. 
In materials such as silicon and germanium, excitons originate from an indirect phonon-assisted mechanism, resulting in the formation of indirect excitons. These indirect excitons can be generated via phonon mediation, as specified in Equation~\ref{ch1.eq9}. In the fourth chapter of this thesis, we will examine phonon-mediated indirect excitonic emission in a semiconductor with an indirect bandgap.
\subsection{Bright and dark excitons}
Based on the optical selection rules, the excitons can be classified as dark excitons and bright excitons~\cite{dark_exciton_TMD}. An exciton with the same spin of its conduction electron and valance hole can be termed a bright exciton because the optical selection rule allows it to absorb a photon with zero momentum, as represented in Figure~\ref{ch1.fig3} (a). In contrast, it is called a spin-forbidden dark exciton if electrons and holes have opposite spin [see Figure~\ref{ch1.fig3} (b)]. Optical excitation is prohibited between a conduction band with the same spin but situated in different valleys within momentum space. These states remain inaccessible to light due to the absence of the necessary momentum transfer and spin-flip, as elucidated by Mueller in 2018~\cite{Mueller2018}.
\section{Excitons in two-dimensional materials}
The discovery of two-dimensional (2D) materials has sparked a huge interest in their potential in next-generation electronic and optoelectronic devices\cite{Novoselov2004_graphene1,Bernardi2013-graphene-light,Butler2013,Mounet2018}. Starting with graphene, an atomically thin semimetal, the 2D family has expanded to include metals like NbSe$_2$, semiconductors such as MoS$_2$, and insulators like hexagonal boron nitride (hBN). These materials offer unique properties like large charge mobilities, metallic nature, and excellent light absorption\cite{Novoselov2004_graphene1,Bernardi2013-graphene-light}. The strong light-matter interaction in these materials, especially in TMDs with formula MX$_2$ (where M = Mo, W and X = S, Se, Te), provides a platform of exceptionally large light absorption and emission useful in optoelectronic applications\cite{Mounet2018}. 
\begin{figure}[t]
	\centering
	\includegraphics[width =.9\linewidth]{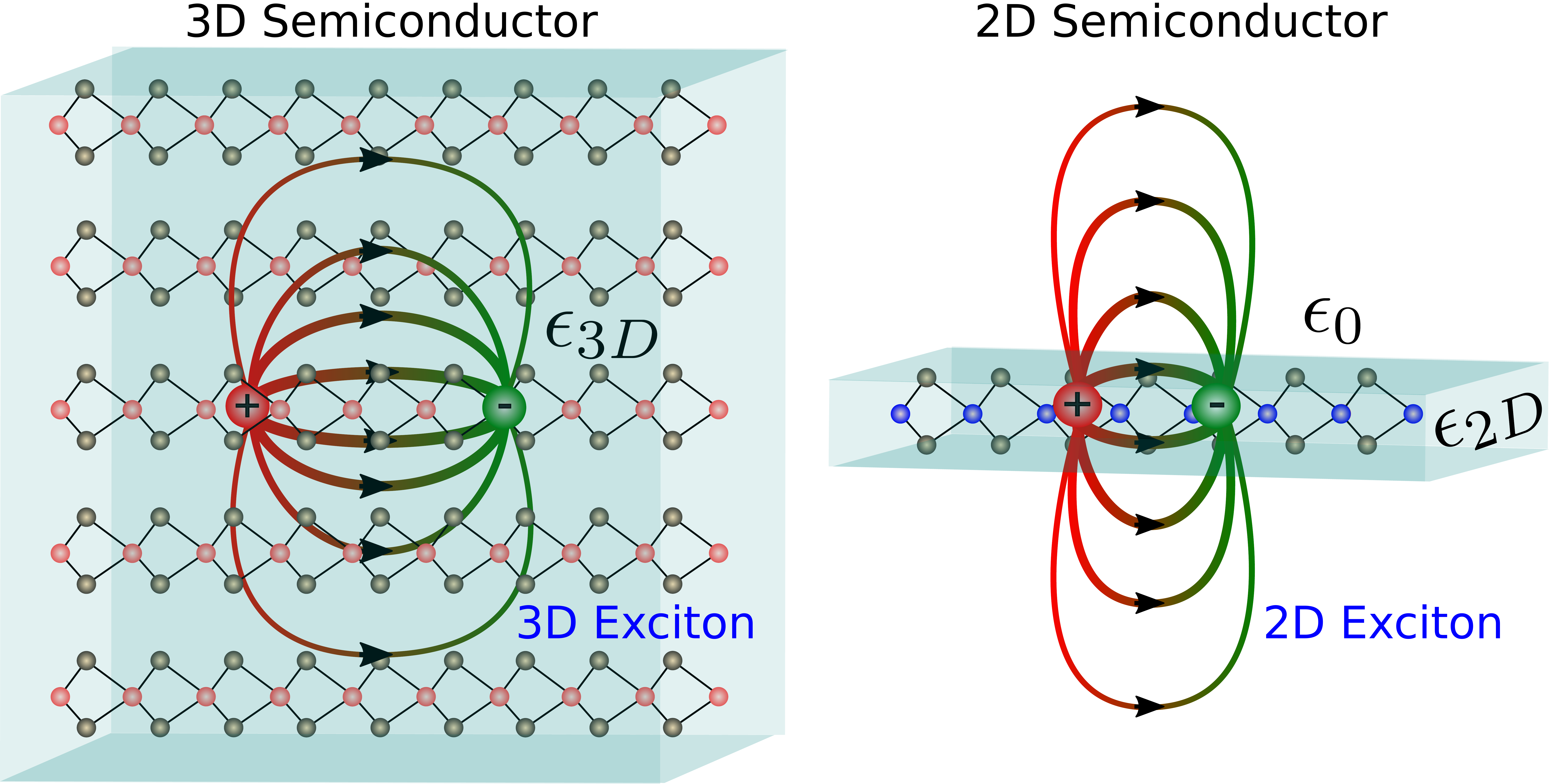}
	\caption {A schematic representation of exciton formation in 3D (left) and 2D (right) dielectric material. The electric field lines between the quasi-electron and quasi-hole in a 3D system are screened by the dielectric environment of the material, However, in a 2D system, the electric field lines are relatively very less screened by the material's dielectric environment. This leads to a strong EHI between the quasiparticles in 2D materials and large binding energy of the excitons.}    
	\label{ch1.fig4}
\end{figure} 
As discussed in the previous section, the exciton energies of a material depend on its dimensionality. Within a hydrogenic approach, it is evident from Equation~\ref{ch1.eq11} that the description of excitons in 2D does not follow the Rydberg model. Even if we ignore the dielectric constant of the medium,  the exciton energy of the lowest state in a 2D system is four times larger than that of a bulk system. A schematic representation of excitons in three-dimensional (3D) and 2D materials is shown in Figure~\ref{ch1.fig4}. As presented in Figure~\ref{ch1.fig4}, in 3D materials, the electric field lines between optically excited negatively charged electrons and positively charged holes are confined within the dielectric environment of the material. Due to large screening effects, the exciton binding energy of bulk semiconductors lies in the order of tens of meV, making them hard to sustain at room temperature~\cite{Thygesen2017}. In 2D materials, the field lines experience reduced screening due to the atomically thin dielectric environment~\cite{TMD-BE}. This leads to strong EHI and, consequently, the magnitude of exciton binding energy in the order of hundreds of meV, making them stable at room temperature. Furthermore, by stacking 2D semiconductors, an electron from one layer can interact with the hole in another layer and form interlayer excitons~\cite{Rivera2015}. Such heterostructures are helpful in studying charge separation and recombination mechanisms in 2D crystals.

Recently, another interesting feature of excitons in 2D insulating materials were reported where the exciton spectrum deviates from it's standard Hydrogenic solution. These studies reveal that the band geometric quantities (berry curveture and quantum metric) can lead to the spliting of exciton energy levels, with opposite angular momentum ($l$), for $l\ne 0$~\cite{Srivastava2015,Zhou2015}.

\subsection{Intralayer and interlayer excitons}
\begin{figure}[t]
	\centering
	\includegraphics[width =.95\linewidth]{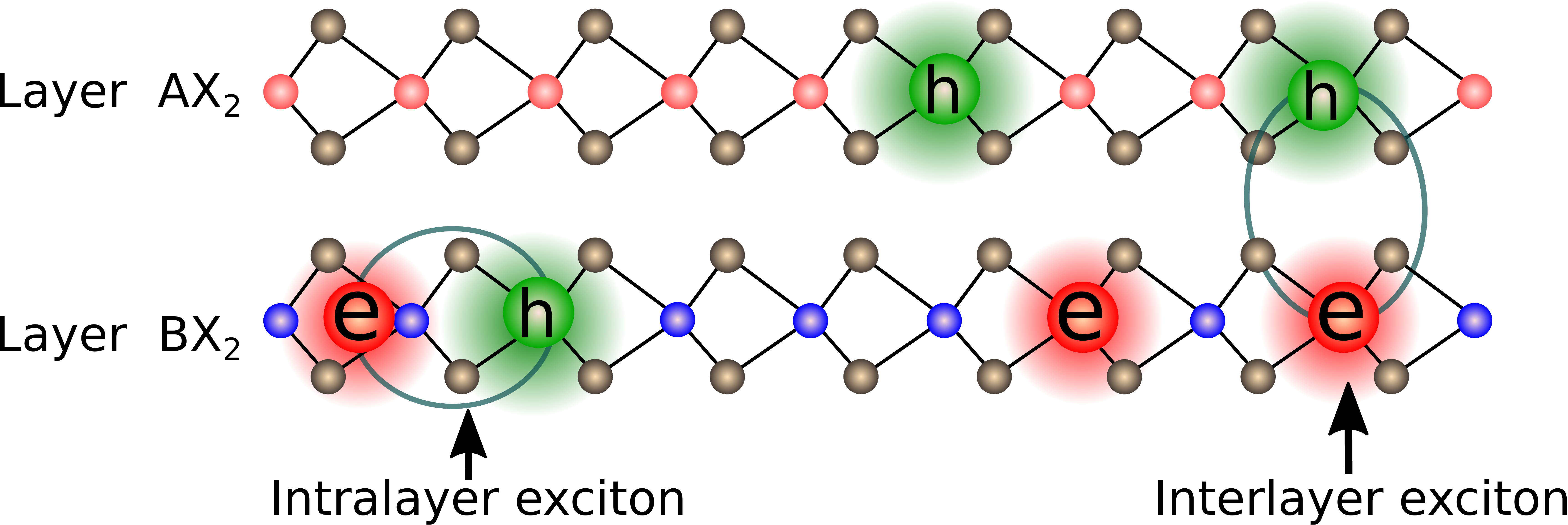}
	\caption {Schematic representation of intralayer exciton (localized in the Layer BX$_2$), and interlayer exciton with the electron being localized in one layer (Layer BX$_2$), and the hole localized in another layer (Layer AX$_2$).}
	\label{ch1.fig5}
\end{figure}

Upon optical excitations, if the electron-hole pairs in a bound pair belong to the same layer of a 2D material, then they are called intralayer excitons. In Figure~\ref{ch1.fig5}, the intralayer exciton formation is shown in the 'Layer BX$_2$', where the electron-hole pairs from the same layer are coupled. However, there can be a situation when an electron in one layer ('Layer BX$_2$' in Figure~\ref{ch1.fig5}) can interact with a hole of another layer ('Layer AX$_2$' in Figure~\ref{ch1.fig5}) in the heterostructure. The excitons in such a scenario are called interlayer excitons. Usually, interlayer excitons have longer lifetimes due to large spacial separation between electron-hole pair~\cite{Rivera2015,Palummo2015} in comparison to intralayer excitons. This makes interlayer excitons ideal to observe exotic many-particle states such as Bose-Einstein condensates\cite{Fogler2014,Thygesen2017}. In this thesis, we will be mainly dealing with the intralayer excitons in 2D materials.
\subsection{Exciton complexes}
Beyond excitons, semiconducting materials also exhibit electron-hole complexes that are made up of more than two particles. In doped semiconductors, an electron or a hole can couple with neutral excitons and form a three-particle complex called trions~\cite{kheng1993observation,mak2013tightly}. The binding energies of trions in 2D TMDs are about 20–30 meV, while in phosphorene, it is about 150 meV~\cite{Mak2013,Berkelbach2013}. Due to the large binding energies of trions in 2D materials, they are predicted to be stable at room temperature, which can be utilized in the physics of nanoscale semiconductors~\cite{Huard2000}.\\
\begin{figure}[t]
	\centering
	\includegraphics[width =.95\linewidth]{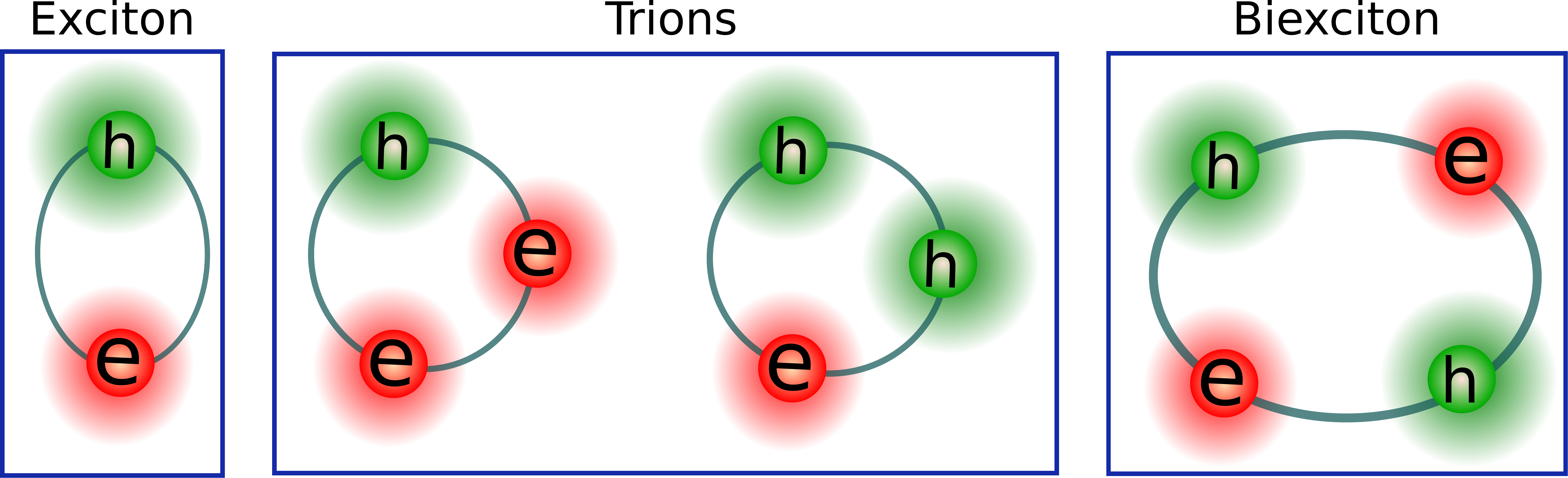}
	\caption {Schematic representation of exciton complexes in 2D materials. An exciton, trions, and biexciton are shown in the left, center, and right boxes respectively. Trions result from interactions between free electrons (in n-doped regions) or holes (in p-doped regions) and neutral excitons. In the rightmost box, a pair of excitons are coupled to form a biexciton.}
	\label{ch1.fig6}
\end{figure}
Apart from trions, exciton-exciton interaction leads to four-particle complexes called biexcitons. In photoluminescence (PL) experiments, the biexcitons are identified through a distinct state at high exciton density with larger binding energy than conventional quantum-well structures~\cite{klingshirn2012semiconductor}. Theoretical calculation based variational principle hints that the large biexciton binding energy is due to reduced and non-local dielectric screening with strong carrier confinement ~\cite{kim1994thermodynamics,You2015}.\\

Theoretical formulations discussed so far regarding the optical property ignore the impact of lattice vibration and the effect of temperature. However, most of the optical measurements are performed at room temperature. Traditional theoretical methods use broadening parameters to fit experimental data observed at a finite temperature. $Ab-initio$ based theoretical methods have made significant progress in explaining the microscopic understanding of temperature effects on optical and electronic properties in materials. Below, we discuss the impact of temperature variations on excitonic properties in semiconducting materials. \\

\section{Temperature dependent excitonic structure}
In semiconductors, the absorption line position, width, and intensity depend on temperature~\cite{Lauten1985}. The theoretical models for solving the exciton problems at zero temperature within the frozen atom condition fail to explain the finite lifetime of excitons. Ignoring temperature effects means overlooking significant features, including the relative magnitude of excitonic peaks and their broadening. Both aspects are directly linked to temperature variations and the nonradiative exciton relaxation time~\cite{Molina2016}. The estimated absorption spectra are commonly modeled with an artificial, \textit{ad hoc} numerical broadening function chosen for the best agreement with the experiment. Theoretical formulation of excitonic effects within the many-body perturbation theory (MBPT) generally neglects the effect of lattice vibrations and temperature effects.

In the broader context, temperature exerts a dominant influence on the electronic and optical properties of semiconductors, shaping their suitability for optoelectronic applications~\cite{Wolpert2011}. It is well-established that temperature renormalizes electronic bandstructure~\cite{Lautenschlager1985} and induces changes in the position and width of optical absorption and emission peaks~\cite{Lautenschlager1987}. One of the major factors that dictate the temperature-dependent optical response of a material is the strength of electron-phonon coupling in that material. 

\subsection{Exciton-Phonon Coupling}
\label{EPC}
As a result of zero-point motion (at absolute zero temperature), the lattice vibrations produce a finite phonon population in materials. Therefore, lattice vibrations and electron-phonon coupling in materials are very fundamental and intrinsic phenomena. The impact of phonons on electronic energies is as important as electron-electron correlations. Through phonon absorption and emission mechanisms, the optically bright excitons can turn into optically dark excitons with finite momentum. The impact of the electron-phonon interactions on optical excitations can be studied in terms of exciton-phonon coupling, particularly in low-dimensional materials exhibiting strongly bound excitons. The exciton-phonon interaction leads to the phonon-induced term of the exciton self-energy, which gives the temperature dependence of the exciton energies and their lifetime broadening. In general, the phonon population in materials would increase with an increase in temperature~\cite{Marini2008,Antonius2022}. The impact of phonons on the band structure and carrier mobility as functions of temperature have been in development recently~\cite{giannozzi2005density,Bernardi2014}. However, the exploration of the impact of phonons on optical excitations is still in the early stages~\cite{Marini2008,Antonius2022}. 

Our interest in temperature-dependent optical excitations is motivated by the recent theoretical development within the {\it ab-initio} method. This approach has been successful in the examination of bulk hexagonal boron nitride, showing strong luminescence in the ultraviolet regime~\cite{Watanabe2004}. The origin of this luminescence has been widely discussed, but only recently has a clear signature of phonon-mediated light emission emerged in the experiments~\cite{Cassabois2016}. Apart from the phonon-mediated luminescence emission, the inclusion of exciton-phonon interaction in calculating the exciton energies and light-matter interaction has been essential in understanding optical measurements performed at finite temperatures~\cite{Paleari2019,Chen2020,Marini2008}.

\subsection{Exciton lifetime}
From the electron-phonon interaction picture, it is established that the inclusion of the lattice vibration corrections transforms the electron-hole interacting Hamiltonian to a non-hermitian Hamiltonian, due to which the excitonic energy eigenvalues become complex quantities. In such a case, the real part of the exciton eigenvalue gives the exciton energy, and the imaginary part defines the excitonic nonradiative lifetime~\cite{Marini2008}. Hence, excitons typically serve as fundamental excitations in condensed matter, exhibiting finite lifetimes. Although longer lifetimes prove beneficial for exciton condensation, this may not always be advantageous, particularly in the context of solar cells. In solar cells, the process involves sunlight absorption to generate excitons, followed by the dissociation of these electron-hole pairs through an electric field. Subsequently, the separated electrons and holes migrate to opposite electrodes for collection, completing the conversion of light to electricity~\cite{Lee2012,Frost2014,Giebink2011}. The efficiency of this process relies on maintaining an optimal exciton lifetime. If the lifetime is excessively prolonged, a stronger electric field becomes necessary for exciton dissociation. Conversely, if the lifetime is too brief, electrons and holes recombine and radiate before reaching their respective electrodes, hindering the photoelectric conversion. Therefore, the regulation of exciton lifetimes stands as a subject of both scientific and practical interest.\\
\section{Density dependent exciton dynamics}
Recent experiments on 2D semiconductors highlight that a controll over photoexcited electron-hole pair density can lead to interesting optical phenomena in the nonequilibrium regime. These include exciton-exciton atom-like interaction, EHP, and EHL phase~\cite{amit-exciton-2,Sei,Arp2019,MoS2-EHL-EXP-ACS-Nano-2019-Yu,Dey-MoS2-2023}. However, the theoretical understanding of these exciting experiments in the nonequilibrium regime is yet to be explored. In this thesis, we will explore the exciton dynamics from equilibrium to nonequilibrium regime in semiconducting solids.
\begin{figure}[t]
	\centering
	\includegraphics[width =0.9\linewidth]{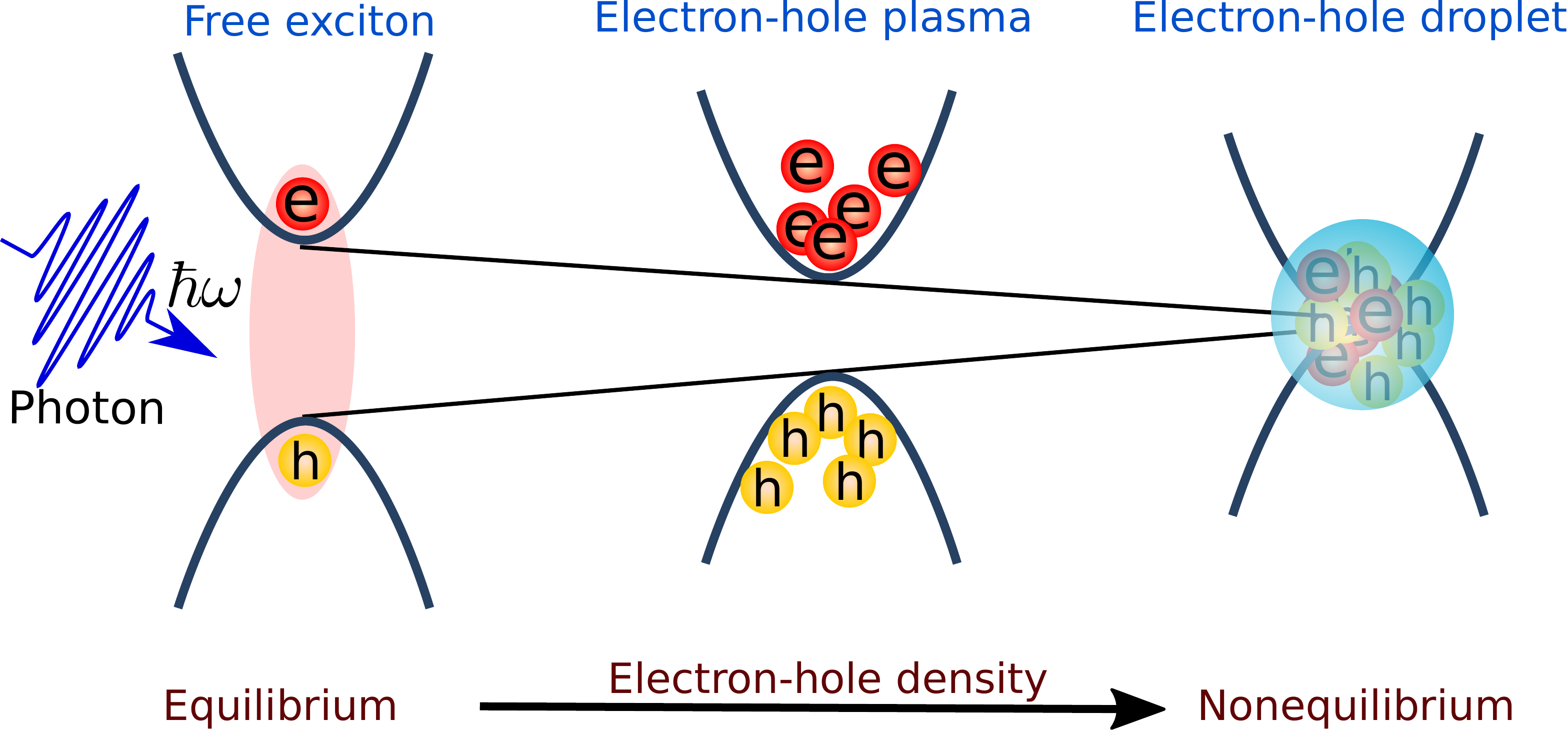}
	\caption {Schematic showing the formation of the electron–hole liquid from photo-excited electrons and	holes. The free excitons dissociate on increasing photo-excited carrier density and form the electron–hole plasma state. In both of these phases, the constituents interact weekly with each other and can be treated as a gaseous state. Further increase in the exciton density leads to the formation of electron–hole droplets or the EHL phase, with the particles interacting collectively}
	\label{ch1.fig7}
\end{figure}
Recently, some experiments have reported exciton-exciton interactions by increasing the pump fluence or photo-excited carrier density, bringing the system into a nonequilibrium regime~\cite{amit-exciton-2,Sei}. Within the nonequilibrium regime, the electron-hole system has been reported to exhibit EHP and EHL phase in the density-temperature parameter space\cite{keldysh1968,keldysh1986,rice-conference}. EHL is a macroscopic quantum state arising from the condensation of electrons, and holes dissociated from excitons at a large photo-excited electron-hole pair density~\cite{keldysh1968,asnin1969, Keldysh-Silin-75, Wolfe-75, Beni-Rice-76,walker87, Jeffries}. At low densities, the excitons interact weakly with each other via van der Waals forces~\cite{Rice1978book} and can be treated as non-interacting exciton gas. These non-interacting excitons are also known as free-excitons (FE). However, at high electron-hole pair densities, ongoing to enhanced screening, the Coulomb attraction between the electron-hole pair is reduced and leads to the band gap renormalization (see Figure~\ref{ch1.fig7}). Hence, the bound excitons start to lose their individuality, leading to EHP state formation~\cite{MOS2-EHP-Bataller2019}. Further increments in the electron-hole pair density give rise to the collective behavior of electrons and holes~\cite{Brinkman-Rice-73, Beni-Rice-78} leading to the formation of electron-hole droplets and hence a phase transition from the FE/EHP to EHL phase.

The emergence of the EHL phase at higher temperatures is restricted by low EHL binding energy in any insulating material. It is observed that the electronic band structures with multiple electron and hole valleys~\cite{Brinkman-Rice-73}, mass anisotropies~\cite{Beni-Rice-78}, and electron-phonon interaction in polar semiconductors~\cite{Beni-Rice-76} contribute toward the stability of the EHL phase. However, the room temperature EHL phase remains challenging to observe in three-dimensional insulating solids because of their small exciton binding energies. However, in 2D semiconducting materials, where the EHI is much stronger than in bulk materials, the binding energies of excitons are large~\cite{Barun-exciton-17}, and these can support the EHL phase.
\section{The role and significance of \textit{ab-initio} theory}
\begin{figure}[t]
	\centering
	\includegraphics[width =0.9\linewidth]{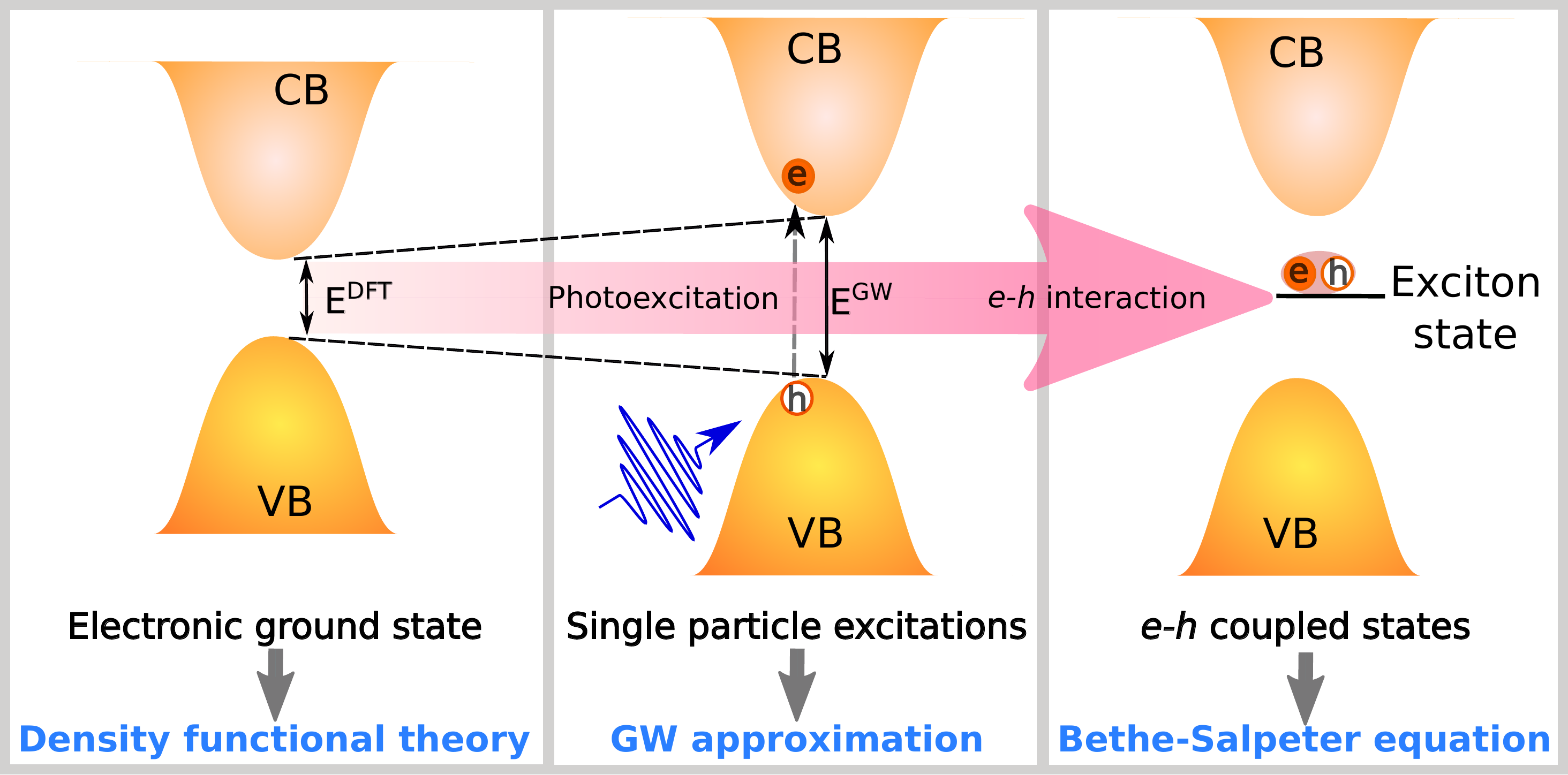}
	\caption{Schematic workflow of \textit{ab-inito} calculation for excitons. (left) DFT, the framework for calculating ground state electronic band structure of materials. (center) Excited state single particle energies calculated by incorporating dynamical screening effects in excited states using the GWA. The electronic band gap at the GWA ($i.e.$,~$E^{GW}$) is larger than the electronic band gap calculated at the DFT level ($i.e.$,~$E^{DFT}$). (right) The electron-hole ($e-h$) coupling is incorporated in the two-body equation of motion called BSE.}
	\label{ch1.fig8}
\end{figure}

To understand the electronic and optical characteristics of quantum materials, \textit{ab-initio} techniques serve as indispensable tools. \textit{Ab-initio} methods, based on quantum mechanical principles, offer a systematic and accurate approach to understand the properties of materials without relying on empirical parameters. These techniques provide a bridge between theoretical predictions and experimental observations in the study of quantum materials. A schematic workflow for investigating optical excitation within the \textit{ab-initio} theory is shown in Figure~\ref{ch1.fig8}. Below, we discuss the significance of \textit{ab-initio} techniques, with a specific focus on density functional theory (DFT), the GW approximation, and the Bethe-Salpeter Equation (BSE) method.\\

\textit{Density Functional Theory} - The DFT emerges as a powerful and versatile theoretical framework within quantum mechanics, offering a comprehensive means of investigating the electronic structure, thermal, magnetic, and other ground state properties of diverse materials. By focusing on the electron density distribution rather than explicitly considering individual electron wave functions, DFT provides a conceptually reliable and computationally efficient approach to describe the behavior of complex systems ranging from atoms and molecules to solids and surfaces~\cite{Hohenberg1964,Kohn-Sham}. A more detailed many-electron theory and DFT formalism are reviewed in Chapter~\ref{chap2}. The efficacy of DFT is extensively documented for numerous ground-state properties across diverse materials. However, a drawback lies in the fact that the Kohn-Sham eigenvalues within the theory do not represent quasiparticle energies. The band gap in DFT, defined by the Kohn-Sham gap $E_g$, typically falls 30—50\% below the value observed in the optical spectrum, even when employing the exact exchange-correlation functional~\cite{Hybertsen-GW2,DFT_GWA_Exciton-GW3}. A major reason for this is that DFT does not include the dynamical screening effects. Therefore, one must look for a theory that can include dynamical screening effects and predict single-particle energies in excited states.\\

\textit{GW Approximation} - While DFT works well for the electronic ground-state properties, the GW Approximation (GWA), based on single-particle Green's function approaches, provides highly successful results for electron quasiparticle spectra by including dynamical screening effects. These quasiparticle spectra are obtained from the poles of the single-particle Green's function~\cite{HedinGW1}. This general formulation has been essential to developing theories of quasiparticle energies in semiconductors and insulators~\cite{Hybertsen-GW2}. The GWA formulation is reviewed in detail in Chapter 2. Despite their efficacy, neither DFT nor GWA adequately captures optical spectra or other charge-neutral excitations as neither of these include interactions between electrons and holes. Incorporating the EHI makes it a two-body problem that is solved using BSE. The BSE has been successful in explaning the optical spectrum obtained in experimental observations~\cite{DFT_GWA_Exciton-GW3}.\\

\textit{Bethe-Salpeter equation} -  Within this theory, an effective two-body Hamiltonian is constructed for studying optical electron-hole excitations. This involves evaluating the two-body Green’s function based on the one-body Green’s function (described by the GWA) and solving the BSE to yield correlated electron-hole excitation states. By considering optical transition matrix elements, the entire linear optical spectrum of a material can be comprehensively evaluated~\cite{BSE-1-Strinati,BSE-2-Strinati,BSE-3-Onida,DFT_GWA_Exciton-GW3}. A detailed review of the BSE method is discussed in Chapter 2.\\
\section{Main theme and outline of the thesis}
In this thesis, firstly, we demonstrate the presence of bound excitons in atomically thin two-dimensional materials majorly from the first-principle approach and MBPT. We estimate the single-particle energies using the GWA theory and incorporate the electron-hole correlations using BSE. The fundamentals of excitons have been studied in the low excitation density, i.e., in the equilibrium limit. With increasing photo-excited carrier density, we show that the lowest energy exciton in monolayer MoSi$_2$Z$_4$ (Z = N, P, As) follows the redshift-blueshift crossover. This is a consequence of renormalized exciton binding energy. In the nonequilibrium regime, it reveals an atom-like interaction among the excitons, and the exciton-exciton interaction potential follows the form of Lennard–Jones potential for atom-atom interaction. Further, we study temperature-dependent optical absorption and PL phenomenon in monolayer aluminum nitride (AlN) in the equilibrium regime. We find that electron-phonon interactions significantly modify the electronic energy, excitonic absorption, emission spectrum, and exciton lifetime. At low temperatures, the PL emission signals exhibit full excitonic thermalization with pronounced phonon replica below the indirect exciton energy. Finally, we combined the impact of temperature and electron-hole density and investigated a photo-excited electron-hole system in the nonequilibrium regime. We find that photo-excited electrons and holes in insulators above a critical density and below a critical temperature can condense to form an electron–hole liquid (EHL) phase. However, observing the EHL phase at room temperature is extremely challenging. Here, we have proposed the monolayer MoSi$_2$Z$_4$ (Z = N, As, P) series of 2D materials as a promising platform for observing the EHL phase at room temperature. Increased effective Coulomb interactions in two dimensions promote the stability of the EHL phase, resulting in higher binding energy, transition temperature, and the formation of strongly bound excitons. This suggests that exploring two-dimensional semiconductors further could lead to the realization of the EHL phase at high temperatures, offering potential for optoelectronic applications.\\

\section{Organization of the thesis}
\subsection{Chapter 1: Introduction (current Chapter)}
In chapter~\ref{chap1}, we introduce the exciton quasiparticles generated upon optical excitation in insulating materials and its impact on different physical observables in condensed matter systems investigated in this thesis. We start by discussing exciton formation upon photoexcitation in finite bandgap materials. A brief discussion about the effect of dimensionality and screening on exciton and binding energies follows this. Next, we classify excitons based on optical selection rules as bright and dark excitons. Excitons are also classified based on their interaction with other charged particles or excitons, such as trion and biexciton complexes. In the context of 2D materials, the Chapter highlights the unique optical responses and excitonic properties of materials like TMDs and hBN, emphasizing intralayer and interlayer excitons. 
A brief discussion about the impact of temperature and exciton density on the optical response of semiconducting solids is introduced, and current research problems are discussed. 
\subsection{Chapter 2: \textit{Ab-initio} based theoretical methods for excited state phenomena}
In chapter~\ref{chap2}, we discuss the formalism used to explore the electronic structures and optical excitations within {\it ab-initio} approach. We start with introducing the many-body Hamiltonian and its solution within the DFT. Further, to estimate the single particle excited state energies, we have adopted the GWA. This is followed by a discussion on the interaction between optically excited electrons and holes leading to the exciton formation and a calculation of the exciton energies and coupled electron-hole wavefunction utilizing the BSE. Furthermore, to understand the effect of lattice vibrations on optical excitations, we estimate the electron-phonon interaction within the DFPT and obtain electron-phonon matrix elements. The electron-phonon interaction effects are finally applied to the quasiparticles (electron, hole, and excitons) energy correction, and their impact on the optical spectrum is analyzed.
\subsection{Chapter 3: Fluence dependent exciton dynamics in monolayer MoSi$_2$Z$_4$ (Z = N, As, P) family of semiconductors}
In Chapter~\ref{chap3}, we predict the excited state properties within the {\formula} series. In monolayer MoSi$_2$N$_4$, MoSi$_2$As$_4$, and MoSi$_2$P$_4$, we investigate their potential to host strongly bound excitons and their role in defining the optical absorption. Starting from DFT, we obtain the ground state electronic band structure. Furthermore, we calculate the excited state quasiparticle energies of electrons and holes using the GWA. We estimate the exciton structure within the GWA-BSE level and report the presence of multiple bright excitons below the QP bandgap with significant contribution to the optical spectrum. We extend our analysis to the optical spectrum for higher pump-fluence value and investigate the nonequilibrium dynamics using the time-dependent BSE. Through this, we unraveled a redshift-blueshift crossover in the exciton binding energy with increasing electron-hole pair density via pump fluence. This dynamic behavior followed atom-like interactions among excitons and can be understood by a Lennard-Jones-like interaction potential between atoms.
\subsection{Chapter 4: Exciton-phonon coupling and indirect photon emission in monolayer aluminum nitride (AlN)}
In Chapter~\ref{chap4}, we utilize DFT, DFPT, and many-body perturbation theory and investigate temperature-dependent optical properties of wide bandgap AlN monolayer. Our investigation demonstrated the profound effect of electron-phonon (el-ph) interactions, leading to the renormalization of quasiparticle energies. Specifically, the calculated optical absorption spectra showcases a redshift and decreased dipole oscillator strengths due to the coupled electronic and vibrational states. We observe phonon-assisted indirect PL emission within the excitonic framework in the ultraviolet regime. Interestingly, the PL emission sustains at higher temperatures ($\ge$ 500). Furthermore, the highlighted role of exciton-phonon interactions extends beyond AlN, emphasizing their importance in group-III nitrides and other two-dimensional materials. 
\subsection{Chapter 5: Electron-hole liquid phase in monolayer MoSi$_2$Z$_4$ (Z = N, As, P) family}
In Chapter~\ref{chap5}, we establish potential observation of the EHL phase at room temperature within the monolayer {\formulaN} family. Because of the dimensional confinement and reduced dielectric constant, the monolayer {\formulaN} family host strongly bound excitons with exciton binding energy of the order of 1 eV, as discussed in Chapter~\ref{chap1}. Since the critical temperature of the EHL phase depends directly on the exciton binding energy ($T_c \sim 0.1E_b$), the {\formulaN} family hosts EHL phase above a critical exciton density ($n_c$). This prediction marks a significant departure from conventional limitations associated with cryogenic conditions, introducing the observation of the EHL phase in these materials at room temperature. The higher impact of the Coulomb interactions in two dimensions helps these monolayers support the EHL phase with an increased EHL binding energy and transition temperature, along with strongly bound excitons. Our findings motivate further exploration of {\formula} monolayers for realizing the EHL phase at high temperatures to harness collective phenomena for optoelectronic applications. 
\subsection{Chapter 6: Conclusions and future work}
Finally, we summarize the main findings of our work in chapter~\ref{chap6}, followed by a discussion of the future scope.

%% file: chap2.tex

\chapter{{\it \textbf{Ab-initio}} based theoretical methods for excited state phenomena}
\label{chap2}
\pagestyle{fancy}

For over three decades, {\it ab-initio} based theories have been utilized to explore various material properties. Among these, the density-functional theory (DFT) has emerged as a powerful method for investigating electronic ground-state properties~\cite{Hohenberg-Kohn-1964}. However, the accurate representation of optical excitations poses a challenge for DFT. Therefore, for the single-particle energies of electrons and holes, approaches based on one-body Green’s functions using the GW approximation (GWA) [where G stands for single particle Green's function and W is the screened Coulomb potential] have demonstrated reasonable accuracy for predicting quasiparticle (QP) properties~\cite{Hedin-GW1, Aryasetiawan1998}. Nonetheless, neither standard DFT nor GWA can accurately evaluate optical spectra. In the independent-quasiparticle picture, the calculated optical spectrum often exhibits significant deviations from experimental results. Characteristic peak positions may be inaccurately determined, and peak amplitudes can deviate from experimental value. The most notable deficiency of the independent-particle spectrum lies in its inability to explain bound exciton states, which play a dominant role in systems with reduced dimensions. This discrepancy is resolved by incorporating the Coulomb interactions between the photo-excited electrons and holes. To calculated the coupled electron-hole problem, the Bethe-Salpeter equation (BSE) is solved~\cite{BSE-1-Strinati, BSE-2-Strinati, DFT_GWA_Exciton-GW3}. The BSE gives a solution for the bound pair of electron and hole called excitons. 

To investigate the optical excitations accurately, in this chapter, we briefly review three methods: DFT for ground state properties, the GWA for quasiparticle (QP) energies, and the GW-BSE method for optical properties. We simplify the many-particle Hamiltonian by the Kohn-Sham theorem in Section~\ref{MBT}, which forms the foundation of DFT in Section~\ref{DFT}. We then delve into a comprehensive review of different approximations for calculating exchange-correlation energy functionals within DFT and the plane-wave basis and pseudopotential methods utilized in DFT calculationsin in Section~\ref{LDA_GGA}. Furthermore, in Secton~\ref{GWA}, we adopt the GWA and to calculate the QP energies. To incorporate the electron-hole correlation, we focus on the BSE in Section~\ref{BSE}. 

\section{A many-body Hamiltonian}
\label{MBT}
To understand the microscopic properties of a many-body system, one can write the total Hamiltonian considering all electrons and nuclei while ignoring certain relativistic effects. In the context of crystalline materials, the Hamiltonian describes the interactions between electrons and ions, which can be written as,
\be
H = T_e + T_I + V_{e-e} + V_{I-I} + V_{e-I}~.
\label{ch2.eq1}
\ee
Here, $T_e$ and $T_I$ denote the kinetic energy part for electrons and ions, respectively. The Coulomb interaction terms for the electron-electron, ion-ion,
and electron-ion are denoted by $V_{e-e},~V_{I-I}$, and $V_{e-I}$, respectively. The explicit form of these terms are given by
\be
T_e = \sum_i \frac{\textbf{p}_{i}^{2}}{2m}~,
\label{ch2.eq2}
\ee

\be
T_I = \sum_n \frac{\textbf{P}_{n}^{2}}{2M}~,
\label{ch2.eq3}
\ee

\be
V_{e-e} = \frac{1}{2}\sum_{\substack{i,i' \\i \ne i'}} \frac{1}{4\pi \epsilon_0}\frac{e^2}{|\textbf{r}_i-\textbf{r}_{i'}|}~,
\label{ch2.eq4}
\ee

\be
V_{I-I} = \frac{1}{2}\sum_{\substack{n,n' \\n \ne n'}} \frac{1}{4\pi \epsilon_0}\frac{(Ze)^2}{|\textbf{R}_n-\textbf{R}_{n'}|}~,
\label{ch2.eq5}
\ee

\be
V_{e-I} = -\frac{1}{2}\sum_{\substack{i,n}} \frac{1}{4\pi \epsilon_0}\frac{Ze^2}{|\textbf{r}_i-\textbf{R}_{n}|}~.
\label{ch2.eq6}
\ee
Here, the index $i(n)$ refers to the electrons (ions) in the system, and their corresponding position and momentum are denoted by $\textbf{r}_{i} (\textbf{R}_n)$ and $\textbf{p}_i(\textbf{P}_n)$, respectively. The ionic and electronic charges are represented by $Ze$ and $e$, while the masses are denoted as $m$ and $M$, respectively.

The many-body Hamiltonian defined for the electron-ion system can be represented by the time-independent Schrodinger equation of
the whole interacting system as,
\be
H \Psi(r_1, r_2, ....,r_N; R_1, R_2, ...., R_{N'}) = E\Psi (r_1, r_2, ....,r_N; R_1, R_2, ...., R_{N'})~,
\label{ch2.eq7}
\ee
Here, $E$ and $\Psi$ represent the eigenvalues and the many-body wavefunction of the ion-electron system having the total of $N$ electrons and $N'$ ions. For a system having a large number of electrons and ions ($10^{23}$, in real materials), solving this equation is extremely difficult within today’s computational constraints. Hence,  need to deploy certain approximations to address these complexities.

To solve Equation \ref{ch2.eq7}, the Born-Oppenheimer approximation has been introduced~\cite{Born-Oppen}, in which the electron and ion motions in solids and molecules can be separated. Due to the relatively higher mass of ions, their movements are slow compared to electrons. Therefore, their equation of motions can be decoupled, and the many-body wavefunction ($\Psi$) defined for Equation~\ref{ch2.eq7}, can be written as a product of electron wavefunction $\psi_e(\textbf{r},\textbf{R})$ and ion wavefunctions $\psi_I(\textbf{r},\textbf{R})$ as,

\be
\Psi = \psi_e(\textbf{r},\textbf{R})\psi_I(\textbf{r},\textbf{R})~.
\label{ch2.eq8}
\ee
This leads to the following separate Schrodinger equation for the electrons,
\be
[T_e + V_{e-e}+ V_{e-I}]\psi_e(\textbf{r},\textbf{R}) = E_e(\textbf{R})\psi_e(\textbf{r},\textbf{R})~.
\label{ch2.eq9}
\ee
Here, $E_e(\textbf{R})$ is the electron's energy eigenvalue for the fixed ionic coordinates (\textbf{R}). The Schrodinger equation for the ionic part of the Hamiltonian can be written as,
\be
[T_I + V_{I-I}]\psi_I(\textbf{R}) = [E_{tot}(\textbf{R})-E_e(\textbf{R})]\psi_I(\textbf{r, R})~.
\label{ch2.eq10}
\ee
Here, $\psi_I(\textbf{r},\textbf{R})$ referes to the ionic wavefunction. Next, we ignore the lattice vibrations; the electronic part of the Hamiltonian can be written as,
\bea
H &= &T_e + V_{e-e}+ V_{e-I} \nonumber \\
&=& \sum_i \left[ \frac{\textbf{p}_{i}^{2}}{2m} + V_{e-I}\right] + \frac{1}{2}\sum_{\substack{i,i' \\i \ne i'}} \frac{1}{4\pi\epsilon_0}\frac{e^2}{|\textbf{r}_i-\textbf{r}_{i'}|}~.
\label{ch2.eq11}
\eea
Here, $V_{e-I}$ (as defined in Equation~\ref{ch2.eq6}) represents the electrostatic potential experienced by the electrons in the presence of ions in the crystal. The last term in Equation~\ref{ch2.eq11}, which incorporates the electron-electron interaction, is not solvable exactly as it still is a many-body problem. To simplify this, we first review the Hartree theory, which treats the electrons as an independent particle~\cite{hartree_1928}. The many-particle wavefunction of the electronic Hamiltonian can be written as a product of single-particle wavefunction,
\be
\psi_e = \psi_{1} (\textbf{r}_1, \sigma_1)\psi_{2} (\textbf{r}_2, \sigma_2).....\psi_{N} (\textbf{r}_N, \sigma_N)~,
\label{ch2.eq12}
\ee
where $\sigma$ denotes the spin of the electrons. The single-particle Schrodinger equation within the Hartree approximation can be written as,
\be
\left[ \frac{\textbf{p}_{i}^{2}}{2m} + V_{e-I}(\textit{\textbf{r}}) + V_{H}(\textit{\textbf{r}})\right]\psi_i(\textit{\textbf{r}}) = E_i\psi_i(\textit{\textbf{r}})~.
\label{ch2.eq13}
\ee
Here $V_{H}(\textit{\textbf{r}})$ is the Hartree potential which is the average electrostatic potential experienced by the $i^{th}$ electron in the presence of all the other electrons in the system we have, 
\be
V_H (\textit{\textbf{r}}) = e^2 \int \frac{d^3\textit{\textbf{r'}}}{|\textit{\textbf{r}}-\textit{\textbf{r'}}|}\sum_{j\ne i}|\psi_j(\textit{\textbf{r'}})|^2~.
\label{ch2.eq14}
\ee
To find the energy eigenvalues and wavefunctions for individual electrons, Equation~\ref{ch2.eq13} must be solved self-consistently.

Although the Hartree theory simplifies the many-electron problem into a single-electron problem, it still needs to account for the electron-electron interaction accurately. Further, it ignores the anti-symmetric nature of the electronic wavefunction. To account for the anti-symmetric nature of the electronic wavefunction, the Harthree-Fock theory represents the many-body wavefunction as,
\be
\psi_e(\textbf{r}_1, \sigma_1; \ldots; \textbf{r}_N, \sigma_N) = \frac{1}{\sqrt{N!}} \sum_P (-1)^P \psi_{P(1)}(\textbf{r}_1, \sigma_1) \psi_{P(2)}(\textbf{r}_2, \sigma_2) \cdots \psi_{P(N)}(\textbf{r}_N, \sigma_N)~.
\label{ch2.eq15}
\ee
In this representation, the wave function is a summation over all possible permutations ($P$) of electron labels, multiplied by the corresponding combined spatial and spin wavefunctions for each electron. This summation with the antisymmetrization operator ensures that the many-electron wavefunction is properly antisymmetrized to obey the Pauli exclusion principle. In a Slater determinant form, Equation~\ref{ch2.eq15} can be expressed as 
\be
\psi_e(\textbf{r}_1, \sigma_1;\ldots; \textbf{r}_N, \sigma_N) = \frac{1}{\sqrt{N!}}
\begin{vmatrix}
	\psi_1(\textbf{r}_1, \sigma_1) & \psi_2(\textbf{r}_1, \sigma_1) & \cdots & \psi_N(\textbf{r}_1, \sigma_1) \\
	\psi_1(\textbf{r}_2, \sigma_2) & \psi_2(\textbf{r}_2, \sigma_2) & \cdots & \psi_N(\textbf{r}_2, \sigma_2) \\
	\vdots & \vdots & \ddots & \vdots \\
	\psi_1(\textbf{r}_N, \sigma_N) & \psi_2(\textbf{r}_N, \sigma_N) & \cdots & \psi_N(\textbf{r}_N, \sigma_N)
\end{vmatrix}.
\label{ch2.eq16}
\ee
The Hartree-Fock method represents a significant advancement over the conventional Hartree method by incorporating the exchange interaction. This addresses the intrinsic anti-symmetry exhibited by the electronic wave function of many-particle systems. However, the Hartree-Fock method overlooks the Coulomb correlation effect, resulting in an overestimation of the exchange term's contribution. Although these methods demonstrate success when applied to atoms and small molecules, yet their applicability needs to improve when dealing with complex systems such as large molecular structures and biomolecules. As the scale of the system increases, both the Hartree and Hartree-Fock methodologies become computationally demanding and impractical, highlighting the necessity for more sophisticated approaches in these cases.
The state-of-the-art method to deal with such a large system is DFT.

\section{Density functional theory} 
\label{DFT}

The DFT has emerged as a powerful and versatile theoretical framework, offering a comprehensive means of investigating the electronic structure and properties of diverse materials. By focusing on the electron density distribution rather than explicitly considering individual electron wave functions, DFT provides an elegant and computationally efficient approach to describe the behavior of complex systems ranging from atoms and molecules to solids and surfaces. This concept hinges on the transition from a 3N variable (where N represents the number of electrons and three accounts for the three spatial dimensions) problem associated with the wave function to a mere three variable problem, represented by the coordinates specifying the electron density. This transformation yields a remarkable advantage, significantly reducing the computational complexity and making DFT a viable method for studying large systems. This simplicity, however, does not compromise accuracy, enabling predictions of a wide range of properties and phenomena across various disciplines such as chemistry, physics, materials science, and biology. 

Hohenberg and Kohn's seminal work on DFT in 1964 treats the many-body ground state wave function as a function of the electron density, allowing the ground state energy to be minimized based on the electron density~\cite{Hohenberg-Kohn-1964}. This obviates the need to explicitly consider many-body wave functions, offering a route to simplifying  the many-body problem. Kohn and Sham in 1965 introduced an iterative approach, utilizing a non-interacting Hamiltonian matched to the true system's density, facilitating practical implementation~\cite{Kohn-Sham}. Two fundamental theorems underpinning DFT are as follows:
\begin{enumerate}
	\item  The external potential, $V_{ext}(r)$, within an interacting many-electron system, can be exclusively represented as a functional derived from the ground state density, $n(r)$. Consequently, the ground state energy of the system is also a distinct function of $n(r)$, denoted as $E = E[n]$.
	
	\item The total energy functional, $E[n]$, possesses a minimum value that aligns with the ground state energy attributed to the corresponding ground state density, $n(r)$.
\end{enumerate}

\subsection{Kohn-Sham Equation}
\label{KSE}
Within the Kohn-Sham formalism of DFT, the interacting electrons are mapped onto an auxiliary non-interacting system designed to represent the same density distribution. Through this trick, Kohn-Sham orbitals $[\psi_{i}^{KS}(\textbf{r})]$ has been introduced — this reproduces the density of real electron arrangement in the system, defined as 
\be
n(\textbf{r}) = \sum_{i=1}^{N}|\psi_{i}^{KS}(\textbf{r})|^2.
\label{ch2.eq17}
\ee
These Kohn-Sham orbitals can be treated as the eigenstates of the Kohn-Sham equation,
\be
\left[ -\frac{\hbar^2}{2m}\nabla^2 + V_{eff}(\textbf{r}) \right] \psi_{i}^{KS}(\textbf{r}) = E_i \psi_{i}^{KS}(\textbf{r})~. 
\label{ch2.eq18}
\ee
Here, $E_i$ are the  Kohn-Sham eigenvalues. The external potential, \(V_{\text{eff}}(\textbf{r})\)  here is a functional of electron density $n$ and it consists of three terms:
\be
V_{eff}(\textbf{r}) = V_{ext}^{[n]}(\textbf{r}) + V_{H}^{[n]}(\textbf{r}) + V_{XC}^{[n]}(\textbf{r})~.
\label{ch2.eq19}
\ee
Since $V_{eff}^{[n]}(\textbf{r})$ is a functional of density and therefore Equation~\ref{ch2.eq18} needs to be solved self-consistently to find the true ground state energy and density. In Equation~\ref{ch2.eq19}, the $V_{H}^{[n]}(\textbf{r})$ represent the Hartree potential and can be written as,

\be
V_{H}^{[n]}(\textbf{r}) = e^2\int d^3r'\frac{n(\textbf{r})}{|\textbf{r}-\textbf{r'}|}~,
\label{ch2.eq20}
\ee
The exchange-correlation potential, $V_{XC}^{[n]}(\textbf{r})$, in Equation~\ref{ch2.eq19}, is defined as,
\be
V_{XC}^{[n]}(\textbf{r}) = \frac{\partial E_{XC}^{[n]}}{\partial n(\textbf{r})}~,
\ee
where $E_{XC}^{[n]}$ is the exchange-correlation part of the total ground state energy. Unfortunately, the exact analytic form of the exchange-correlation energy is unknown. Therefore, several approximations have been proposed for calculating the exchange-correlation part of the potential. Among them, the most adopted are the local-density approximation (LDA) and generalized gradient approximations (GGA). These approximations serve as foundational pillars in understanding the behavior of electrons within a material, with each approach addressing specific aspects of electron density variation.

\subsection{LDA and GGA}
\label{LDA_GGA}
Initially proposed by Kohn and Sham in 1965, the LDA rests on the fundamental assumption that the electron density, denoted as $n(\mathbf{r})$, varies smoothly across space, displaying homogeneity within small local volumes. Within this framework, the exchange-correlation energy functional, $E_{xc}[n]$, takes the form:

\be
E_{xc}^{LDA}[n] = \int \epsilon_{xc}[n(\mathbf{r})] n(\mathbf{r}) d^3r~.
\label{ch2.eq22}
\ee
Here, $\epsilon_{xc}[n(\mathbf{r})]$ represents the exchange-correlation function, and the integral term signifies the exchange-correlation energy for an interacting homogeneous electron gas with density $n(\mathbf{r})$ and volume $d^3r$. The corresponding exchange-correlation potential, $v_{xc}(\mathbf{r})$, is derived as:

\be
v_{xc}(\mathbf{r}) = \frac{\partial E^{LDA}_{xc}[n(\mathbf{r})]}{\partial n(\mathbf{r})} \equiv \mu_{xc}(n(\mathbf{r}))~,
\label{ch2.eq23}
\ee
where, $\mu_{xc}(n(\mathbf{r}))$ denotes the exchange-correlation contribution to the chemical potential of the uniform electron gas. The total energy functional within LDA can thus be expressed as:
\be
E^{LDA}[n] = \sum_{i} \epsilon_i - \frac{e^2}{2} \int \int \frac{n(\mathbf{r}) n(\mathbf{r}^{\prime})}{|\mathbf{r}-\mathbf{r}^{\prime}|}d^3rd^3r^{\prime} + \int \{ \epsilon_{xc}(n(\mathbf{r}))-\mu_{xc}(n(\mathbf{r}))\}  n(\mathbf{r}) d^3r~.
\label{ch2.eq24}
\ee
LDA has proven highly effective in predicting the ground state energy for numerous systems. Nevertheless, it exhibits limitations when dealing with systems characterized by rapidly varying electron densities, necessitating the incorporation of gradient corrections.
Building upon the LDA, the GGA enhances the accuracy of the calculation of electronic states. GGA accounts for local variations in electron density by extending the exchange-correlation functional to be dependent not only on the electron density, $n(\mathbf{r})$, but also on its spatial gradient, $\mathbf{\nabla}n(\mathbf{r})$. The GGA exchange-correlation energy functional takes the form:
\be
E_{xc}[n(\mathbf{r})] = \int \epsilon_{xc}[n(\mathbf{r}), \mathbf{\nabla}n(\mathbf{r})] n(\mathbf{r}) d^3r.
\label{ch2.eq25}
\ee
There as several implementations of GGA, such as the widely used Perdew, Burke, and Ernzerhof (PBE)~\cite{GGA-PBE} and the optimized PBE for solids (PBE-sol)~\cite{PBE_sol}. Compared to LDA, GGA offers improved agreement with experimental electronic structure data. However, similar to LDA, it underestimates the bandgap in various cases due to its inherent "local density" approximation and the omission of long-range Coulomb interactions.

Hence, a more advanced method is required to accurately predict the individual energy levels of electrons and holes, which is crucial for understanding how charge carriers behave in materials. This is where the GWA for the electron self-energy becomes essential. Now, we review the GWA, a more accurate method for calculating the energy levels of electrons and holes.

\section{The GW approximation}
\label{GWA}
The GWA, rooted in one-body Green's function methods, has proven to be highly successful for the single-particle energies of the quasi-electrons and quasi-holes in excited states~\cite{Hedin-GW1, Hedin1970, SGLGW1985, Godby1986}. Hedin's GW theory, as introduced by Hedin in his seminal work \cite{Hedin-GW1}, offers a vital solution to address the inherent challenge in DFT, which struggles to account for dynamical electron-electron correlations accurately. In many-body interactions, the GW theory can be viewed as an energy convolution process involving two key components: the interacting electronic Green's function $(G)$ and the dynamic electron-electron screened potential denoted as $W$. The construction of these crucial components involves a series of convolutions.

We begin with the non-interacting single-particle Green's function, represented as ${G}^0(\textbf{r},\textbf{r}^{\prime};\tau)$. Using ${G}^0(\textbf{r},\textbf{r}^{\prime};\tau)$, we calculate the polarization function within the random phase approximation,
\be
P\left(\textbf{r},\textbf{r}^{\prime\prime};\tau\right)=-i{G}^{0}\left(\textbf{r},\textbf{r}^{\prime};\tau\right){G}^{0}\left(\textbf{r}^{\prime},\textbf{r};-\tau\right)~.
\label{ch2.eq26}
\ee
In this equation, $\tau= t-t^{\prime}$, represent the temporal evolution of the propagator from time $t$ to time $t^{\prime}$. In the Kohn-Sham eigenstates basis, the polarization function can be expainded as a summation over the occupied and unoccupied states. In the Fourier space, it is given by,
\be
\begin{aligned}
	{P}\left(\textbf{r},\textbf{r}^{\prime\prime};\omega\right) = \sum_{v}^{occ}\sum_{c}^{unocc}&\psi_{c}^{KS}\left(\textbf{r}\right)\psi_{v}^{KS\ast}\left(\textbf{r}\right)\psi_{c}^{KS \ast}\left(\textbf{r}^{\prime}\right)\psi_{v}^{KS}\left(\textbf{r}^{\prime}\right)\\
	&\times\left[\frac{1}{\omega+\varepsilon_{v}^{KS}-\varepsilon_{c}^{KS}+i\eta}-\frac{1}{\omega-\varepsilon_{v}^{KS}+\varepsilon_{c}^{KS}-i\eta}\right]~.
\end{aligned}
\label{ch2.eq27}
\ee
Here, $\psi_{v}^{KS}$ and $\varepsilon_{c}^{KS}$ correspond to the wave functions and energies of the Kohn-Sham equation for the electrons in occupied and unoccupied states, respectively. This approach by Hedin's GW theory, captures the dynamical electron-electron correlations that DFT often struggles to account for.

To calculate the QP energies within the GW theory, first we need to construct the microscopic dielectric function, denoted as $\epsilon(\mathbf{r},\mathbf{r}^\prime;\omega)$, which characterizes the response of electrons to an external perturbation at frequency $\omega$. This dielectric function is obtained by integrating the polarization function (Equation~\ref{ch2.eq27}) with the bare Coulomb potential $v(\mathbf{r},\mathbf{r}^{\prime\prime})$,
\be
\epsilon(\mathbf{r},\mathbf{r}^\prime;\omega) = \delta(\mathbf{r}-\mathbf{r}^\prime) - \int {P}(\mathbf{r},\mathbf{r}^{\prime\prime};\omega) v(\mathbf{r},\mathbf{r}^{\prime\prime}) d^{3}\mathbf{r}^{\prime\prime}~. 
\label{ch2.eq28}
\ee
Here, $\delta(\mathbf{r}-\mathbf{r}^\prime)$ represents the Kronecker delta function. Next, it utilizes the inverse of this dielectric function to construct the renormalized dynamic screened potential, 

\be
W(\mathbf{r},\mathbf{r}^\prime;\omega) = \int \epsilon^{-1}(\mathbf{r},\mathbf{r}^{\prime\prime};\omega) v(\mathbf{r},\mathbf{r}^{\prime\prime}) d^{3}\mathbf{r}^{\prime\prime}~.
\label{ch2.eq29}
\ee
Further, the electron self-energy is evaluated, which involves a convolution over all frequencies between the non-interacting electronic Green's function and the dynamic screening function $W(\mathbf{r},\mathbf{r}^\prime;\omega)$:

\be
\Sigma^{GW}(\mathbf{r},\mathbf{r}^\prime;\omega) = \frac{i}{2\pi}\int_{-\infty}^{+\infty}G_{0}(\mathbf{r},\mathbf{r}^\prime;\omega+\omega^{\prime})W(\mathbf{r},\mathbf{r}^\prime;\omega^{\prime})~ e^{i\omega^{\prime}\eta}d\omega^{\prime}~. 
\label{ch2.eq30}
\ee
Here, $\Sigma^{GW}{(\mathbf{r},\mathbf{r}^\prime;\omega)}$ represents a non-local and energy-dependent self-energy operator. The self-energy operator plays a crucial role in determining the difference between the energy of a quasiparticle and that of a non-interacting, particle. Consequently, it encapsulates the intricate many-body interactions within the system. The electron-self-energy in Equation~\ref{ch2.eq30} has two distinct components: a pure exchange contribution and a correlational self-energy component:

\be
\textstyle \sum^{x}\left(\textbf{r},\textbf{r}^{\prime};\omega\right) = \frac{i}{2\pi}\int_{-\infty}^{+\infty}G^{0}\left(\textbf{r},\textbf{r}^{\prime};\omega+\omega^{\prime}\right)v\left(\textbf{r},\textbf{r}^{\prime}\right)e^{i\omega^{\prime}\eta}d\omega^{\prime},
\label{ch2.eq31}
\ee
and
\be
\textstyle \sum^{c}\left(\textbf{r},\textbf{r}^{\prime};\omega\right) = \frac{i}{2\pi}\int_{-\infty}^{+\infty}{G}^{0}\left(\textbf{r},\textbf{r}^{\prime};\omega+\omega^{\prime}\right)\left[\textrm{W}\left(\textbf{r},\textbf{r}^{\prime};\omega\right)-v\left(\textbf{r},\textbf{r}^{\prime}\right)\right]d\omega^{\prime},
\label{ch2.eq32}
\ee
respectively. Finally, we can write the QP energy equation,

\be
\varepsilon_{n\textbf{k}}^{QP} = \varepsilon_{n\textbf{k}}^{KS} + \langle n\textbf{k}| \Sigma^{GW} (\varepsilon_{n\textbf{k}}^{QP}) - V_{xc}|n\textbf{k}\rangle~.
\label{ch2.eq33}
\ee
This approximation holds well when the DFT eigenvectors closely resemble the GW eigenvectors, a common scenario for moderately correlated systems. 

\section{The Bethe-Salpeter equation}
\label{BSE}
The GWA predicts a very accurate QP energies, whereas it fails to capture the Coulomb correlation between the photo-excited electron-hole pair, and hence, it does not predict accurate optical absorption and other excited state properties~\cite{DFT_GWA_Exciton-GW3,yambo2019,DESLIPPE20121269}. Therefore, our next objective is to review the formulation that incorporates electron-hole correlations using a two-particle Green's function and the subsequent solution of its equation of motion, commonly referred to as the BSE. The BSE gives us the energy eigenvalues and eigenfunctions of the correlated electron-hole pair called excitons within a bound state or resonant. We can write the exciton states as a linear combination of quasi-electron and quasi-hole states, 
\bea
|\varPsi^{s}_{\textbf{Q}}\rangle = \sum_{vc\textbf{k}}A^{S}_{vc\textbf{k}}|{v\textbf{k}}\rangle \otimes |c\textbf{k+Q}\rangle,
\label{ch2.eq34}
\eea
where $S$ represents the exciton state, \textbf{Q} denotes the center-of-mass momentum of the exciton, and $A^S_{vc\textbf{k}}$ represents the amplitude of the free quasi-electron and quasi-hole pair with an electron in the state $|c\textbf{k + Q}\rangle$ and a hole from the state $|v\textbf{k}\rangle$.
To incorporate the electron-hole interaction, a two-particle correlation function can be defined as~\cite{BSE-1-Strinati, BSE-2-Strinati},
\be
L(1,2;1',2') = -G_2(1,2;1',2') + G(1,1')G(2,2')~.
\label{ch2.eq35}
\ee
Here, $G_2$ denotes the two-particle Green's function. The variables (1) involve position, spin, and time coordinates: (1) = (\textbf{x}$_1$, t1) = (\textbf{r}$_1$,$\sigma_1$, t1). The function $L$ relies on four time variables, corresponding to two creation processes (electron and hole) and two annihilation processes. The electron-hole correlation function can be represented as Dyson equation.
\be
L(12;1'2') = L_{0}(12;1'2') + \int d(3456) L_0(14;1'3)K(35;46)L(62;52')~.
\label{ch2.eq36}
\ee
Equation~\ref{ch2.eq36} is known as the BSE.
$L(12; 1'2')$ represents the electron-hole correlation function, and $K(35; 46)$ is the electron-hole interaction kernel, as discussed below. $L_0(12; 1'2') = G_1(1,2')G_1(2,1')$ corresponds to free electron-hole pairs with the interaction $K$ switched off. \\
The BSE (Equation~\ref{ch2.eq36}) can be written as a generalized eigenvalue problem in the matrix form
\begin{equation}
	H^{BS}(\mathbf{Q}) = (\varepsilon_{c\mathbf{k+Q}}^{QP}-\varepsilon_{v\mathbf{k'}}^{QP})\delta_{\mathbf{k+Q},\mathbf{k'}} + \begin{pmatrix}
		K^{AA}(\mathbf{Q}) & K^{AB}(\mathbf{Q}) \\
		K^{BA}(\mathbf{Q}) & K^{BB}(\mathbf{Q})
	\end{pmatrix}~.
	\label{ch2.eq37}
\end{equation}
The kernels in the matrix are calculated based on single-particle wavefunctions as follows,
\be
K^{AA}_{vc\textbf{k},v^{\prime}c^{\prime}\textbf{k}^{\prime}}(\mathbf{Q}) = i\int d(3456)\phi_{v\textbf{k}}(\mathbf{x}_4)\phi^{\star}_{c\textbf{k+Q}}(\mathbf{x}_3)K(3,5;4,6)\phi^{\star}_{v^{\prime}\mathbf{k}^{\prime}}(\mathbf{x_5})\phi_{c^{\prime}\textbf{k}^{\prime}+\textbf{Q}} (\mathbf{x}_6)~,
\label{ch2.eq38}
\ee  
\be
K^{AB}_{vc\textbf{k},v^{\prime}c^{\prime}\textbf{k}^{\prime}}(\mathbf{Q}) = i\int d(3456)\phi_{v\textbf{k}}(\mathbf{x}_4)\phi^{\star}_{c\textbf{k+Q}}(\mathbf{x}_3)K(3,5;4,6)\phi^{\star}_{v^{\prime}\mathbf{k}^{\prime}}(\mathbf{x_6})\phi_{c^{\prime}\textbf{k}^{\prime}+\textbf{Q}} (\mathbf{x}_5)
\label{ch2.eq39},
\ee
\be  
K^{BB} = -K^{AA*}~,
\label{ch2.eq40}
\ee
\be
K^{BA} = -K^{AB*}~.
\label{ch2.eq41}
\ee
The equation presented above (Equation ~\ref{ch2.eq37}) exhibits a block-matrix structure, with diagonal blocks representing energy differences $(E_c - E_v)$ and interaction matrix elements $K^{AA}$ and $K^{BB}$. Meanwhile, off-diagonal blocks, namely $K^{AB}$ and $K^{BA}$, are typically negligible in determining excitation energies. Consequently, one can set $K^{AB} = K^{BA} =0$ if the QP gap is small compared to the electron-hole interaction. In that case, Equation~\ref{ch2.eq37} reduces to the Bethe-Salpeter Hamiltonian $H^{BS}$ within the Tamm-Dancoff approximation (TDA), 
\begin{equation}
	H^{BS}(\mathbf{Q}) = (\varepsilon_{c\mathbf{k+Q}}^{QP}-\varepsilon_{v\mathbf{k'}}^{QP})\delta_{\mathbf{k+Q},\mathbf{k'}} + K^{AA}(\mathbf{Q})
	\label{ch2.eq42}~,
\end{equation}
or, in terms of the eigenvalue problem, it can be written as,
\begin{equation} 
	\left(\varepsilon_{c\textbf{k}}^{QP}-\varepsilon_{v\textbf{k}}^{QP}\right)A_{vc\textbf{k}}^{s}+
	\sum_{v^{\prime}c^{\prime}\textbf{k}^{\prime}}\left\langle vc\textbf{k}\left|K^{AA}_{vc\textbf{k},v^{\prime}c^{\prime}\textbf{k}^{\prime}}\right|v^{\prime}c^{\prime}\textbf{k}^{\prime}\right\rangle A_{v^{\prime}c^{\prime}\textbf{k}^{\prime}}^{s}=\mathcal{E}^{X}_{s}A_{vc\textbf{k}}^{s}~.
	\label{eq2.43}
\end{equation}
The electron-hole interaction kernel $K^{AA}$, denoted as $K(35;46)$, is defined as:
\be
K(35;46) =    \frac{\delta[V_{Coul}(3)\delta(3,4)+\Sigma(3,4)]}{\delta G_1 (6,5)}
\label{ch2.eq44}~.
\ee
Assuming that the derivative of the $W$ with respect to $G_1$ can be disregarded, one can derive:
\bea
K(35;46) &= &i \delta (3,4)\delta (5^{-},6)v(3,6)+i\delta (3,6)\delta (4,5)W(3^{+},4)\\
&=& K^{x}(35;46) + K^{d}(35;46)~. 
\label{ch2.eq46}
\eea
$K^{x}$ is known as the exchange kernel due to the bare Coulomb interaction, and $k^{d}$ represents the screened Coulomb interaction and is known as the direct kernel. In cases where the exciton binding energies $(\varepsilon_{c\textbf{k}}^{QP}-\varepsilon_{v\textbf{k}}^{QP}-\mathcal{E}_{s}^{X})$ are significantly less than the plasmon frequency, we can neglect the frequency-dependence screening in $W$. Consequently, within the quasi-electron basis, the matrix elements of the BSE kernel for both the exchange and direct contributions are given as,
\be
\langle vc\textbf{k}|K^{x}|v^{\prime}c^{\prime}\textbf{k}^{\prime}\textbf{Q}\rangle =\int d\textbf{x}d\textbf{x}^{\prime} \phi^{*}_{c\textbf{k+Q}}(\mathbf{x})\phi_{v\textbf{k}}(\mathbf{x})v(\textbf{r,r'})\phi^{*}_{v^{\prime}\mathbf{k}^{\prime}}(\mathbf{x'})\phi_{c^{\prime}\textbf{k}^{\prime}+\textbf{Q}} (\mathbf{x'})~,
\label{ch2.eq47}
\ee
and
\be
\langle vc\textbf{k}|K^{d}|v^{\prime}c^{\prime}\textbf{k}^{\prime}\textbf{Q}\rangle =-\int d\textbf{x}d\textbf{x}^{\prime} \phi^{*}_{c\textbf{k+Q}}(\mathbf{x})\phi_{c'\textbf{k'+Q}}(\mathbf{x})W(\textbf{r,r'};\omega=0)\phi^{*}_{v^{\prime}\mathbf{k}^{\prime}}(\mathbf{x'})\phi_{v\textbf{k}} (\mathbf{x'})~.
\label{ch2.eq48}
\ee
After obtaining the BSE solutions, we can utilize them to find out the physical observables of the interest, i.e., the optical absorption and conductivity, and both of them are defined from the imaginary part of the macroscopic dielectric function ($\epsilon_{M}$).

To determine the macroscopic dielectric function, we diagonalize the matrices in Equation \ref{ch2.eq34} under the condition of the long-wavelength limit ($\textbf{q}\rightarrow 0$). This results in the following expression for the macroscopic dielectric function $\epsilon_{M}(\omega)$:
\begin{eqnarray}    
	\epsilon_{M}\left(\omega\right)&=&\mathrm{\underset{\textbf{q}\rightarrow0}{Lim}}\frac{1}{\epsilon_{\textbf{G}=0,\textbf{G}^{\prime}=0}^{-1}\left(\omega\right)}\nonumber \\ 
	&=&1-{\textstyle {\displaystyle \mathrm{\underset{\textbf{q}\rightarrow0}{lim}}}}\frac{8\pi}{\left|\textbf{q}\right|^{2}\Omega}\sum_{s}\frac{\left|\sum_{cv\textbf{k}}\left\langle v\textbf{k}-\textbf{q}\left|e^{-i\textbf{q}.\textbf{r}}\right|c\textbf{k}\right\rangle A_{cv\textbf{k}}^{s}\right|^{2}}{\omega-\mathcal{E}^{X}_{s}+i\eta}~.
	\label{ch2.eq49}
\end{eqnarray}
In this expression, $\Omega$ represents the cell volume, and $\left\langle v\textbf{k}-\textbf{q}\left|e^{-i\textbf{q}.\textbf{r}}\right|c\textbf{k}\right\rangle$ denotes the dipole oscillator strength. To evaluate the exciton wavefunction in real space lattice, we can write Equation~\ref{ch2.eq34} in terms of single particle wavefunction of electron and hole:
\begin{equation}
	\left|\varPsi^{s}\right\rangle = \sum_{cv\textbf{k}}A_{cv\textbf{k}}^{s}\phi_{v\textbf{k}}(\textbf{r}_{e})\phi_{c\textbf{k}}(\textbf{r}_{h})~.
	\label{ch2.eq50}
\end{equation}
Here, $\textbf{r}_e$ and $\textbf{r}_h$ represent the electron and hole coordinates, respectively. It is worth noting that the evaluation of this wavefunction involves six coordinates.

Additionally, the temperature dependency of the absorption spectra is calculated within the framework developed by Marini\cite{Marini2008}. This involves analyzing the zero-point energy and the excitonic Hamiltonian's temperature dependency. These calculations yield valuable information about how the material's optical properties change with temperature.

\section{Temperature-dependent Bethe-Salpeter equation}
\label{temperatureBSE}
The {\it ab-initio} calculation of optical spectrum in semiconducting solids has been proven to be quantitatively accurate within the framework of many-body perturbation theory (MBPT). Within this approach, the GWA and BSE deployed to calculate the quasiparticle energies and the optical spectrum are well-established state-of-the-art methods~\cite{HedinGW1, Hybertsen-GW2,spin-orbit-MoS2-PRB}. Different from the standard GW+BSE implementation to study the excitonic effects, it is insufficient to accurately predict the optical properties because it does not account for the lattice vibrations. Incorporating the lattice vibrations, even at zero temperature, modifies the optical properties significantly. The exciton-driven optical spectrum obtained within a frozen-atom condition gets modified at zero and finite temperatures, and excitons acquire a nonradiative lifetime. The electron-phonon (el-ph) interaction in optical study plays a crucial role in determining exciton nonradiative dynamics~\cite{Purz2022}, decoherence times~\cite{moody2015intrinsic,dey2016optical} exciton linewidths~\cite{selig2016excitonic,cadiz2017excitonic}. 
In particular, the presence of lattice vibrations introduces intriguing physics into optical absorption via excitons, as the energy eigenvalues now become complex, encompassing both energy broadening and lifetime ~\cite{Marini2008}. In the optical limit ($q \rightarrow 0$), optical absorption was calculated at different temperatures using the BSE method, considering the influence of el-ph interactions on the coupled electron-hole BSE Hamiltonian.
In the frozen-atom (FA) case, from the BS Hamiltonian $H_{FA}$ (Equation~\ref{ch2.eq42}), the excitonic states and energies are denoted as $|\varPsi^{s}\rangle$ and $\mathcal{E}_{s}^{X}$, respectively, in the electron-hole basis as defined in Equation~\ref{ch2.eq50}. The absorption spectrum that is the imaginary part of the macroscopic dielectric function defined in Equation~\ref{ch2.eq49} can be written as,
\be
\epsilon_{2} = -\frac{8\pi}{\Omega} \sum_s  \left|A^{FA}_{s}\right|^2 \mathrm{Im} \left[(\omega-\mathcal{E}_{X}^{s}+i \eta)^{-1}\right]~.
\label{ch2.eq51}
\ee
In the finite-temperature regime, the  single particle energy levels $E_i$ have an explicit dependence on the temperature:
\be
\varepsilon_i (T)=\varepsilon_i + \Delta \varepsilon_i(T),
\label{ch2.eq52}
\ee
with 
\be
\Delta \varepsilon_i(T) = \Delta \varepsilon^{el-ph}_i(T) + \Delta \varepsilon^{TE}_i(T),
\label{ch2.eq53}
\ee
where $\varepsilon_i^{TE}(T)$ is the thermal expansion (TE) contribution~\cite{CARDONA20053,Paszkowicz2002,Marini2008}. $\Delta E_{el-ph}$ represents the correction originating from the el-ph interaction and is a complex number. To calculate the $\Delta E_{el-ph}$, we have adopted the Heine, Allen, and Cardona approach~\cite{Allen1976}, where it is defined in terms of Eliashberg function $g^2 F_i(\omega)$:
\be
\Delta E_{el-ph} = \int d\omega g^2 F_i(\omega) [N(\omega, T) +1/2]~,
\label{ch2.eq54}
\ee 
where $N(\omega, T)$ represents the Bose occupation function, $g$ being the el-ph matrix elements which is defined as,
\begin{equation}
	g^{\textbf{q}\nu}_{n',n\textbf{q}} = \sum_{\xi l} \langle n, \textbf{k}| \nabla_{\xi l}\phi|n', \textbf{k-q}\rangle \times \sum_{\textbf{q} \nu} \left ( \frac{1}{2M_{l}\omega_{\textbf{q}\nu}}\right)^{\frac{1}{2}}e^{-iq \cdot \tau_{l}}\varepsilon^{*}\left( \frac{\textbf{q}\nu}{l} \right)~.
	\label{ch2.eq55}
\end{equation}
This expression referres to the first-order el-ph matrix element. The self-consistent potential $\phi$ is computed using the charge density obtained within the Kohn-Sham DFT. The symbols $\xi$ represent atomic displacements, and $\tau_l$ indicates the position of $l$-th atomic species having mass $M$, $\omega_{\mathbf{q}}$ refers to the phonon frequency at lattice momentum ($\mathbf{q}$), $\nu$ represents the index for the phonon mode. Polarization vectors denoted as $\varepsilon^{*}( \frac{\textbf{q}\nu}{l})$. 
It is important to note that including el-ph interaction in the self-energy term transforms the BSE Hamiltonian into a non-Hermitian temperature-dependent Hamiltonian. 

\be
H_{BS}(T) = H_{BS}^{FA} + [\Delta E_{e}(T)-\Delta E_{h}(T) ]\delta_{eh,e'h'}~.
\label{ch2.eq56}
\ee
Here, $H_{BS}^{FA}$ represents the BS Hamiltonian for frozen atoms as defined in Equation~\ref{ch2.eq42}. Equation~\ref{ch2.eq56} can be solved as a standard eigenvalue problem 
\begin{equation}
	H_{BS}(T) |\varPsi_s(T)\rangle =\mathcal{E}_{s}^{X}(T) |\varPsi_s(T)\rangle~.
	\label{ch2.eq57}	
\end{equation}
The eigenstates $\varPsi_s(T)$ are a linear combination of electron-hole (e-h) pairs,
\begin{equation}
	|\varPsi_s(T)\rangle=\sum_{eh} A_{eh}^{s}(T)|eh\rangle~,
\label{ch2.eq58}
\end{equation}
where, $A^{s}_{eh}=\langle eh|\varPsi_s\rangle$. Within this formalism, we can get the temperature-dependent exciton energy eigenvalues as,
\begin{equation}
	\mathcal{E}_{s}^{X}(T) = \langle \varPsi_s(T)|H_{BS}^{FA}|\varPsi_s(T)\rangle +\sum_{eh} |A_{eh}^{s}(T)|^2[\Delta E_{e}(T)-\Delta E_{h}(T)]~.
\label{ch2.eq59}
\end{equation}
By omitting the TE contribution from Equation~\ref{ch2.eq53}, we obtain the following expressions:
\begin{equation}
	\mathrm{Re}[\Delta\mathcal{E}_{s}(T)] = \langle \varPsi_s(T)|H_{BS}^{FA}|\varPsi_s(T)\rangle -\langle \varPsi_s|H_{BS}^{FA}|\varPsi_s\rangle + \int d\omega\mathrm{Re}[g^2F(\omega,T)][N(\omega,T)+1/2]~.
\label{ch2.eq60}
\end{equation}
Additionally,
\begin{equation}
	\mathrm{Im}[\mathcal{E}_{s}(T)] =  \int d\omega ~ \mathrm{Im}[g^2F(\omega,T)][N(\omega,T)+1/2]~.
	\label{ch2.eq61}
\end{equation}
Here, $\Delta \mathcal{E}_s(T) = \mathcal{E}_s(T)-\mathcal{E}_s^{FA}(T)$, with the exciton-phonon coupling function defined as:
\begin{equation}
	g^2F_{s}(\omega, T) =\sum_{eh}|A^{s}_{eh}(T)|^2[g^2F_e(\omega)-g^2F_h(\omega)]~.
	\label{ch2.eq62}	
\end{equation}
We can now define the nonradiative excitonic lifetime ($\tau^s$) using the imaginargy part of the exciton eigenvalue as defined in Equation~\ref{ch2.eq61},
\begin{equation}
	\tau^s = \frac{1}{2\mathrm{Im}[\mathcal{E}_s(T)]}~.
	\label{ch2.eq63}
\end{equation}
The nonradiative excitonic lifetime is otherwise infinite in the FA approximation. Notably, the dielectric function now explicitly depends on temperature $T$ and is expressed as:

\begin{equation}
	\epsilon_{2}(\omega, T) = - (8\pi/\Omega)\sum_s|A_{s}(T)|^2\mathrm{Im}\{[\omega - \mathcal{E}_{s}(T)]^{-1}\}~.
	\label{ch2.eq64} 
\end{equation}
Importantly, there is no longer a need for a damping parameter in this context.\\

In this chapter, we have outlined our methodology for investigating excitonic properties in 2D semiconductors. The theoretical framework involves employing DFT for ground state electronic structure calculations, the GW approximation (GWA) for excited state single-particle energies, and the Bethe-Salpeter Equation (BSE) for excitonic structure calculations in Chapter~\ref{chap3}, Chapter~\ref{chap4}, and Chapter~\ref{chap5}.

Furthermore, we explore temperature-dependent excitonic properties of monolayer AlN in Chapter~\ref{chap4} using the temperature-dependent BSE, as discussed in Section~\ref{temperatureBSE}. Detailed computational parameters for DFT, GW, and BSE calculations, including kinetic energy cutoff, k-grid sampling, and relevant settings, are expounded upon in their respective chapters, providing a comprehensive insight into our methodology.

%% file: chap3.tex
\chapter{Fluence dependent dynamics of excitons in monolayer MoSi$_2$Z$_4$ (Z = N, As, P)}
\label{chap3}
\pagestyle{fancy}

The interplay of light-matter interactions and Coulomb interactions in two dimensional (2D) materials gives rise to very exciting physics such as emergence of strongly bound excitons with exceptionally large binding energies~\cite{Thygesen2017,Wang2018,Mounet2018,amit-nanoscale-2018}. Coupled with the rapidly growing family of stable 2D materials, this has accelerated the exploration of novel optical responses and their potential use for next-generation  optoelectronic
device applications~\cite{Novoselov2004_graphene1,Bernardi2013-graphene-light,Butler2013}. In particular,  2D monolayers of transition metal di-chalcogenides (TMDs) exhibit prominent excitonic effects~\cite{optical-MoS2-PRL,strain-exciton-Mos2-PRB,Dark-Exciton-PhysRevMaterials.2.014002,exciton-TMD}, due to the reduced dielectric screening and enhanced Coulomb interactions. These excitonic effects significantly modify the optical absorption spectrum and can be tuned by various external stimuli, opening up new possibilities for optoelectronic applications~\cite{Butler2013,Bernardi2013-graphene-light}.

In recent experimentals investigation conducted under non-equilibrium conditions, 2D TMDs  have been identified to show a signature of exciton-exciton interaction. As the excitation density increased, a distinct redshift in the exciton resonance energy was identified, succeeded by an unexpected blueshift~\cite{Sei,amit-exciton-2,Schmit1985}. The observation suggests a connection to plasma effects and an attraction-repulsion crossover in exciton-exciton interactions, akin to the behavior observed in the Lennard-Jones potential between atoms. These experimental inquiry raises several questions about the many-particle effects in a mixture of Fermionic (electrons and holes) and Bosonic (excitons) system in nonequilibrium regime. This invites for a further theoretical understanding of the exciton-exciton interaction upon large photo-excitation. 

Recently, {\formulaN} class of monolayer materials, has been added to the family of 2D materials. These are synthetic materials with no known naturally occurring three-dimensional counterparts~\cite{Hong670}. In contrast to the monolayers of naturally occurring crystals, these synthetic 2D materials were designed using a bottom-up approach, and a monolayer of MoSi$_2$N$_4$ was grown using chemical vapor deposition. Additionally, several other materials of the same family of compounds, such as {\formulaAs}, $\mathrm{WSi}_{2}{\mathrm{N}}_{4}$, $\mathrm{WSi}_{2}{\mathrm{V}}_{4}$, were predicted to be dynamically stable~\cite{Hong670}. 
The {\formula} series of materials has been shown to have interesting electrical~\cite{Wu2021-APL,Cao2021-APL,Guo_2020-IOP,Bafekry2020MoSi2N4SA,keshari}, thermal~\cite{Yu_2021}, optical~\cite{Yao2021-Nanomaterials,SHG-PRB}, valley~\cite{Valley-1,Yang_valley-valley2,ai2021theoretical-valley3}, and spin dependent~\cite{Rajibul-Barun-spin} properties. However, the physics of excitons in this series of materials is relatively less explored~\cite{exciton-1-wu2021mosi2n4}. 

In this chapter [\footnote{This chapter is adapted from the following paper:
	\\
	Fluence dependent dynamics of excitons in monolayer MoSi$_2$Z$_4$ (Z = pnictogen), \href{https://doi.org/10.1088/1361-648X/acc43f}{J. Phys.: Condens. Matter \textbf{35} 235701 (2022)} by Pushpendra Yadav, Bramhachari Khamari, Bahadur Singh, K V Adarsh,
	and Amit Agarwal.}], we investigate the equilibrium and non-equilibrium optical properties of the monolayer {\formula} series of compounds (with Z $=$ N, As, or P), focusing on the excitonic effects. We start by calculating the electronic ground state, quasipartical energies, and optical absorption of the monolayer {\formula} series of compounds (with Z $=$ N, As, or P), focusing on the excitonic effects within the first principle approach as discussed in Chapter~\ref{chap2}. Within the low intensity limit (the linear and equilibrium range) we predict the the presence of strongly bound excitons with the exciton energies of the order of 1 eV. Going beyond the equilibrium properties~\cite{exciton-1-wu2021mosi2n4}, we study the fluence-dependent optical spectra and the emergence of exciton-exciton interaction in the non-equilibrium regime for MoSi$_2$N$_4$ for the first time in our work. We reveal the renormalization of the exciton binding energy (BE) of the A and B exciton peaks with increasing photo-generated carrier density in the {\formulaN} monolayer. The exciton BE shows a redshift with increasing pump fluence or photo-excited charge carrier density. This decrease in the BE arises from screening the excitonic Coulomb potential by the photo-excited charge carriers. However, on further increase in the pump fluence or exciton density, the exciton binding energies show a blueshift crossover. This establishes that 
excitons in the {\formulaN} series display an atom-like attractive and repulsive interaction depending on the inter-exciton separation. 
Furthermore, for the direct band gap {\formulaAs} and {\formulaP} monolayers, we predict the fundamental quasiparticle band gap and optical band gap with their detailed excitonic character for the first time.

\label{CMCH1}
We use the density functional theory (DFT) calculations to obtain electronic properties. We include the quasiparticle (QP) self-energy corrections using quantum many-body perturbation theory (MBPT) following the GW approximation (GWA). 
\begin{figure}[t]
		\centering
	\includegraphics[width=0.95\textwidth]{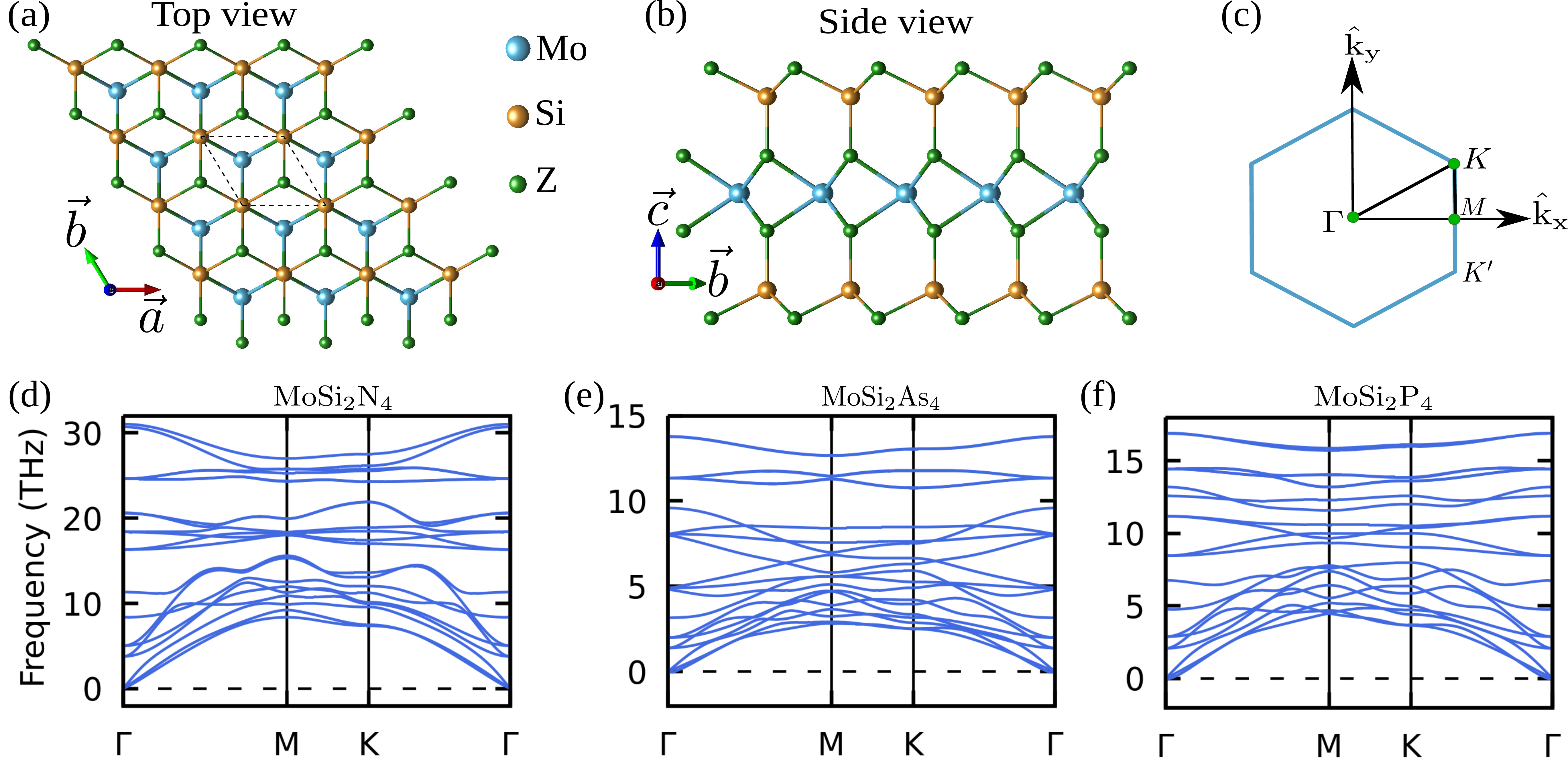}
	\caption {(a) The top and (b) side view of the monolayer crystal structure of {\formula} (Z= N, As or P). The  Z-Si-Z-Mo-Z-Si-Z arrangement of atoms along the $c$-axis can be clearly seen in panel (b). (c) The 2D hexagonal Brillouin zone (BZ). (d)-(e) The phonon dispersion of the {\formulaN}, {\formulaAs}, and {\formulaP} monolayer respectively. These three {\formula} series monolayers display no negative frequency over the entire BZ and are mechanically stable.
		\label{ch3.fig1}}
\end{figure}
To study the excitonic resonances and their impact on the optical absorption spectrum, we include electron-hole correlations on top of the QP states evaluated within the GWA, and using the Bethe-Salpeter equation (BSE). 
Our calculations show that in contrast to two excitonic peaks in the QP gap region found in MoS$_2$~\cite{optical-MoS2-PRL,spin-orbit-MoS2-PRB,amit-exciton-1,amit-exciton-2}, the {\formula} series of materials host three or more strongly bound bright excitonic peaks in the bandgap region~\cite{exciton-1-wu2021mosi2n4}. Compared to other 2D materials, the lowest energy exciton peak in all three monolayers has a very high binding energy (BE) of 1 eV or more. 
\par
Finally, to understand pump fluence's impact on the exciton BE's renormalization, we solve the time-dependent BSE (td-BSE). We find that the exciton-exciton interactions in the \formula series of monolayers mimic the attractive and repulsive behavior of atom-atom interactions ~\cite{amit-exciton-2}. We show that the exciton-exciton interactions can be modeled via a Lennard-Jones like potential. Our study establishes the {\formula} series of compounds as an exciting platform for i) exploring the physics of strongly bound exciton in 2D materials and ii) exploring optoelectronic applications in the infrared (MoSi$_2$As$_4$ and MoSi$_2$P$_4$ with an optical bandgap of $\sim 0.7$ eV) and visible regime (MoSi$_2$N$_4$ with an optical bandgap of 2.35 eV). 

\section{Crystal structure and computational methods}
\label{CSCM}
Experimentally synthesized monolayer {\formulaN} crystal structure consists of seven atomic layers in the sequence of N-Si-N-Mo-N-Si-N as shown in Figure~\ref{ch3.fig1} (a)-(b). These individual atomic layers are held together by strong covalent bonds~\cite{Yu_2021}. The {\formulaAs} and {\formulaP} monolayers with different lattice parameters share the same structure. For the bi-layers, the AB stacking is energetically the most favorable structure~\cite{Rajibul-Barun-spin, Zhong2021}. 
The monolayer {\formula} has a hexagonal lattice structure with space group $P\bar{6}m2$ (No. 187) [see Figure~\ref{ch3.fig1} (a) and (b)], which breaks the space inversion symmetry~\cite{Hong670, Valley-1}. For our $ab-initio$ calculations, we have done the lattice structure relaxation, starting from the reported lattice parameters by Hong {\it et al.} in Ref.~\cite{Hong670}. The lattice parameters reported in Ref.~\cite{Hong670} and after our relaxation are summarized in Table-\ref{ch3.table1}. To avoid spurious inter-layer interactions, we add a $27$ {\AA} vacuum along the out-of-plane axis. 
\begin{table}[h]
		\centering
	\caption{ The lattice constants for all the three {\formula} monolayer structures, before and after the relaxation. Starting from the given lattice constants in the literature, we again relaxed the structures for our calculations. There is hardly any perceptible difference ($<$ 0.06 $\%$) between the reported and our relaxed values for all three structures.}
	\begin{center}
		\begin{tabular}{c c c c}
			\hline \hline \vspace{1 mm}
			Structure \hsp & $\mathrm{a}$ ($\mathrm{\AA}$)  \cite{Hong670} \hsp & $\mathrm{a}$ ($\mathrm{\AA}$) (This work) \hsp &  $\Delta$a ($\%$)  \\
			\hline
			{\formulaN}  \hsp & 2.909 \hsp & 2.910 \hsp & 0.03 \\
			{\formulaAs} \hsp & 3.621 \hsp & 3.622 \hsp & 0.03 \\
			{\formulaP}  \hsp & 3.471 \hsp & 3.473 \hsp & 0.06 \\
			\hline
		\end{tabular}
	\end{center}
\label{ch3.table1}
\end{table}

To confirm the dynamical stability of the monolayer {\formula} structure, we have performed the phonon calculations with a $2\times 2\times 1$ supercell. For this, we have used the first-principle calculations based on density functional theory (DFT) (see Section~\ref{DFT} for details), as implemented in the Vienna \textit {ab-initio} simulation package (VASP)~\cite{PhysRevB.54.11169, PhysRevB.59.1758}. The exchange-correlation effects are treated within the generalized gradient approximation (GGA)~\cite{GGA-1, GGA-PBE}. 
An energy cutoff of 500 eV for the plane-wave basis set and tolerance of $10^{-7} $ eV is used for electronic energy minimization. After calculating the ground state charge density self-consistently, we use the Phonopy code package~\cite{phonon} to get the phonon dispersion of all three {\formula} monolayers. The phonon dispersion of the monolayer {\formulaN}, {\formulaAs}, and {\formulaP} are shown in Figure~\ref{ch3.fig1} (d)-(f). These three {\formula} series monolayers display no negative frequency over the entire Brillouin zone (BZ) and are mechanically stable.
As an additional check for the GGA band structure, we use the relaxed structure and validate our DFT calculations using the Quantum ESPRESSO (QE) package~\cite{QE} with fully relativistic norm-conserving pseudo-potential. The QE and VASP codes produce similar band structures for all three studied materials. An energy cutoff of 50 Ry for the plane-wave basis set is used after a convergence test. To perform the BZ integration, we used a $\Gamma$-centered $12\times 12 \times 1$ Monkhorst $k$ mesh~\cite{monkhorst}.
To simulate the QP energy and the optical excitation calculations, we have used quantum MBPT following the implementation in the YAMBO package~\cite{yambo20091392,yambo2019}. The ground-state Kohn-Sham~\cite{Kohn-Sham} electronic structure data from the GGA+SOC calculations are used as the initial input by the MBPT. 

For incorporating the QP self-energy corrections in the electronic structure, we have used the GWA implemented in the YAMBO package~\cite{yambo20091392,yambo2019}. To evaluate the diagonal elements of exchange self-energy, we have used $10^6$ random points in our calculation with an energy cutoff of 50 Ry after a convergence test (see Appendix~\ref{convergence}). This integral has been evaluated using a Monte Carlo scheme, known as the random integration method~\cite{Pulci1998-RIM, Rozzi2006-Coulomb-truncation}. The numerical integral has been defined within a box-type geometry of $50.01$ {\AA} on either side of the monolayer \formulaN. We have used three hundred forty bands and an energy cutoff of 14 Ry after a convergence test to calculate the polarization function within the random-phase approximation. The convergence results are shown in Figure~\ref{conv_band} [see Appendix~\ref{convergence}]. Following this, we used a plasmon-pole approximation~\cite{PPA} to calculate the inverse of the microscopic dynamic dielectric function. A self-consistent GWA on eigenvalues only (evGW) is adopted for the QP self-energy calculations (see Section~\ref{GWA} for details). We have calculated the optical spectra BSE as discussed in Section~\ref{BSE}. The linear response optical spectrum was converged with the top eight valence and lowest eight conduction bands. We have used a 12$\times$12$\times$1 k-grid for the excited state calculations, based on the fact that our simulations of the optical band gap are in good agreement with the experimentally reported value optical bandgap as discussed in Section~\ref{ch3.sec2.2}. A recent report~\cite{Sun2022}, utilizing a k-grid 24$\times$24$\times$1 in this family of monolayers, reports approximately the same optical band gap for {\formulaN}. 
To probe the impact of pump fluence on the excitons binding energies, we study the non-equilibrium carrier dynamics of the system. We perform real-time simulation using the td-BSE equation~\cite{Marini} and the non-equilibrium Green's functions (NEGFs) technique as implemented in the YAMBO  code. 
The non-equilibrium population of the photo-excited electronic states in the presence of the pump laser pulse is obtained by following the time evolution of the density matrix. The equation of motion for the density matrix is projected onto 20 bands. To account for the dissipative effects in the dynamics, a relaxation term with different scattering timescale for the population relaxation and dephasing is added to the propagation equation for the Green's function [Equation (11) of Ref.~\cite{Marini}]. We have chosen 650 $fs$ as the scattering timescale of the perturbed electronic population and 100 $fs$ for the dephasing rate. The pump field is simulated as a sinusoidal time-dependent external potential (of a specific frequency) convoluted with a Gaussian function in time. 
We have chosen the full-width at half maximum (FWHM) to be 100 $fs$. The intensity of the applied field is varied from $(6-70) \times 10^5$ ($kW/cm^2$). We have chosen a smearing of 60 meV to obtain the fluence-dependent optical absorption spectrum.
\section{Equilibrium properties}
\label{ch1.sec1}
To explore the light-matter interaction and optical absorption, we start from the ground state electronic bandstructure of the monolayer {\formula} series. We note that for accurate calculation of the exchange interaction in both the evGW and the BSE, it is essential to use the semi-core ($4s$ and $4p$) orbitals for the Mo atoms~\cite{spin-orbit-MoS2-PRB}.
\begin{figure}[t]
	\centering
	\includegraphics[width =0.95\textwidth]{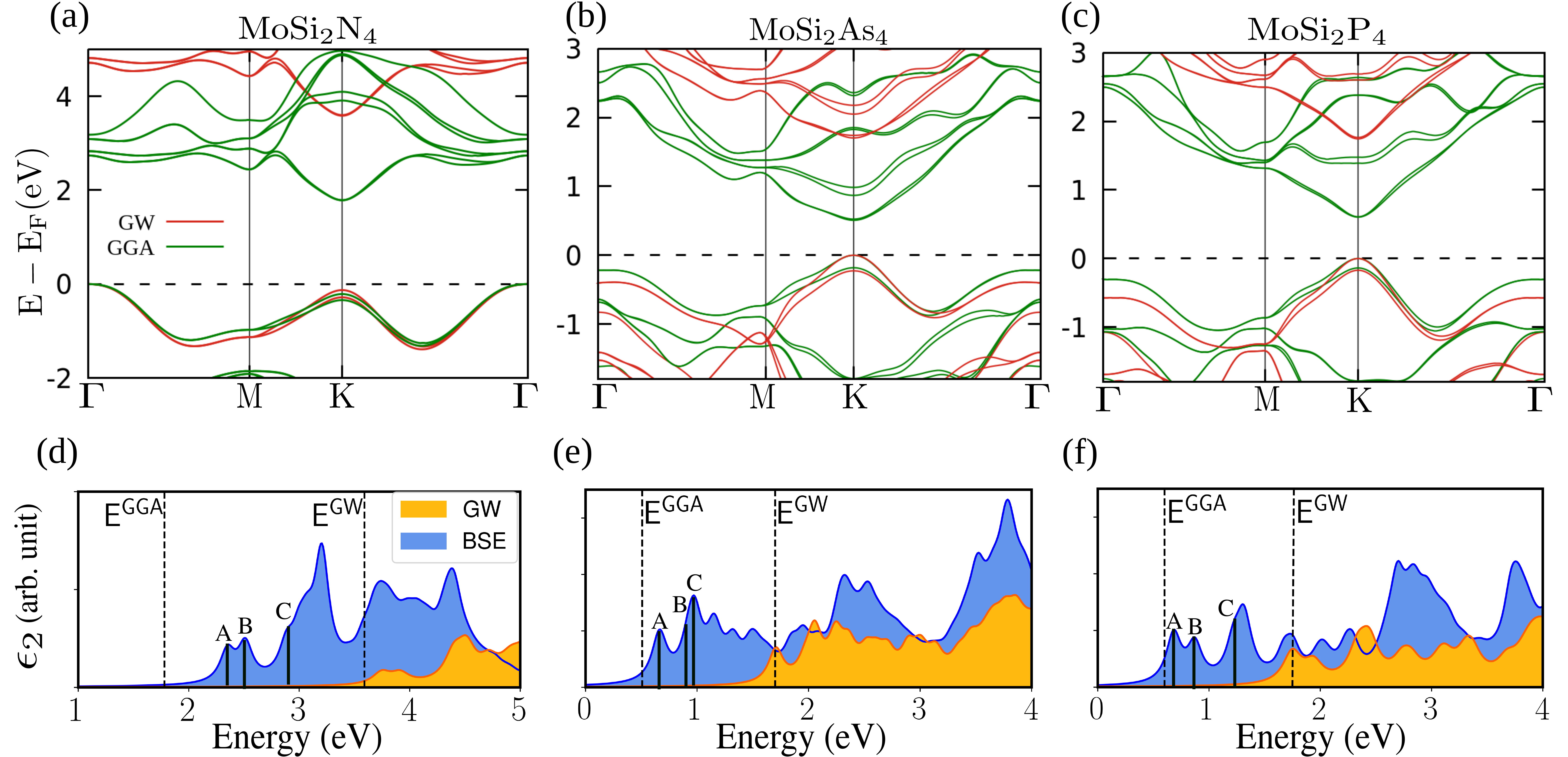}
	\caption {The electronic bandstructure calculated in the presence of spin-orbit coupling (SOC) using GGA (green color) and the QP  bandstructure calculated using evGW (red color) for monolayer (a) {\formulaN}, (b) {\formulaAs}, and (c) {\formulaP}. The corresponding optical absorption spectra, calculated using the Bethe-Salpeter equation on top of the evGW bandstructure (with SOC) for the three monolayers, are shown in (d), (e), and (f), respectively. The bandgap calculated within the GGA approximation and the evGW scheme is marked by dashed vertical lines. The BSE optical spectrum includes the two-particle electron-hole interactions (EHI) and captures the excitonic resonances, manifesting as several prominent absorption peaks below the evGW bandgap. The location of the first three prominent bright exciton peaks, similar to the A, B, and C peaks in monolayer MoS$_2$, is marked by black lines for all three synthetic monolayers. In contrast to monolayer MoS$_2$, which hosts two prominent exciton peaks in the bandgap region, the {\formula} monolayers host several (more than three) bright exciton peaks in the electronic bandgap.}
	\label{ch3.fig2}
\end{figure}

\subsection{Quasiparticle bandstructure}
\label{ch3.sec.2.1}
Within the GGA approximation, including spin-orbit coupling (SOC), we find that monolayer {\formulaN} is an indirect bandgap semiconductor with a bandgap of $1.78$ eV. In contrast,  the monolayers {\formulaAs}  and {\formulaP} are direct bandgap semiconductors with a bandgap of $0.51$ eV and $0.60$ eV at the $K$ point of the 2D hexagonal BZ. 
To improve the estimation of the electronic bandgap, we include the QP self-energy corrections on top of the GGA-based DFT calculations using the evGW method. 
\begin{table}
		\centering
	\caption{The electronic bandgap for the three {\formula} monolayers with spin-orbit coupling (SOC), calculated within the GGA approximation ($\mathrm{E_{g}^{GGA}}$), the QP G$_0$W$_0$ bandgap ($\mathrm{E_g^{G_0W_0}}$) and evGW bandgap ($\mathrm{E_g^{evGW}}$), the optical bandgap, and the direct/indirect nature of the bandgap. All three synthetic monolayers support strongly bound excitons with 1 eV or more binding energies, as listed in Table~\ref{ch3.table3}.}	
	\vspace{0.15 cm}
	\begin{tabular}{c c c c c}
		
		\hline \hline \vspace{1 mm}
		Structure \hsp & $\mathrm{E_g^{GGA}}$(eV) \hsp & $\mathrm{E_g^{G_0W_0}}$ (eV) \hsp& $\mathrm{E_g^{evGW}}$ (eV) \hsp   &Optical bandgap (eV)\hsp  \\
		\hline
		{\formulaN}  \hsp & 1.78 \hsp  & 3.30 \hsp & 3.58 \hsp & 2.35\\
		{\formulaAs} \hsp & 0.51 \hsp & 1.58 \hsp & 1.70 \hsp & 0.66 \\
		{\formulaP}  \hsp & 0.60 \hsp & 1.60 \hsp & 1.74 \hsp & 0.68 \\
		\hline \hline
	\end{tabular}
	\label{ch3.table2}
\end{table}

At the self-consistent eigenvalue GW approximation (evGW) level, our calculation indicates that the monolayer {\formulaN} has an indirect bandgap of 3.58 eV (3.30 eV), while {\formulaAs} and {\formulaP} have a direct bandgap of 1.70 eV (1.58 eV), and 1.74 eV (1.60 eV), respectively (see Table~\ref{ch3.table2}). 
We present the band structure of these three monolayers in Figure~\ref{ch3.fig2} (a)-(c). The bandgap for all three materials obtained from different approximations are summarized in Table-\ref{ch3.table2}. 

A subtle feature of the electronic band structure is that the combined effect of the absence of inversion symmetry and the strong SOC of the Mo-$d$ orbitals breaks the valence band edge degeneracy at the $K$ point. This is similar to the valence band splitting in TMD  monolayers~\cite{SOC-effect-MoS2-PRL}. 
We find that the SOC splits the valence band maxima (VBM) at the K point in the evGW (GGA) calculations by 154 (129), 226 (182), and 171 (138) meV in monolayer {\formulaN}, {\formulaAs}  and {\formulaP},  respectively.

Having obtained the electronic spectrum, we now focus on the optical absorption spectrum. Owing to the reduced screening of the Coulomb interactions in 2D materials, these synthetic monolayers have enhanced electron-hole interaction (EHI). This can give rise to several excitonic peaks below the electronic bandgap, similar to that found in MoS$_2$ monolayers. Thus, it is essential to consider the attraction between the quasi-electrons and quasi-holes by solving the BSE to obtain reasonably good optical absorption spectra~\cite{strain-exciton-Mos2-PRB}.

\subsection{Optical absorption spectra and excitons}
\label{ch3.sec2.2}
To include the Coulomb interactions between the electrons and holes in our calculations, we use the quantum MBPT (see Section~\ref{BSE} of Chapter~\ref{chap2}). 
\begin{figure}[t]
		\centering
		\includegraphics[width =0.95\textwidth]{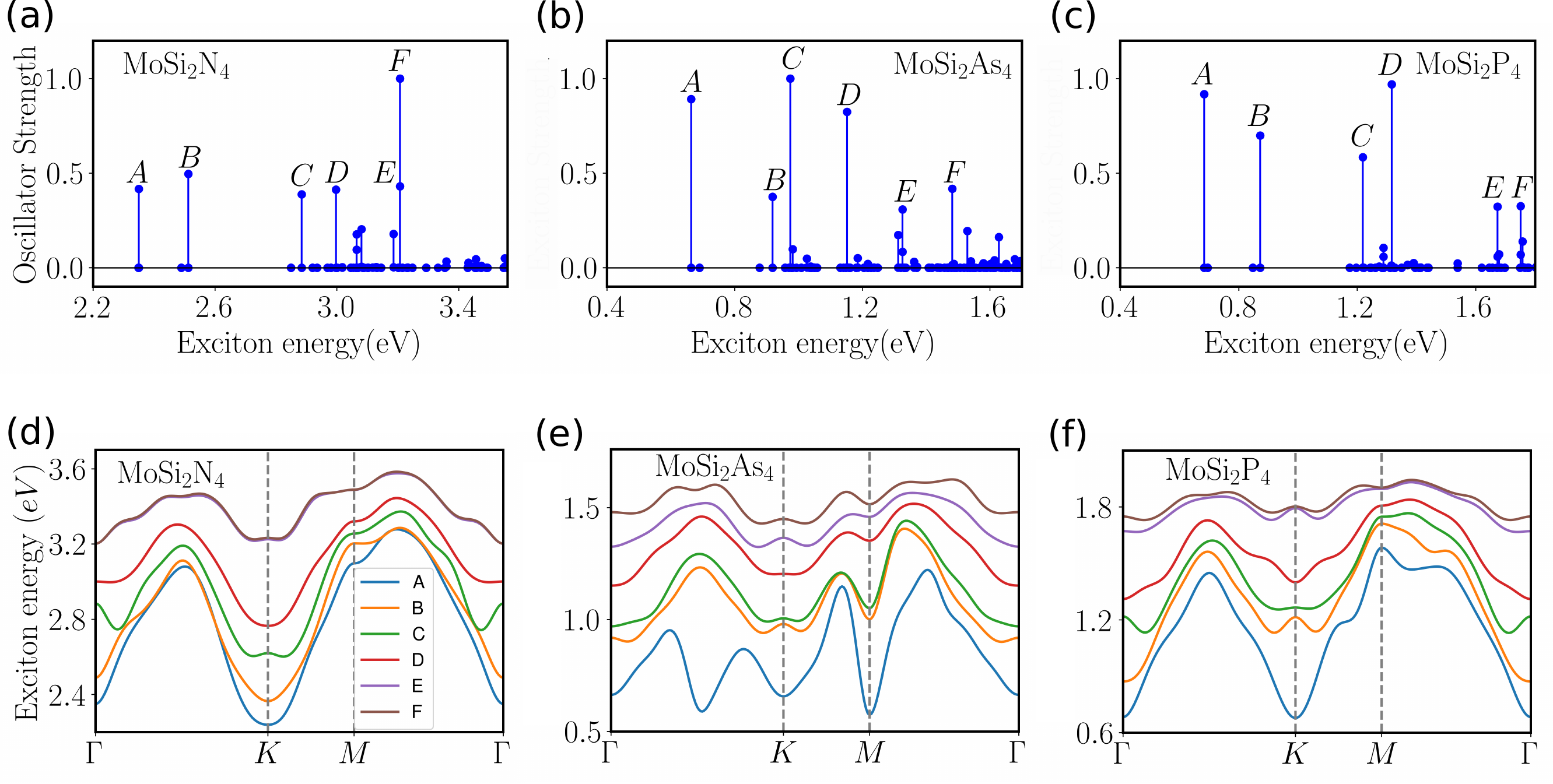}
		\caption {The exciton oscillator strength and the energy location of all the excitons in monolayers of a) \formulaN, b) \formulaAs, and c) \formulaP. Large dipole oscillator strength is generally indicative of the brightness of an exciton. Each of the three \formula~ monolayers supports at least six prominent bright excitons in the electronic bandgap region. Finite 
		center-of-mass momentum exciton bandstructure for the six bright excitons listed in Table~\ref{ch3.table3} for the monolayers (d) {\formulaN}, (e) {\formulaAs}, and (f) {\formulaP}, respectively.}  
		\label{ch3.fig3}
\end{figure}
\begin{table}
	\centering
	\caption{The exciton energy and the dipole oscillator strength of the most prominent exciton peak shown in  Figure~\ref{ch3.fig3} (a)-(c) for the monolayers of {\formulaN}, {\formulaAs}, and {\formulaP} respectively. We have normalized the oscillator strength for each monolayer with their maximum oscillator strength (0.22131, 0.20692, and 0.24570 for {\formulaN}, {\formulaAs}, and {\formulaP}), respectively. The binding energies are calculated for the $A$ and $B$ excitons. In all three monolayers, there are six excitons (with multiple degeneracies shown in Figure~\ref{ch3.fig3}) found below the minimum of the non-interacting QP bandgap and are strongly bound with a BE of $~1.0$ eV or more.}
	\vspace{0.15 cm}
	
	\begin{tabular}{c| c|c|c}
		\hline \hline
		Exciton & {\formulaN} & {\formulaAs} & {\formulaP}\\
		\hline
		\begin{tabular}{c}
			 \\ 
			A \\
			B\\ 
			C\\ 
			D\\
			E\\
			F\\
		\end{tabular}&\begin{tabular}{c c}
			E (eV) & BE (eV)\\ 
			\hline
			2.35 &  1.35\\
			2.51 &  1.35\\ 
			2.88 &  \\ 
			2.99 &  \\
			3.21 &  \\ 
			3.21 &  \\ 
		\end{tabular} & \begin{tabular}{c c}
			E (eV) & BE (eV)\\ 
			\hline
			0.66 &  1.04\\
			0.92 &  1.01\\ 
			0.97 &  \\ 
			1.15 &  \\
			1.33 &  \\ 
			1.48 &  \\ 
		\end{tabular} & \begin{tabular}{c c}
			E (eV)  & BE (eV)\\ 
			\hline
			0.68 &  1.06\\
			0.87 & 1.01\\ 
			1.22 &  \\ 
			1.32 &  \\
			1.67 &  \\ 
			1.74 &  \\ 
		\end{tabular} \\ 
		\hline \hline
	\end{tabular}
	\label{ch3.table3}
\end{table}

\begin{figure}[t]
		\centering
		\includegraphics[width =0.95\textwidth]{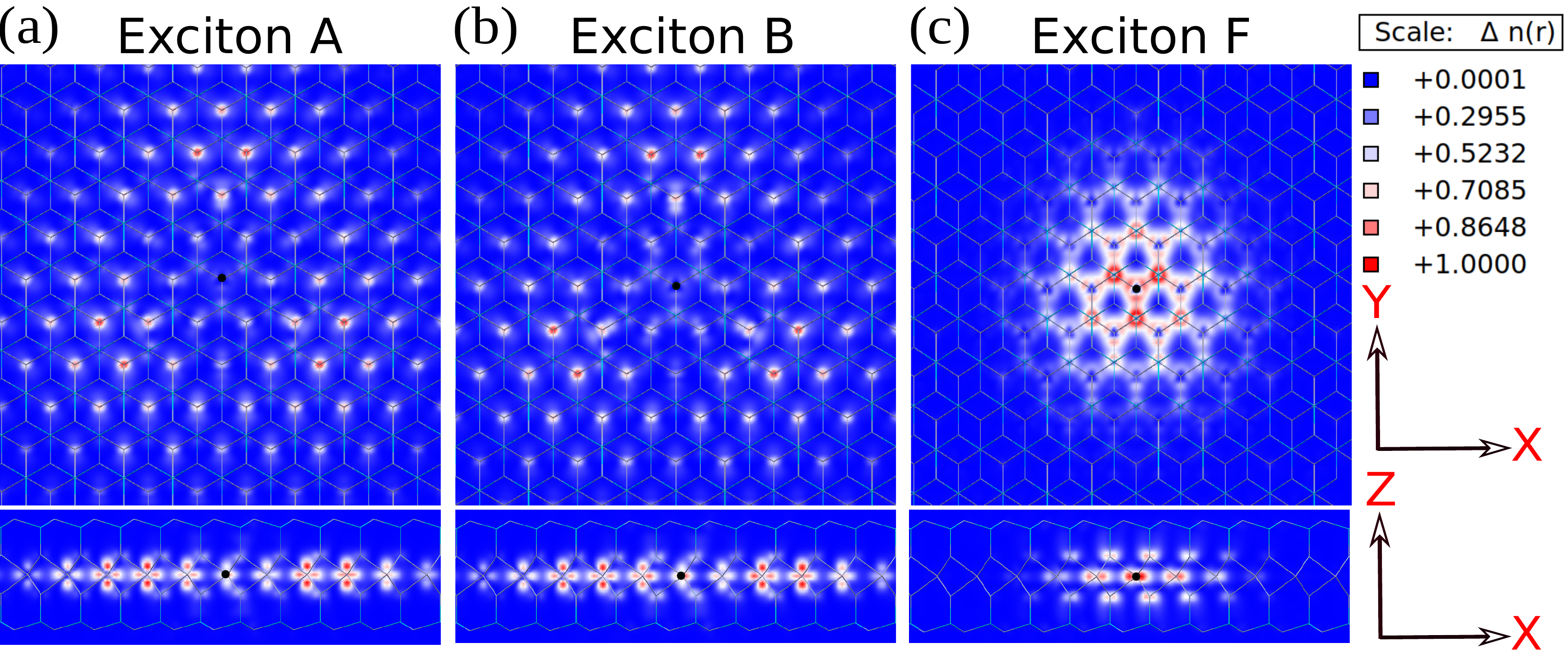}
		\caption {The real space probability density of the {\formulaN} (a) A exciton, (b) B exciton, and (c) F exciton wavefunction. In all figures, i) the top panel shows the 2D $x-y$ plane, ii) the bottom panel shows the out-of-plane ($x-z$)  view, and iii) the hole has been placed on the Mo atom (black dot) in the center. The A and the B exciton have a similar probability distribution, indicating that they have a similar origin. Analyzing the momentum-resolved oscillator strength for these excitons confirms that they arise from the direct transitions from the spin split valence bands at the $K$ and $K'$ points of the BZ.}
		\label{ch3.fig4}
	\end{figure}
	We present the calculated optical absorption spectrum for monolayer {\formulaN}, {\formulaAs}, and {\formulaP}, including the EHI (blue color) and excluding the EHI (orange color), in Figure~\ref{ch3.fig2} (d)-(f). The calculated optical bandgap (presented in Table~\ref{ch3.table2}) is in good agreement with the previously reported theoretical~\cite{exciton-1-wu2021mosi2n4,Bafekry2020MoSi2N4SA,Sun2022} and experimental~\cite{Hong670} values. It is evident from Figure~\ref{ch3.fig2} that the inclusion of excitonic effects changes the optical spectrum significantly and reduces the optical bandgap by almost 1 eV in all three {\formula} monolayers.
	The excitonic absorption spectrum also shows multiple prominent excitonic peaks, even in the QP bandgap region. We have explicitly marked the location of the first three bright exciton peaks as $A$, $B$, and $C$ exciton peaks lying in the QP bandgap region in Figure~\ref{ch3.fig2} (d)-(f). 
	
	We analyze the eigenvalues obtained from the BSE to identify all the excitonic states in the three \formula~ monolayers (see Equation~\ref{ch2.eq42}).
	The energy location of all the different exciton states and their oscillator strength are presented in Figure~\ref{ch3.fig3} (a)-(c) for all three {\formula} monolayers. We find that the two of the lowest energy excitons, the $A$ and $B$ exciton peaks, in all three monolayers are doubly degenerate. 
	 Monolayer \formula~ possesses the valley degeneracy at the $K$ and $K^\prime$ points of the hexagonal 2D BZ. We have applied a linearly polarized light to excite our system. This leads to two possible transitions at the $K/K^\prime$ points for each spin-splitted bands. For example, at the $K$ point, if the transition is happening between two same-spin bands then only left circularly polarized light will lead to a bright exciton however, the right circularly polarized light will lead to a dark exciton. Therefore, for the linearly polarized light (which can be represented by the linear combination of left- and right-circularly polarised lights) as our external perturbation, we observe both bright and dark exciton states at the same energy, but with different oscillator strengths. This argument validates the presence of both bright and dark states corresponding to different exciton states.
	Apart from the $A$ and $B$ exciton peaks, We find several bright exciton peaks in the QP bandgap region of these monolayers. Six prominent bright excitons, with the largest oscillator strength, are explicitly marked by alphabetic letters $A-F$ in Figure~\ref{ch3.fig3} (a)-(c). The properties of these six prominent bright excitons are summarized in  Table~\ref{ch3.table3}. Specifically, 
	Each of these doubly degenerate excitons consists of a bright exciton (with a large oscillator strength) and a dark exciton (with a vanishingly small oscillator strength), as shown in Figure~\ref{ch3.fig3} (a)-(c). The excitonic peaks $A$ and $B$ correspond to electron-hole pairs arising from the direct transition from the SOC split valence bands to the conduction band at the $K$ point of the BZ. These features are similar to the $A$ and $B$ exciton peaks reported in monolayer MoS$_2$~\cite{optical-MoS2-PRL,spin-orbit-MoS2-PRB}.

	The $A$ and $B$ excitons are strongly bound with a BE $\sim 1.35$ eV for monolayer {\formulaN}. In comparison, the BE of the lowest energy exciton~\cite{optical-MoS2-PRL, TMD-BE} in MoS$_2$, WS$_2$ and WSe$_2$ is 0.96 eV, 0.83 eV, and 0.79 eV, respectively.
	Interestingly, the ferromagnetic monolayer CrBr$_3$ with almost the same quasi-particle band gap (3.8 eV) as in monolayer {\formulaN} shows a large binding energy of 2.3 eV~\cite{CrBr3-PRM2022}. 
	
	\formulaAs and \formulaP also display similar excitonic peaks in their absorption spectrum shown in Figure~\ref{ch3.fig2} (d)-(f). The $A$ and $B$ exciton of the monolayer {\formulaAs} and {\formulaP} also have a  large BE of around 1 eV. 
	
	Finite-momentum excitons are optically dark but essential in hot carrier relaxation and valley dynamics~\cite{Wang2013, Exciton_band_mos2_PRB}. The finite momentum exciton bandstructure can be probed via momentum-resolved electron energy loss spectroscopy and nonresonant inelastic x-ray scattering~\cite{exciton_BS_C3N}. To calculate the exciton band structure, we solve the BSE for finite 
	center-of-mass (CoM) momentum at the irreducible momentum points of the CoM BZ. We interpolate the exciton bandstructure along the high-symmetry direction in the CoM BZ of the {\formula} monolayer. We find that several excitons (degenerate dark and bright excitons at the Gamma point or $q=0$) preserve their degeneracy at the $K$ and $M$ high-symmetry points. However, these degenerate exciton bands split at some generic $q$ points in the CoM BZ. We present the exciton dispersion of the six prominent excitons listed in Table~\ref{ch3.table3}, for monolayer {\formulaN}, {\formulaAs}, and {\formulaP} in Figure~\ref{ch3.fig3} (d)-(f). We find that the lowest energy excitons have almost equal energies at the $\Gamma$ and the $K$ points, similar to that found in MoS$_2$~\cite{Exciton_band_mos2_PRB} and hexagonal boron nitride~\cite{exciton_BS_hBN_PRB}. The lowest bright excitons ($A$) have an almost linear dispersion around the $\Gamma$ point, similar to that observed in other 2D materials~\cite{exciton_BS_PRM_Daniele} but a quadratic behavior around the $K$ point in {\formula} monolayers~\cite{exciton_BS_PRL}. 
	
	To visualize the spatial distribution of the excitons for {\formulaN}, we plot the exciton probability density for the $A$, $B$, and $F$ exciton peaks in Figure~\ref{ch3.fig4}. We fix the hole position at the top of a Mo atom at the center of each panel (black dot). 
	The exciton wavefunctions of the $A$ and $B$ excitons are almost identical, suggesting a similar origin. This is also expected because the $A$ and $B$ exciton peaks arise from direct transitions from the spin split valence bands at the $K/K'$ point of the BZ. 
	The exciton wavefunction of the $A$ and $B$ exciton peaks are spread over 4-5 unit cells in the real lattice structure, suggesting that these excitons are of Wannier–Mott type ~\cite{optical-MoS2-PRL,exciton_wf_ws2}. In contrast, the real space wavefunction of the brightest exciton peak $F$ shown in Figure~\ref{ch3.fig4} (c) has a relatively localized wavefunction. The analysis of the momentum-resolved exciton oscillator strength shows that the $F$ exciton originates from  
	several direct band transitions around the minimum QP bandgap from different k-points in the BZ~\cite{exciton-1-wu2021mosi2n4,optical-MoS2-PRL}. 
	All three excitons in Figure~\ref{ch3.fig4} show strong in-plane confinement of the exciton wavefunction, similar to that observed in other 2D materials. This 2D confinement indicates reduced dielectric screening in the out-of-plane direction 
	~\cite{hBN_excitonWF,hBN_excitonWF2}.
	Having explored the excitonic equilibrium properties of the monolayer MoSi$_2$Z$_4$ series, we now focus on the renormalization of the exciton binding energies induced by increasing the pump fluence. 

\section{Carrier dynamics and its effect on absorption spectrum}
\begin{figure}[!t]
		\begin{centering}
			\includegraphics[width=0.95\textwidth]{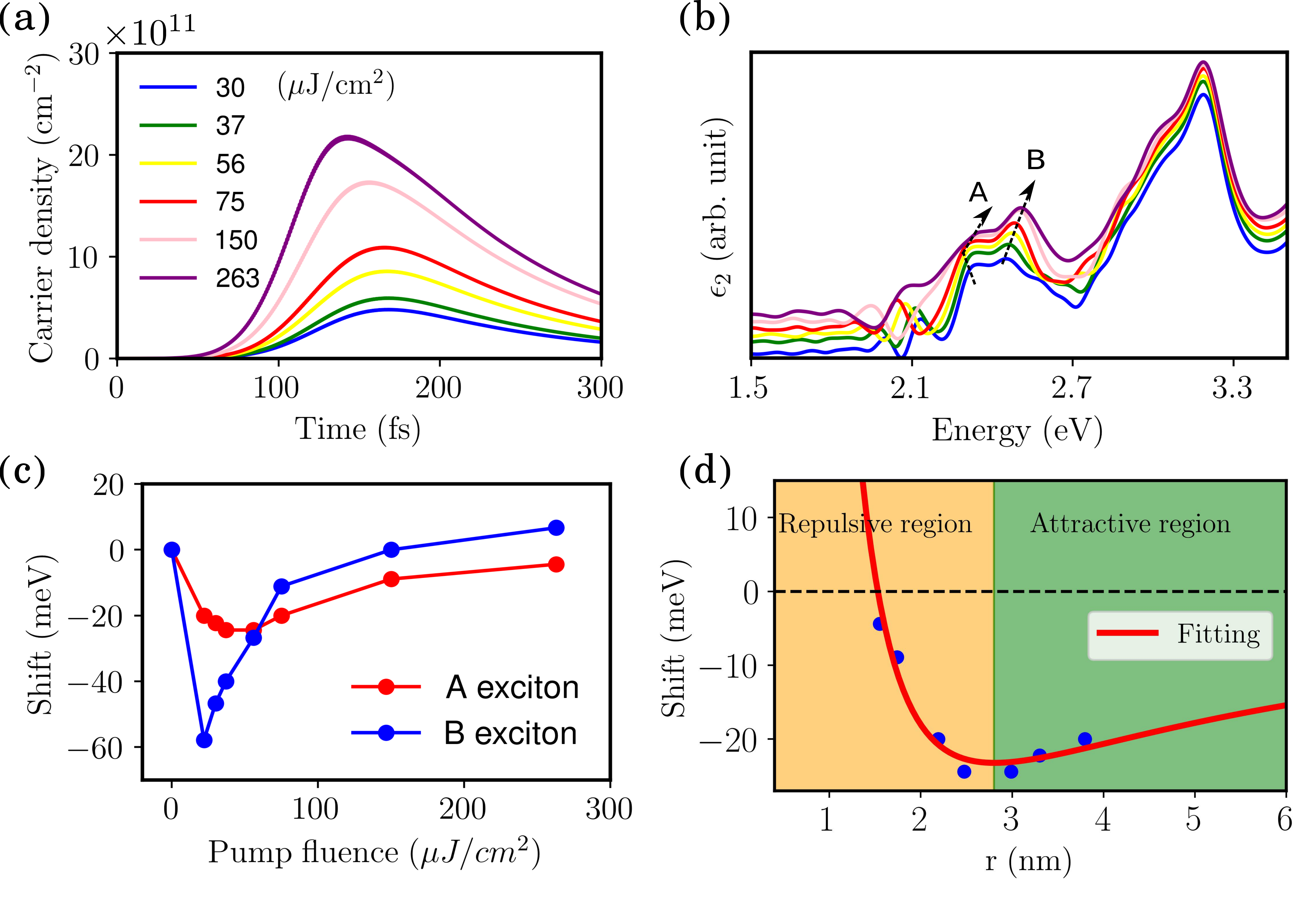}
			\caption{(a) The evolution of photo-excited carriers density (per unit surface area) in {\formulaN} with time, for different values of pump fluence ranging from 30-263 $\mu$J/cm$^2$. The incident beam is a Gaussian pump beam with a full width at half maximum of 100 $fs$. The excitation frequency of the pump beam is set at the energy of the first excitonic peak ($\omega_{pump}$= 2.35 eV) in \formulaN. (b) The pump-fluence dependent optical absorption spectrum, with the $A$ and the $B$ exciton peaks marked. The color of the curves represents a different value of the pump fluence, as indicated in (a). (c) Pump-fluence induced shift in the binding energies of the $A$ and $B$ exciton peaks. The redshift-blueshift crossover in the BE of the $A$ and the $B$ exciton with increasing pump fluence can be clearly seen. (d) The redshift-blueshift crossover in the BE highlights that the exciton-exciton interactions mimic the atom-atom interactions captured by a Lennard-Johnes like potential. Here, the blue dots capture the calculated BE shift for the $A$ exciton in c), while the red line is the fit to a Lennard-Johnes like potential of Equation~\ref{ch3.eq1}.}
	\label{ch3.fig5}	
	\end{centering}
\end{figure}

To study the non-equilibrium optical properties of monolayer {\formula} series, we use the td-BSE framework as implemented in the YAMBO code~\cite{Roth, Marini, Pedio}.  
In our real-time simulation, we apply a pump electric field with a frequency locked to the equilibrium location of the $A$ exciton peak in \formulaN. 
At resonance, we excite the system and as we increase the pump fluence, there are electron-hole pairs generated which remain in a dynamical equilibrium with the exciton gas. Furthermore, at room temperature, there are always free carriers generated due to thermal energy ($k_BT$) until an electron-hole pair interacts via a long range Coulomb interaction. Increasing the pump fluence value increases the excitation density as observed in Figure~\ref{ch3.fig5} (a). The carrier profile for different fluence values shows different exciton density {\it i.e.} with increasing fluence, excitation density increases.
More pump fluence generates more photo-excited carriers over the entire pulse duration. For a particular value of the pump fluence, the number of carriers first increases with time. Then, it decreases as expected for a pump pulse with a Gaussian temporal profile. \\

The variation of the non-equilibrium absorption spectrum for different values of the pump fluence is shown in Figure~\ref{ch3.fig5} (b).  
We present the renormalization of the BE of the $A$ and $B$ exciton peaks with increasing pump fluence in Figure~\ref{ch3.fig5} (c). The exciton BE first shows a redshift with increasing pump fluence. This decrease in the BE arises from screening the excitonic Coulomb potential induced by the photo-excited charge carriers. Initially, when electrons and holes are photoexcited in semiconductors, they exist as free charge carriers up to a specific density. This occurs till the inter-carrier separation becomes small enough for the Coulomb energy scale to become comparable to the temperature energy scale ~\cite{keldysh1986}. On increasing the density of photo-excited carriers, the Coulomb interaction effects dominate the thermal energy. Excitons start to form as the photo-excited carrier density continues to rise, 
a state is reached in which both photo-excited charge carriers and excitons coexist. During this stage, the free carriers can counteract the attractive Coulomb interactions in the system, reducing the binding energy of the excitons~\cite{keldysh1986}.\\

The density of the excitons and the free charge carriers keep increasing with further increase in the photo-excited charge carriers. With increasing exciton density, exciton-exciton interactions come into play. For large photoexcited carrier densities, the exciton-exciton repulsion starts dominating the physics. Combined with the exciton screening, this reflects in a redshift to blueshift crossover of the exciton BE with increasing pump fluence, as seen in Figure~\ref{ch3.fig5} (c). A similar redshift to blueshift crossover of the exciton BE with increasing pump fluence obtained here for MoSi$_2$N$_4$ has been recently demonstrated experimentally for MoS$_2$~\cite{amit-exciton-2}, and also for WS$_2$~\cite{Sei}. \\

The redshift to blueshift crossover in the exciton BE is similar to the atom-atom interaction with changing inter-atomic separation. To show this explicitly, we fit the exciton BE to the inter-atomic interaction potential energy specified by
\begin{equation}
	\delta E =A\left[\left(\frac{r_{0}}{r}\right)^{p}-\left(\frac{r_{0}}{r}\right)^{q}\right]~.
	\label{ch3.eq1}
\end{equation}
Here, $r$ is the separation between the interacting atoms or excitons. In Equation~\ref{ch3.eq1} $r_0$, $p$, $q$ and $A$ are treated as fitting parameters. Assuming all the photo-excited carriers form excitons, $r$ can be related to the photo-excited carrier density ($n$) via the condition, $\pi r^{2} n = 1$. 
Fitting yields, the Bohr radius $r_0 = 1.53$ nm, $p = 2.30$ and $q = 1.13$. 
Note that in WS$_2$ the Bohr radius is $r_0=2.6$ nm~\cite{Sei}, while in MoS$_2$ we have $r_0= 0.53$ nm \cite{amit-exciton-2}. 
The minima of the exciton-exciton interaction potential occurs at $r_{\rm eq} = 2.9$ $nm$.  
We show in Figure~\ref{ch3.fig5} (d) that the exciton BE fits well to the interaction potential specified by Equation~\ref{ch3.eq1}. 
\section{Summary}
\label{ch3.sec4}
We have demonstrated that the {\formula} series of monolayers hosts very strongly bound excitons with potential for optoelectronic applications in the infrared and the visible regime. By explicitly including the QP self-energy corrections in evGW based bandstructure calculations, we find that monolayer MoSi$_2$N$_4$ hosts an indirect bandgap of 3.58 eV with a comparable direct bandgap. In contrast, the monolayer MoSi$_2$As$_4$ and MoSi$_2$P$_4$ have a direct bandgap of 1.70 eV and 1.74 eV, respectively. Starting from the evGW based QP bandstructure calculations and including the two-particle electron-hole correlations, we show the existence of several (around six) strongly bound bright excitons within the QP bandgap region in this series of materials. The binding energies of the prominent $A$ and $B$ exciton peaks in all three monolayers are greater than 1 eV, and they can be as large as 1.35 eV in {\formulaN}.

Exploring beyond the equilibrium properties, we solve the td-BSE to study the fluence-dependent renormalization of the excitonic BE. We unveil a redshift-blueshift crossover of the exciton BE of the A and B exciton peaks with increasing exciton density in the {\formulaN} monolayer. At low density, the exciton BE shows a redshift. This decrease in the exciton BE  arises from screening the long-range attractive excitonic Coulomb potential induced by the photo-excited charge carriers. The blueshift in the exciton BE at high density reflects the short-range exciton-exciton repulsion. This redshift-blueshift crossover in the exciton BE with increasing exciton density indicates atom-like interactions between the excitons. We show that the density dependence of the excitonic interactions can be modeled as a Lennard-Jones-like interaction potential between atoms. Our study not only enhances our understanding of exciton dynamics in non-equilibrium scenarios but also establishes a meaningful analogy between excitons and atoms in their interparticle interactions. Consequently, these findings open avenues for further exploration, particularly in unraveling the anticipated liquid and crystalline phases of excitons in two-dimensional materials.
Our study establishes the monolayers of synthetic {\formula} series to be an exciting platform for exploring the physics of strongly bound excitons and their non-equilibrium dynamics.\\

In this chapter, we studied the optical properties of \formula~ materials and investigated exciton-exciton interaction as a function of exciton density in the nonequilibrium regime. In our study, we have ignored the impact of lattice variation and temperature effects on excitonic properties. However, the exciton energy, absorption amplitude, and linewidth in semiconductors depend on temperature~\cite{Lauten1985}. The theoretical models for solving the exciton problems at zero temperature within the frozen atom condition fail to explain the finite lifetime of excitons and do not capture the optical absorption and emission spectrum either qualitatively or quantitatively. Therefore, in the next chapter (\textit{i. e.} Chapter~\ref{chap4}), we investigate temperature-dependent electronic and excitonic properties of monolayer AlN and address the importance of electron-phonon interaction in the indirect emission process.

\section{Appendix}
\subsection{Numerical convergence test for QP energy calculation}
\label{convergence}
\begin{figure}[t]
	\centering
	\includegraphics[width =0.95\textwidth]{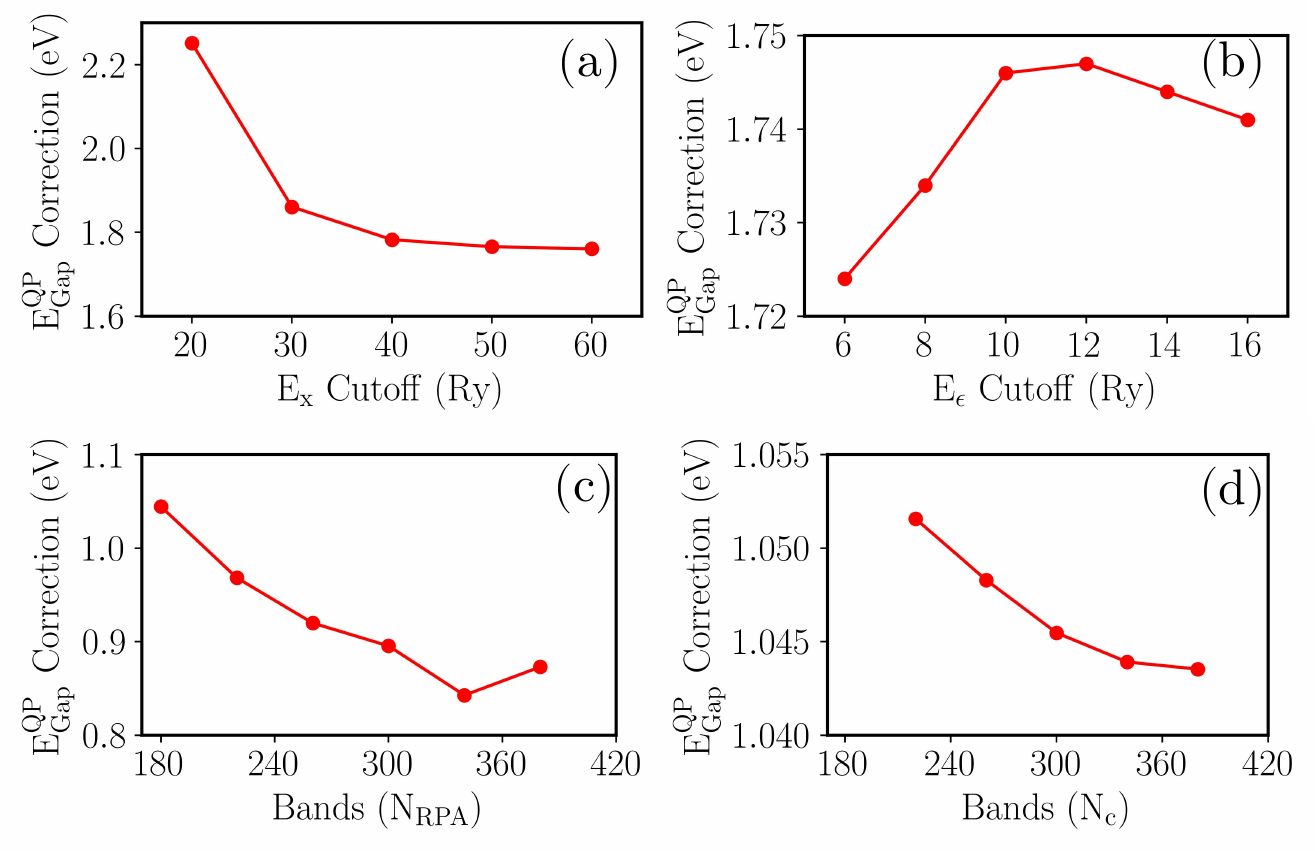}
	\caption{The convergence test for different parameters used in our first principle calculations. The QP bandgap ($\rm{E^{QP}_{Gap}}$) corrections to the GGA bandgap with (a) the energy cutoff for the exchange self-energy, (b) the energy cutoff to incorporate the local field effects in the dielectric screening for the calculation of the response function for monolayer {\formulaN}. The variation in the band gap for the last four values is less than 0.2\%. (c) The convergence checks for the number of bands entering the sum over states in the RPA response function. (d) The number of bands used for the correlation part of the self-energy ($\rm{N_c}$) for monolayer {\formulaAs}. The QP corrections to the band gap are calculated at the $K$-point of the 2D BZ.}
	\label{conv_band}
\end{figure}

The QP energy calculation is done after various numerical convergence tests and presented in Figure~\ref{conv_band}. 
For the exchange self-energy for GW calculations, an energy cutoff of 50 Ry gives reasonable results, as shown in Figure~\ref{conv_band} (a). 
The energy cutoff used to incorporate the local field effects in the dielectric screening of the response function used for GW calculations, which gives reasonable convergence, is 14 Ry, as shown in Figure~\ref{conv_band} (b). The convergence test for the number of bands used in the sum over states in the RPA response function ($\rm{N_{RPA}}$) suggests that considering three hundred forty converged bands yields reasonable results, see  Figure~\ref{conv_band} (c) for {\formulaN} and (d) for {\formulaAs}.

\begin{figure}[t]
	\centering
	\includegraphics[width =\textwidth]{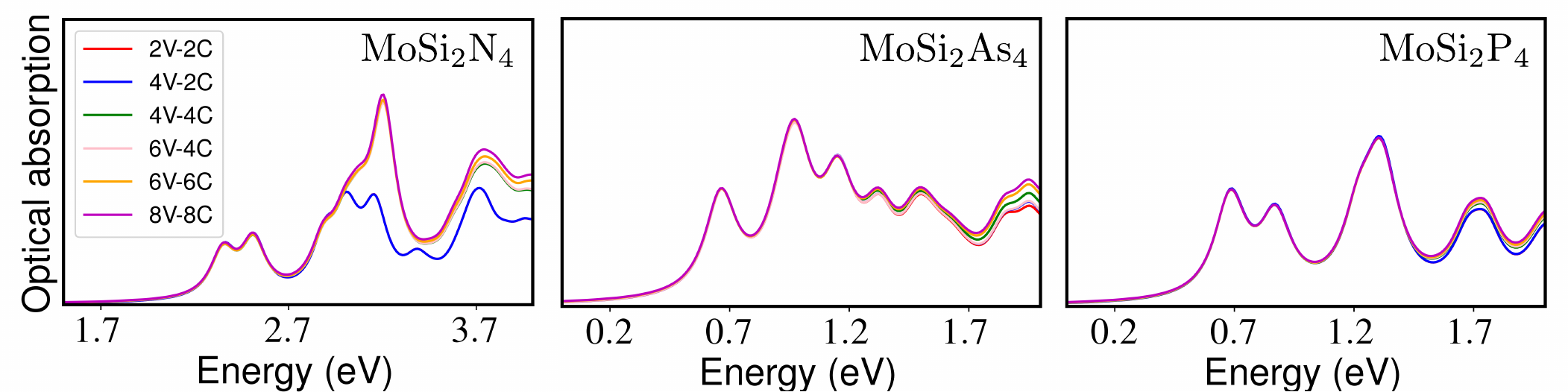}
	\caption{ The optical absorption spectrum of the MoSi2Z4 series of compounds with different sets of valence and conduction bands selected for calculating the optical excitations using the Bethe-Salpeter equation. We started the convergence test with two valence bands and two conduction bands (represented as 2V-2C) in the inset of the top panel. We calculated several other combinations, such as (4v-2C, 4V-4C, 6V-4C, 6V-6C, and 8V-8C). We find that a minimum of eight valence and eight conduction bands give well-converged absorption spectra for all three monolayers.}
	\label{conv_bse}
\end{figure}

\subsection{Numerical convergence test for BSE absorption spectrum}
Another important convergence test in our work is done before calculating the BSE optical absorption. We select a set of valence and conduction bands (required for optical excitations) and calculate the excitonic energies and corresponding optical absorption for the {\formula} monolayers to construct the BSE matrix kernel. The optical absorption spectrum of the {\formula} series of compounds with different sets of valence and conduction bands selected for calculating the optical excitations using the
Bethe-Salpeter equation. Starting with two valence bands and two
conduction bands (represented as 2V-2C) in the inset of the top panel of Figure~\ref{conv_bse} and did the calculation with several other sets of combinations such as (4V-2C, 4V-4C, 6V-4C, 6V-6C, and 8V-8C). We found that eight valence and eight conduction bands (8V-8C) are sufficient for generating a numerically acceptable optical absorption spectrum in these monolayers.

%% file: chap4.tex
\chapter{Exciton-phonon coupling and indirect photon emission in monolayer aluminum nitride (AlN)}
\label{chap4}
\pagestyle{fancy}
The remarkable optical characteristics of two-dimensional (2D) materials have fueled extensive research into the light-matter interaction and exciton physics~\cite{Mounet2018,Thygesen2017,Wang2018}. This includes ongoing efforts to comprehend the fundamental phenomenon of photoluminescence (PL) in atomically thin materials, where unexpectedly intense light emissions are observed. Optical experiments on 2D WSe$_2$ and WS$_2$~\cite{Brem2020} reveal robust emissions at lower temperatures despite being assisted by phonons. Contrary to initial assumptions about the direct gap in bulk hexagonal boron nitride (h-BN)~\cite{Watanabe2004}, subsequent studies identified its indirect gap, leading to phonon-assisted emission in the deep ultraviolet (UV) region~\cite{Cassabois2016}. Interestingly, experiments on h-BN reveal distinct excitonic absorption and emission spectrum. While the former involves a direct exciton transition, the latter requires phonons through an indirect process. However, a quantitatively accurate description of such phonon-assisted phenomena is still under development~\cite{Marini2008,Monserrat2014, Shen2020,Cannucciaarxive}. Additionally, 2D materials exhibiting such potential optical properties in the ultraviolet regime need a comprehensive theoretical framework.

Recently, there has been a growing interest in understanding the electro-optic characteristics of III-V materials in their 2D crystal phases~\cite{Zhuang13, Michael2018, Jiongyao2017}. Among these materials, aluminum nitride (AlN), exhibiting a graphene-like 2D structure~\cite{Bourret1998, Qi2021}, has shown potential for nanoscale optoelectronics. This interest is primarily motivated by its remarkable physical properties, which include high chemical stability, thermal conductivity, and exceptional mechanical characteristics~\cite{AlN-exciton1,AlN-optical-1,AlN-optical-2, Luo2020, Strite1992, Jain2000}. 2D AlN has been successfully synthesized as high-purity films using conventional growth techniques~\cite{Jain2000, Zhang2007, Yan2010, Alizadeh2017}. These group-III nitrides, when confined to their 2D limit, have found extensive applications in various optoelectronic devices~\cite{Dahal2009, Jiang2017-GaN, Jiang2016, Zhao2017, VanHove1997, Morko1998, Nakamura2009}. They are crucial in solid-state optical applications, solar energy, and electronic power~\cite{Wang2021-JMCC}.
Nevertheless, the optical excitations in 2D AlN, particularly in the presence of lattice vibrations, still need to be adequately understood. 
\begin{figure}[t] 
	\centering
	\includegraphics[width=0.95\linewidth]{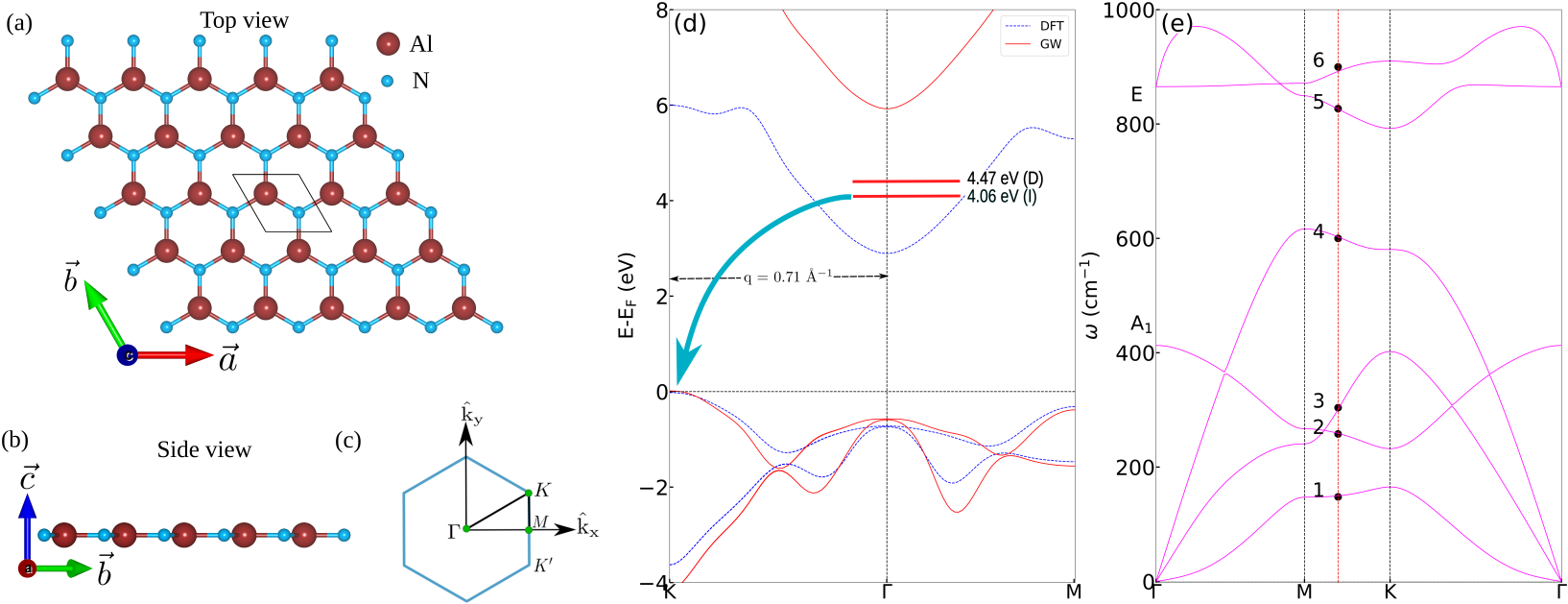}
	\caption{(a) A top view of the lattice structure of monolayer AlN, (b) a side view showing an atomically thin structure, (c) the corresponding hexagonal Brillouin zone. (d) DFT (blue) and excited state (red) electronic band structure of monolayer AlN. The blue (curved) arrow shows a schematic indirect electron-hole recombination assisted by phonons. The direct (D) and indirect (I) optical gaps are mentioned as levels. The top of the valence band in both cases is set to zero. (e) Corresponding disperson showing degenerate LO-TO mode at $\Gamma$. The black dots show the phonon modes assisting the PL emission process at finite transferred momentum, $\textbf{q}$ = 0.71 \AA$^{-1}$.} 
	\label{ch4.fig1}
\end{figure}

In this chapter [\footnote{This chapter is adapted from the following work:
	\\
	Phonon-assisted photoluminescence and exciton recombination in monolayer aluminum nitride \href{https://iopscience.iop.org/article/10.1088/2053-1583/adb8be}{2D Mater. \textbf{12} 025022 (2025)}, by Pushpendra Yadav, Amit Agarwal, and Sitangshu Bhattacharya.}], we investigate temperature-dependent optical excitations and photoluminescence emission in monolayer AlN. PL emission in AlN is supported by phonon assistance.

\section{Structural properties and electronic ground state}
\label{sm_sec1}
\begin{figure}[t]
	\centering
	\includegraphics[width=0.95\columnwidth]{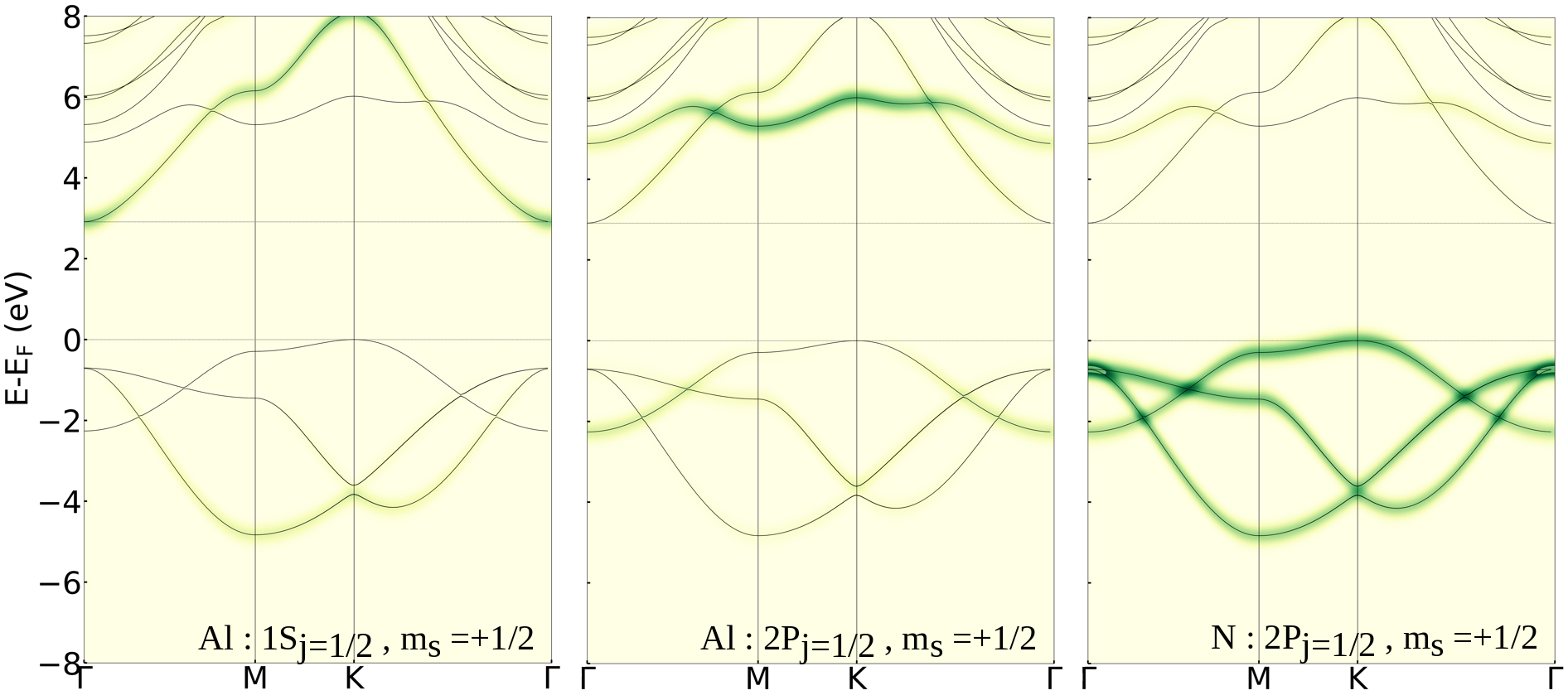}
	\caption{The orbital resolved electronic band structure at the GGA level. (a) The lowest energy unoccupied bands are majorly populated with the $s$-orbital (m$_s$ = +1/2) and $p$-orbital (m$_s$ = +1/2) of the Al atom. In contrast, the highest occupied (valence) bands in (b) are dictated by the $p$-orbitals (m$_s$ = +1/2) of the N atom.} 
	\label{ch4.fig2}
\end{figure}

Monolayer AlN is a group-III nitride wide-bandgap semiconductor. We report the electronic, mechanical, and dynamical stability of the planer AlN monolayer structure from the DFT calculations. For electronic stability, we find that the enthalpy of the individual Al and N atoms are -4.38 Ry and -20.17 Ry per atom, respectively, while the enthalpy of the AlN compound is -25.51 Ry. This leads to the formation energy of -0.96 Ry/atom for the monolayer AlN. The electronic structure relaxation confirms that the monolayer AlN structure is a stable planer hexagonal structure [see Figure~\ref{ch4.fig1} (a) and (b)], and the corresponding Brillouin zone (BZ) in Figure~\ref{ch4.fig1} (a). The crystal structure belongs to the $C_{3v}$ point group symmetry with lattice constant a = 3.12 \AA, consistent with the previous reports~\cite{ZhuangAlN, AlN-exciton1}. To confirm the mechanical and dynamical stability of the planer structure, we have computed the phonon energies within density function perturbation theory (DFPT); see Section~\ref{methods-elph} for computational details. The absence of negative phonon frequency in the phonon dispersion [see Figure~\ref{ch4.fig1} (e)] confirms the stability of the planer structure in the monolayer limit. Further, from the electronic structure calculation within density functional theory (DFT), we find that the monolayer AlN has an indirect and wide band gap of 2.91 eV along the $\Gamma-K$ path and a direct band gap of 3.61 eV at the $\Gamma$ point of the BZ within the generalized gradient approximation (GGA), see Section~\ref{methods-GS} for computational details. Further, to get more insight into the orbital character of the bands participating in the photoexcited phenomena, we have calculated the orbital projected bands as shown in Figure~\ref{ch4.fig2}. It is evident from the orbital projections on the band structure that the valence bands are mainly occupied by the $P$ orbital with $s_z = 1/2$ of the nitrogen atom. In contrast, the conduction bands are filled with the $P$ and $S$ orbitals with majority spin $s_z = 1/2$ of the aluminum atom. Specifically, the effect of spin-orbit splitting (SOS) can be seen in the valence band at $\Gamma$. Due to the three-fold $C_{3v}~(3m)$ crystalline symmetry and time-reversal symmetry, the top valence state at $\Gamma$ is a Kramer's doublet ($\Gamma_{5v}, $$\Gamma_{6v}$) signifying a heavy-hole. The next lower state at ($\Gamma_{4v}$) is the light hole with a spin-orbit split energy energy of 16 meV.
\section{Excitonic resonances and optical absorption}
All ground and excited state quasi-particle energy calculations are conducted using the open-source DFT code Quantum Espresso~\cite{QE} and the many-body perturbation theory (MBPT) code Yambo~\cite{yambo2019}, respectively. In the single-particle framework, quasi-particle energies of electrons and holes are computed within the GWA, starting from Kohn-Sham energy eigenvalues. Subsequently, the band structure is evaluated along high-symmetry BZ directions, as illustrated in Figure~\ref{ch4.fig1} (a). In 2D AlN, we observe large quasi-particle direct (6.30 eV) and indirect (5.73 eV) gaps. The direct gap is at $\Gamma$, while the indirect gap lies between $\mathrm{K}$ and $\Gamma$. The influence of spin-orbit splitting (SOS) is evident in the valence band at $\Gamma$. 
Table~\ref{ch4.table1} summarizes the 2D AlN gaps calculated at DFT and the GWA level.	
\begin{figure}[h]
	\centering
	\includegraphics[width=0.7\columnwidth]{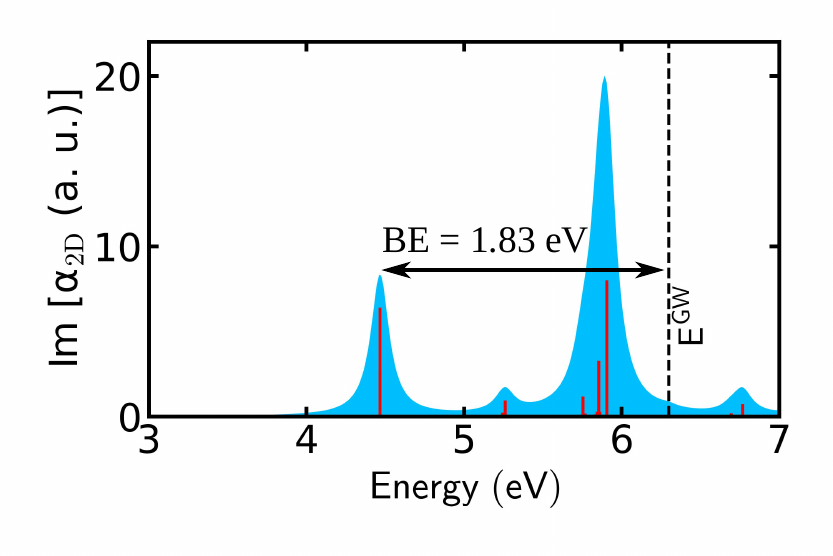}
	\caption{The optical absorption calculated within GW-BSE theory. The first exciton peak appears at 4.47 eV ($E^{OP}$), below the single-particle band gap ($E^{GW}$ = 6.30), leading to the binding energy of 1.83 eV (= $E^{GW} - E^{OP}$) for the corresponding exciton.}
	\label{ch4.fig3}
\end{figure}
\begin{table}[t]
	\centering
	\caption{The electronic band gap is calculated with different theoretical methods: the GGA, GGA (0 K/ 300K), i.e., GGA  at zero and three-hundred Kelvin, and GW methods. The indirect band gap values at the $\Gamma-K$ point and the direct band gap values at the $\Gamma$ point are provided. The units for the band gap values are in electron volts (eV).}
	\setlength{\tabcolsep}{10pt} 
	\begin{tabular}{c c c c c}
		\hline
		\hline
		Bandgap Type  & GGA & GGA at 0 K & GGA at 300 K  & GW \\
		& (eV) &  (eV) & (eV)  &  (eV)\\
		
		\hline 
		Indirect ($\Gamma-K$) &   2.91    & 2.73  &  2.91  & 5.73 \\
		Direct  ($\Gamma$)  &   3.61    & 3.14  &  3.44  &  6.30  \\
		\hline
		\hline
	\end{tabular}     
	\label{ch4.table1}
\end{table}
We use the BSE discussed in Section~\ref{BSE} of Chapter~\ref{chap2} to calculate the excitonic spectrum.
\begin{figure}[t]
	\begin{centering}
	\includegraphics[width=0.8\linewidth]{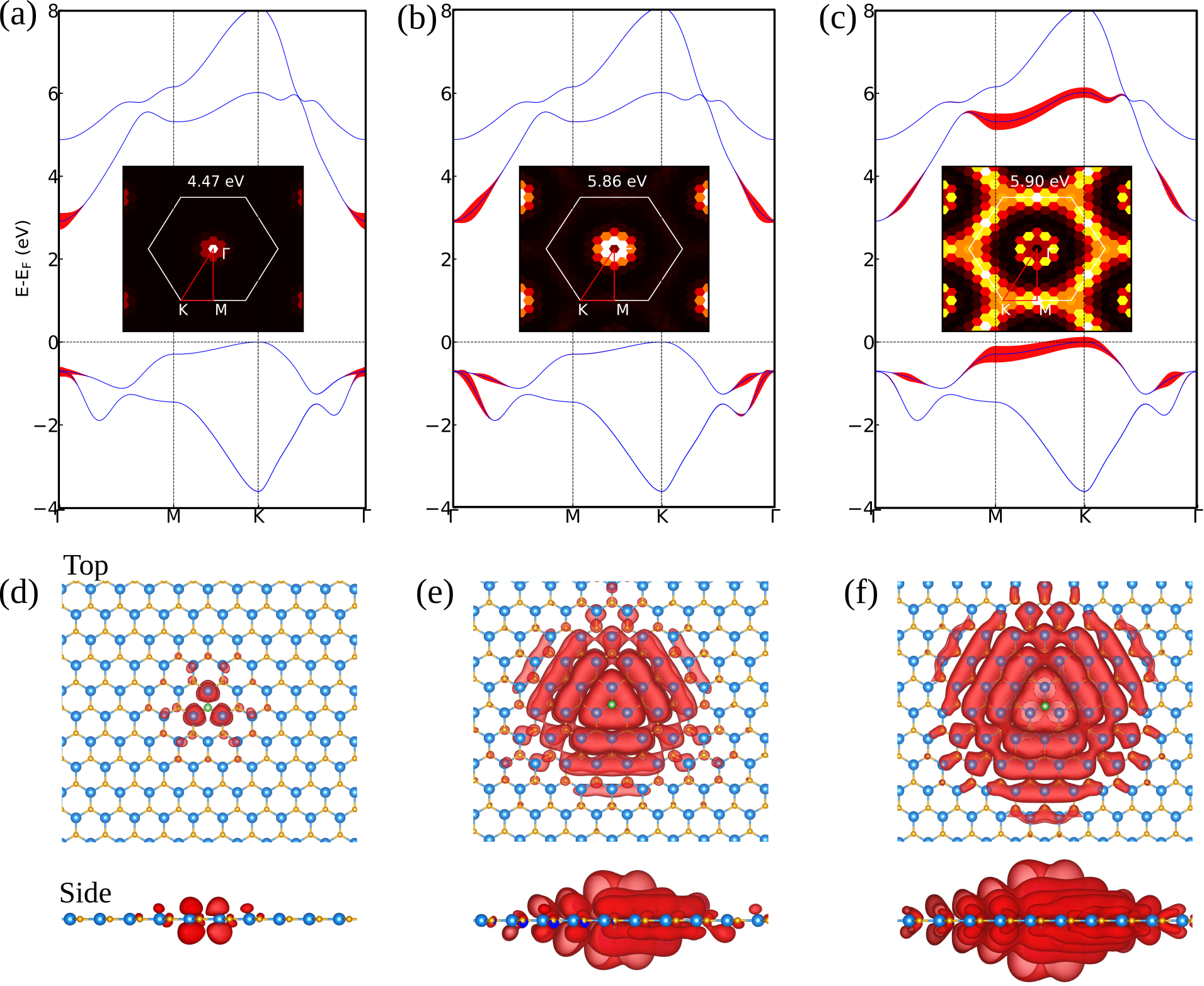}
	\caption{Excitonic structure of prominent bright excitonic peaks: (a)-(c) Exciton weight mapped on electronic band structure, showing the first three bright excitons at 4.47 $eV$, 5.86 $eV$, and 5.90, respectively $eV$. Inset: Exciton wavefunction in momentum space shown in the BZ. The corresponding exciton wavefunction in real space for the first bright exciton is shown in (a)-(c). The magnitude of the electronic wavefunction (in red) centered around the hole (green ball) shows the spread over the lattice. The first bright exciton shown in (a) indicates the localized nature of the lowest energy bright exciton. (e) and (f) exciton wavefunctions in real space for the second and third bright excitons are relatively spread over several unit cells.}
	\label{ch4.fig4}
		\end{centering}
\end{figure}
The optical absorbance, given as,  $A=1-\textrm{exp}\left(-\frac{\Im\varepsilon\left(\omega\right)\mathcal{E_{X}^{\mathrm{s}}}L}{\hbar c}\right)$, is shown in Figure~\ref{ch4.fig3}. The absorbance is calculated using the 2D microscopic polarizability $\alpha_{2D}\left(\omega\right)$, and the $\alpha_{2D}\left(\omega\right)$ is obtained from the imaginary dielectric function, $\Im [\varepsilon(\omega)]=\frac{4\pi\alpha_{2D}\left(\omega\right)}{L}$. Here, $L$ is the monolayer thickness, and $c$ is the speed of light. The absorbance spectrum shows major bound excitonic peaks in the 4.0-6.0 eV range. The first exciton peak, at 4.47 eV, corresponds to the fundamental optical band gap. The calculated exciton binding energy ($\mathcal{E}_b$ for the lowest energy exciton is 1.83 eV, indicating a strong electron-hole correlation. In the absence of this correlation [in the independent particle (IP) limit], the absorbance blue-shifts at the GW direct gap and significantly reduces in magnitude. Two more prominent peaks appear at 5.86 (shoulder peak) and 5.90 eV below the quasi-particle direct band gap. These three peaks consist of doubly degenerate pairs, each composed of one optically bright (large exciton oscillator strength) and one optically dark (small exciton oscillator strength) exciton. In Figure~\ref{ch4.fig4} (a)-(c), we illustrate the optical transitions of these three bound excitons mapped on the electronic dispersion and in reciprocal lattice space. The lowest exciton is localized around $\Gamma$, while the next two excitons are more delocalized, signifying a less relatively bounded nature. The orbital spin occupancy of the electronic transitions is detailed in Figure~\ref{ch4.fig2}. The lowest conduction band near $\Gamma$ is mainly occupied by the 1S$_{j=+1/2}$ orbital of Al atom with spin m$_s$=+1/2. Whereas, along the $\mathrm{M - K}$ path of the BZ, the occupancy is primarily by the 2P$_{j=+1/2}$ orbital of N atom with spin m$_s$=+1/2. The top three valence bands are mostly occupied by the 2P$_{j=+1/2}$ orbital of the N atom with spin m$_s$=+1/2. The lowest exciton involves a transition from 2P$_{j=+1/2}$ orbital of the N atom with spin m$_s$=+1/2 to 1S$_{j=+1/2}$ orbital of the Al atom with spin m$_s$=+1/2. Similarly, the exciton at 5.90 eV has a transition from 2P$_{j=+1/2}$ orbital of the N atom with spin m$_s$=+1/2 to 1S$_{j=+1/2}$ orbital of the Al atom with spin m$_s$=+1/2 near $\Gamma$, while 2P$_{j=+1/2}$ with of the to Al atom spin m$_s$=+1/2 near $\mathrm{M - K}$. The maximum strength is observed for the exciton at 5.90 eV (deep UV region), with light absorbance as large as 30$\%$. In contrast, the absorbance corresponding to the lowest exciton at 4.26 eV (near UV) is approximately 10$\%$. Analyzing exciton wavefunctions in real space provides insight into the structure of these three prominent bright excitonic peaks. 

The excitonic wavefunction in real space can be expressed as $\psi^{\mathrm{s}}\left(\textbf{r}_{e},\textbf{r}_{h}\right)=\sum_{cv\textbf{k}}A_{vc\textbf{k}}^{\mathrm{s}}\phi_{v\textbf{k}}^{*}\left(\textbf{r}_{h}\right)\phi_{c\textbf{k}}\left(\textbf{r}_{e}\right)$, where the electron and hole coordinates are ${\textbf{r}_e}$ and ${\textbf{r}_h}$, respectively. This formulation allows the unveiling of the corresponding excitonic character, distinguishing between Mott-Wannier or Frenkel-type excitons. The probability distribution $\left|\psi^{\mathrm{s}}\left(\textbf{r}_{e},\textbf{r}_{h}\right)\right|^{2}$ is computed and projected onto the 2D AlN lattice space, providing the probability of finding electron localization around a hole fixed in the lattice ($\textbf{r}_{e}=\mathbf{r}$ and $\textbf{r}_{h}=\mathbf{\overline{r}}_{h}$) and shown in Figure~\ref{ch4.fig4}. Here, we fix a hole on the top of the N-atom ($\sim$1\AA) within the cell, as most valence band maxima consist of nitrogen orbitals. The exciton wavefunctions in real space shown in Figure~\ref{ch4.fig4} (d)-(f) illustrate the symmetry of the excitonic wavefunction for the aforementioned bound excitons. The 2D AlN possesses a C$_{3v}$(3m) point group symmetry with three irreducible representations: A$_1$, A$_2$, and $E$. The former two are mono-dimensional with even and odd $\sigma_{v}$ reflection symmetries, respectively, while the last irreducible representation $E$ is two-dimensional. In this case, all three bound states, being double-degenerate, are assigned $E$-type symmetry. The exciton wavefunctions shown in Figure~\ref{ch4.fig4} show no significant charge densities observed in the hole site. The lattice trigonal excitonic character is well-preserved, reflecting an overall C$_{3v}$(3m) symmetry. Moreover, all three bound excitons maintain a double-degenerate $E$-type symmetry. The lowest bound exciton is tightly bound and spread out only to the nearest sites, suggesting a Frenkel exciton character. In contrast, the next two excitons exhibit characteristics similar to Mott-Wannier excitons. A small peak near 5.26 eV is also observed, composed of a bright and dark degenerate pair. Excitonic textures along the out-of-plane direction are also presented. All excitons proliferate along the aperiodic direction, increasing their volumes as we approach higher energies. This behavior becomes non-trivial for excitonic interactions between bilayers or hetero-structures, leading to additional Davydov splittings in the absorption spectrum~\cite{Paleari2018, Henrique2017}.
\begin{figure}[t]
	\centering 
	\includegraphics[width=0.95\columnwidth]{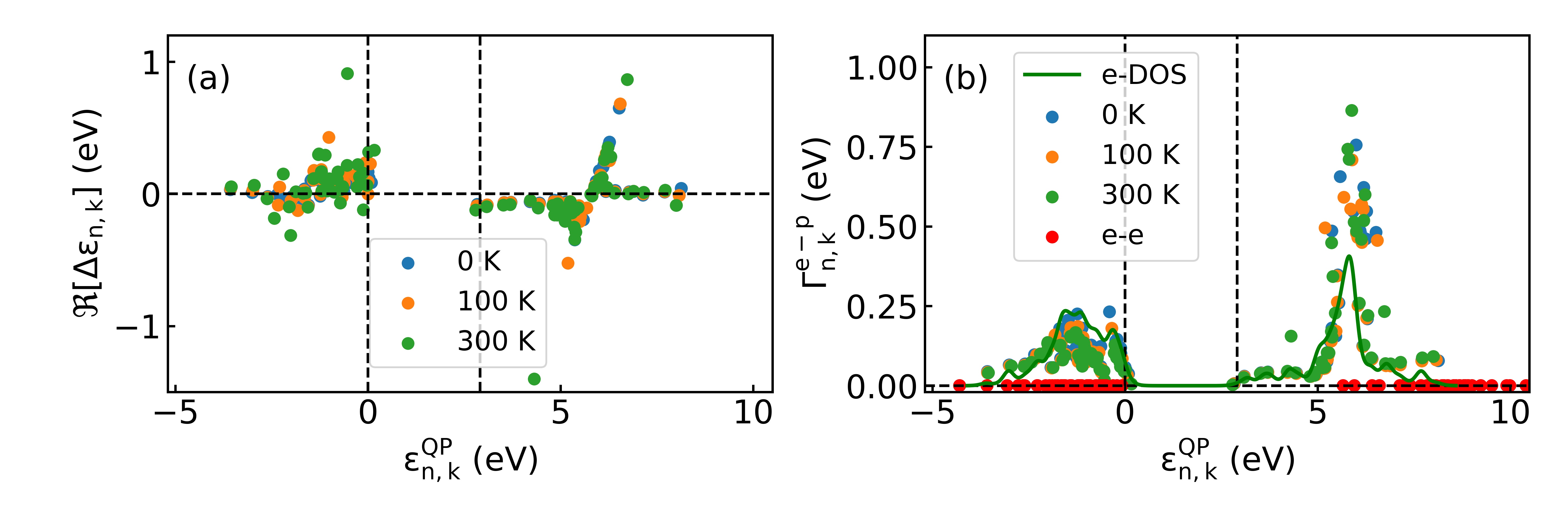}
	\caption{(a) Effect of Fan and DW correction to the electron-phonon self energies at various temperatures exhibiting the strength of electron-phonon couplings. (b) Electronic linewidth in the presence of electron-phonon interaction at various temperatures. The corresponding phonon-assisted lifetime is $\hbar\left[2\Im\sum_{n\textbf{k}}^{\mathrm{Fan}}\left(\omega,T\right)\right]^{-1}$. The solid curve is the electronic DOS without lattice vibrations (scaled to fit), whereas the red dotted symbols are the GW corrected linewidths (scaled). This clearly highlights the role of electron-phonon interaction in dictating the exciton lifetime.}  
	\label{ch4.fig5}
\end{figure}

\begin{figure}[t]
	\centering
	\includegraphics[width=0.95\linewidth]{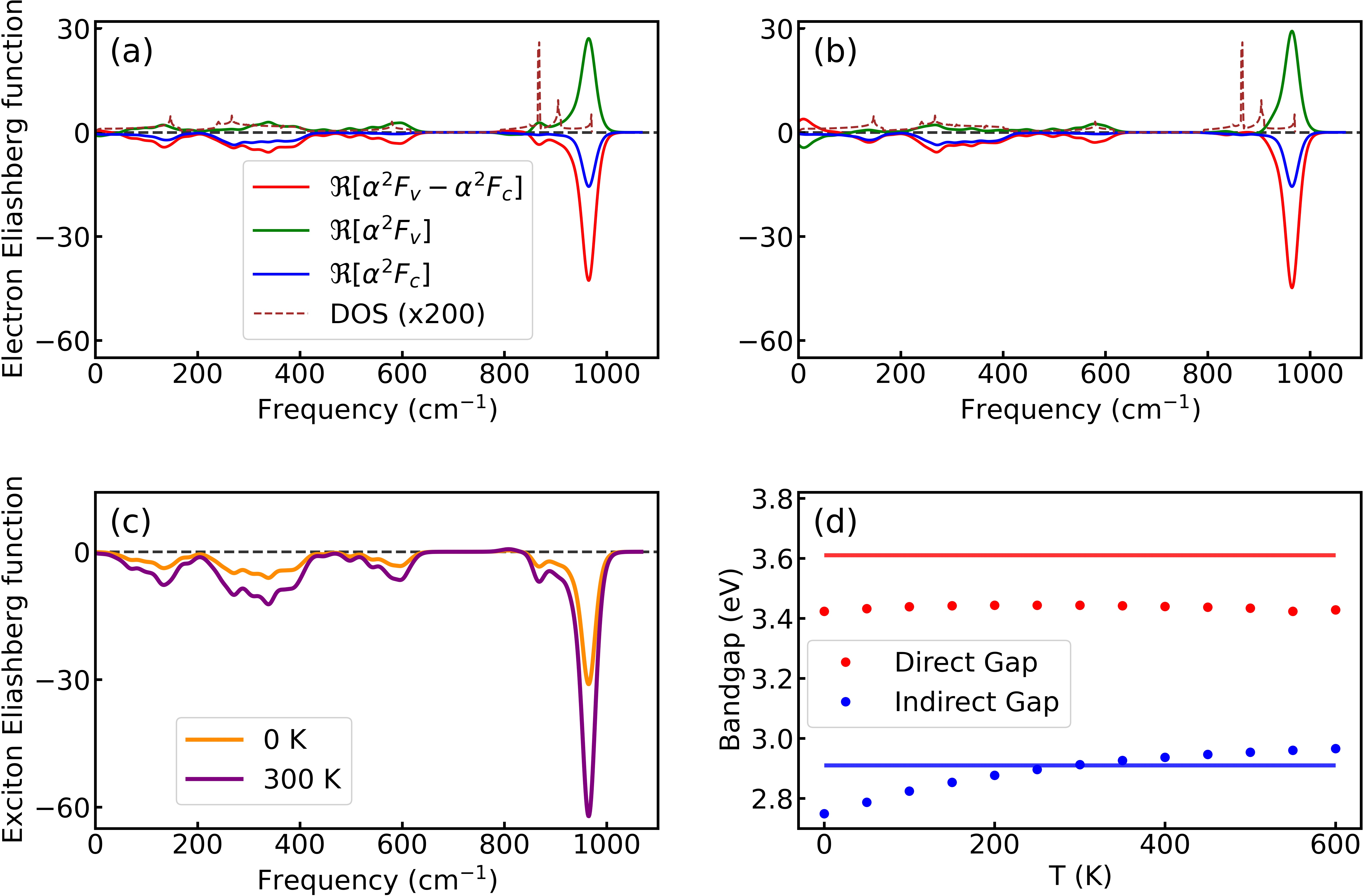}
	\caption{The electron Eliashberg functions for the valence band maxima (green), conduction band minima (blue), and their difference (red) corresponding to the band edges for the (a) direct and (b) indirect band gap. (c) The exciton Eliashberg function at 0 K and 300 K. (d) The temperature-dependent electronic direct bandgap (red dotted curve) and indirect bandgap (blue dotted curve) for the frozen atom condition direct band gap (red horizontal line) and indirect band gap (blue horizontal line). The direct bandgap appears to be temperature-independent in the 0-600 K range, while the indirect bandgap shows a sub-linear increment with temperature.} 
	\label{ch4.fig6}
\end{figure}

	\section{Electron-phonon interaction}
	\noindent Absorption and PL emission experiments are typically conducted at finite temperatures ~\cite{Cassabois2016, Park2018}. Therefore it is essential to comprehend the role of electron-phonon interaction from a microscopic theoretical perspective~\cite{Allen1976, Allen1983,Fan1950, Zollner1992,Cannucciaarxive}. 

The electron-phonon matrix elements are evaluated directly from the density functional perturbation theory (DFPT) through the first order Taylor's expansion of the change in self-consistent charge density with respect to atomic displacements. The Fan self-energy~\cite{Fan1950, Mahan2014} $\left(\textstyle \sum_{n\textbf{k}}^{Fan}\left(\omega,T\right)\right)$ of an electronic state $\left|n,\textbf{k}\right\rangle$ is then obtained by converting the convoluted Matsubara frequency sum of electron and phonon Green’s propagator into the retarded self-energy.
By introducing the electron-phonon interaction, the quasiparticle energy eigenvalues become a complex quantity which depends on both temperature and frequency. The real portion modifies the bare energy in the pole, while the imaginary portion adds to a finite polaronic lifetime.

Since, Fan self-energy includes only first order term in the electron-phonon Hamiltonian, therefore, a more accurate description of the electron-phonon interaction is provided by the Debye-Waller (DW) correction to the self-energies $\left(\textstyle \sum_{n\textbf{k}}^{DW}\left(T\right)\right)$~\cite{Zollner1992, Cannucciaarxive}. The DW self-energy calculation requires the expensive second-order electron-phonon coupling matrix elements and are not provided by the DFPT. Instead they are obtained by virtue of crystal translation invariance property within the MBPT approach. These elements are then transformed similar to the first order product-like Fan terms. A Sternheimer's approach~\cite{Sternheimer1954} may then be used to compute the sum-over empty states, which however the MBPT code Yambo does not use.
	Nevertheless, the DW self-energy is dependent only on temperature through Bose occupation factor $n_{B}$, therefore, it is completely real.\\	
\begin{figure}[t]
		\centering
		\includegraphics[width=0.95\linewidth]{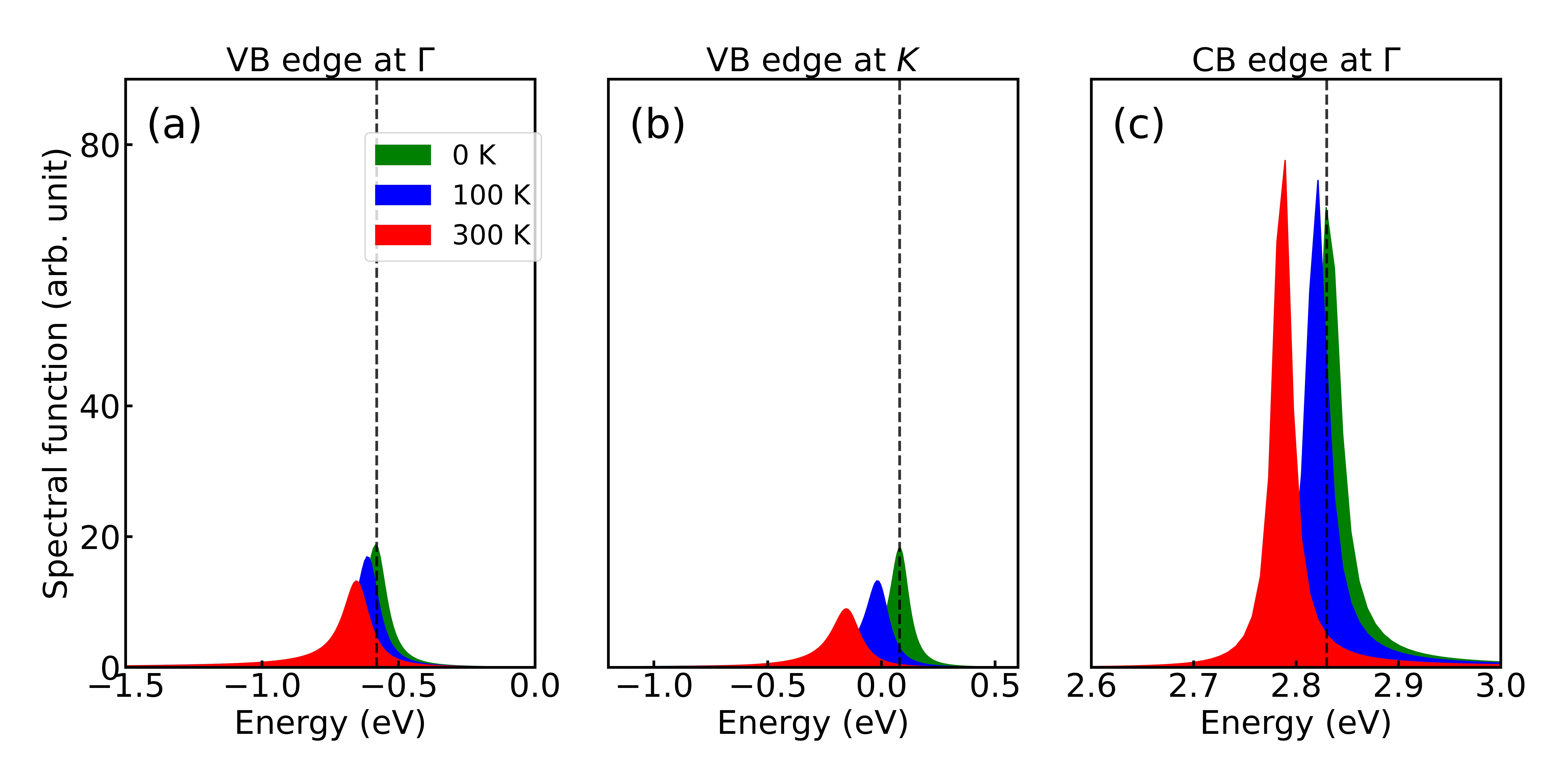}
		\caption{variation of the spectral functions with temperature for the direct valence edge at $\Gamma$ in (a) and indirect valence band edge in (b), and conduction state in (c). The vertical dash lines represent the band edges without el-ph couplings.}
		\label{ch4.fig7}
\end{figure}
In Figure~\ref{ch4.fig5} we have shown the impact of electron-phonon correlations compared to the corresponding electron-electron interactions. The panel (a) demonstrate the corrections from both the Fan and DW self-energies and the panel (b) exhibits the phonon mediated electronic linewidths at different temperature, which is also proportional to the electronic density-of-states (e-DOS). 
The corrections to the quasiparticle energies are also significantly larger to the corresponding linewidths found by the GW corrections (in few tens of meV). In fact several studies in past taking into account the impact of lattice vibrations on electronic band-gaps in variety of semiconductors such as diamond-like~\cite{Lautenschlager1985, Lautenschlager1987, Zollner1992}, GaAs~\cite{Lautenschlager1987}, bulk h-BN~\cite{Marini2008}, and atomically thin MoS$_2$~\cite{Chiara2020, Chan23}, WSe$_2$\cite{HimaniWSe2}, h-BN~\cite{Mishra2019}, NP~\cite{Kolos2021}, etc. have been reported. The impact of zero-point motion (ZPM) effect was found to significantly renormalize the electronic and fundamental optical band gaps, exciton binding energies and linewidths, and absorption and PL emission lines.
	
After obtaining the self-energies, we finally evaluate the pole of the single-particle Green’s function $G_{n,\textbf{k}} (\omega, T)$. The real part of poles of the $G_{n,\textbf{k}} (\omega, T)$ is the renormalized the QP energy, and the imaginary part defines the QP lifetime. The imaginary part of interacting electron's Green propagator gives us the spectral functions, $A_{n,\textbf{k}} (\omega, T) = -(1/\pi)\Im |G_{n,\textbf{k}} (\omega, T)|$. The spectral functions calculated for monolayer AlN for the bottom conduction (at $\Gamma$) and top valence (at $\Gamma$ and \textbf{K}) states are shown in Figure~\ref{ch4.fig7}. The spectral functions are Lorentzian-type with FWHM representing the phonon mediated electronic linewidths. The strength of the electron-phonon correlation can be understood from the shape of the Lorentzian plots. An asymmetric and dwarf spectral function indicates a strong correlation, whereas a sharp and symmetric shape corresponds to a weak correlation. We find that the quasiparticle renormalized weight factor ~\cite{Mahan2014, cannuccia2011effect} $Z$ (bounded between 0 and 1), in this case to be about 83$\%$ and 80$\%$ for the valence states at $\Gamma$ and \textbf{K}, whereas 92$\%$ for the conduction state at $\Gamma$ at 0 K respectively. The weight $Z$ carries the finger-print of quasiparticle interaction. A $Z$ value close to 1 would imply a true quasi-particle state, while a strongly correlated state has $Z$ near 0. At a higher temperature of 300 K, these reduces to about 70$\%$, 55$\%$ and 81$\%$ respectively, showing an increased impact of electron-phonon interaction. We also see as temperature increases, the Lorentzians becomes small in magnitude and broadens. The renormalized electronic energies appears to shift from the respective DFT energy edges,even at T=0. This shift in the band edge even at T = 0 is the known as the zero point renormalization (ZPR). This ZPR is eventually responsible for the experimentally observed red and blue shifts observed in dielectric function~\cite{Lautenschlager1987, Zollner1992, Marini2008}. 

Our calculations shows that at 0 K, the valence states at $\Gamma$ and \textbf{K} blue-shifts by about 111 meV and 85 meV. In contrast, the conduction state at $\Gamma$ red-shifts by about 79 meV from their respective DFT band-edges. This results in a net gap-shrinkage of about 42 and 6 meV at the direct ($\Gamma$) and indirect ($\Gamma$-\textbf{K}) points. Figure \ref{ch4.fig6}~(d) shows this dependency of the electronic direct and indirect band-gaps with other non-zero temperatures. We observe that the direct gap in the lower temperature has a small incremental tendency, which beyond 200 K starts decreasing. In case of indirect gap, we observe a slight increasing behaviour. These shifts are also captured from the corresponding spectral functions shown in Figure~\ref{ch4.fig7}.

To understand the temperature dependent electronic band gap, due to the electron-phonon interaction, we calculate the electronic Eliashberg function using polaronic quasiparticle energy ${\Delta\varepsilon_{n\textbf{k}}}$. This complex quantity can be written as~\cite{Marini2008} 
	\begin{equation}\label{eliash}
		{g^{2}F_{n\textbf{k}}\left(\omega\right)=\sum_{\textbf{q}\nu}\frac{\partial\varepsilon_{n\textbf{k}}}{\partial n_{B}\left(\omega_{\textbf{q}\nu}\right)}\delta\left(\omega-\omega_{\textbf{q}\nu}\right)~,
		}
	\end{equation}
	and is summed over all phonon modes $\textbf{q}\nu$. Here $g$ represents the matrix elements while $F$ is the e-DOS. Figures~\ref{ch4.fig6} (a) and (b) demonstrate how the real portion of $g^{2}F_{n\textbf{k}}\left(\omega\right)$ evaluated for valence state at $\Gamma$ and $\mathrm{K}$ and at conduction state $\Gamma$, changes with phonon frequencies. In conventional semiconductors, $\Re \left[g^{2}F_{v\textbf{k}}\left(\omega\right)\right]$ is a positive quantity at valence band, whereas $\Re \left[g^{2}F_{c\textbf{k}}\left(\omega\right)\right]$ is a negative quantity at conduction band. Thus their difference $\Re \left[g^{2}F_{c\textbf{k}}\left(\omega\right)-g^{2}F_{v\textbf{k}}\left(\omega\right)\right]$ becomes negative. This leads to bandgap shrinkage, as evident in Figure~\ref{ch4.fig6} (d). Most of the area under this envelope is negative in magnitude, thus justifying the decrease of the gap with an increase in temperature. Additionally, we also see the limiting behavior of this difference. The area diminishes at both lower frequency and Debye frequency, signifying the validity of the crystal translation invariance property. In case of indirect band gap between $\Gamma - \mathrm{K}$, the increment tendency could be due to positive portion of $\Re \left[g^{2}F_{c\textbf{k}}\left(\omega\right)-g^{2}F_{v\textbf{k}}\left(\omega\right)\right]$ (small in magnitude near 800 cm$^{-1}$ frequency). Since the electronic Eliashberg is proportional to the phonon-DOS $\delta\left(\omega-\omega_{\textbf{q}\nu}\right)$, we can now identify the phonon modes responsible for band gap widening. We observe that the electrons and holes are mainly coupled with the optical branches of phonons around 965 cm$^{-1}$ frequency. The small shoulder peak near 800 cm$^{-1}$ in Figure~\ref{ch4.fig6} (a) is due to the in-plane longitudinal and transverse optical $\mathrm{E}$-type modes. In Figure~\ref{ch4.fig6} (b), we observe that there are frequencies where $\Re \left[g^{2}F_{c\textbf{k}}\left(\omega\right)-g^{2}F_{v\textbf{k}}\left(\omega\right)\right]$ becomes positive, such as near 900 cm$^{-1}$ and near lower acoustic modes 10 cm$^{-1}$. These are the major modes responsible for the incremental of the indirect band gap at lower and higher temperatures. Nevertheless, these outcomes are derived from a cascaded computational process and are quite delicate, with corrections in the order of a few milli-electron volts. Therefore, there are inevitable instances of spurious computational inaccuracies that we cannot ignore. It is only up to the carefully conducted experiments add more insight to this behavior, along with the effect of substrate, ambient conditions, etc. 

After understanding the impact of electron-phonon interaction on single-particle energies, we estimate the optical absorption and emission in monolayer AlN within the interacting electron-hole (excitonic) framework.

\section{Temperature-dependent absorbance}
The role of lattice vibrations in investigating the optical absorption and excitonic characteristics in \textit{ab - initio} calculations has been very successful~\cite{Marini2008}. Including the polaronic corrections transforms the hermitian Bethe-Salpeter Hamiltonian (in the absence of lattice vibrations) to become non-Hermitian as a concequence of the excitonic energy eigenvalues become complex quantities. In such cases, the imaginary part of the exciton eigenvalues is proportional to the excitonic non-radiative lifetime, as defined in Equation~\ref{ch2.eq64}.

\begin{table}
	\footnotesize
	\centering
	\caption{The excitonic properties of monolayer AlN are calculated using the GW+BSE method, with and without exciton-phonon interaction. The properties include the optical bandgap (lowest energy bright exciton), binding energy, dipole oscillator strength of the corresponding doubly degenerate excitons, and excitonic linewidth at zero and three-hundred Kelvin temperatures.}
	\setlength{\tabcolsep}{10pt} 
	\begin{tabular}{c c c }
		\hline
		\hline
		Theoretical     & Optical band gap & Excitonic linewidth \\
		method & (eV) & (meV) \\
		\hline
		GW+BSE  &   4.47      & \\ 
		GW+BSE (0 K)  &   4.27   &  92.69  \\
		GW+BSE (300 K)  &   4.25   &  87.97  \\
		\hline
		\hline
	\end{tabular}     
	\label{ch4.table2}
\end{table}

\begin{figure}[t]
	\begin{centering}
	\includegraphics[width=0.8\linewidth]{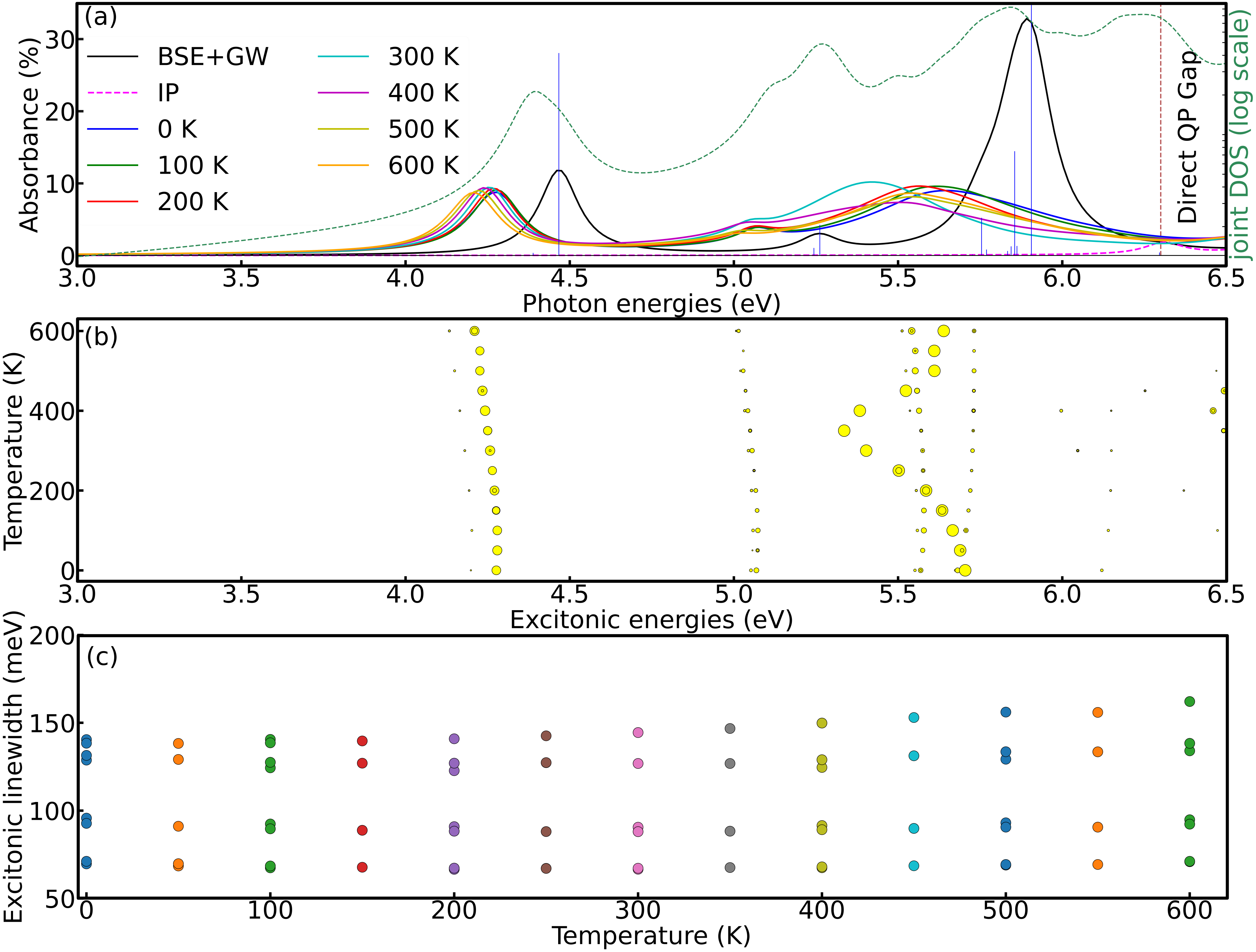}
	\caption{The impact of lattice vibrations on the absorption, excitonic structure, and excitonic non-radiative linewidths. (a) The absorbance in the frozen atomic configurations and at different temperatures. Due to the lattice vibrations, the el-ph interactions modify the excitonic resonances significantly at 0 K, whereas the changes at finite temperatures compared to 0 K are small. (b) The exciton dipole oscillator strength, and (c) the excitonic linewidth as a function of temperature. The dipole oscillator strength for the first exciton (4.47 $eV$ ) increases with temperature, but the second bright exciton (5.26 $eV$) becomes a dark exciton after 200 $K$. Interestingly, the third bright exciton (at 5.90 $eV$) shows a redshift-blueshift crossover across the different temperature ranges.} 
		\label{ch4.fig8}
\end{centering}

\end{figure}

The temperature-dependent exciton eigenvalues are calculated using Equation~\ref{ch2.eq59}. The real part of the energy eigenvalue, $\Re\Delta \mathcal{E}_{\mathcal{X}}^{\mathrm{s}}\left(T\right)$, [see Equation~\ref{ch2.eq60}], represents the coherent interaction between the exciton and the phonon where the excitonic amplitudes are dependent on their interaction strength. Finally, to obtain the temperature-dependent optical absorbance, we calculate the temperature-dependent dielectric function given by Equation~\ref{ch2.eq64}. The optical absorbance calculated at different temperatures, along with the absorbance and the joint density of states within the frozen atom condition, is shown in Figure~\ref{ch4.fig8} (a). The optical spectrum is significantly red-shifted with respect to the frozen atom GW+BSE solution. Because of the '$\frac{1}{2}$' factor in $\Re\Delta \mathcal{E}_{\mathcal{X}}^{\mathrm{s}}\left(T\right)$, the fundamental peak at 0 K is red-shifted by an about of 200 meV thus signifying an intrinsic spatial uncertainty of atoms. 
In addition, the presence of an imaginary part of the excitonic energies does not need any more \textit{ad-hoc} damping. As a result, the full width at half maximum (FWHM) is their phonon-assisted broadening. The red-shifting of the absorbance can then be understood from the sign of $g^{2}F{_s}\left(\omega, T\right)$. The net cancellation between the coherent and incoherent terms can result in lowering or increasing of $\Re\Delta \mathcal{E}_{\mathcal{X}}^{\mathrm{s}}\left(T\right)$. 

This excitonic Eliashberg function is shown in Figure~\ref{ch4.fig6} (c), where we observe that the incoherent interaction with the phonons makes this function mostly negative at all frequencies. The major peak is near 900 cm$^{-1}$, corresponding to the optical phonon branches. As temperature increases, the phonon number density increases, with the most intense interaction coming from branches near 900 cm$^{-1}$. These are the modes where the torsional motion of atoms appears. This leads to the stretching and compressing of the 2D sheet, consequently relaxing and reinforcing the excitonic interactions with the lattice. These combined effects make the excitons red-shifts in energy with temperature.

Figure~\ref{ch4.fig8} (b) shows the exciton energies weighted by their oscillator strengths as a function of temperature. The effect of temperature is more evident in the states of the excitons. Corresponding to the respective temperature-dependent absorbance peak locations, we observe a slow red-shifting of the fundamental bound exciton near 4.25 eV. We consider only those excitons whose strengths are more than 10$\%$ of the maximum and are considered bright. This reveals that, throughout the entire temperature range, the fundamental exciton always remains bright (as indicated by the size of the circle). Similarly, the other bound exciton near 5.06 eV undergoes a slower red-shifting with comparatively less strength throughout the range and therefore falls on the darker side. A relatively large pace in the red-shifting of the exciton energies with temperature is observed for the group of bound excitons in the energy range 5.25-5.75 eV. Interestingly, there are a few excitons (near 5.06 and 5.25 eV) that share their oscillator strengths when other excitons are in close proximity. It is clear that their strength increases when two excitons are close by, whereas when they separate, they both lose their strengths and become dark. This exchange of optical strengths is entirely due to the contribution from coherent interactions, similar to Si and h-BN~\cite{Marini2008}.

The excitonic linewidths of first few bound excitons with oscillator strengths more than 10$\%$ are shown in Figure~\ref{ch4.fig8} (c). At 0 and 600 K, the linewidths are between 60-150 meV and 80-170 meV, respectively. Since, the interactions are mainly from the optical branches, we use the empirical equation for the non-radiative (NR) linewidth\cite{Selig2016},
\begin{equation}
\gamma_{NR}\left(T\right)=\gamma_{0}+\gamma_{\textrm{op}}\left[\textrm{exp}\left(\frac{\Lambda}{k_{B}T}\right)-1\right]^{-1}~.	
\end{equation}
Here, $\gamma_{0}$ denotes the 0 K residual linewidth, $\gamma_{\textrm{OP}}$ represents the interaction strength between excitons and optical phonons and $\Lambda$ denotes the phonon frequency of the coupled phonon mode. Our result indicate a $\gamma_{0}$ = 89.21 meV, and $\gamma_{op}$ = 13.09 meV. This is a strong exciton-optical phonon interaction strength
Due to a small acoustic interaction, these linewidths increase significantly with temperature after about 350 K. This makes $\gamma_{NR}$(300 K)$\sim\gamma_{0}$.

In order to compare the phonon-assisted non-radiative lifetime with the intrinsic radiative recombination lifetime $\tau_{R}$, we follow Palummo \textit{et al.}~\cite{Palummo2015}
\begin{equation}
 \ensuremath{\gamma_{R}^{\mathrm{s}}=\left[\tau_{R}^{\mathrm{s}}\right]^{-1}=\frac{8\pi e^{2}\mathcal{E}^{\mathrm{s}}}{\hbar^{2}c}\frac{\mu_{\mathrm{s}}^{2}}{A_{uc}}}~.
\end{equation}
 Here, $c$ represents the speed of light, $A_{uc}$ is the primitive cell area, and $\mu_{\mathrm{s}}^{2}$ denotes the square modulus of the transition dipole divided by the sampled momenta. 
	\begin{table}[h]
		\centering
		\caption{Excitonic radiative and non-radiative lifetimes in AlN monolayer at 0K and 300K.}
			\setlength{\tabcolsep}{5pt} 
		\begin{tabular}{c c c c c c}
			\hline
			\hline
			Excitons & $\tau_{NR}$ at 0 K  & $\tau_{NR}$ at 300 K & $\tau_{R}$ at 0 K& $\left\langle \tau_{R}\right\rangle$ at 300 K& $\langle \tau_{R}^{eff}\rangle$ at 300 K\\
			(in eV)	& (in fs) & (in fs) & (in ps) & (in ns) & (in ns) \\
			\hline
			4.47  &~ 2.33 &~ 2.33 & 10.11 & 1.00 & 1.00 \\
			5.86  &~ 1.64 &~ 1.63 & 29.19 & 1.70 & 1.69 \\
			5.90  &~ 0.61 &~ 0.58 & 4.90 & 0.30 & 0.31 \\
	\hline
\hline	
	\end{tabular}
	\label{ch4.table3}
\end{table}
At low temperatures, a thermally averaged radiative lifetime can be formulated as 
\begin{equation}
 \ensuremath{\ensuremath{\ensuremath{\left\langle \tau_{R}^{\mathrm{s}}\right\rangle =\tau_{R}^{\mathrm{s}}\frac{3}{4}k_{B}T\left(\frac{\left(\mathcal{E}^{\mathrm{s}}\right)^{2}}{2m_{\mathrm{s}}c^{2}}\right)^{-1}}}},
 \label{ch4.eq5}
\end{equation} 
under the parabolic exciton dispersion assumption and it is proportional to temperature. In this case, $m_{\mathrm{s}}$ is the effective exciton mass obtained by adding the effective electron and hole masses at the transition point. Moving to higher temperatures, an effective lifetime is calculated using
\begin{equation}
\left\langle \tau_{R}^{eff}\right\rangle ^{-1}=\frac{\sum_{\mathrm{s}}\left\langle \tau_{R}^{\mathrm{s}}\right\rangle ^{-1}e^{-\mathcal{E}^{\mathrm{s}}/k_{B}T}}{\sum_{\mathrm{s}}e^{-\mathcal{E}^{\mathrm{s}}/k_{B}T}}~.
\end{equation}
We find that at 0 K, the intrinsic lifetime of the three prominent bound excitons ranges from 4.90 to 29.19 picoseconds (ps). However, this dramatically increases to the order of nanoseconds (ns) at higher temperatures, primarily due to large oscillator strength, and exciton energies. These lifetimes are comparable to those observed in 2D TMDCs~\cite{Palummo2015}. When comparing to the $\tau_{NR}$, we observe that it is the phonon-assisted non-radiative processes which are significantly faster (in fs), even at 300 K, leading to a substantial ratio $\frac{\tau_{R}}{\tau_{NR}}\sim10^{6}$. Such ultrafast non-radiative lifetimes are also observed experimentally in various 2D TMDCs~\cite{Selig2016}. We summarize these lifetimes in Table~\ref{ch4.table3}.

\section{Phonon-assisted indirect emission}
After solving the BSE for exciton energies, the first optically bright exciton appears at 4.47 eV in the optical limit ($\mathrm{\textbf{Q}}\rightarrow0$). Further, we identify three additional pairs of degenerate, optically dark excitons (with $E$-type symmetry) located at 4.38. Since the single-particle band gap is indirect gap, therefore understanding the emission spectra requires going beyond $\mathrm{\textbf{Q}}\rightarrow0$ to calculate excitonic energies at non-zero exciton momentum ($\mathrm{\textbf{Q}}\neq0$). 
\begin{figure}[t]
	\centering
	\includegraphics[width=0.7\linewidth]{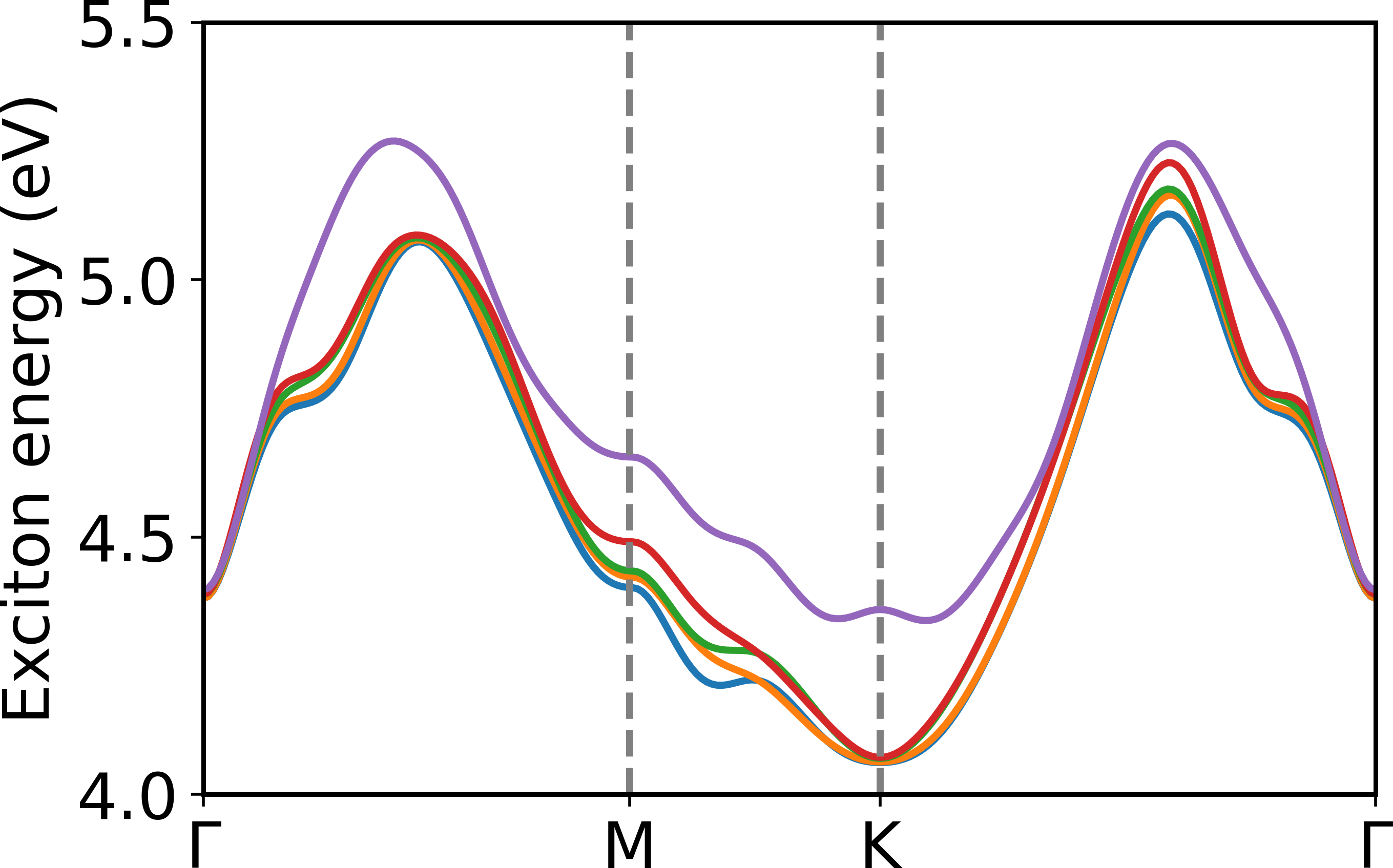}
	\caption{Finite momentum exciton dispersion in 2D AlN. The band-structure is shown for the lowest five excitons. The exciton energies at $\Gamma$ are the optical limit excitons.} 
	\label{ch4.fig9}
\end{figure}
The finite monetum exciton energies calculated for a first five excitons along the high-symmetry path in BZ are shown in Figure~\ref{ch4.fig9}. The $\Gamma$ point represents the $\mathrm{\textbf{Q}}\rightarrow0$ excitonic energies utiized in the absorbance spectrum in Figure~\ref{ch4.fig3}. Along the $\Gamma$-\textbf{K} direction, we observe that the degeneracy in the exciton energies is lifted due to non $E$-type symmetry. The exciton dispersion reveals a minimum energy of 4.06 eV at the \textbf{K} point ($|\mathrm{\textbf{Q}}| = 0.71$ \AA),  an indirect optical gap resulting from a degenerate dark exciton pair, called indirect exciton ($i$X) below the direct exciton (D at 4.47 eV). These $i$X pairs are formed by an electron-hole pair coupled around the $\Gamma$ and \textbf{K} points of the BZ [see Figure~\ref{ch4.fig1} (d)]. 

The exciton energy at $Q = K$ is 4.06 eV identified as indirect exciton ($iX$) and at $Q=\Gamma$ it is 4.47 eV (the direct exciton), and the difference leads to 0.41 eV. Such differences are close to the crystal Debye energy ($\sim$0.12 eV), indicating assistance from optical phonons during recombination. This configuration breaks the dipolar approximation for light-matter coupling, necessitating phonon scattering to maintain momentum conservation during photon emission. We therefore now identify the phonon modes responsible for this assistance during emission [see Figure~\ref{ch4.fig1} (d) and (e)]. These are shown as dots in phonon dispersion plot in Figure~\ref{ch4.fig1} (e). These branches are between \textbf{M}-\textbf{K} with an exactly phonon momentum of 0.71 \AA. The lowest two branches (denoted by 1 and 2) corresponds to out of plane acoustic vibrations, while the next higher branch (denoted by 3) represents an acoustic in-plane longitudinal motion. The mid-frequency (mode denoted by 4) corresponds to longitudinal optical in-plane vibrations, whereas the modes 5 and 6 corresponds to in-plane circular vibrations, stretching and compressing the layer along longitudinal and transverse directions.
\begin{figure}[h]
	\centering
	\includegraphics[width=0.95\linewidth]{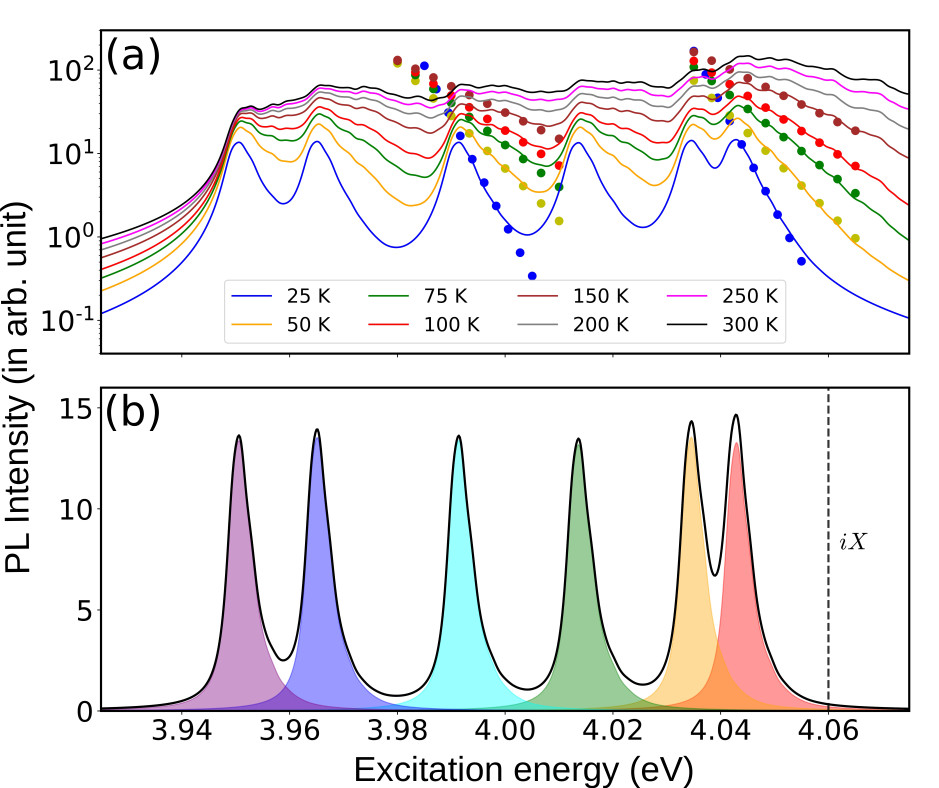}
	\caption{(a) PL emission spectra at various temperature in 2D AlN. The sharp ripples at lower temperatures corresponds to phonon replica due to various modes. The dotted symbols corresponds to the falling edge of the exponential dependency of the exciton thermalization. (b) PL emission at 25 K resolving individual phonon mode assistance. The vertical line is the indirect exciton \textit{iX} located at 4.06 eV.} 
	\label{ch4.fig10}
\end{figure}
The phonon-assisted density of states for the PL emission at various temperatures is shown in Figure \ref{ch4.fig10}. Computation of these spectra involves considering excitonic states at \textbf{Q}$\neq$0 for first few excitons. Subsequently, these excitons are correlated with phonon frequencies, computed through DFPT and utilizing electron-phonon matrix elements. The phonon-assisted DOS can be written as ~\cite{BSE-4-Paleari}, 
\begin{equation}
\varrho\left(\omega,T\right)=\sum_{\mathrm{s},Q,\lambda}\left[1+n_{B}\left(\omega_{\textbf{Q}}^{\lambda}\right)\right]\exp\left[-\frac{\mathcal{E}_{\textbf{Q}}^{\mathrm{s}}-\mathcal{E}_{min}}{k_{B}T}\right]\delta\left(\mathcal{E}_{\textbf{Q}}^{\mathrm{s}}-\omega_{\textbf{Q}}^{\lambda}-\omega\right)~.
\end{equation}
here, $\mathcal{E}_{\textbf{Q}}^{\mathrm{s}}$ is the energy of $\mathrm{s}^{th}$ exciton with momentum \textbf{Q} and $\mathcal{E}_{min}$ is the minimum excitonic energy. We note here that the phonon-assisted spectra presented do not involve rigorous consideration of exciton-phonon matrix elements and dipoles due to the severe complexities associated with such calculations. Instead, $\varrho\left(\omega,T\right)$ serves to pinpoint the emission peak locations based on rigorous electron-phonon matrix elements. 

The luminescence spectra obtained using $\varrho\left(\omega,T\right)$ is known to accurately capture the phonon modes associated with emission processes and align with experimental results ~\cite{Cassabois2016}. In Figure \ref{ch4.fig10} (a), the indirect emission processes is illustrated. Two distinct sets of bands near 4.02-4.05 and 3.94-4.00 eV emerge as the PL emission lines. These lines exhibit at a lower energy compared to the indirect exciton at 4.06 eV, indicating the necessity of phonon assistance for the emission process. Furthermore, examining the high-energy tails near the 4.00 and 4.05 eV lines, we observe an exponential fall-off rate, depicted by dotted symbols for \{25, 50, 175, 100, 150\} K temperatures.
\begin{figure}[t]
	\centering
	\includegraphics[width=0.7\linewidth]{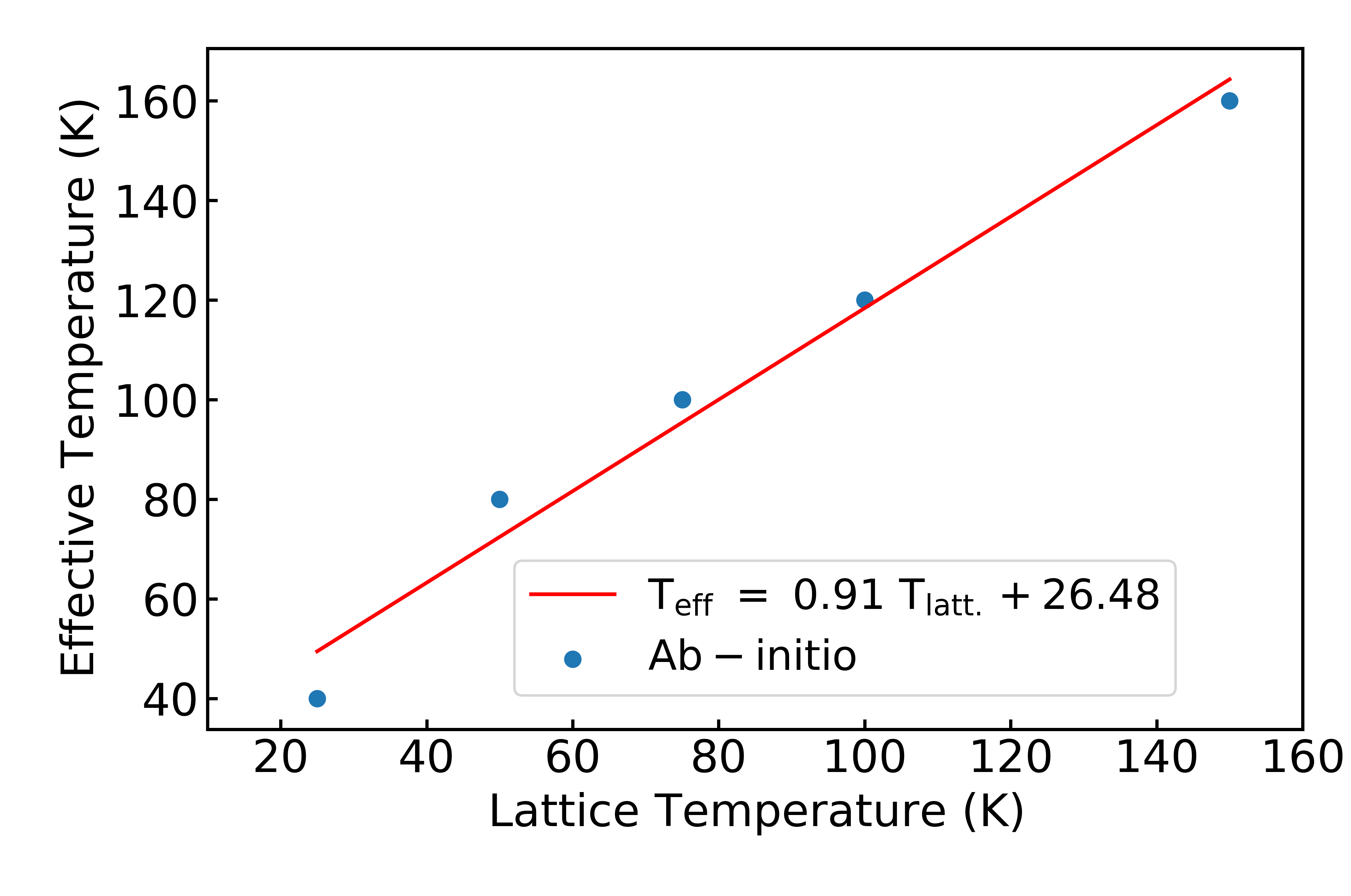}
	\caption{Exciton thermalization in 2D AlN. The dots corresponds to the effective excitonic temperature against the lattice temperature. The straight line is the best fit with slope 0.91.} 
	\label{ch4.fig11}
\end{figure}
As the temperature increases, we observe a reduction in the corresponding slope, although this trend is not evident on their lower energy sides. 

To comprehend the thermalization of excitons, we employ a fitting approach akin to that demonstrated by Cassabois $et$ $al$.~\cite{Cassabois2016}, applying an exponential Boltzmann factor with an effective temperature $\mathrm{T}_{\mathrm{eff}}$ to the declining edge of the spectra. Analysis of the plot correlating $\mathrm{T}_{\mathrm{eff}}$ and lattice temperature $\mathrm{T}_{\mathrm{lat}}$ as shown in Figure \ref{ch4.fig11}.  This suggests that exciton thermalization with the crystal occurs at more than 26 K, similar to a recently reported emission spectra from monolayer h-BN ~\cite{Cassabois2016}.

The phonon-assisted emission becomes evident through the presence of phonon replicas, as shown in Figure~\ref{ch4.fig10} (b).  We observe that the rightmost emission lines at 4.04 eV, result from the out-of-plane ZA mode. The set of replicas below 4.00 eV can be attributed to the optical branches, with a peak corresponding to the optical ZO mode. As the temperature increases, the replicas become less distinct, overshadowed by a large number of phonon densities. Since the optical-limit absorption spectrum of the lowest exciton situated at 4.47 eV is at higher energy level by 0.41 eV with respect to the indirect exciton at 4.06 eV. This shows the emission process is indirect and occurs through phonon assistance. Similar physics was demonstrated in experiments conducted on the emission spectra of tungsten-based WSe$_2$ and WS$_2$~\cite{Brem2020}. Empirical observations suggest that these materials exhibit a diverse range of emission peaks, particularly at low temperatures.

\newpage
\section{Conclusions}
\label{conclusion}
Using first-principles calculations, we investigate temperature dependent electronic and optical properties of planar AlN monolayer. The electronic band structure exhibits an indirect and wide bandgap of 6.73 eV between the $\Gamma$ and $K$ point of the BZ at the GWA level. The optical absorption spectrum reveals an optical bandgap of 4.47 eV and exciton binding energy of 1.83 eV. We identify sevral bright and dark bound excitons below the QP band gap among which three bright excitons contribute to the prominent absorption peaks at 4.47, 5.26, and 5.90 eV. Furthermore, we explore the effect of lattice vibrations and temperature on electronic and optical properties such as single particle band gap, optical absorption, and PL emission. The electron-phonon interaction leads to the renormalization of quasiparticle energies. The calculated optical absorption spectra demonstrate the redshift and changes in dipole oscillator strengths due to the electron-phonon interactions. We observe the phonon-assisted indirect PL emission with excitonic effects. The results obtained from our study offer a deeper understanding of the temperature-dependent optical characteristics of hexagonal AlN monolayers. The significant role of exciton-phonon interactions in shaping absorption and emission behaviors highlights their importance in group-III nitrides and other two-dimensional materials. These findings contribute to advancing 2D material-based optoelectronic devices and solid-state optical applications.\\
 
In this chapter, we thoroughly explored the temperature-dependent electronic and excitonic properties of monolayer AlN. Our focus was on understanding the role of electron-phonon interaction in the indirect emission process. In the previous chapter, we investigated the optical properties of {\formula} materials, with a specific emphasis on exciton-exciton interaction as a function of exciton density. From these individual projects, we gained insights into the impact of exciton density and lattice temperature on the excitonic properties of different 2D semiconductors. Moving forward, in the next chapter, we integrate these crucial parameters \textit{i.e.} exciton density and lattice temperature. To investigate the electron-hole system in the nonequilibrium regime at finite temperatures, this approach allows us to build a comprehensive understanding of different exotic phases of electron-hole systems, such as electron-hole plasma and electron-hole liquid in the nonequilibrium regime at different temperatures.
\section{Appendix}
\subsection{Computational details}
\subsubsection{Ground state electronic structure}
\label{methods-GS}
The ground state properties for the ML AlN are calculated using the Quantum ESPRESSO (QE) code~\cite{QE} the DFT level. The fully relativistic norm-conserving pseudopotentials, including nonlinear core corrections with the state frozen in the core, is 1$s$. Valence states that are included in the pseudopotential are 2$s$ and 2$p$ for aluminium and nitrogen, the states shifted to the core are 1$s$ and 2$s$, and those included in pseudopotential are 3$s$, and 3p \cite{hamann2013optimized}. We have used the \mbox{Perdew – Burke – Ernzerhof} (PBE)  \mbox{exchange-correlation} functionals for the ground state calculations. A cutoff of 80 Ry for the kinetic energy was sufficient for achieving energy convergence.
Energy minimization was performed using a plane-wave basis set, employing a $\Gamma$-centred 12$\times$12$\times$1 k-grid and ensuring convergence of force and energy in the order of 10$^{-5}$ Ry and 10$^{-8}$ Ry, respectively. A two-spinor wavefunction and noncollinear \mbox{spin-orbit} interactions were incorporated for the charge densities in self-consistent calculations. Our electronic stability calculation shows that the ML AlN's formation energy is -0.96 eV/atom. Further, we optimized the lattice structure within the DFT and used the minimum energy structure for our ground state and excited state studies. To confirm the mechanical and thermal stability of the ML structure, we have computed the lattice vibration within the density function perturbation theory (DFPT). 

\subsubsection{El-ph coupling calculations}
\label{methods-elph}
We performed lattice vibration calculations using the PHonon package of the QE. To ensure accuracy, we selected a phonon \textbf{q}-grid of size 8 $\times$ 8 $\times$ 1, applying a self-consistent energy threshold of 10$^{-16}$ Ry.
We randomly sampled the irreducible Brillouin zone (BZ) to calculate the el-ph couplings, generating a fine \textbf{q}-grid with $12\times 12 \times 1$ phonon momenta. We computed the perturbed potentials and dynamical matrices by utilizing the the self-consistent charge densities. Subsequently, we conducted a non-self-consistent calculation on these randomly generated grids, resulting in the el-ph corrected electronic states across the BZ. Through this final step, we evaluated the el-ph matrix elements using the first-order Fan and second-order Debye-Waller perturbation theory after constructing the initial states.

\subsubsection{Excited state calculation}
\label{methods-GW-BSE}
We employed the MBPT open-source code package Yambo~\cite{Sangalli2019} to perform excited state corrections, ensuring accurate capture of optical transitions. We have calculated the QP energies of the electron and hole quasiparticles within the self-consistent GW method~\cite{Hedin-GW1, Hybertsen-GW2, DFT_GWA_Exciton-GW3}, where both the non-interacting Green's function and dynamic screening were updated at each iteration. For 2D systems where the Coulomb potential ($V_q\sim\frac{1}{q}$) leads to the electron-self integrals towards numerical divergence for ($q\rightarrow 0$). To make the density matrices a continuous function of momenta ($q$), we have integrated them them over the irreducible BZ space. To evaluate the diverging integrals, we have employed Monte Carlo techniques with $10^7$ random points across the irreducible BZ, using an energy cutoff of 3 Ry. Additionally, we applied a Coulomb truncation technique to reduce interactions between repeated monolayer images. In order to accurately sum up the irreducible polarization response function, a total of 150 bands were included, combining the occupied and unoccupied bands after a convergence test. We have chosen the random-phase approximation (RPA) kernel to incorporate the local field effects with an energy cutoff of 10 Ry after a convergence test. To tackle the challenge posed by poles around the real axis in the inverse microscopic dielectric function, we employed the Godby and Needs approach of plasmon-pole model~\cite{Godby1989}. This model effectively utilized a pole at zero frequency and another at the plasmon frequency. We have applied GW corrections to the five valence and five conduction bands for an enhanced quantitative understanding of the excited state properties. To obtain the absorption spectra, we solved the time-independent BSE. The equal energy cutoffs used in the GW calculation were employed to construct the electron-hole attractive and repulsive BSE kernels. To generate a Lorentzian-shaped spectrum with an in-plane electric field, as the perturbing field, we have applied a broadening of 0.007 eV. 

Furthermore, we incorporated the quasiparticle energy corrections and static screening obtained from the previous GW calculation. In the presence of lattice vibrations, we implemented corrections corresponding to el-ph interactions on the energy bands. We computed the exciton line widths using el-ph matrix elements, eliminating the need for external broadening. To surpass the limitations of the standard Tamm-Dancoff approximation~\cite{Cannuccia2011, DFT_GWA_Exciton-GW3}, we extended the BSE Hamiltonian to include both resonant and anti-resonant electron-hole matrix elements. We determined the poles representing the transition energies through diagonalization of the BSE matrix.
\clearpage

%% file: chap5.tex
\chapter{Room temperature electron-hole liquid in semiconductors}
\label{chap5}
\begin{figure}[t]
		\centering
		\includegraphics[width =0.7\linewidth]{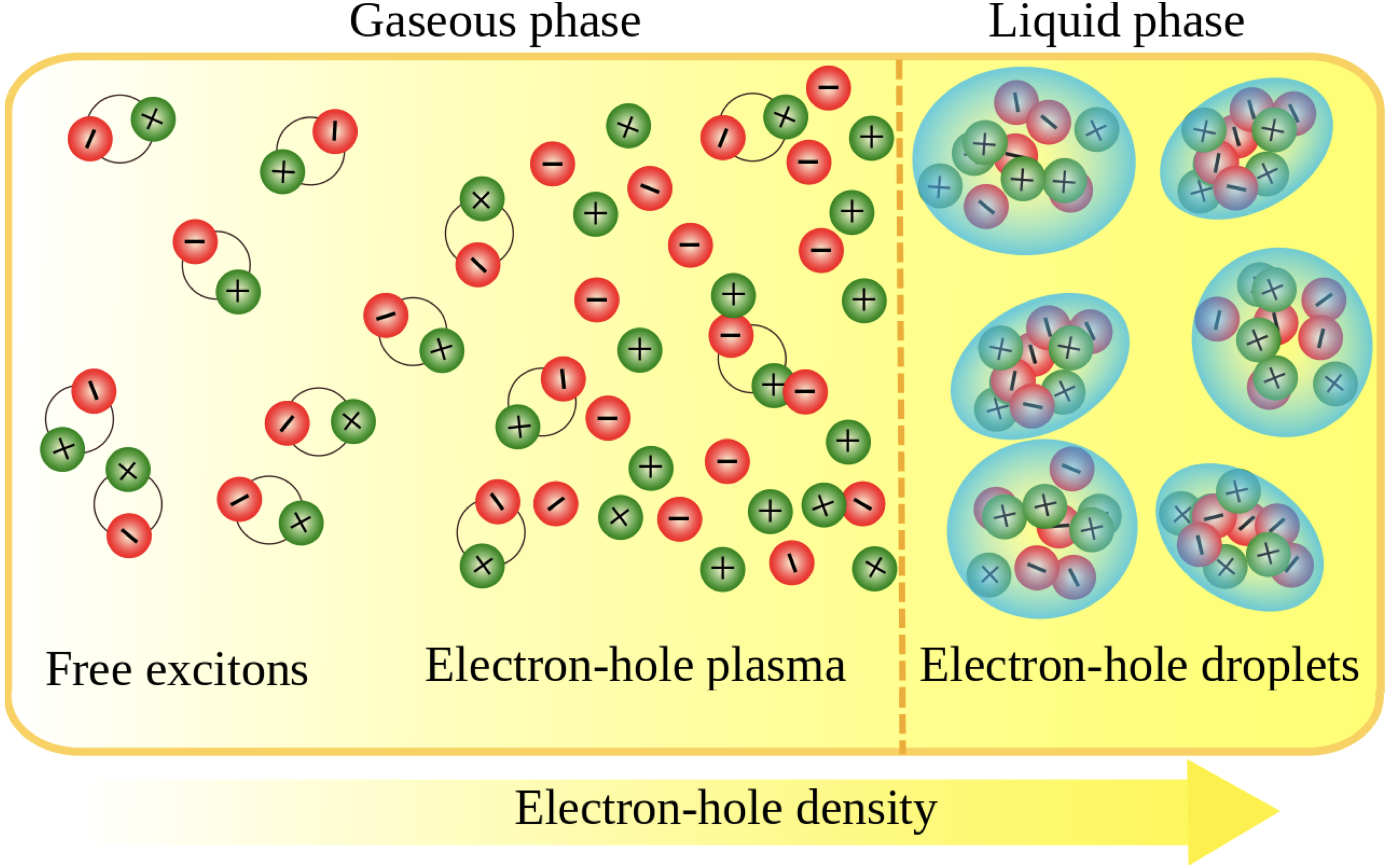}
		\caption {Schematic showing the formation of the electron-hole liquid from photo-excited electrons and holes. The free excitons dissociate on increasing photo-excited carrier density and form the electron-hole plasma state. In both phases, the constituents interact weakly with each other and can be treated as a gaseous state. Further increase in the exciton density leads to the formation of electron-hole droplets or the EHL phase, with the particles interacting collectively.} 
		\label{ch5.fig1}
\end{figure}

\pagestyle{fancy}
The electron-hole liquid (EHL) state is an exciting example of a phase in the non-equilibrium regime of photo-excited carrier~\cite{keldysh1968}. It arises from the condensation of quasi-electrons and quasi-holes in an insulating system at high carrier densities~\cite{Arp2019, Keldysh-Silin-75}. This condensation transforms the electron-hole pairs into a metallic, degenerate Fermi liquid state. At low electron-hole carrier densities and temperatures, bound pairs of excitons are formed, and they interact weakly with each other, forming a non-interacting free-exciton gas~\cite{walker87}. As carrier densities increase, the pairwise Coulomb attraction between electrons and holes is screened, and excitons dissociate into electrons and holes, resulting in an electron-hole plasma (EHP) state~\cite{MOS2-EHP-Bataller2019}. Further increase in the carrier density strengthen the collective interaction between electrons and holes, leading to their condensation into droplets (see Figure~\ref{ch5.fig1}) with a rich phase diagram~\cite{Brinkman-72, Beni-Rice-78, landau, Shah-1977, Simon-1992, Vashishta-PRL-1974, Berciaud2019}. 
Unfortunately, the EHL phase generally occurs at extremely low temperatures. 
This is dictated by the EHL binding energy, which is typically one-tenth of the exciton binding energy \textit{i.e.}, $k_BT_c < 0.1~E_{\rm ex}$~\cite{landau,MoS2-nano-lett}, where $k_B$ is the Boltzmann constant, $T_c$ is the critical temperature of the EHL phase and, $E_{\rm ex}$ is the exciton binding energy. The exciton binding energies in 3D semiconductors range from $0.1-0.001$ eV resulting in $T_c < 20$ K~\cite{Beni-Rice-78, Vashishta-PRL-1974}. However, their 2D counterparts have exciton binding energies of 100s of meV \cite{SOC-effect-MoS2-PRL, TMD-BE, MoS2-nano-lett, MoS2-EHL-EXP-ACS-Nano-2019-Yu}, owing to the reduced dielectric screening of the Coulomb interaction. For example, monolayer MoS$_2$ has an exciton binding energy of  $E_{\rm ex} \approx 0.6$ eV. Due to this, 2D semiconductors offer an ideal platform to observe the EHL phase at room temperature. In addition to the exciton binding energy, other necessary conditions must be satisfied to observe the EHL phase experimentally. These include i) the presence of long-lived photoexcited carriers and ii) a lower ground state energy for the EHL phase compared to the exciton/plasma mixture. Stringent experimental conditions such as high crystal purity and suspended monolayers (effectively eliminating the influence of substrates on exciton dynamics) are essential to meet these requirements. Heterostructures of 2D materials are known to increase the lifetime of the photoexcited carriers and can support the EHL phase.

Recently, the room temperature EHL phase has been probed in the van der Waal heterostructure of MoTe$_2$ and graphene using the technique of multi-parameter dynamic photoresponse microscopy~\cite{Arp2019}. The study observed a distinctive ring-like interlayer photoresponse, indicating the formation of an EHL. As far as intrinsic 2D materials are concerned, electron-hole plasma~\cite{MOS2-EHP-Bataller2019, Pekh2020,pekh2021phase-1} and the EHL phase has been investigated in TMDs such as MoS$_2$,  theoretically as well as experimentally~\cite{MoS2-EHL-EXP-ACS-Nano-2019-Yu, Pekh2020,pekh2021phase-1}. Specifically, a combination of power-dependent photoluminescence experiments and transient differential absorption spectroscopy has helped to demonstrate the EHL phase in MoS$_2$~\cite{MoS2-EHL-EXP-ACS-Nano-2019-Yu, Dey-MoS2-2023}. Ultrafast transient absorption spectroscopy allows for direct probing of excited state dynamics and detecting EHL state's formation, decay, and relaxation processes with high temporal resolution. These techniques provide insights into the aggregation and behavior of electron-hole pairs, shedding light on the collective properties of the EHL state~\cite{amit-exciton-1,amit-exciton-2, Arp2019, MOS2-EHP-Bataller2019, MoS2-EHL-EXP-ACS-Nano-2019-Yu}. This has instigated our search for other 2D systems supporting the room-temperature EHL phase, which can open new avenues for exploring non-equilibrium phase transitions without the limitation of cryogenics.

In this chapter [\footnote{This chapter is adapted from the following paper:
	\\
	Room temperature electron-hole liquid phase in monolayer MoSi$_2$Z$_4$ (Z = pnictogen), \href{https://doi.org/10.1088/2053-1583/ace83b}{2D Mater. 10 (2023) 045007} by Pushpendra Yadav, K V Adarsh,
	and Amit Agarwal.}], we predict that monolayer {\formula} series (Z = N, P, or As) can host the EHL phase for temperatures above room temperature. This is facilitated by the strongly bound excitons in the {\formula} series having binding energies of up to 1 eV as per our study in Chapter~\ref{chap1}. The binding energy ($E_{\rm ex}$) of the free exciton is a significant factor in determining the critical temperature for the EHL phase, which follows the empirical relation, T$_c$ $\sim$ 0.1 $E_{\rm ex}$. Therefore, the monolayers of {\formula} are promising platforms for observing the EHL phase at higher temperatures and experimentally accessible exciton density. 

We calculate the ground state energy and the phase diagram of the EHL phase in the {\formula} series, taking into account the electron-hole pairs' kinetic, exchange, and correlation energy. Our phase diagram predicts that monolayer {\formulaN} can sustain the EHL phase below a critical temperature of $T_c \sim 415$ K and for photo-excited carrier densities higher than $n_c \sim 10^{11}$ cm$^{-2}$. In addition to our prediction for the room temperature EHL phase in these synthetic monolayers, we systematically explore the impact of the variation of effective thickness and the background dielectric constant on the phase diagram of the electron-hole system. Furthermore, we have explored the layer-dependent EHL phase diagram and report the EHL phase in bilayer {\formula}. 
Additionally, we show that our results for the EHL phase are consistent with the Saha ionization formula~\cite{MoS2-nano-lett,saha}. Our findings open new avenues for exploring the non-equilibrium quantum many-body EHL state in monolayer {\formula} series for potential quantum technology and high-power laser applications.
\begin{table}
	\begin{center}
		\def \hsp{\hspace{10cm}}
		\caption{The lattice parameter $a$ in Angstrom (\AA) for the monolayers of {\formula} (Z = N, As, P) and their electronic bandgap (E$_g$) in electron-Volt (eV), calculated within GW method. The effective masses of electrons ($m^*_e$) and holes ($m^*_h$) are listed in terms of electron mass ($m_e$). The number of electron and hole valleys are represented as $\nu_e$ and $\nu_h$ for the lowest unoccupied conduction band minima and the highest occupied valence band maxima at the $K/K^{\prime}$ of the 2D hexagonal Brillouin zone.}	
		\vspace{0.15 cm}
		\begin{tabular}{c c c c c c c c c }
			\hline \hline \vspace{1 mm}
			Compound  & $a$ (\AA) & E$_{g}$ (eV) & $m^*_e/m_e$   & $m^*_h/m_h$ &  $\nu_e$  & $\nu_h$ \\
			\hline
			\formulaN  & 2.909  & 3.58  & 0.407{\cite{Valley-pseudospin-PRB}} & 0.554{\cite{Valley-pseudospin-PRB}} & 2 & 2 \\
			\formulaAs & 3.621  & 1.70  & 0.499{\cite{nanotech-MoSi2As4}} & 0.419{\cite{nanotech-MoSi2As4}} & 2 & 2 \\
			\formulaP  & 3.471  & 1.74 & 0.325{\cite{nanotech-MoSi2As4}} & 0.393{\cite{nanotech-MoSi2As4}} & 2 & 2 \\
			\hline
		\end{tabular}
		\label{ch5.table1}
	\end{center}
\end{table}	

\section{Ground state energy of the electron-hole system}
\label{ch5.sec1}
The total ground state energy of a system with interacting electrons and holes comprises of kinetic, exchange, and correlation energies. The total kinetic energy of  a 2D electron-hole system can be approximated as,  
\begin{equation}
	E_{\text{kin}} = \sum_{i=e,h}\nu_i \sigma_i \sum_{k<k_F^i}\frac{\hbar^2 k^2}{2m_i}~ = ~\sum_{i=e,h} \frac{1}{2}E_F^{i}~.
	\label{ch5.eq1}
\end{equation}
Here, $i = e/h$ represents electrons/holes, $\sigma_i$ ($\nu_i$) is the spin (valley) degeneracy of the bands, and $k_F^i$ is the corresponding Fermi wavevector. In Equation~\ref{ch5.eq1}, the Fermi wavevector for each species is given by $k_F^{i} = k_F/\sqrt{\nu_{i}}$ where, $k_F=(2\pi n)^{1/2}$ and $n$ represents electron-hole pair density. The corresponding Fermi energy is specified by $E^{i}_F = \hbar^2(k_F^{i})^2/2m_{i}$.	The valley structure of the {\formula} series of monolayers can be seen from Figure~\ref{ch3.fig2} of Chapter~\ref{chap3}. We find that for all three compounds, the optically relevant band extrema occurs at the $K$ and $K'$ points of the BZ. This implies $\nu_e = \nu_h = 2$ for all the three studied monolayers. The effective masses of electrons and holes near these band extrema are obtained from the $ab-initio$ band structure calculations for the monolayer {\formula} and tabulated in Table~\ref{ch5.table1}.
\begin{figure}[t]
	\centering
	\includegraphics[width =\linewidth]{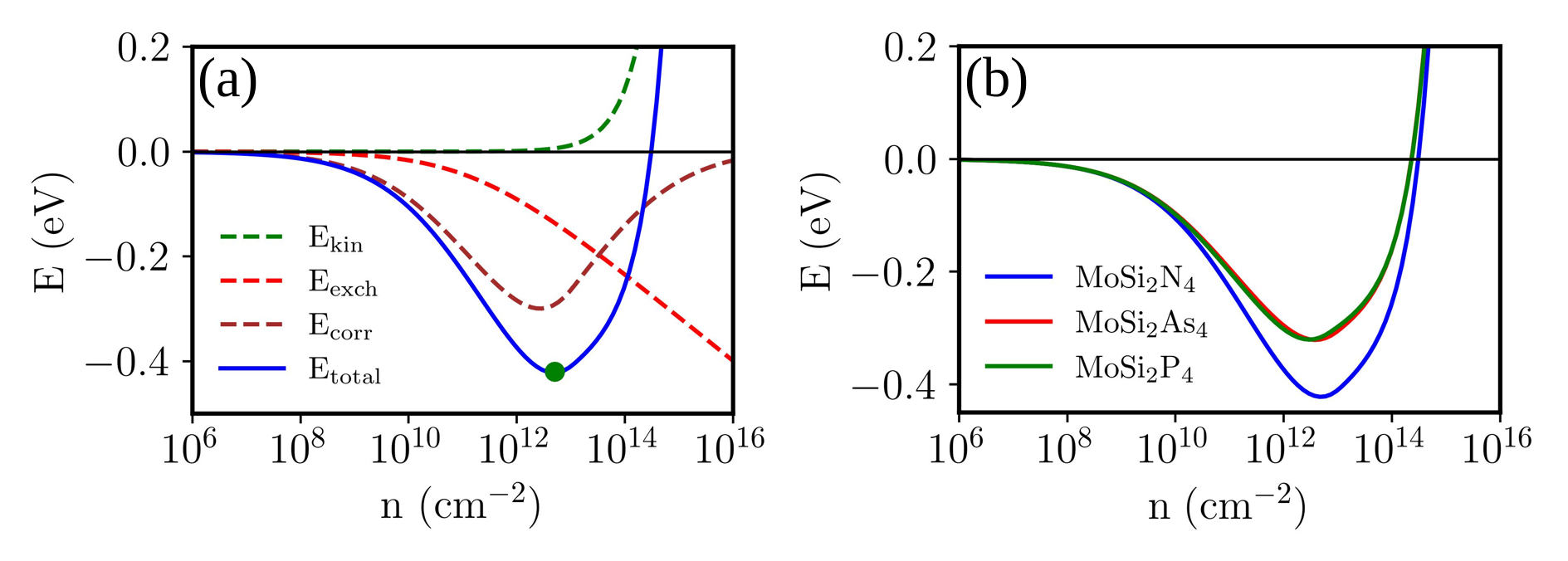}
	\caption {The total ground state energy as a function of photo-excited carrier density $n$ for the monolayer {\formula} series. (a) The kinetic, exchange,  correlation, and total ground state energy (dashed green, red, brown, and solid blue curve, respectively) for monolayer {\formulaN}. The kinetic energy is prominent in the high-density limit, while the correlation energy dominates in the low-density regime. (b) The total ground state energy of the three monolayers. The photo-excited electrons and holes in monolayer {\formulaN} have the lowest ground state energy. 
		\label{ch5.fig2}}
\end{figure}
The impact of coulomb interactions can be split into the exchange and the correlation contributions. The exchange contribution is captured within the first-order perturbation theory by calculating the expectation value of the Coulomb interaction Hamiltonian using the multi-particle eigenstates of the non-interacting Hamiltonian~\cite{Brinkman-Rice-73, Giuliani-Giovanni-Vignale, Bergersen_1975}. 
For a generic Coulomb potential specified by $V_k$ (in the momentum space), it has the following form~\cite{MoS2-nano-lett, Pekh2020,pekh2021phase-1},  
\begin{equation}
	E_{\text{exch}} = -\sum_{i=e,h}\frac{\nu_i \sigma_i}{2L^2} \sum_{k,q<k_{F,i}}V_{k-q}~,
	\label{ch5.eq2}
\end{equation}
where $L^2$ is the area of the 2D system. For a freestanding 2D system of zero thickness, $V_k = 2 \pi e^2/k$ gives the unscreened Coulomb potential. However, most 2D crystalline systems have a finite width and can be encapsulated on both sides by substrates of different dielectric constants. These effects are captured by the Keldysh potential~\cite{keldysh1986} which has the form, 
\begin{equation}
	V_k = \frac{2\pi e^2}{\epsilon'k(1+r_0k)}~.
	\label{ch5.eq3}
\end{equation}
Here, $\epsilon^{\prime}$=($\epsilon_1$ + $\epsilon_2$)/2 with $\epsilon_1$ ($\epsilon_2$) being the dielectric constant of the top (bottom) substrate and $r_0$ is the effective thickness of the 2D system~\cite{Angel-Rubio2011_r0, MacDonald_r0}. The Keldysh potential is known to be more accurate for calculating the exciton binding energy of the monolayer transition metal dichalcogenides~\cite{MoS2-nano-lett}. It also captures the change in the impact of the Coulomb interaction across dimensional crossover. The Keldysh potential reduces to the unscreened 2D Coulomb potential in the $r_0 \to 0$ limit. For large $r_0$ values, it mimics the 3D Coulomb potential with $V_k \propto 1/k^2$ [see Figure~\ref{ch5.fig5} (d)-(e)]. The effective thickness $r_0$ in monolayer {\formula} series is calculated using the relation $r_0 \propto d/\epsilon$ where $d$ is the layer thickness, and $\epsilon$ is the dielectric constant of the material. For monolayer MoS$_2$, we have $d$ = 3.1 \AA, and $r_0/d = 14.16$ \cite{MacDonald_r0,Angel-Rubio2011_r0,MoS2-nano-lett}. Using this $r_0/d$ value for monolayer {\formulaN},  {\formulaAs}, and {\formulaP} we estimate their $r_0$ to be 99.15 \AA, 140.65 {\AA}, and 132.12 \AA, respectively.

In contrast to the exchange energy for unscreened Coulomb potential~\cite{Pekh2020,pekh2021phase-1}, the analytical form of the exchange energy for the Keldysh potential (Equation~\ref{ch5.eq3}) is not known. Therefore,  we calculate it numerically. The dependence of the exchange energy for monolayer {\formulaN} is shown by the red dashed curve in Figure~\ref{ch5.fig2} (a). Together, the kinetic and exchange energy contributions form the Hartree-Fock energy. The remaining correlation energy contributions from the Coulomb interaction includes the second and higher order terms of the perturbation series~\cite{Combescot_1972, Rice1978book}. These are typically captured by the Random phase approximation (RPA)~\cite{Combescot_1972}, which sums the infinite series of the bubble diagrams. Within the RPA scheme, the correlation energy has the following form,
\begin{equation}
	E_{\rm{corr}} = \sum_{q}\int_{0}^{\infty} \frac{\hbar d\omega}{2\pi}\left[ \tan^{-1}\left (\frac{-B_q(\omega)}{1-A_q(\omega)}\right )+B_q(\omega) \right]~.
	\label{ch5.eq4}
\end{equation}
Here, $A_q(\omega)$ and $B_q(\omega)$ are the frequency-dependent real and imaginary parts of $V_q\chi(q,\omega)$, or $A_q(\omega) + iB_q(\omega) = V_q\chi(q,\omega)$. $\chi(q,\omega)$ is the sum of the Lindhard susceptibility of the electrons and holes~\cite{Giuliani-Giovanni-Vignale}, and $V_q$ is the Keldysh potential defined in Equation~\ref{ch5.eq3}. To understand the role of dimensionality in interaction effects, we calculate the correlation energies for different $r_0$ values. We find that a smaller $r_0$ value in the Keldysh potential yields the known correlation energy for the 2D case with unscreened Coulomb interactions. On increasing the $r_0$ value, the correlation energy decreases. This highlights that correlation effects become more pronounced when reducing the dimensions of the systems [see Figure~\ref{ch5.fig5} (d)-(e)].

Our numerical calculations of the total energy show that for low densities of the photo-excited carriers, both the exchange and correlation energies dominate over the kinetic energy contribution. With a gradual increase in the photo-excited carrier density, the kinetic energy dominates the correlation and exchange terms. This leads to a non-monotonic behavior in the total energy curve with a minimum at the equilibrium density [see Figure~\ref{ch5.fig2} (a)] for monolayer {\formulaN}. We find that the ground state energy for monolayer {\formulaN} is larger in magnitude than the ground state energies of {\formulaAs} and {\formulaP} [see Figure~\ref{ch5.fig2} (b)]. The relatively smaller effective thickness (or $r_0$) of {\formulaN} compared to {\formulaAs} and {\formulaP} make the exchange-correlation effects stronger, lowering its ground state energy. We show below that this makes the EHL phase in monolayer {\formulaN} relatively more stable with a higher temperature.

\section{Thermodynamics of the EHL phase}
\label{ch5.sec2}
\begin{figure}[t]
	\centering
	\includegraphics[width =0.95\linewidth]{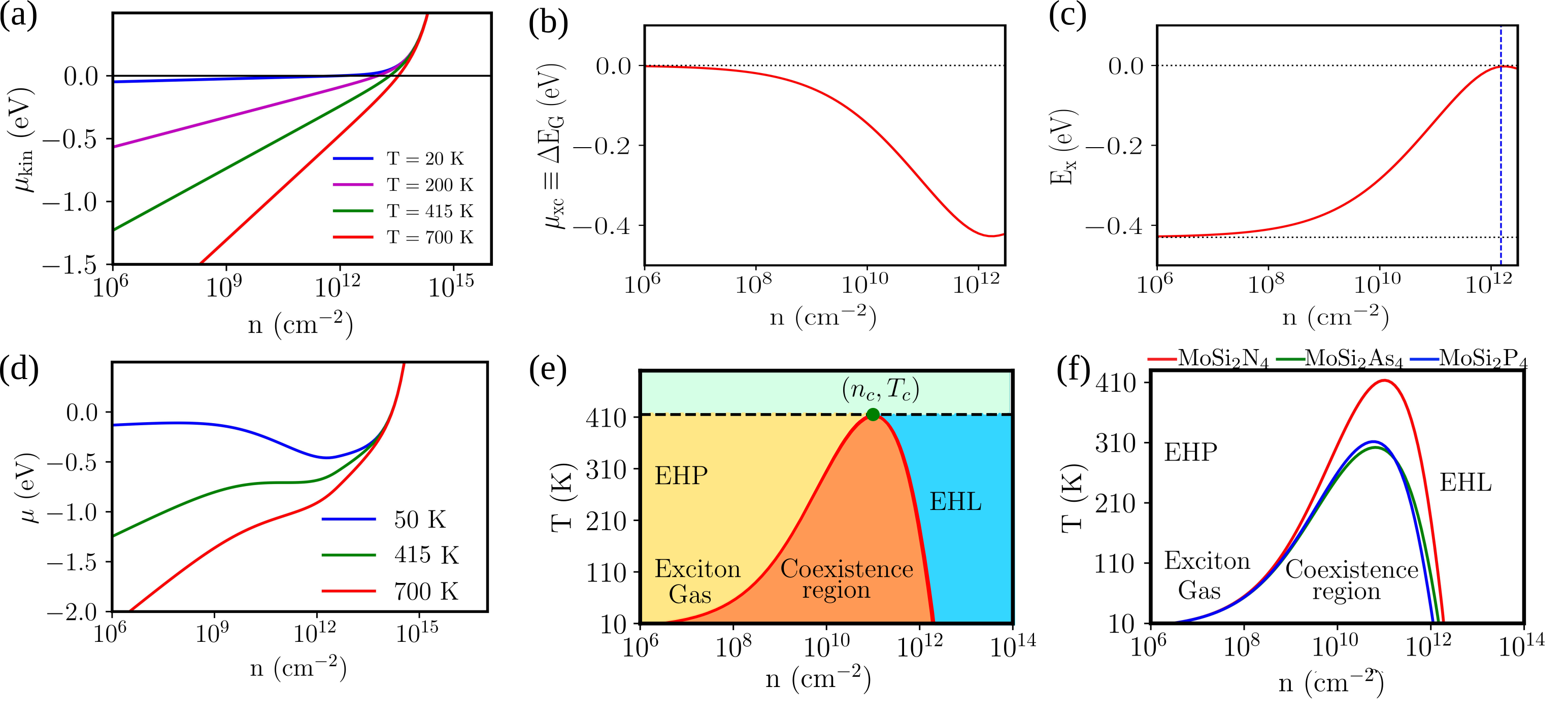}
	\caption {(a) The density variation of the chemical potential contribution from the kinetic energy term at different temperatures. (b) The exchange-correlation part of the chemical potential as a function of carrier density and the corresponding bandgap renormalization leading to the renormalized binding energy as a function of the photo-excited carrier density $n$ in (c). For monolayer {\formulaN}, we find the Mott density [the blue dashed vertical line in (c)] to be 1.5 $\times$ 10$^{12}$ cm$^{-2}$. (d) Represents the total chemical potential for the interacting electron-hole system. Including the exchange and correlation terms, the chemical potential becomes a non-monotonic density function. Beyond a critical temperature $(T_c)$, the system can have two different densities at the same $\mu$, indicating the coexistence of the electron-hole gas and the condensed electron-hole liquid phase. (e) The thermodynamic phase diagram in the $n-T$ plane shows regions of coexistence of an electron-hole gas and electron-hole liquid phase. (f)) The phase boundaries and the coexistence region for the three {\formula} monolayers in the $n-T$ plane. 
		\label{ch5.fig3}}
\end{figure}

The EHL droplet formation is a first-order phase transition similar to the gas-liquid transition. The EHL droplet condensation happens when a supersaturated electron-hole system simultaneously exhibits a high-density liquid phase and a low-density gaseous phase. This occurs at a critical density ($n_c$) and a critical temperature ($T_c$). The excitons lose their individuality above the critical density and below the critical temperature, and electrons and holes condense into a droplet~\cite{Simon-2002}. To study the thermodynamics of the photo-excited electron-hole pairs, we calculate the free energy and derive the chemical potential. The free energy per particle can be expressed as $F (n, T) = F_0(n, T) + F_{\mathrm{xc}}(n, T).$ Here, $F_0(n, T)$ is the free energy of the non-interacting electrons and holes, and $F_{\mathrm{xc}}(n, T)$ is the Coulomb interaction induced exchange and correlation contribution. At a high density of electrons and holes, the EHL is known to be metallic~\cite{Brinkman-72, Thomas-73,thomas-PRL-1974}. In the metallic regime, we can safely ignore the explicit $T$ dependence of the interaction part of the free energy and express $F_{\mathrm{xc}}(n, T) = F_{\mathrm{xc}}(n)$. Accordingly, the chemical potential is given by
\begin{equation}
	\mu = \left( \frac{\partial F}{\partial N}\right)_{T,V} = \mu_{\rm{kin}} + \left( \frac{\partial F_{\rm{xc}}}{\partial N}\right )_{T,V}~.
	\label{ch5.eq5}
\end{equation}
The first term on the right-hand side of Equation~\ref{ch5.eq5} refers to kinetic energy contribution to the chemical potential, while the second term refers to the contribution of the exchange and correlation energy to the chemical potential ($\mu_{\rm xc}$). These can be calculated from ~\cite{thomas-PRL-1974}, 
\bea
\mu_{\rm{kin}} & = & \frac{1}{\beta}\left( \ln[e^{\beta E^e_F} -1] + \ln[e^{\beta E^h_F} -1] \right)~, \nonumber \\ 
\mu_{\rm{xc}} & = & E_{\rm{xc}} + n \frac{\partial E_{\rm{xc}} }{\partial n}~.
\label{ch5.eq6}
\eea
Here, $\beta = 1/(k_BT)$ is the Boltzmann constant, and $E_{\rm{xc}}$ is the sum of exchange and correlation energies. We present the temperature and carrier density dependence of the $\mu_{\rm kin}$ for the monolayer {\formulaN} in Figure~\ref{ch5.fig3} (a). 

\textit{Bandgap renormalization:} The dependence of $\mu_{\rm xc}$ on the photoexcited carrier density is presented in Figure~\ref{ch5.fig3} (b). One important aspect of the variation in the exciton binding energy with photoexcited carrier density is the band gap renormalization (BGR). The density-dependent exchange-correlation chemical potential captures the BGR~\cite{BGR-Kalt-1992, MoS2-nano-lett} with changing density of photo-excited carriers. BGR captures the impact of the correlation and exchange energy in reducing the single particle energy bandgap. The variation in the chemical potential induced by the exchange and correlation energy terms gives an approximate estimation of the BGR as a function of the number density of photoexcited electron-hole pairs. We have calculated the change in the bandgap, $\Delta E_g \approx \mu_{xc}$, where $\mu_{xc}$ is defined by Equation~\ref{ch5.eq6}. We present the variation of the exchange-correlation part of the chemical potential and the BGR in {\formulaN} in Figure~\ref{ch5.fig3} (b) and Figure~\ref{ch5.fig3} (c), respectively. We find that beyond a critical density of photoexcited carriers (referred to as the Mott density), the excitons merge with the continuum single-particle excitations~\cite{MoS2-nano-lett, BGR-Kalt-1992}. For \formulaN, we find the Mott density to be approximately 1.5 $\times 10^{12}$ cm$^{-2}$. From the $\mu_{\rm xc}(n)$ plot for monolayer {\formulaN}, we find that on including the effective thickness of the layer ($r_0$ = 99.15 \AA), the free exciton binding energy becomes 0.432 eV. Similarly, the free exciton binding energy calculated for the monolayer {\formulaAs} and {\formulaP} are summarized in Table~\ref{ch5.table2}. These exciton binding energies in Table~\ref{ch5.table2} are significantly lower than those calculated from first principles, which do not include the impact of effective layer thickness in the Coulomb interactions. 

\textit{EHL Phase diagram:} The total chemical potential calculated using Equation~\ref{ch5.eq5} is shown in Figure~\ref{ch5.fig3} (d). Depending on the temperature, there are regions with the possibility of having two different densities at the same chemical potential, a clear indication of the coexistence of two phases. For $T > T_c$, the chemical potential increases monotonically with exciton density. On reducing the temperature, we reach a critical temperature $T =T_c$, for which the slope of the chemical potential curve goes to zero at a critical density ($n = n_c$). This inflection point ($n_c, T_c$) marks the onset of the EHL phase transition. The boundary of the coexistence region of the liquid and the gas phase is determined by $\partial_{n} \mu = 0$, and the critical point is obtained from~\cite{Rice1978book, Tikhodeev_1985}, or 
\bea
 \frac{\partial^2\mu}{\partial n^2} \bigg |_{(n_c,T_c)} = 0~.
\label{ch5.eq7}
\eea	
We present the boundary of the coexistence region and the critical point for the {\formulaN} monolayer in the temperature-density plane in Figure~\ref{ch5.fig3} (e). The phase diagram clearly shows the `gas region' supporting free excitons and electron-hole plasma, the coexistence region, and the region with the EHL. Our calculations suggest that the critical temperature $T_c = 415$ $K$, and the critical density $n_c = 1.0\times 10^{11}$ cm$^{-2}$  for monolayer {\formulaN}. Monolayers {\formulaAs} and {\formulaP} have a qualitatively similar phase diagram as shown in Figure~\ref{ch5.fig3} (f). The critical density and critical temperature for the EHL phase transition for all three materials are summarized in Table~\ref{ch5.table2}.
\begin{table}
	\begin{center}
		\def \hsp{\hspace{10cm}}
		\caption{{The free exciton binding energy ($E_{\rm{x}}$), critical temperature ($T_c$), and the corresponding critical electron-hole pair density ($n_c$) from our calculations.}}	
		\vspace{0.15 cm}
		\begin{tabular}{c c c c}
			\hline \hline \vspace{1 mm}
			Compound \hsp    &  $E_x$ (eV) \hsp   & $T_c$ (K)  \hsp   &  $n_c$ ($\times 10^{11}$ $\mathrm{cm^{-2}})$ \\
			\hline
			\formulaN &  0.432 & 415 & 1.0 \\
			\formulaAs & 0.466 &  302 & 0.7 \\
			\formulaP  & 0.433 &  312 & 0.6 \\
			\hline
		\end{tabular}
		\label{ch5.table2}	
	\end{center}	
\end{table}

\textit{EHL phase via Saha ionization formula:} As an independent check, we investigate the different phases of the electron-hole system using the Saha ionization equation~\cite{saha, MoS2-nano-lett}. It describes the thermodynamics of the ionization of atoms in a gas and is typically used to understand the behavior of ionized gases, where atoms lose or gain electrons. In the context of electron-hole systems, it captures the equilibrium between the electron-hole plasma and the excitons. We use it to check the consistency of our results for the formation of electron-hole plasma and the EHL. The Saha ionization formula is specified by, 
\begin{equation}
	\frac{\alpha^2}{1-\alpha} = \frac{g_eg_h}{g_{\rm ex}} \frac{1}{n \lambda_T^2}~\exp\left ( \frac{-|E_{\rm ex}(n)|}{k_BT}\right )~.
	\label{ch5.eq8}
\end{equation}
Here, $\alpha$ is the ionization ratio for the exciton gas, $g_e$ = 2, $g_h$ = 2, $g_{\rm ex}$ = 4, are the spin and valley degeneracy factors, and  $k_B$ is the Boltzmann constant, and $E_{\rm ex} (n)$ is the binding energy of exciton as a function of photoexcited carrier density. The variation of $E_{\rm ex}(n)$ with the carrier density is calculated and presented in Figure~\ref{ch5.fig3} (c). In Equation~\ref{ch5.eq8}, $\lambda_{T}$ (= $h/\sqrt{2\pi \mu_{r}k_BT}$) is the thermal de Broglie wavelength of the electron-hole pair, with $\mu_r$ being the reduced electron-hole mass.
\begin{figure}[h]
	\centering
	\includegraphics[width =.6\linewidth]{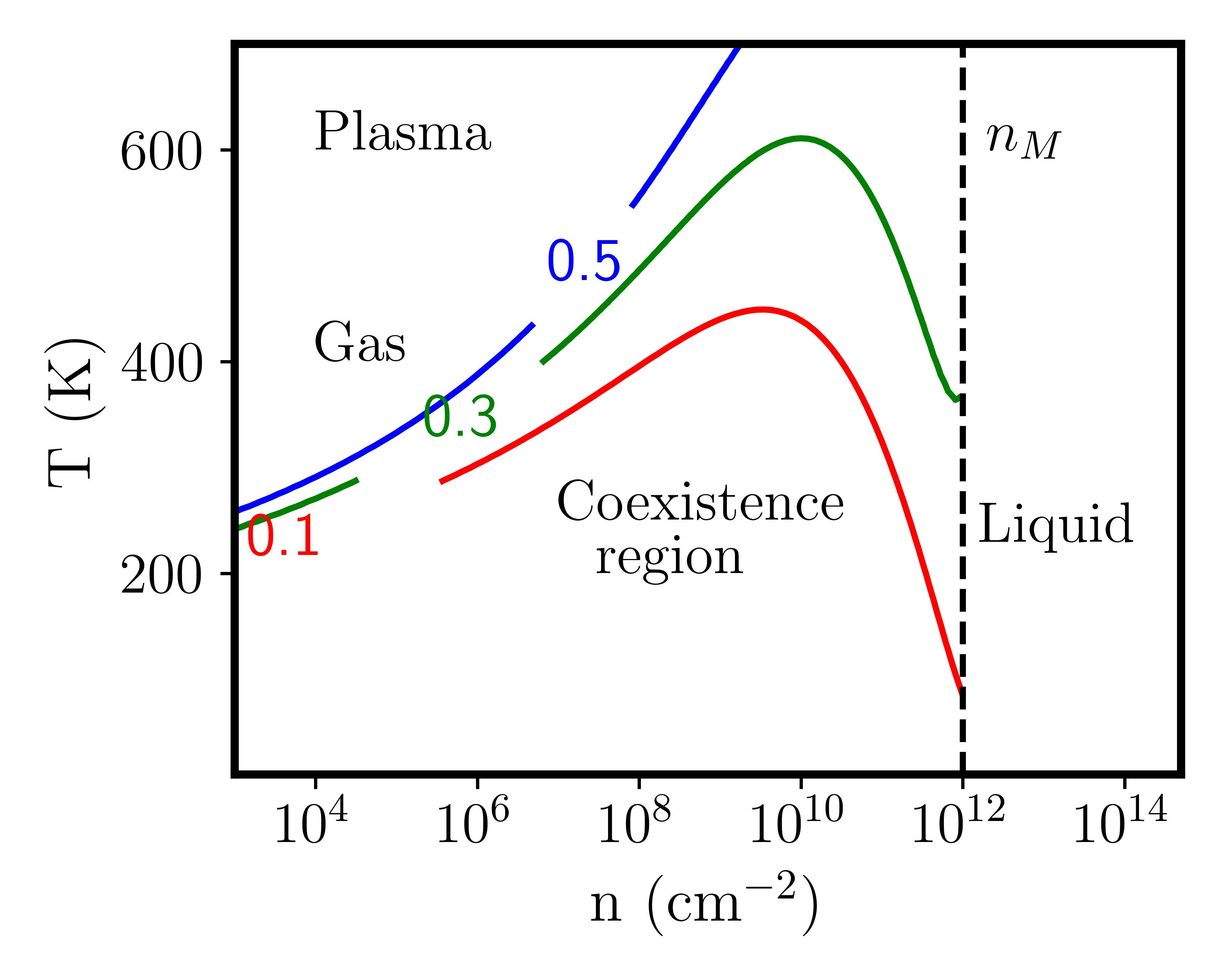}
	\caption {Ionization ratio lines obtained by solving the Saha ionization equation (Equation~\ref{ch5.eq8}) for different ionization ratios [$\alpha$ = 0.1, 0.3, 0.5] for monolayer {\formulaN}. The black vertical line represents the Mott density ($n_M$) at which all the excitons are dissociated. Beyond the Mott density, the exciton gas condenses into an EHL.  
			\label{ch5.fig4}}
\end{figure}	
Using the Saha ionization equation, we find that the gaseous and the plasma phases are well-defined in the temperature-density plane for different ionization ratios [$\alpha$ = 0.1, 0.3, 0.5]. In Figure~\ref{ch5.fig4}, we see that at low temperatures and low density, the excitons partially dissociate into electron-hole plasma and coexist. However, with increasing density, the exciton gas and the electron-hole plasma coexist with the liquid phase. Beyond the critical density of the Mott transition, the exciton gas ceases to exist, and the plasma phase can coexist with the liquid phase at higher temperatures. 
\section{Impact of dielectric constant and layer thickness on the EHL phase}
\label{ch5.sec3}
\begin{figure}[t]
	\centering
	\includegraphics[width =.95\linewidth]{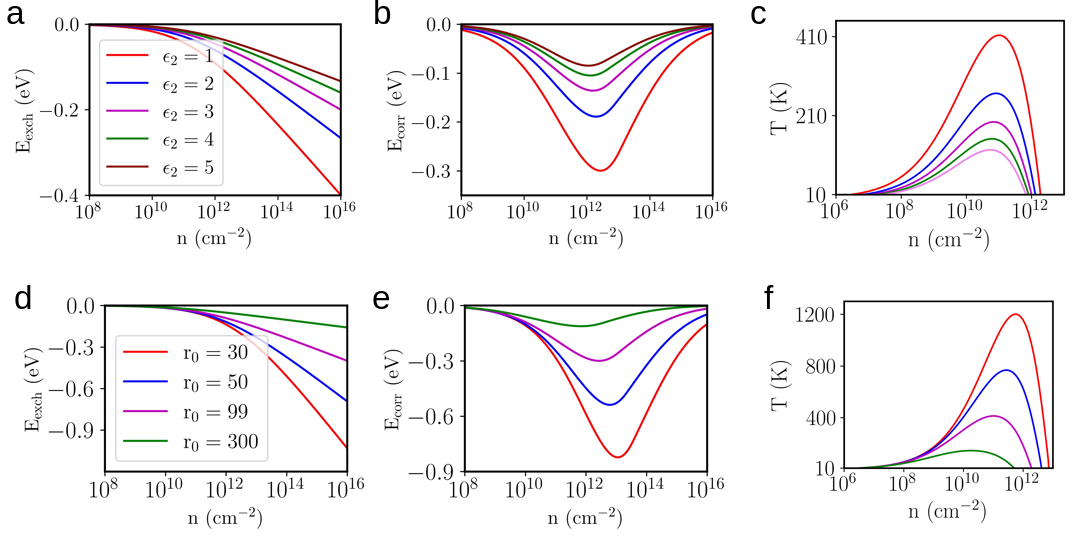}
	\caption {The sensitivity of the EHL transition temperature on the dielectric constant and the effective thickness. The variation of the (a) exchange energy, (b) correlation energy, and (c) the phase boundary of monolayer {\formulaN} with varying dielectric constant of the substrate. Similarly, the effect of the effective layer thickness $r_0$ on the (d) exchange energy, (e) correlation energy, and (f) the phase diagram of the monolayer {\formulaN}.
		\label{ch5.fig5}}
\end{figure}

\begin{figure}[t]
	\centering
	\includegraphics[width =\linewidth]{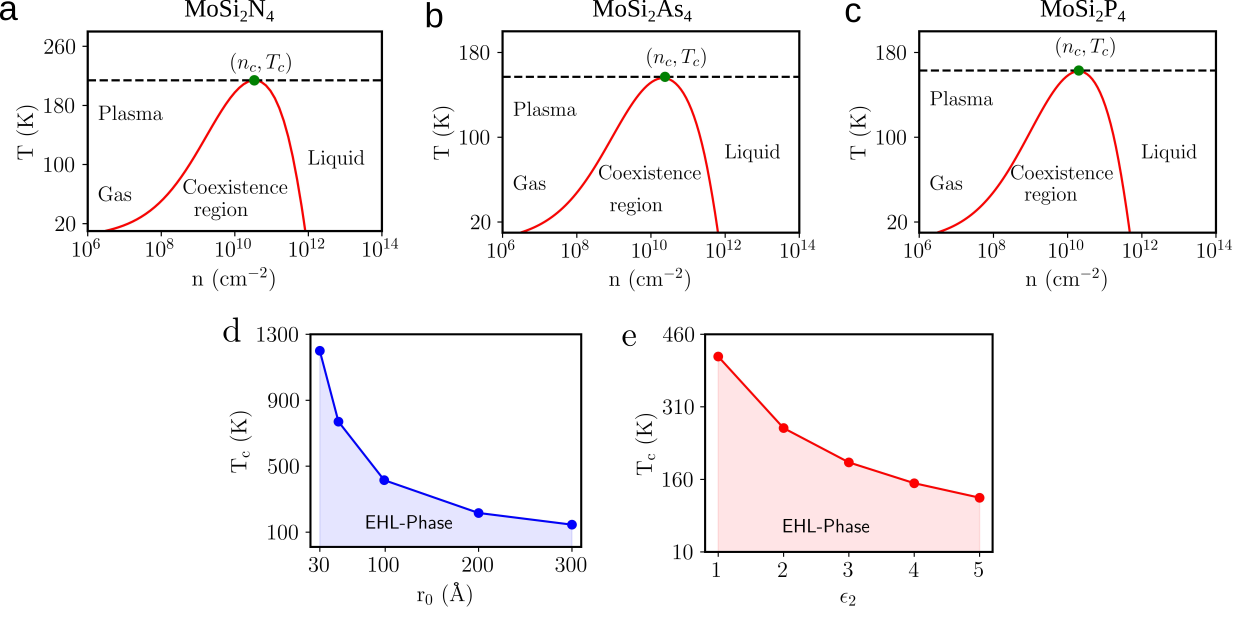}
	\caption {(a) The thermodynamic phase diagram of the electron-hole system in the density-temperature plane for bilayer {\formulaN}, (b) bilayer {\formulaAs}, and (c) bilayer {\formulaP}. The coexistence of an electron-hole gas and electron-hole liquid phase is evident in bilayer {\formulaAs} with critical temperature 157 K and critical density $2.4~\times~10^{10}~cm^{-2}$, and in bilayer {\formulaP} with critical temperature 163 K and critical density $2.0~ \times~10^{10}~cm^{-2}$.The coexistence of an electron-hole gas and electron-hole liquid phase can be seen in bilayer {\formulaN} with critical temperature 214 K and critical density 3.4 $\times$ $10^{10}~cm^{-2}$. The impact of the effective 2D layer thickness on the transition temperature of the EHL phase can also be captured by increasing $r_0$. (d) The variation of the EHL transition temperature with increasing $r_0$ and (e) the $T_c$ variation with the dielectric constant of the substrate. Increasing both parameters ($r_0$ and $\epsilon$) reduces the Coulomb interaction strength [see Equation~\ref{ch5.eq3}], decreasing the critical temperature. 
		\label{ch5.fig6}}
\end{figure}

Figure~\ref{ch5.fig4} demonstrates the possibility of room temperature EHL phase in the monolayers of the {\formula} series. However, our calculations rely on the specific choice of $r_0$ and the dielectric constant. For 2D materials, the substrate's dielectric constant can also significantly impact its optical properties and EHL phase [see Equation~\ref{ch5.eq3}]. The increase in the effective dielectric constant ($\epsilon'$) decreases the strength of the Coulomb potential. This results in a reduction of the exchange and correlation energy or the magnitude of the total ground state energy of the electron-hole system with increasing dielectric constant. To quantify this variation, we show the dependence of the exchange energy, correlation energy, and the EHL phase boundary on the dielectric constant in Figure~\ref{ch5.fig5} (a), (b), and (c), respectively. As expected, an increase in the dielectric constant pushes the EHL phase boundaries towards lower temperatures. This becomes even more evident in Figure~\ref{ch5.fig6} (e), which shows the decrease of the $T_c$ with increasing dielectric constant. However, even with a dielectric constant of the substrate $\epsilon_2=5$, {\formulaN} has a $T_c$ of more than 100 K for the EHL phase. 

Furthermore, to understand the thickness dependence on the EHL phase, we calculated the EHL phase diagram of the bilayer {\formula}. The phase diagram for bilayer {\formulaN} is shown in Figure~\ref{ch5.fig6} (a), and for the {\formulaAs}, and {\formulaP} bilayers, it is shown in Figure~\ref{ch5.fig6} (b)  and ~\ref{ch5.fig6} (c), respectively. As expected, the increase in layer thickness increases the screening of the Coulomb interaction, decreasing its impact. This leads to a  decrease in the exciton binding energy and the critical temperature for the formation of EHL.
\label{section-sm3}

We find that all three monolayers can support room-temperature EHL phase. Amongst the three monolayers, {\formulaN} has the lowest $r_0$ and the highest $T_c$. A higher $r_0$ value decreases the strength of the effective Coulomb interactions and the exchange-correlation energy [see Figure~\ref{ch5.fig5} (d) and (e)] leading to modified phase diagrams in Figure~\ref{ch5.fig5} (f) and a lowering of the $T_c$ which is demonstrated in Figure~\ref{ch5.fig6}~(d). 

Increasing the dielectric constant reduces the strength of the Coulomb interactions. This pushes the EHL phase boundaries towards lower temperatures. Figure~\ref{ch5.fig6} {d} shows that the $T_c$ decreases with increasing dielectric constant. 
We also check the sensitivity of the $T_c$ with the effective thickness, $r_0$, in the Keldysh potential defined in Equation~\ref{ch5.eq3}. $r_0$ captures the dimensional crossover of the Coulomb potential from 2D to an effective 3D. Increasing $r_0$  decreases the strength of the Coulomb interaction and the $T_c$ of the EHL phase. To observe the quantitative impact of $r_0$, we show the variation of the exchange and correlation energies of the electron-hole system with increasing $r_0$ in Figure~\ref{ch5.fig5} (d) and Figure~\ref{ch5.fig5} (e), respectively. The exchange and correlation energies increase in magnitude as $r_0$ decreases. This reflects in the lowering of the $T_c$ with increasing $r_0$, as shown in Figure~\ref{ch5.fig5} (f) and a quantitative variation of $T_c$ with $\epsilon_2$ is shown in Figure~\ref{ch5.fig6}). 
We investigate the electron-hole liquid phase for bilayer {\formula} to understand the thickness-dependent EHL phase. We find that the bilayers of {\formula} exhibit the EHL phase at lower temperatures (214 K for {\formulaN}, 157 K for {\formulaAs}, and 163 K for {\formulaP}) in comparison to their critical temperatures for corresponding monolayers. The phase diagrams are presented in Figure~\ref{ch5.fig6}. However, the EHL phase critical temperatures for the bilayers are still much higher than the those observed from bulk conventional semiconductors~\cite{Jeffries,Keldysh-86}. The increase in layer thickness increases the screening of the Coulomb interaction, decreasing its impact. This leads to decreased exciton binding energy and the critical temperature for forming electron-hole liquid.

More interestingly, we find that the critical density needed to achieve the EHL phase in all three monolayers is easily achievable in experiments \cite{amit-exciton-2}. Amongst the three monolayers, the critical density for the EHL phase is lowest in {\formulaP}. This is a consequence of the lower (electron and hole) effective masses in {\formulaP} (see Table~\ref{ch5.table1}). The qualitative criteria for EHL formation is $n > a_{ex}^{-2}$, where $a_{ex}$ is the exciton Bohr radius which depends on the exciton effective mass ($m^{*}$) and the dielectric constant of the material ($\epsilon$) as $a_{ex} \approx {\epsilon \hbar^2/(m^{*} e^2)}$. Here, $\hbar$ and $e$ are the reduced Planck's constant and electronic charge, respectively. This allows the electrons and holes in {\formulaP}  to condense into a macroscopic EHL phase at a relatively lower density. 

However, in some extreme cases, it should also be possible to completely destabilize the EHL phase, or make it stronger in some cases. Along with using a more accurate lattice model, or more accurate the electron-electron interactions , there are several other factors such as effective masses of electrons and holes, valley degeneracy of the bands that could also play an important role in stabilizing the EHL phase~\cite{Brinkman-72}.

The EHL is known to be a metallic state with large mobility. This is typically reflected in a sharp rise in the photocurrent measurements once the threshold density for EHL formation has been breached~\cite{Arp2019}. The high mobility of an EHL is attributed to the reduction of friction with the crystal lattice caused by the presence of Fermi degenerate charge carriers. This characteristic allows the EHL to move through crystal lattices in the presence of nonuniform deformation, strain, electric/magnetic fields, thermal gradients, radiation pressure, and phonon wind~\cite{MoS2-EHL-EXP-ACS-Nano-2019-Yu, Hensel1978}. The droplet mobility ($\mu_d$) can be interpreted from a phenomenological theory, {\it, i.e.}, $\mu_d$ = $\tau_p/M$, for the droplet mass $M$ and scattering time $\tau_p$, that allows us to establish the relation with microscopic collision processes. The relaxation time and, hence, the mobility increases significantly with decreasing temperature. Considering that the electron and hole represent a degenerate Fermi liquid behavior in the EHL, we can make a very simplifying approximation by approximating the $\tau_p$ with the scattering time for constituent particles or electrons and holes in the metallic state~\cite{Hensel1978}. First-principle-based transport calculations for individual electrons and holes have shown that the electron and hole mobilities in the {\formula} class of materials are relatively large in comparison to the much explored 2D TMDs~\cite{Hong670, Rajibul-Barun-spin, Yin2023}.
\section{Summary}
\label{ch5.sec5}
In summary, we predict the possibility of observing the EHL phase at room temperature in the monolayer {\formula} family. Our calculations show that monolayer {\formulaN}, {\formulaAs}, and {\formulaP} are capable of supporting a stable EHL phase at room temperature with easily achievable photo-excited carrier densities. This is due to the more prominent exchange and correlation effects in 2D systems compared to 3D systems, which helps to stabilize the EHL phase at higher temperatures. The synthetic {\formula} series opens up new avenues for exploring non-equilibrium phase transitions without the limitation of cryogenics. In addition to this prediction, we systematically explore the impact of the variation of effective thickness and the background dielectric constant on the phase diagram of the electron-hole system. 
This highlights the robustness of the EHL phase in these materials. We have also shown the consistency of our results with the Saha ionization formula. This exciting possibility can be experimentally verified through photoluminescence experiments, transient differential absorption spectroscopy~\cite{MOS2-EHP-Bataller2019, MoS2-EHL-EXP-ACS-Nano-2019-Yu} or by photocurrent spectroscopy experiments \cite{Arp2019}. This motivates a more detailed study of potential applications of EHL phase in synthetic monolayer {\formula} series for quantum technology and high-power photonic applications. Given that only a few experimental works explore the EHL phase in 2D MoS$_2$ and other heterostructures, the exploration, search, and systematic study of other 2D materials supporting room temperature EHL phase and their applications has become crucial.

%% file: chap6.tex
\chapter{Summary and outlook}
\label{chap6}
\pagestyle{fancy}
	
In this thesis we have addressed two key problems in the photo-excited insulating solids - (i) the fluence-dependent exciton-exciton interactions, and it's consequence in the form of an EHL phase at room temperature and (ii) temperature-dependent optical response and indirect exciton emission in aluminium nitride monolayer. The theoretical approach adopted by us involved advanced \textit{ab-initio} techniques, including DFT, GW approximation, BSE for electron-hole interactions, and time-dependent BSE for non-equilibrium exciton dynamics. Further, for the optical properties including the electron-phonon interaction using DFPT and indirect emission processs using the photoluminescence theory.  These methods provided a solid foundation for unraveling the intricate interplay between light-matter interactions, electronic structures, and emergent phenomena in quantum materials.
\section{Summary}
Our work in this thesis has twofold objectives. Firstly, to deepen our understanding of fundamentals of the excited state phenomena in quantum materials, specifically in the {\formula} series and planar AlN monolayers. Secondly, to explore the potential applications of these materials in optoelectronics, quantum technology, and high-power laser applications. Theoretical predictions and insights from this research could guide experimental efforts and contribute to the development of novel materials for technological advancements. Next, we briefly summarize our key findings from each chapter of the thesis.\\
	
In Chapter~\ref{chap3}, we have briefly discussed the foundational understanding of the excited state properties within the {\formula} series. Our study focused on monolayers MoSi$_2$N$_4$, MoSi$_2$As$_4$, and MoSi$_2$P$_4$, revealing their potential to host strongly bound excitons with promising applications in optoelectronics. To achieve a more accurate electronic structure, we incorporated QP self-energy corrections in the GWA bandstructure calculations. The inclusion of QP self-energy corrections not only provided distinct bandgaps compared to those obtained from DFT but also highlighted the importance of accounting for screening effects in accurately characterizing the excited states. Taking a step further and considering electron-hole interactions at the GW-BSE level, we validated the existence of multiple bright excitons. This confirmation aligns with and reinforces existing experimental results, underscoring the reliability of our theoretical framework.

We extended our analysis to non-equilibrium dynamics using the time dependent BSE. Through this, we unraveled a redshift-blueshift crossover in the exciton binding energy with increasing electron-hole pair density via pump-fluence. This dynamic behavior was found to follow atom-like interactions among excitons, enhancing our understanding of exciton physics in 2D materials.\\
	
In Chapter~\ref{chap4}, we have utilized first-principle calculations and conducted an in-depth exploration of the electronic and optical properties of a planar AlN monolayer. Our investigation at the GW level revealed a distinctive electronic band structure characterized by an indirect and wide bandgap of 6.73 eV between the $\Gamma$ and $K$ points of the Brillouin Zone (BZ). In tandem, the optical absorption spectrum unveiled an optical bandgap of 4.47 eV, accompanied by a notably large exciton binding energy of 1.83 eV. The analysis of excitonic effects delved into the presence of several bright excitons below the quasiparticle bandgap, manifesting in three prominent absorption peaks at 4.47, 5.26, and 5.90 eV. To comprehensively understand the influence of external factors, we extended our study to explore the impact of lattice vibrations and thermal energy on electronic and optical properties, including optical absorption and Photoluminescence (PL) emission. Our investigation demonstrated the profound effect of electron-phonon (el-ph) interactions, leading to the renormalization of quasiparticle energies. Specifically, the calculated optical absorption spectra showcased a redshift and decrease in dipole oscillator strengths, underscoring the intricate interplay between electronic and vibrational states. Notably, we observed phonon-assisted indirect PL emission within the excitonic framework. Furthermore, the highlighted role of exciton-phonon interactions extends beyond AlN, emphasizing their importance in group-III nitrides and other two-dimensional materials. This work not only advances our understanding of fundamental physical phenomena but also holds implications for the development of 2D material-based optoelectronic devices and solid-state optical applications.

In Chapter~\ref{chap5}, we established potential observation of the EHL phase at room temperature within the monolayer {\formula} family. This prediction marks a significant departure from conventional limitations associated with cryogenic conditions, introducing the exciting prospect of observing non-equilibrium phase transitions in these materials under more practical circumstances. A key highlight of this chapter lies in the stability of the EHL phase at room temperature, and systematic exploration of the robustness of the EHL phase under variations in effective thickness and dielectric enviorment. This analysis contributes a nuanced understanding of the factors influencing the stability of the EHL phase. The predictions made in this chapter offer insights into fundamental quantum phenomena of optically excited electron-hole systems.

\section{Outlook}
Understanding the response of materials under an external perturbation is the key to designing a device for practical applications. Specifically, the photoexcitation processes in insulating solids reveal the characteristics of light absorption, reflection, and transmission. In this thesis, we focused on temperature and exciton density as parameters to investigate the light-matter interactions and nature of exciton dynamics at absolute zero and finite temperatures. We have predicted an atom-like exciton-exciton interactions in the higher photoexcited carrier density limit. However, the strength of exciton-exciton interactions in different semiconducting materials needs to be studied for a deeper understanding of excitonic interactions in other materials~\cite{amit-exciton-2,Sei,PhysRevB.96.115409}. Additionally, the impact of external perturbations such as electric and magnetic fields, strain, and varying dielectric environments can be used to tune the amplitude of exciton-exciton interactions~\cite{PhysRevB.103.045426}. 

Our {\it ab-initio} based calculations have been limited to the monolayers of the discussed 2D materials. However, the fabrication of monolayer materials is challenging, and it is hard to keep them stable for a longer time. Therefore, first-principle excited state calculation for a few-layer or heterostructures of 2D materials could be more helpful to study optical excitations and design optoelectronic devices for application~\cite{Tartakovskii2019}. Heterostructures are also advantageous in observing interlayer excitons, holding promise for the advancement of excitonic integrated circuits. These circuits serve as counterparts to electronic integrated circuits helpful in integrating optical communication and signal processing~\cite{Jiang2021}. Another interesting experimental observation on photoexcited electrons and holes ignored in this thesis is the formation of multiparticle excitations like trions and biexcitons. These multiparticle excitations, exciton-exciton interactions, and excitons coupled with free carriers drive the charge and energy transfer mechanisms in optoelectronic and photovoltaic materials~\cite{Li2018,Plechinger2015}. Since trions and biexcitons are three-body and four-body quantum mechanical problems, respectively, their {\it ab-initio} based calculations are challenging and need methodological development for their study. Recently, a parametrized model Hamiltonian to simulate excitonic complexes has been proposed~\cite{Cho2021}.

The collective interactions among the excitons lead to exotic phases such as EHL, which we studied in one of the projects. The EHL phase represents the merged electron and hole Fermi liquids and excitons, which do not hold the bound characteristic in the liquid phase. For conventional semiconductors, at a very low temperature, the EHL phase was achieved by applied strain ~\cite{Brinkman-72}, whereas, the EHL phase in low-dimensional semiconductors has been recently reported at room temperatures ~\cite{Arp2019,MoS2-EHL-EXP-ACS-Nano-2019-Yu,Dey-MoS2-2023}, and can further be investigated under magnetic fields and strain for their stability at higher temperatures. The impact of electron-phonon interaction on the EHL phase was recently studied by us in  CsPbBr$_3$, CsPbI$_3$, and CsPbBrI$_2$ bulk semiconductor, in a collaborative study in Ref.~\cite{Poonia2023}. However, the role of electron-phonon interaction in observing the EHL phase in 2D materials needs to be explored. Furthermore, in the study of the EHL phase, we have adopted the many-body perturbation theory, where the electronic responce function was treated within the Random Phase approximation. However, the time-dependent BSE can be utilized to incorporate the electron-hole interactions, which can predict accurate exciton-binding energy renormalization as a function of exciton density as studied in Chapter 3 and Ref.~\cite{amit-exciton-2}. 

The stable collective phases, such as the EHL phase, hold the potential for exceptionally high mobility, which may find practical applications in devices for generating and detecting high-power, high-frequency terahertz signals. These devices could be manipulated using electronic and optical control~\cite{Arp2019,MoS2-EHL-EXP-ACS-Nano-2019-Yu,Dey-MoS2-2023}.

Furthermore, since excitons are the bound electron-hole pairs, they are bosonic quasiparticles and exhibit Bose-Einstein condensation~\cite{Morita2022}. Both theoretical and experimental understanding of such charge-neutral excitations and their quantum phases are new avenues for today's research focus. Experimental methods for such phenomena are still in their developing stage~\cite{Kasprzak2006}. They will help elucidate the behaviors of relevant excitations in real materials and aid in the design of new energy materials.

In addition to the exciton-exciton interactions and EHL in 2D ~\formula~ semiconductors, they are also promising candidates for nonlinear optical properties such as second, third, and even higher-order harmonic generations~\cite{SHG-PRB}. The nonlinear phenomena have made an advancement via time-resolved optical spectroscopy. By leveraging time-resolved techniques, one can uncover the temporal evolution of these materials' properties, gaining a deeper understanding of excitation dynamics~\cite{Mogi2022}. These insights contribute to the fundamental understanding of quantum materials and drive advancements in lasers, spectroscopy, and optoelectronic devices, enabling applications in high-speed communication, medical imaging, and quantum technologies. Their precise control of light-matter interactions is pivotal in these fields.

The impact of electron-phonon interaction studied in this thesis was limited to the phonon-mediated emission mechanism with excitonic effects ~\cite{Marini2008,Chen2020}. However, phonon-assisted absorption is still in its early stages of theoretical and computational developments~\cite{Zhang2023}. Specifically, it opens the path to studying microscopic insights into excitonic thermal and dynamical processes ~\cite{Cassabois2016}.

In this thesis, we have reported the finite momentum exciton energies, i.e., the exciton bandstructure, for different 2D semiconductors, mainly to understand the indirect emission processes. However, exciton band structures, which can be probed experimentally through momentum-resolved electron energy-loss spectroscopy~\cite{Hong2020}, need to be studied in detail for exciton transport. Exciton band structures are also promising in understanding the crystal dimensionality and symmeteies~\cite{Qiu2021}, and exciton characteristics ~\cite{exciton_BS_PRL}. 

%% file: thesis_revised.bbl
\begin{thebibliography}{100}

\bibitem{Mueller2018}
T.~Mueller and E.~Malic.
\newblock Exciton physics and device application of two-dimensional transition
  metal dichalcogenide semiconductors.
\newblock {\em npj 2D Materials and Applications}, 2(1):29,  2018.

\bibitem{walker87}
G.~Walker.
\newblock Excitonic Processes in Solids.
\newblock {\em Journal of Modern Optics}, 34(1):2--2, 1987.

\bibitem{Hanke1979-PRL}
W.~Hanke and L.~J. Sham.
\newblock Many-Particle Effects in the Optical Excitations of a Semiconductor.
\newblock {\em Phys. Rev. Lett.}, 43:387--390,  1979.

\bibitem{Kasprzak2006}
J.~Kasprzak, M.~Richard, S.~Kundermann, A.~Baas, P.~Jeambrun, J.~M.~J. Keeling,
  F.~M. Marchetti, M.~H. Szymańska, R.~André, J.~L. Staehli, V.~Savona, P.~B.
  Littlewood, B.~Deveaud, and L.~S. Dang.
\newblock Bose–Einstein condensation of exciton polaritons.
\newblock {\em Nature}, 443(7110):409–414,  2006.

\bibitem{Eisenstein2004}
J.~P. Eisenstein and A.~H. MacDonald.
\newblock Bose--Einstein condensation of excitons in bilayer electron systems.
\newblock {\em Nature}, 432(7018):691--694,  2004.

\bibitem{Knox1983}
R.~S. Knox.
\newblock {\em Introduction to Exciton Physics}, pages 183--245.
\newblock Springer US, Boston, MA, 1983.

\bibitem{Man2021}
M.~K.~L. Man, J.~Madéo, C.~Sahoo, K.~Xie, M.~Campbell, V.~Pareek, A.~Karmakar,
  E.~L. Wong, A.~Al-Mahboob, N.~S. Chan, D.~R. Bacon, X.~Zhu, M.~M.~M.
  Abdelrasoul, X.~Li, T.~F. Heinz, F.~H. da~Jornada, T.~Cao, and K.~M. Dani.
\newblock Experimental measurement of the intrinsic excitonic wave function.
\newblock {\em Science Advances}, 7(17),  2021.

\bibitem{Quintela2022}
M.~F. C.~M. Quintela, J.~C.~G. Henriques, L.~G.~M. Tenório, and N.~M.~R.
  Peres.
\newblock Theoretical Methods for Excitonic Physics in 2D Materials.
\newblock {\em physica status solidi (b)}, 259(7),  2022.

\bibitem{Diana2013}
D.~Y. Qiu, F.~H. da~Jornada, and S.~G. Louie.
\newblock Optical Spectrum of ${\mathrm{MoS}}_{2}$: Many-Body Effects and
  Diversity of Exciton States.
\newblock {\em Phys. Rev. Lett.}, 111:216805,  2013.

\bibitem{Bernardi2013-graphene-light}
M.~Bernardi, M.~Palummo, and J.~C. Grossman.
\newblock Extraordinary Sunlight Absorption and One Nanometer Thick
  Photovoltaics Using Two-Dimensional Monolayer Materials.
\newblock {\em Nano Letters}, 13(8):3664--3670,  2013.

\bibitem{Feldmann1987}
J.~Feldmann, G.~Peter, E.~O. G\"obel, P.~Dawson, K.~Moore, C.~Foxon, and R.~J.
  Elliott.
\newblock Linewidth dependence of radiative exciton lifetimes in quantum wells.
\newblock {\em Phys. Rev. Lett.}, 59:2337--2340,  1987.

\bibitem{BUTOV20172}
L.~Butov.
\newblock Excitonic devices.
\newblock {\em Superlattices and Microstructures}, 108:2--26, 2017.

\bibitem{High07}
A.~A. High, A.~T. Hammack, L.~V. Butov, M.~Hanson, and A.~C. Gossard.
\newblock Exciton optoelectronic transistor.
\newblock {\em Opt. Lett.}, 32(17):2466--2468,  2007.

\bibitem{Alex2008}
A.~A. High, E.~E. Novitskaya, L.~V. Butov, M.~Hanson, and A.~C. Gossard.
\newblock Control of Exciton Fluxes in an Excitonic Integrated Circuit.
\newblock {\em Science}, 321(5886):229--231, 2008.

\bibitem{Grosso2009}
G.~Grosso, J.~Graves, A.~T. Hammack, A.~A. High, L.~V. Butov, M.~Hanson, and
  A.~C. Gossard.
\newblock Excitonic switches operating at around 100 K.
\newblock {\em Nature Photonics}, 3(10):577--580,  2009.

\bibitem{Stier2016}
A.~V. Stier, K.~M. McCreary, B.~T. Jonker, J.~Kono, and S.~A. Crooker.
\newblock Exciton diamagnetic shifts and valley Zeeman effects in monolayer WS2
  and MoS2 to 65{\thinspace}Tesla.
\newblock {\em Nature Communications}, 7(1):10643,  2016.

\bibitem{GonzalezMarin2019}
J.~F. Gonzalez~Marin, D.~Unuchek, K.~Watanabe, T.~Taniguchi, and A.~Kis.
\newblock MoS2 photodetectors integrated with photonic circuits.
\newblock {\em npj 2D Materials and Applications}, 3(1):14,  2019.

\bibitem{Lee2016}
H.~S. Lee, D.~H. Luong, M.~S. Kim, Y.~Jin, H.~Kim, S.~Yun, and Y.~H. Lee.
\newblock Reconfigurable exciton-plasmon interconversion for nanophotonic
  circuits.
\newblock {\em Nature Communications}, 7(1):13663,  2016.

\bibitem{Ciarrocchi2022}
A.~Ciarrocchi, F.~Tagarelli, A.~Avsar, and A.~Kis.
\newblock Excitonic devices with van der Waals heterostructures: valleytronics
  meets twistronics.
\newblock {\em Nature Reviews Materials}, 7(6):449--464,  2022.

\bibitem{lampert1958mobile}
M.~A. Lampert.
\newblock Mobile and immobile effective-mass-particle complexes in nonmetallic
  solids.
\newblock {\em Physical Review Letters}, 1(12):450, 1958.

\bibitem{kheng1993observation}
K.~Kheng, R.~Cox, M.~Y. d’Aubign{\'e}, F.~Bassani, K.~Saminadayar, and
  S.~Tatarenko.
\newblock Observation of negatively charged excitons X- in semiconductor
  quantum wells.
\newblock {\em Phys. Rev. Lett.}, 71(11):1752, 1993.

\bibitem{mak2013tightly}
K.~F. Mak, K.~He, C.~Lee, G.~H. Lee, J.~Hone, T.~F. Heinz, and J.~Shan.
\newblock Tightly bound trions in monolayer MoS2.
\newblock {\em Nature materials}, 12(3):207--211, 2013.

\bibitem{Sei}
E.~J. Sie, A.~Steinhoff, C.~Gies, C.~H. Lui, Q.~Ma, M.~Rösner, G.~Schönhoff,
  F.~Jahnke, T.~O. Wehling, Y.-H. Lee, J.~Kong, P.~Jarillo-Herrero, and
  N.~Gedik.
\newblock Observation of Exciton Redshift–Blueshift Crossover in Monolayer
  WS2.
\newblock {\em Nano Letters}, 17(7):4210--4216, 2017.

\bibitem{MoS2-EHL-EXP-ACS-Nano-2019-Yu}
Y.~Yu, A.~W. Bataller, R.~Younts, Y.~Yu, G.~Li, A.~A. Puretzky, D.~B. Geohegan,
  K.~Gundogdu, and L.~Cao.
\newblock Room-Temperature Electron–Hole Liquid in Monolayer MoS2.
\newblock {\em ACS Nano}, 13(9):10351--10358, 2019.

\bibitem{amit-exciton-2}
S.~K. Bera, M.~Shrivastava, K.~Bramhachari, H.~Zhang, A.~K. Poonia, D.~Mandal,
  E.~M. Miller, M.~C. Beard, A.~Agarwal, and K.~V. Adarsh.
\newblock Atomlike interaction and optically tunable giant band-gap
  renormalization in large-area atomically thin ${\mathrm{MoS}}_{2}$.
\newblock {\em Phys. Rev. B}, 104:L201404,  2021.

\bibitem{Arp2019}
T.~B. Arp, D.~Pleskot, V.~Aji, and N.~M. Gabor.
\newblock Electron{\textendash}hole liquid in a van der Waals heterostructure
  photocell at room temperature.
\newblock {\em Nature Photonics}, 13(4):245--250,  2019.

\bibitem{Dey-MoS2-2023}
P.~Dey, T.~Dixit, V.~Mishra, A.~Sahoo, C.~Vijayan, and S.~Krishnan.
\newblock Emergence and Relaxation of an e–h Quantum Liquid Phase in
  Photoexcited MoS2 Nanoparticles at Room Temperature.
\newblock {\em Advanced Optical Materials}, n/a(n/a):2202567.

\bibitem{optical-MoS2-PRL}
D.~Y. Qiu, F.~H. da~Jornada, and S.~G. Louie.
\newblock Optical Spectrum of ${\mathrm{MoS}}_{2}$: Many-Body Effects and
  Diversity of Exciton States.
\newblock {\em Phys. Rev. Lett.}, 111:216805,  2013.

\bibitem{Davide}
A.~Molina-Sánchez, D.~Sangalli, L.~Wirtz, and A.~Marini.
\newblock Ab Initio Calculations of Ultrashort Carrier Dynamics in
  Two-Dimensional Materials: Valley Depolarization in Single-Layer WSe2.
\newblock {\em Nano Letters}, 17(8):4549--4555, 2017.

\bibitem{Molina2013}
A.~Molina-S{\'a}nchez, D.~Sangalli, K.~Hummer, A.~Marini, and L.~Wirtz.
\newblock Effect of spin-orbit interaction on the optical spectra of
  single-layer, double-layer, and bulk MoS$_2$.
\newblock {\em Phys. Rev. B}, 88:045412, 2013.

\bibitem{Galvani2016}
T.~Galvani, F.~Paleari, H.~P.~C. Miranda, A.~Molina-S{\'a}nchez, L.~Wirtz,
  S.~Latil, H.~Amara, and F.~Ducastelle.
\newblock Excitons in boron nitride single layer.
\newblock {\em Phys. Rev. B}, 94:125303, 2016.

\bibitem{Hanbickia2015}
A.~Hanbickia, M.~Curriea, G.~Kioseogloub, A.~Friedmana, and B.~Jonker.
\newblock Measurement of high exciton binding energy in the monolayer
  transition-metal dichalcogenides WS$_2$and WSe$_2$.
\newblock {\em Solid State Commun.}, 203:16, 2015.

\bibitem{Hill2015}
H.~M. Hill, A.~F. Rigosi, C.~Roquelet, A.~Chernikov, T.~C.Berkelbach, D.~R.
  Reichman, M.~S. Hybertsen, L.~E. Brus, and T.~F. Heinz.
\newblock Observation of excitonic Rydberg states in monolayer MoS$_2$and
  WS$_2$ by photoluminescence excitation spectroscopy.
\newblock {\em Phys. Rev. B}, 15:2992, 2015.

\bibitem{Robert2016}
C.~Robert, R.~Picard, D.~Lagarde, G.~Wang, J.~P. Echeverry, F.~Cadiz,
  P.~Renucci, A.~H\"ogele, T.~Amand, and X.~M. \textit{et al.}
\newblock Excitonic properties of semiconducting monolayer and bilayer
  MoTe$_2$.
\newblock {\em Phys. Rev. B}, 94:155425, 2016.

\bibitem{Thygesen2017}
K.~S. Thygesen.
\newblock Calculating excitons, plasmons, and quasiparticles in 2D materials
  and van der Waals heterostructures.
\newblock {\em 2D Materials}, 4(2):022004,  2017.

\bibitem{Marini2008}
A.~Marini.
\newblock Ab initio finite-temperature excitons.
\newblock {\em Phys. Rev. Lett.}, 101:106405, 2008.

\bibitem{Molina2015}
A.~Molina-S{\`a}nchez, M.~Palummo, A.~Marini, and L.~Wirtz.
\newblock Temperature-dependent excitonic effects in the optical properties of
  single-layer MoS$_2$.
\newblock {\em Phys. Rev. B}, 93:155435, 2016.

\bibitem{Pflug2021}
T.~Pflug, M.~Olbrich, J.~Winter, J.~Schille, U.~L\"{o}schner, H.~Huber, and
  A.~Horn.
\newblock Fluence-Dependent Transient Reflectance of Stainless Steel
  Investigated by Ultrafast Imaging Pump–Probe Reflectometry.
\newblock {\em The Journal of Physical Chemistry C}, 125(31):17363–17371,
  2021.

\bibitem{DRESSELHAUS195614}
G.~Dresselhaus.
\newblock Effective mass approximation for excitons.
\newblock {\em Journal of Physics and Chemistry of Solids}, 1(1):14--22, 1956.

\bibitem{Slater_exciton}
J.~C. Slater and W.~Shockley.
\newblock Optical Absorption by the Alkali Halides.
\newblock {\em Phys. Rev.}, 50:705--719,  1936.

\bibitem{Wannier_exciton}
G.~H. Wannier.
\newblock The Structure of Electronic Excitation Levels in Insulating Crystals.
\newblock {\em Phys. Rev.}, 52:191--197,  1937.

\bibitem{Mott1938}
N.~F. Mott.
\newblock Conduction in polar crystals. II. The conduction band and
  ultra-violet absorption of alkali-halide crystals.
\newblock {\em Transactions of the Faraday Society}, 34:500, 1938.

\bibitem{DFT_GWA_Exciton-GW3}
M.~Rohlfing and S.~G. Louie.
\newblock Electron-hole excitations and optical spectra from first principles.
\newblock {\em Phys. Rev. B}, 62:4927--4944,  2000.

\bibitem{SHIAU2023169431}
S.-Y. Shiau and M.~Combescot.
\newblock A fresh view on Frenkel excitons: Electron–hole pair exchange and
  many-body formalism.
\newblock {\em Annals of Physics}, 458:169431, 2023.

\bibitem{Frenkel1931}
J.~Frenkel.
\newblock On the Transformation of light into Heat in Solids. I.
\newblock {\em Phys. Rev.}, 37:17--44,  1931.

\bibitem{Ng2020}
K.~Ng, M.~Webster, W.~P. Carbery, N.~Visaveliya, P.~Gaikwad, S.~J. Jang,
  I.~Kretzschmar, and D.~M. Eisele.
\newblock Frenkel excitons in heat-stressed supramolecular nanocomposites
  enabled by tunable cage-like scaffolding.
\newblock {\em Nature Chemistry}, 12(12):1157–1164,  2020.

\bibitem{Combescot2015}
M.~Combescot and S.-Y. Shiau.
\newblock {108Frenkel Excitons}.
\newblock In {\em {Excitons and Cooper Pairs: Two Composite Bosons in Many-Body
  Physics}}. Oxford University Press,  2015.

\bibitem{AGRANOVICH2003317}
V.~Agranovich, V.~Yudson, and P.~Reineker.
\newblock Hybridization of Frenkel and Wannier–Mott Excitons in
  Organic-Inorganic Heterostructures. Strong Coupling Regime.
\newblock In {\em Electronic Excitations in Organic Nanostructures}, volume~31
  of {\em Thin Films and Nanostructures}, pages 317--353. Academic Press, 2003.

\bibitem{LAROCCA200397}
G.~La Rocca.
\newblock Wannier–Mott Excitons in Semiconductors.
\newblock In {\em Electronic Excitations in Organic Nanostructures}, volume~31
  of {\em Thin Films and Nanostructures}, pages 97--128. Academic Press, 2003.

\bibitem{dark_exciton_TMD}
E.~Malic, M.~Selig, M.~Feierabend, S.~Brem, D.~Christiansen, F.~Wendler,
  A.~Knorr, and G.~Bergh\"auser.
\newblock Dark excitons in transition metal dichalcogenides.
\newblock {\em Phys. Rev. Mater.}, 2:014002,  2018.

\bibitem{Novoselov2004_graphene1}
K.~S. Novoselov, A.~K. Geim, S.~V. Morozov, D.~Jiang, Y.~Zhang, S.~V. Dubonos,
  I.~V. Grigorieva, and A.~A. Firsov.
\newblock Electric Field Effect in Atomically Thin Carbon Films.
\newblock {\em Science}, 306(5696):666--669, 2004.

\bibitem{Butler2013}
S.~Z. Butler, S.~M. Hollen, L.~Cao, Y.~Cui, J.~A. Gupta, H.~R. Guti{\'e}rrez,
  T.~F. Heinz, S.~S. Hong, J.~Huang, A.~F. Ismach, E.~Johnston-Halperin,
  M.~Kuno, V.~V. Plashnitsa, R.~D. Robinson, R.~S. Ruoff, S.~Salahuddin,
  J.~Shan, L.~Shi, M.~G. Spencer, M.~Terrones, W.~Windl, and J.~E. Goldberger.
\newblock Progress, Challenges, and Opportunities in Two-Dimensional Materials
  Beyond Graphene.
\newblock {\em ACS Nano}, 7(4):2898--2926,  2013.

\bibitem{Mounet2018}
N.~Mounet, M.~Gibertini, P.~Schwaller, D.~Campi, A.~Merkys, A.~Marrazzo,
  T.~Sohier, I.~E. Castelli, A.~Cepellotti, G.~Pizzi, and N.~Marzari.
\newblock Two-dimensional materials from high-throughput computational
  exfoliation of experimentally known compounds.
\newblock {\em Nature Nanotechnology}, 13(3):246--252,  2018.

\bibitem{TMD-BE}
A.~Hanbicki, M.~Currie, G.~Kioseoglou, A.~Friedman, and B.~Jonker.
\newblock Measurement of high exciton binding energy in the monolayer
  transition-metal dichalcogenides WS2 and WSe2.
\newblock {\em Solid State Communications}, 203:16--20, 2015.

\bibitem{Rivera2015}
P.~Rivera, J.~R. Schaibley, A.~M. Jones, J.~S. Ross, S.~Wu, G.~Aivazian,
  P.~Klement, K.~Seyler, G.~Clark, N.~J. Ghimire, J.~Yan, D.~G. Mandrus,
  W.~Yao, and X.~Xu.
\newblock Observation of long-lived interlayer excitons in monolayer
  MoSe2–WSe2 heterostructures.
\newblock {\em Nature Communications}, 6(1),  2015.

\bibitem{Srivastava2015}
A.~Srivastava and A.~Imamoglu.
\newblock Signatures of Bloch-Band Geometry on Excitons: Nonhydrogenic Spectra
  in Transition-Metal Dichalcogenides.
\newblock {\em Phys. Rev. Lett.}, 115:166802,  2015.

\bibitem{Zhou2015}
J.~Zhou, W.-Y. Shan, W.~Yao, and D.~Xiao.
\newblock Berry Phase Modification to the Energy Spectrum of Excitons.
\newblock {\em Phys. Rev. Lett.}, 115:166803,  2015.

\bibitem{Palummo2015}
M.~Palummo, M.~Bernardi, and J.~C. Grossman.
\newblock Exciton Radiative Lifetimes in Two-Dimensional Transition Metal
  Dichalcogenides.
\newblock {\em Nano Lett.}, 5:2794, 2015.

\bibitem{Fogler2014}
M.~M. Fogler, L.~V. Butov, and K.~S. Novoselov.
\newblock High-temperature superfluidity with indirect excitons in van der
  Waals heterostructures.
\newblock {\em Nature Communications}, 5(1),  2014.

\bibitem{Mak2013}
K.~F. Mak, K.~He, C.~Lee, G.~H. Lee, J.~Hone, T.~F. Heinz, and J.~Shan.
\newblock Tightly bound trions in monolayer MoS2.
\newblock {\em Nature Materials}, 12(3):207--211,  2013.

\bibitem{Berkelbach2013}
T.~C. Berkelbach, M.~S. Hybertsen, and D.~R. Reichman.
\newblock Theory of neutral and charged excitons in monolayer transition metal
  dichalcogenides.
\newblock {\em Phys. Rev. B}, 88:045318,  2013.

\bibitem{Huard2000}
V.~Huard, R.~T. Cox, K.~Saminadayar, A.~Arnoult, and S.~Tatarenko.
\newblock Bound States in Optical Absorption of Semiconductor Quantum Wells
  Containing a Two-Dimensional Electron Gas.
\newblock {\em Phys. Rev. Lett.}, 84:187--190,  2000.

\bibitem{klingshirn2012semiconductor}
C.~F. Klingshirn.
\newblock {\em Semiconductor optics}.
\newblock Springer Science \& Business Media, 2012.

\bibitem{kim1994thermodynamics}
J.~Kim, D.~Wake, and J.~Wolfe.
\newblock Thermodynamics of biexcitons in a GaAs quantum well.
\newblock {\em Physical Review B}, 50(20):15099, 1994.

\bibitem{You2015}
Y.~You, X.-X. Zhang, T.~C. Berkelbach, M.~S. Hybertsen, D.~R. Reichman, and
  T.~F. Heinz.
\newblock Observation of biexcitons in monolayer WSe2.
\newblock {\em Nature Physics}, 11(6):477--481,  2015.

\bibitem{Lauten1985}
P.~Lautenschlager, P.~B. Allen, and M.~Cardona.
\newblock Temperature dependence of band gaps in Si and Ge.
\newblock {\em Phys. Rev. B}, 31:2163--2171,  1985.

\bibitem{Molina2016}
A.~Molina-S\'anchez, M.~Palummo, A.~Marini, and L.~Wirtz.
\newblock Temperature-dependent excitonic effects in the optical properties of
  single-layer ${\mathrm{MoS}}_{2}$.
\newblock {\em Phys. Rev. B}, 93:155435,  2016.

\bibitem{Wolpert2011}
D.~Wolpert and P.~Ampadu.
\newblock {\em Temperature Effects in Semiconductors}, page 15–33.
\newblock Springer New York,  2011.

\bibitem{Lautenschlager1985}
P.~Lautenschlager, P.~B. Allen, and M.~Cardona.
\newblock Temperature dependence of band gaps in Si and Ge.
\newblock {\em Phys. Rev. B}, 31:2163, 1985.

\bibitem{Lautenschlager1987}
P.~Lautenschlager, M.~Garriga, L.~Vina, and M.~Cardona.
\newblock Temperature dependence of the dielectric function and interband
  critical points in silicon.
\newblock {\em Phys. Rev. B}, 36:4821, 1987.

\bibitem{Antonius2022}
G.~Antonius and S.~G. Louie.
\newblock Theory of exciton-phonon coupling.
\newblock {\em Phys. Rev. B}, 105:085111,  2022.

\bibitem{giannozzi2005density}
P.~Giannozzi and S.~Baroni.
\newblock Density-functional perturbation theory.
\newblock In {\em Handbook of Materials Modeling}, pages 195--214. Springer,
  2005.

\bibitem{Bernardi2014}
M.~Bernardi, D.~Vigil-Fowler, J.~Lischner, J.~B. Neaton, and S.~G. Louie.
\newblock Ab Initio Study of Hot Carriers in the First Picosecond after
  Sunlight Absorption in Silicon.
\newblock {\em Phys. Rev. Lett.}, 112:257402,  2014.

\bibitem{Watanabe2004}
K.~Watanabe, T.~Taniguchi, and H.~Kanda.
\newblock Direct-bandgap properties and evidence for ultraviolet lasing of
  hexagonal boron nitride single crystal.
\newblock {\em Nat. Mater.}, 3:404, 2004.

\bibitem{Cassabois2016}
G.~Cassabois, P.~Valvin, and B.~Gil.
\newblock Hexagonal boron nitride is an indirect bandgap semiconductor.
\newblock {\em Nature Photonics}, 10(4):262--266,  2016.

\bibitem{Paleari2019}
F.~Paleari, H.~P.~C.~Miranda, A.~Molina-S\'anchez, and L.~Wirtz.
\newblock Exciton-Phonon Coupling in the Ultraviolet Absorption and Emission
  Spectra of Bulk Hexagonal Boron Nitride.
\newblock {\em Phys. Rev. Lett.}, 122:187401,  2019.

\bibitem{Chen2020}
H.-Y. Chen, D.~Sangalli, and M.~Bernardi.
\newblock Exciton-Phonon Interaction and Relaxation Times from First
  Principles.
\newblock {\em Phys. Rev. Lett.}, 125:107401,  2020.

\bibitem{Lee2012}
M.~M. Lee, J.~Teuscher, T.~Miyasaka, T.~N. Murakami, and H.~J. Snaith.
\newblock Efficient Hybrid Solar Cells Based on Meso-Superstructured
  Organometal Halide Perovskites.
\newblock {\em Science}, 338(6107):643–647,  2012.

\bibitem{Frost2014}
J.~M. Frost, K.~T. Butler, F.~Brivio, C.~H. Hendon, M.~van Schilfgaarde, and
  A.~Walsh.
\newblock Atomistic Origins of High-Performance in Hybrid Halide Perovskite
  Solar Cells.
\newblock {\em Nano Letters}, 14(5):2584–2590,  2014.

\bibitem{Giebink2011}
N.~C. Giebink, G.~P. Wiederrecht, M.~R. Wasielewski, and S.~R. Forrest.
\newblock Thermodynamic efficiency limit of excitonic solar cells.
\newblock {\em Phys. Rev. B}, 83:195326,  2011.

\bibitem{keldysh1968}
L.~V. Keldysh.
\newblock Proceedings of the Ninth International Conference on the Physics of
  Semiconductors, Moscow.
\newblock 1968.
\newblock edited by S. M. Ryvkin and V. V. Shmastsev Nauka, Leningrad (1968).

\bibitem{keldysh1986}
L.~V. Keldysh.
\newblock The electron-hole liquid in semiconductors.
\newblock {\em Contemporary Physics}, 27(5):395--428,  1986.

\bibitem{rice-conference}
T.~M. Rice.
\newblock The electron-hole liquid in semiconductors, 1986.

\bibitem{asnin1969}
V.~M. Asnin and P.~R. A.~A.~Rogachev, Zh. Eksp. Teor.~Fiz.
\newblock {\em JETP Lett.}, 9:415, 1969.

\bibitem{Keldysh-Silin-75}
L.~V. Keldysh and A.~P. Silin.
\newblock Electron-hole fluid in polar semiconductors.
\newblock {\em Zh. Eksp. Teor. Fiz.}, 69:1053--1057,  1975.

\bibitem{Wolfe-75}
J.~P. Wolfe, W.~L. Hansen, E.~E. Haller, R.~S. Markiewicz, C.~Kittel, and C.~D.
  Jeffries.
\newblock Photograph of an Electron-Hole Drop in Germanium.
\newblock {\em Phys. Rev. Lett.}, 34:1292--1293,  1975.

\bibitem{Beni-Rice-76}
G.~Beni and T.~M. Rice.
\newblock Electron-Hole Liquids in Polar Semiconductors.
\newblock {\em Phys. Rev. Lett.}, 37:874--877,  1976.

\bibitem{Jeffries}
C.~D. Jeffries.
\newblock Electron-Hole Condensation in Semiconductors.
\newblock {\em Science}, 189(4207):955--964, 1975.

\bibitem{Rice1978book}
F.~S. Henry~Ehrenreich and D.~Turnbull.
\newblock {\em Solid State Physics-32 (Eds.)}.
\newblock Academic Press, 1978.

\bibitem{MOS2-EHP-Bataller2019}
A.~W. Bataller, R.~A. Younts, A.~Rustagi, Y.~Yu, H.~Ardekani, A.~Kemper,
  L.~Cao, and K.~Gundogdu.
\newblock Dense Electron–Hole Plasma Formation and Ultralong Charge Lifetime
  in Monolayer MoS2 via Material Tuning.
\newblock {\em Nano Letters}, 19(2):1104--1111, 2019.

\bibitem{Brinkman-Rice-73}
W.~F. Brinkman and T.~M. Rice.
\newblock Electron-Hole Liquids in Semiconductors.
\newblock {\em Phys. Rev. B}, 7:1508--1523,  1973.

\bibitem{Beni-Rice-78}
G.~Beni and T.~M. Rice.
\newblock Theory of electron-hole liquid in semiconductors.
\newblock {\em Phys. Rev. B}, 18:768--785,  1978.

\bibitem{Barun-exciton-17}
B.~Ghosh, P.~Kumar, A.~Thakur, Y.~S. Chauhan, S.~Bhowmick, and A.~Agarwal.
\newblock Anisotropic plasmons, excitons, and electron energy loss spectroscopy
  of phosphorene.
\newblock {\em Phys. Rev. B}, 96:035422,  2017.

\bibitem{Hohenberg1964}
P.~Hohenberg and W.~Kohn.
\newblock Inhomogeneous Electron Gas.
\newblock {\em Phys. Rev.}, 136:B864--B871,  1964.

\bibitem{Kohn-Sham}
W.~Kohn and L.~J. Sham.
\newblock Self-Consistent Equations Including Exchange and Correlation Effects.
\newblock {\em Phys. Rev.}, 140:A1133--A1138,  1965.

\bibitem{Hybertsen-GW2}
M.~S. Hybertsen and S.~G. Louie.
\newblock Electron correlation in semiconductors and insulators: Band gaps and
  quasiparticle energies.
\newblock {\em Phys. Rev. B}, 34:5390--5413,  1986.

\bibitem{HedinGW1}
L.~Hedin.
\newblock New Method for Calculating the One-Particle Green's Function with
  Application to the Electron-Gas Problem.
\newblock {\em Phys. Rev.}, 139:A796--A823,  1965.

\bibitem{BSE-1-Strinati}
G.~Strinati.
\newblock Dynamical Shift and Broadening of Core Excitons in Semiconductors.
\newblock {\em Phys. Rev. Lett.}, 49:1519--1522,  1982.

\bibitem{BSE-2-Strinati}
G.~Strinati.
\newblock Effects of dynamical screening on resonances at inner-shell
  thresholds in semiconductors.
\newblock {\em Phys. Rev. B}, 29:5718--5726,  1984.

\bibitem{BSE-3-Onida}
G.~Onida, L.~Reining, and A.~Rubio.
\newblock Electronic excitations: density-functional versus many-body
  Green's-function approaches.
\newblock {\em Rev. Mod. Phys.}, 74:601--659,  2002.

\bibitem{Hohenberg-Kohn-1964}
P.~Hohenberg and W.~Kohn.
\newblock Inhomogeneous Electron Gas.
\newblock {\em Phys. Rev.}, 136:B864--B871,  1964.

\bibitem{Hedin-GW1}
L.~Hedin.
\newblock New Method for Calculating the One-Particle Green's Function with
  Application to the Electron-Gas Problem.
\newblock {\em Phys. Rev.}, 139:A796--A823,  1965.

\bibitem{Aryasetiawan1998}
F.~Aryasetiawan and O.~Gunnarsson.
\newblock The GW method.
\newblock {\em Reports on Progress in Physics}, 61(3):237--312,  1998.

\bibitem{Born-Oppen}
M.~Born and R.~Oppenheimer.
\newblock Zur Quantentheorie der Molekeln.
\newblock {\em Annalen der Physik}, 389(20):457--484, 1927.

\bibitem{hartree_1928}
D.~R. Hartree.
\newblock The Wave Mechanics of an Atom with a Non-Coulomb Central Field. Part
  I. Theory and Methods.
\newblock {\em Mathematical Proceedings of the Cambridge Philosophical
  Society}, 24(1):89–110, 1928.

\bibitem{GGA-PBE}
J.~P. Perdew, K.~Burke, and M.~Ernzerhof.
\newblock Generalized Gradient Approximation Made Simple.
\newblock {\em Phys. Rev. Lett.}, 77:3865--3868,  1996.

\bibitem{PBE_sol}
A.~V. Terentjev, L.~A. Constantin, and J.~M. Pitarke.
\newblock Dispersion-corrected PBEsol exchange-correlation functional.
\newblock {\em Phys. Rev. B}, 98:214108,  2018.

\bibitem{Hedin1970}
L.~Hedin and S.~Lundqvist.
\newblock Effects of Electron-Electron and Electron-Phonon Interactions on the
  One-Electron States of Solids.
\newblock In {\em Solid State Physics}, pages 1--181. Elsevier, 1970.

\bibitem{SGLGW1985}
M.~S. Hybertsen and S.~G. Louie.
\newblock First-Principles Theory of Quasiparticles: Calculation of Band Gaps
  in Semiconductors and Insulators.
\newblock {\em Phys. Rev. Lett.}, 55:1418--1421,  1985.

\bibitem{Godby1986}
R.~W. Godby, M.~Schl\"uter, and L.~J. Sham.
\newblock Accurate Exchange-Correlation Potential for Silicon and Its
  Discontinuity on Addition of an Electron.
\newblock {\em Phys. Rev. Lett.}, 56:2415--2418,  1986.

\bibitem{yambo2019}
D.~Sangalli, A.~Ferretti, H.~Miranda, C.~Attaccalite, I.~Marri, E.~Cannuccia,
  P.~Melo, M.~Marsili, F.~Paleari, A.~Marrazzo, G.~Prandini, P.~Bonf{\`{a}},
  M.~O. Atambo, F.~Affinito, M.~Palummo, A.~Molina-S{\'{a}}nchez, C.~Hogan,
  M.~Grüning, D.~Varsano, and A.~Marini.
\newblock Many-body perturbation theory calculations using the yambo code.
\newblock {\em Journal of Physics: Condensed Matter}, 31(32):325902,  2019.

\bibitem{DESLIPPE20121269}
J.~Deslippe, G.~Samsonidze, D.~A. Strubbe, M.~Jain, M.~L. Cohen, and S.~G.
  Louie.
\newblock BerkeleyGW: A massively parallel computer package for the calculation
  of the quasiparticle and optical properties of materials and nanostructures.
\newblock {\em Computer Physics Communications}, 183(6):1269--1289, 2012.

\bibitem{spin-orbit-MoS2-PRB}
A.~Molina-S\'anchez, D.~Sangalli, K.~Hummer, A.~Marini, and L.~Wirtz.
\newblock Effect of spin-orbit interaction on the optical spectra of
  single-layer, double-layer, and bulk MoS${}_{2}$.
\newblock {\em Phys. Rev. B}, 88:045412,  2013.

\bibitem{Purz2022}
T.~L. Purz, E.~W. Martin, W.~G. Holtzmann, P.~Rivera, A.~Alfrey, K.~M. Bates,
  H.~Deng, X.~Xu, and S.~T. Cundiff.
\newblock {Imaging dynamic exciton interactions and coupling in transition
  metal dichalcogenides}.
\newblock {\em The Journal of Chemical Physics}, 156(21):214704,  2022.

\bibitem{moody2015intrinsic}
G.~Moody, C.~Kavir~Dass, K.~Hao, C.-H. Chen, L.-J. Li, A.~Singh, K.~Tran,
  G.~Clark, X.~Xu, G.~Bergh{\"a}user, et~al.
\newblock Intrinsic homogeneous linewidth and broadening mechanisms of excitons
  in monolayer transition metal dichalcogenides.
\newblock {\em Nature communications}, 6(1):8315, 2015.

\bibitem{dey2016optical}
P.~Dey, J.~Paul, Z.~Wang, C.~Stevens, C.~Liu, A.~Romero, J.~Shan, D.~Hilton,
  and D.~Karaiskaj.
\newblock Optical coherence in atomic-monolayer transition-metal
  dichalcogenides limited by electron-phonon interactions.
\newblock {\em Phys. Rev. Lett.}, 116(12):127402, 2016.

\bibitem{selig2016excitonic}
M.~Selig, G.~Bergh{\"a}user, A.~Raja, P.~Nagler, C.~Sch{\"u}ller, T.~F. Heinz,
  T.~Korn, A.~Chernikov, E.~Malic, and A.~Knorr.
\newblock Excitonic linewidth and coherence lifetime in monolayer transition
  metal dichalcogenides.
\newblock {\em Nature communications}, 7(1):13279, 2016.

\bibitem{cadiz2017excitonic}
F.~Cadiz, E.~Courtade, C.~Robert, G.~Wang, Y.~Shen, H.~Cai, T.~Taniguchi,
  K.~Watanabe, H.~Carrere, D.~Lagarde, et~al.
\newblock Excitonic linewidth approaching the homogeneous limit in MoS 2-based
  van der Waals heterostructures.
\newblock {\em Physical Review X}, 7(2):021026, 2017.

\bibitem{CARDONA20053}
M.~Cardona.
\newblock Electron–phonon interaction in tetrahedral semiconductors.
\newblock {\em Solid State Communications}, 133(1):3--18, 2005.

\bibitem{Paszkowicz2002}
W.~Paszkowicz, J.~Pelka, M.~Knapp, T.~Szyszko, and S.~Podsiadlo.
\newblock Lattice parameters and anisotropic thermal expansion of hexagonal
  boron nitride in the 10-297.5 K temperature range.
\newblock {\em Applied Physics A: Materials Science $\&$ Processing},
  75(3):431--435,  2002.

\bibitem{Allen1976}
P.~B. Allen and V.~Heine.
\newblock Theory of the temperature dependence of electronic band structures.
\newblock {\em J. Phys. C}, 9:2305, 1976.

\bibitem{Wang2018}
G.~Wang, A.~Chernikov, M.~M. Glazov, T.~F. Heinz, X.~Marie, T.~Amand, and
  B.~Urbaszek.
\newblock Colloquium : Excitons in atomically thin transition metal
  dichalcogenides.
\newblock {\em Reviews of Modern Physics}, 90(2),  2018.

\bibitem{amit-nanoscale-2018}
A.~Agarwal, M.~S. Vitiello, L.~Viti, A.~Cupolillo, and A.~Politano.
\newblock Plasmonics with two-dimensional semiconductors: from basic research
  to technological applications.
\newblock {\em Nanoscale}, 10:8938--8946, 2018.

\bibitem{strain-exciton-Mos2-PRB}
H.~Shi, H.~Pan, Y.-W. Zhang, and B.~I. Yakobson.
\newblock Quasiparticle band structures and optical properties of strained
  monolayer MoS${}_{2}$ and WS${}_{2}$.
\newblock {\em Phys. Rev. B}, 87:155304,  2013.

\bibitem{Dark-Exciton-PhysRevMaterials.2.014002}
E.~Malic, M.~Selig, M.~Feierabend, S.~Brem, D.~Christiansen, F.~Wendler,
  A.~Knorr, and G.~Bergh\"auser.
\newblock Dark excitons in transition metal dichalcogenides.
\newblock {\em Phys. Rev. Materials}, 2:014002,  2018.

\bibitem{exciton-TMD}
M.~Koperski, M.~R. Molas, A.~Arora, K.~Nogajewski, A.~O. Slobodeniuk,
  C.~Faugeras, and M.~Potemski.
\newblock Optical properties of atomically thin transition metal
  dichalcogenides: observations and puzzles.
\newblock {\em Nanophotonics}, 6(6):1289--1308, 2017.

\bibitem{Schmit1985}
S.~Schmitt-Rink, D.~S. Chemla, and D.~A.~B. Miller.
\newblock Theory of transient excitonic optical nonlinearities in semiconductor
  quantum-well structures.
\newblock {\em Phys. Rev. B}, 32:6601--6609,  1985.

\bibitem{Hong670}
Y.-L. Hong, Z.~Liu, L.~Wang, T.~Zhou, W.~Ma, C.~Xu, S.~Feng, L.~Chen, M.-L.
  Chen, D.-M. Sun, X.-Q. Chen, H.-M. Cheng, and W.~Ren.
\newblock Chemical vapor deposition of layered two-dimensional MoSi2N4
  materials.
\newblock {\em Science}, 369(6504):670--674, 2020.

\bibitem{Wu2021-APL}
Q.~Wu, L.~Cao, Y.~S. Ang, and L.~K. Ang.
\newblock Semiconductor-to-metal transition in bilayer MoSi2N4 and WSi2N4 with
  strain and electric field.
\newblock {\em Applied Physics Letters}, 118(11):113102, 2021.

\bibitem{Cao2021-APL}
L.~Cao, G.~Zhou, Q.~Wang, L.~K. Ang, and Y.~S. Ang.
\newblock Two-dimensional van der Waals electrical contact to monolayer
  MoSi2N4.
\newblock {\em Applied Physics Letters}, 118(1):013106, 2021.

\bibitem{Guo_2020-IOP}
S.-D. Guo, Y.-T. Zhu, W.-Q. Mu, and W.-C. Ren.
\newblock Intrinsic piezoelectricity in monolayer {MSi}2N4 (M = Mo, W, Cr, Ti,
  Zr and Hf).
\newblock {\em {EPL} (Europhysics Letters)}, 132(5):57002,  2020.

\bibitem{Bafekry2020MoSi2N4SA}
A.~Bafekry, M.~Faraji, D.~M. Hoat, M.~Shahrokhi, M.~M. Fadlallah, F.~Shojaei,
  S.~A.~H. Feghhi, M.~Ghergherehchi, and D.~Gogova.
\newblock {MoSi}2N4 single-layer: a novel two-dimensional material with
  outstanding mechanical, thermal, electronic and optical properties.
\newblock {\em Journal of Physics D: Applied Physics}, 54(15):155303,  2021.

\bibitem{keshari}
K.~Nandan, B.~Ghosh, A.~Agarwal, S.~Bhowmick, and Y.~S. Chauhan.
\newblock Two-Dimensional MoSi2N4: An Excellent 2-D Semiconductor for
  Field-Effect Transistors.
\newblock {\em IEEE Transactions on Electron Devices}, 69(1):406--413, 2022.

\bibitem{Yu_2021}
J.~Yu, J.~Zhou, X.~Wan, and Q.~Li.
\newblock High intrinsic lattice thermal conductivity in monolayer {MoSi}2N4.
\newblock {\em New Journal of Physics}, 23(3):033005,  2021.

\bibitem{Yao2021-Nanomaterials}
H.~Yao, C.~Zhang, Q.~Wang, J.~Li, Y.~Yu, F.~Xu, B.~Wang, and Y.~Wei.
\newblock Novel Two-Dimensional Layered MoSi2Z4 (Z = P, As): New Promising
  Optoelectronic Materials.
\newblock {\em Nanomaterials}, 11(3):559, 2021.

\bibitem{SHG-PRB}
L.~Kang and Z.~Lin.
\newblock Second harmonic generation of
  $\mathrm{Mo}{\mathrm{Si}}_{2}{\mathrm{N}}_{4}$-type layers.
\newblock {\em Phys. Rev. B}, 103:195404,  2021.

\bibitem{Valley-1}
S.~Li, W.~Wu, X.~Feng, S.~Guan, W.~Feng, Y.~Yao, and S.~A. Yang.
\newblock Valley-dependent properties of monolayer
  ${\mathrm{MoSi}}_{2}{\mathrm{N}}_{4}, {\mathrm{WSi}}_{2}{\mathrm{N}}_{4}$,
  and ${\mathrm{MoSi}}_{2}{\mathrm{As}}_{4}$.
\newblock {\em Phys. Rev. B}, 102:235435,  2020.

\bibitem{Yang_valley-valley2}
C.~Yang, Z.~Song, X.~Sun, and J.~Lu.
\newblock Valley pseudospin in monolayer
  $\mathrm{Mo}{\mathrm{Si}}_{2}{\mathrm{N}}_{4}$ and
  $\mathrm{Mo}{\mathrm{Si}}_{2}{\mathrm{As}}_{4}$.
\newblock {\em Phys. Rev. B}, 103:035308,  2021.

\bibitem{ai2021theoretical-valley3}
H.~Ai, D.~Liu, J.~Geng, S.-P. Wang, K.~H. Lo, and H.~Pan.
\newblock Theoretical evidence of the spin-valley coupling and valley
  polarization in two-dimensional MoSi2X4 (X= N, P, and As).
\newblock {\em Physical Chemistry Chemical Physics}, 2021.

\bibitem{Rajibul-Barun-spin}
R.~Islam, B.~Ghosh, C.~Autieri, S.~Chowdhury, A.~Bansil, A.~Agarwal, and
  B.~Singh.
\newblock Tunable spin polarization and electronic structure of bottom-up
  synthesized ${\mathrm{MoSi}}_{2}{\mathrm{N}}_{4}$ materials.
\newblock {\em Phys. Rev. B}, 104:L201112,  2021.

\bibitem{exciton-1-wu2021mosi2n4}
Y.~Wu, Z.~Tang, W.~Xia, W.~Gao, F.~Jia, Y.~Zhang, W.~Zhu, W.~Zhang, and
  P.~Zhang.
\newblock Prediction of protected band edge states and dielectric tunable
  quasiparticle and excitonic properties of monolayer MoSi2N4.
\newblock {\em npj Computational Materials}, 8(1):129,  2022.

\bibitem{amit-exciton-1}
M.~Karmakar, S.~Bhattacharya, S.~Mukherjee, B.~Ghosh, R.~K. Chowdhury,
  A.~Agarwal, S.~K. Ray, D.~Chanda, and P.~K. Datta.
\newblock Observation of dynamic screening in the excited exciton states in
  multilayered ${\mathrm{MoS}}_{2}$.
\newblock {\em Phys. Rev. B}, 103:075437,  2021.

\bibitem{Zhong2021}
H.~Zhong, W.~Xiong, P.~Lv, J.~Yu, and S.~Yuan.
\newblock Strain-induced semiconductor to metal transition in $M{A}_{2}Z{}_{4}$
  bilayers ($M=\mathrm{Ti},\mathrm{Cr},\mathrm{Mo}$; $A=\mathrm{Si}$;
  $Z=\mathrm{N},\mathrm{P}$).
\newblock {\em Phys. Rev. B}, 103:085124,  2021.

\bibitem{PhysRevB.54.11169}
G.~Kresse and J.~Furthm\"uller.
\newblock Efficient iterative schemes for ab initio total-energy calculations
  using a plane-wave basis set.
\newblock {\em Phys. Rev. B}, 54:11169--11186,  1996.

\bibitem{PhysRevB.59.1758}
G.~Kresse and D.~Joubert.
\newblock From ultrasoft pseudopotentials to the projector augmented-wave
  method.
\newblock {\em Phys. Rev. B}, 59:1758--1775,  1999.

\bibitem{GGA-1}
J.~P. Perdew and Y.~Wang.
\newblock Accurate and simple analytic representation of the electron-gas
  correlation energy.
\newblock {\em Phys. Rev. B}, 45:13244--13249,  1992.

\bibitem{phonon}
A.~Togo and I.~Tanaka.
\newblock First principles phonon calculations in materials science.
\newblock {\em Scripta Materialia}, 108:1--5, 2015.

\bibitem{QE}
P.~Giannozzi, O.~Andreussi, T.~Brumme, O.~Bunau, M.~B. Nardelli, M.~Calandra,
  R.~Car, C.~Cavazzoni, D.~Ceresoli, M.~Cococcioni, N.~Colonna, I.~Carnimeo,
  A.~D. Corso, S.~de~Gironcoli, P.~Delugas, R.~A. DiStasio, A.~Ferretti,
  A.~Floris, G.~Fratesi, G.~Fugallo, R.~Gebauer, U.~Gerstmann, F.~Giustino,
  T.~Gorni, J.~Jia, M.~Kawamura, H.-Y. Ko, A.~Kokalj, E.~Kü{\c{c}}ükbenli,
  M.~Lazzeri, M.~Marsili, N.~Marzari, F.~Mauri, N.~L. Nguyen, H.-V. Nguyen,
  A.~O. de-la Roza, L.~Paulatto, S.~Ponc{\'{e}}, D.~Rocca, R.~Sabatini,
  B.~Santra, M.~Schlipf, A.~P. Seitsonen, A.~Smogunov, I.~Timrov,
  T.~Thonhauser, P.~Umari, N.~Vast, X.~Wu, and S.~Baroni.
\newblock Advanced capabilities for materials modelling with Quantum
  {ESPRESSO}.
\newblock {\em Journal of Physics: Condensed Matter}, 29(46):465901,  2017.

\bibitem{monkhorst}
H.~J. Monkhorst and J.~D. Pack.
\newblock Special points for Brillouin-zone integrations.
\newblock {\em Phys. Rev. B}, 13:5188--5192,  1976.

\bibitem{yambo20091392}
A.~Marini, C.~Hogan, M.~Grüning, and D.~Varsano.
\newblock yambo: An ab initio tool for excited state calculations.
\newblock {\em Computer Physics Communications}, 180(8):1392--1403, 2009.

\bibitem{Pulci1998-RIM}
O.~Pulci, G.~Onida, R.~Del~Sole, and L.~Reining.
\newblock Ab Initio Calculation of Self-Energy Effects on Optical Properties of
  GaAs(110).
\newblock {\em Phys. Rev. Lett.}, 81:5374--5377,  1998.

\bibitem{Rozzi2006-Coulomb-truncation}
C.~A. Rozzi, D.~Varsano, A.~Marini, E.~K.~U. Gross, and A.~Rubio.
\newblock Exact Coulomb cutoff technique for supercell calculations.
\newblock {\em Phys. Rev. B}, 73:205119,  2006.

\bibitem{PPA}
R.~W. Godby and R.~J. Needs.
\newblock Metal-insulator transition in Kohn-Sham theory and quasiparticle
  theory.
\newblock {\em Phys. Rev. Lett.}, 62:1169--1172,  1989.

\bibitem{Sun2022}
M.~Sun, M.~R. Fiorentin, U.~Schwingenschl\"{o}gl, and M.~Palummo.
\newblock Excitons and light-emission in semiconducting {MoSi}2X4
  two-dimensional materials.
\newblock {\em npj 2D Materials and Applications}, 6(1),  2022.

\bibitem{Marini}
C.~Attaccalite, M.~Gr\"uning, and A.~Marini.
\newblock Real-time approach to the optical properties of solids and
  nanostructures: Time-dependent Bethe-Salpeter equation.
\newblock {\em Phys. Rev. B}, 84:245110,  2011.

\bibitem{SOC-effect-MoS2-PRL}
D.~Xiao, G.-B. Liu, W.~Feng, X.~Xu, and W.~Yao.
\newblock Coupled Spin and Valley Physics in Monolayers of ${\mathrm{MoS}}_{2}$
  and Other Group-VI Dichalcogenides.
\newblock {\em Phys. Rev. Lett.}, 108:196802,  2012.

\bibitem{CrBr3-PRM2022}
M.~Wu, Z.~Li, and S.~G. Louie.
\newblock Optical and magneto-optical properties of ferromagnetic monolayer
  ${\mathrm{CrBr}}_{3}$: A first-principles $GW$ and $GW$ plus Bethe-Salpeter
  equation study.
\newblock {\em Phys. Rev. Materials}, 6:014008,  2022.

\bibitem{Wang2013}
Q.~Wang, S.~Ge, X.~Li, J.~Qiu, Y.~Ji, J.~Feng, and D.~Sun.
\newblock Valley Carrier Dynamics in Monolayer Molybdenum Disulfide from
  Helicity-Resolved Ultrafast Pump--Probe Spectroscopy.
\newblock {\em ACS Nano}, 7(12):11087--11093,  2013.

\bibitem{Exciton_band_mos2_PRB}
F.~Wu, F.~Qu, and A.~H. MacDonald.
\newblock Exciton band structure of monolayer ${\mathrm{MoS}}_{2}$.
\newblock {\em Phys. Rev. B}, 91:075310,  2015.

\bibitem{exciton_BS_C3N}
M.~Bonacci, M.~Zanfrognini, E.~Molinari, A.~Ruini, M.~J. Caldas, A.~Ferretti,
  and D.~Varsano.
\newblock Excitonic effects in graphene-like ${\mathrm{C}}_{3}\mathrm{N}$.
\newblock {\em Phys. Rev. Materials}, 6:034009,  2022.

\bibitem{exciton_BS_hBN_PRB}
L.~Sponza, H.~Amara, C.~Attaccalite, S.~Latil, T.~Galvani, F.~Paleari,
  L.~Wirtz, and F.~m.~c. Ducastelle.
\newblock Direct and indirect excitons in boron nitride polymorphs: A story of
  atomic configuration and electronic correlation.
\newblock {\em Phys. Rev. B}, 98:125206,  2018.

\bibitem{exciton_BS_PRM_Daniele}
M.~Bonacci, M.~Zanfrognini, E.~Molinari, A.~Ruini, M.~J. Caldas, A.~Ferretti,
  and D.~Varsano.
\newblock Excitonic effects in graphene-like ${\mathrm{C}}_{3}\mathrm{N}$.
\newblock {\em Phys. Rev. Materials}, 6:034009,  2022.

\bibitem{exciton_BS_PRL}
P.~Cudazzo, L.~Sponza, C.~Giorgetti, L.~Reining, F.~Sottile, and M.~Gatti.
\newblock Exciton Band Structure in Two-Dimensional Materials.
\newblock {\em Phys. Rev. Lett.}, 116:066803,  2016.

\bibitem{exciton_wf_ws2}
Z.~Ye, T.~Cao, K.~O'Brien, H.~Zhu, X.~Yin, Y.~Wang, S.~G. Louie, and X.~Zhang.
\newblock Probing excitonic dark states in single-layer tungsten disulphide.
\newblock {\em Nature}, 513(7517):214--218,  2014.

\bibitem{hBN_excitonWF}
B.~Arnaud, S.~Leb\`egue, P.~Rabiller, and M.~Alouani.
\newblock Huge Excitonic Effects in Layered Hexagonal Boron Nitride.
\newblock {\em Phys. Rev. Lett.}, 96:026402,  2006.

\bibitem{hBN_excitonWF2}
S.~Galambosi, L.~Wirtz, J.~A. Soininen, J.~Serrano, A.~Marini, K.~Watanabe,
  T.~Taniguchi, S.~Huotari, A.~Rubio, and K.~H\"am\"al\"ainen.
\newblock Anisotropic excitonic effects in the energy loss function of
  hexagonal boron nitride.
\newblock {\em Phys. Rev. B}, 83:081413,  2011.

\bibitem{Roth}
S.~Roth, A.~Crepaldi, M.~Puppin, G.~Gatti, D.~Bugini, I.~Grimaldi, T.~R.
  Barrilot, C.~A. Arrell, F.~Frassetto, L.~Poletto, M.~Chergui, A.~Marini, and
  M.~Grioni.
\newblock Photocarrier-induced band-gap renormalization and ultrafast charge
  dynamics in black phosphorus.
\newblock {\em 2D Materials}, 6(3):031001,  2019.

\bibitem{Pedio}
G.~D. Tsibidis, L.~Mouchliadis, M.~Pedio, and E.~Stratakis.
\newblock Modeling ultrafast out-of-equilibrium carrier dynamics and relaxation
  processes upon irradiation of hexagonal silicon carbide with femtosecond
  laser pulses.
\newblock {\em Phys. Rev. B}, 101:075207,  2020.

\bibitem{Brem2020}
S.~Brem, A.~Ekman, D.~Christiansen, F.~Katsch, M.~Selig, C.~Robert, X.~Marie,
  B.~Urbaszek, A.~Knorr, and E.~Malic.
\newblock Phonon-Assisted Photoluminescence from Indirect Excitons in
  Monolayers of Transition-Metal Dichalcogenides.
\newblock {\em Nano Letters}, 20(4):2849--2856, 2020.

\bibitem{Monserrat2014}
B.~Monserrat and R.~J. Needs.
\newblock Comparing electron-phonon coupling strength in diamond, silicon, and
  silicon carbide: First-principles study.
\newblock {\em Phys. Rev. B}, 89:214304(1)--(8), 2014.

\bibitem{Shen2020}
T.~Shen, X.-W. Zhang, H.~Shang, M.-Y. Zhang, X.~Wang, E.-G. Wang, H.~Jiang, and
  X.-Z. Li.
\newblock Influence of high-energy local orbitals and electron-phonon
  interactions on the band gaps and optical absorption spectra of hexagonal
  boron nitride.
\newblock {\em Phys. Rev. B}, 102:045117, 2020.

\bibitem{Cannucciaarxive}
E.~Cannuccia and A.~Marini.
\newblock Ab-initio study of the effects induced by the electron-phonon
  scattering in carbon based nanostructures.
\newblock {\em arXiv preprint arXiv:1304.0072}, 2013.

\bibitem{Zhuang13}
H.~L. Zhuang, A.~K. Singh, and R.~G. Hennig.
\newblock Computational discovery of single-layer III-V materials.
\newblock {\em Phys. Rev. B}, 87:165415,  2013.

\bibitem{Michael2018}
M.~C. Lucking, W.~Xie, D.-H. Choe, D.~West, T.-M. Lu, and S.~B. Zhang.
\newblock Traditional Semiconductors in the Two-Dimensional Limit.
\newblock {\em Phys. Rev. Lett.}, 120:086101, 2018.

\bibitem{Jiongyao2017}
J.~Wu, Y.~Yang, H.~Gao, Y.~Qi, J.~Zhang, Z.~Qiao, and W.~Ren.
\newblock Electric field effect of GaAs monolayer from first principles.
\newblock {\em AIP Adv.}, 7:035218, 2017.

\bibitem{Bourret1998}
A.~Bourret, A.~Barski, J.~L. Rouvi{\`{e}}re, G.~Renaud, and A.~Barbier.
\newblock Growth of aluminum nitride on (111) silicon: Microstructure and
  interface structure.
\newblock {\em Journal of Applied Physics}, 83(4):2003--2009,  1998.

\bibitem{Qi2021}
J.~Qi, L.~Wang, Y.~Zhang, X.~Guo, W.~Yu, Q.~Wang, K.~Zhang, P.~Ren, and M.~Wen.
\newblock Amorphous {AlN} nanolayer thickness dependent toughness, thermal
  stability and oxidation resistance in {TaN}/{AlN} nanomultilayer films.
\newblock {\em Surface and Coatings Technology}, 405:126724,  2021.

\bibitem{AlN-exciton1}
C.~Attaccalite, M.~S. Prete, M.~Palummo, and O.~Pulci.
\newblock Interlayer and Intralayer Excitons in AlN/WS2 Heterostructure.
\newblock {\em Materials}, 15(23), 2022.

\bibitem{AlN-optical-1}
M.-I. Choe and K.-H. Kim.
\newblock Second-Order Nonlinear Optical Responses of {AlN} Two-Dimensional
  Monolayer: A Real-Time First-Principles Study.
\newblock {\em {ChemPhysChem}}, 23(13),  2022.

\bibitem{AlN-optical-2}
D.~V. Fakhrabad, N.~Shahtahmasebi, and M.~Ashhadi.
\newblock Optical excitations and quasiparticle energies in the {AlN} monolayer
  honeycomb structure.
\newblock {\em Superlattices and Microstructures}, 79:38--44,  2015.

\bibitem{Luo2020}
N.~Luo, W.~Duan, B.~I. Yakobson, and X.~Zou.
\newblock Excitons and Electron{\textendash}Hole Liquid State in 2D
  $\gamma$-Phase Group-{IV} Monochalcogenides.
\newblock {\em Advanced Functional Materials}, 30(19):2000533,  2020.

\bibitem{Strite1992}
S.~Strite.
\newblock {GaN}, {AlN}, and {InN}: A review.
\newblock {\em Journal of Vacuum Science $\&$ Technology B: Microelectronics
  and Nanometer Structures}, 10(4):1237,  1992.

\bibitem{Jain2000}
S.~C. Jain, M.~Willander, J.~Narayan, and R.~V. Overstraeten.
\newblock {III}{\textendash}nitrides: Growth, characterization, and properties.
\newblock {\em Journal of Applied Physics}, 87(3):965--1006,  2000.

\bibitem{Zhang2007}
X.~Zhang, Z.~Liu, and S.~Hark.
\newblock Synthesis and optical characterization of single-crystalline {AlN}
  nanosheets.
\newblock {\em Solid State Communications}, 143(6-7):317--320,  2007.

\bibitem{Yan2010}
X.~Yan, Y.~Dong, H.~Li, J.~Gong, and C.~Sun.
\newblock Synthesis of aluminum nitride thin films by filtered arc ion plating
  deposition.
\newblock {\em Materials Letters}, 64(11):1261--1263,  2010.

\bibitem{Alizadeh2017}
M.~Alizadeh, B.~T. Goh, and S.~A. Rahman.
\newblock Controlled Growth of Conductive {AlN} Thin Films by Plasma-Assisted
  Reactive Evaporation.
\newblock {\em Metallurgical and Materials Transactions A}, 48(7):3461--3469,
  2017.

\bibitem{Dahal2009}
R.~Dahal, B.~Pantha, J.~Li, J.~Y. Lin, and H.~X. Jiang.
\newblock {InGaN}/{GaN} multiple quantum well solar cells with long operating
  wavelengths.
\newblock {\em Applied Physics Letters}, 94(6),  2009.

\bibitem{Jiang2017-GaN}
C.~Jiang, L.~Jing, X.~Huang, M.~Liu, C.~Du, T.~Liu, X.~Pu, W.~Hu, and Z.~L.
  Wang.
\newblock Enhanced Solar Cell Conversion Efficiency of {InGaN}/{GaN} Multiple
  Quantum Wells by Piezo-Phototronic Effect.
\newblock {\em {ACS} Nano}, 11(9):9405--9412,  2017.

\bibitem{Jiang2016}
L.~Jiang, J.~Liu, A.~Tian, Y.~Cheng, Z.~Li, L.~Zhang, S.~Zhang, D.~Li,
  M.~Ikeda, and H.~Yang.
\newblock {GaN}-based green laser diodes.
\newblock {\em Journal of Semiconductors}, 37(11):111001,  2016.

\bibitem{Zhao2017}
D.~Zhao, J.~Yang, Z.~Liu, P.~Chen, J.~Zhu, D.~Jiang, Y.~Shi, H.~Wang, L.~Duan,
  L.~Zhang, and H.~Yang.
\newblock Fabrication of room temperature continuous-wave operation {GaN}-based
  ultraviolet laser diodes.
\newblock {\em Journal of Semiconductors}, 38(5):051001,  2017.

\bibitem{VanHove1997}
J.~M.~V. Hove, R.~Hickman, J.~J. Klaassen, P.~P. Chow, and P.~P. Ruden.
\newblock Ultraviolet-sensitive, visible-blind {GaN} photodiodes fabricated by
  molecular beam epitaxy.
\newblock {\em Applied Physics Letters}, 70(17):2282--2284,  1997.

\bibitem{Morko1998}
H.~Morko{\c{c}}, R.~Cingolani, W.~Lambrecht, B.~Gil, H.-X. Jiang, J.~Lin,
  D.~Pavlidis, and K.~Shenai.
\newblock Material Properties of {GaN} in the Context of Electron Devices.
\newblock {\em {MRS} Proceedings}, 537, 1998.

\bibitem{Nakamura2009}
S.~Nakamura.
\newblock Current Status of {GaN}-Based Solid-State Lighting.
\newblock {\em {MRS} Bulletin}, 34(2):101--107,  2009.

\bibitem{Wang2021-JMCC}
Z.~Wang, G.~Wang, X.~Liu, S.~Wang, T.~Wang, S.~Zhang, J.~Yu, G.~Zhao, and
  L.~Zhang.
\newblock Two-dimensional wide band-gap nitride semiconductor {GaN} and {AlN}
  materials: properties, fabrication and applications.
\newblock {\em Journal of Materials Chemistry C}, 9(48):17201--17232, 2021.

\bibitem{ZhuangAlN}
H.~L. Zhuang, A.~K. Singh, and R.~G. Hennig.
\newblock Computational discovery of single-layer III-V materials.
\newblock {\em Phys. Rev. B}, 87:165415,  2013.

\bibitem{Paleari2018}
F.~Paleari, T.~Galvani, H.~Amara, F.~Du\c{c}astelle, A.~Molina-S{\'a}nchez, and
  L.~Wirtz.
\newblock Excitons in few-layer hexagonal boron nitride: Davydov splitting and
  surface localization.
\newblock {\em 2D Mater.}, 5:045017, 2018.

\bibitem{Henrique2017}
H.~P.~C. Miranda, S.~Reichardt, G.~Froehlicher, A.~Molina-S{\'a}nchez,
  S.~Berciaud, and L.~Wirtz.
\newblock Excitons in few-layer hexagonal boron nitride: Davydov splitting and
  surface localization.
\newblock {\em Nano Lett.}, 17:2381, 2017.

\bibitem{Park2018}
S.~Park, N.~Mutz, T.~Schultz, S.~Blumstengel, A.~Han, A.~Aljarb, L.-J. Li,
  E.~J.~W. List-Kratochvil, P.~Amsalem, and N.~Koch.
\newblock Direct determination of monolayer MoS$_2$ and WSe$_2$ exciton binding
  energies on insulating and metallic substrate.
\newblock {\em 2D Mater.}, 5:025003, 2018.

\bibitem{Allen1983}
P.~B. Allen and M.~Cardona.
\newblock Temperature dependence of the direct gap of Si and Ge.
\newblock {\em Phys. Rev. B}, 27:4760, 1983.

\bibitem{Fan1950}
H.~Y. Fan.
\newblock Temperature Dependence of the Energy Gap in Monatomic Semiconductors.
\newblock {\em Phys. Rev.}, 6:808, 1950.

\bibitem{Zollner1992}
S.~Zollner, M.~Cardona, and S.~Gopalan.
\newblock Isotope and temperature shifts of direct and indirect band gaps in
  diamond-type semiconductors.
\newblock {\em Phys. Rev. B}, 45:3376, 1992.

\bibitem{Mahan2014}
G.~D. Mahan.
\newblock {\em Many-Particle Physics}.
\newblock Springer International Edition, New York, United States of America, 3
  edition, 2014.

\bibitem{Sternheimer1954}
R.~Sternheimer.
\newblock Temperature dependence of the electronic structure of semiconductors
  and insulators.
\newblock {\em Phys. Rev.}, 96:951, 1954.

\bibitem{Chiara2020}
C.~Trovatello, H.~P.~C. Miranda, A.~Molina-Sanchez, R.~Borrego-Varillas,
  C.~Manzoni, L.~Moretti, L.~Ganzer, M.~Maiuri, J.~Wang, D.~Dumcenco, A.~Kis,
  L.~Wirtz, A.~Marini, G.~Soavi, A.~C. Ferrari, G.~Cerullo, D.~Sangalli, and
  S.~D. Conte.
\newblock Strongly Coupled Coherent Phonons in Single-Layer MoS2.
\newblock {\em {ACS} Nano}, 14(5):5700--5710,  2020.

\bibitem{Chan23}
Y.~hao Chan, J.~B. Haber, M.~H. Naik, J.~B. Neaton, D.~Y. Qiu, F.~H.
  da~Jornada, and S.~G. Louie.
\newblock Exciton Lifetime and Optical Line Width Profile via Exciton-Phonon
  Interactions: Theory and First-Principles Calculations for Monolayer
  {MoS}$_2$.
\newblock {\em Nano Letters}, 23(9):3971--3977,  2023.

\bibitem{HimaniWSe2}
H.~Mishra, A.~Bose, A.~Dhar, and S.~Bhattacharya.
\newblock Exciton-phonon coupling and band-gap renormalization in monolayer
  ${\mathrm{WSe}}_{2}$.
\newblock {\em Phys. Rev. B}, 98:045143,  2018.

\bibitem{Mishra2019}
H.~Mishra and S.~Bhattacharya.
\newblock Giant exciton-phonon coupling and zero-point renormalization in
  hexagonal monolayer boron nitride.
\newblock {\em Phys. Rev. B}, 99:165201, 2019.

\bibitem{Kolos2021}
M.~Kolos, L.~Cigarini, R.~Verma, F.~Karlický, and S.~Bhattacharya.
\newblock Giant Linear and Nonlinear Excitonic Responses in an Atomically Thin
  Indirect Semiconductor Nitrogen Phosphide.
\newblock {\em The Journal of Physical Chemistry C}, 125(23):12738--12757,
  2021.

\bibitem{cannuccia2011effect}
E.~Cannuccia and A.~Marini.
\newblock Effect of the quantum zero-point atomic motion on the optical and
  electronic properties of diamond and trans-polyacetylene.
\newblock {\em Phys. Rev. Lett.}, 107(25):255501, 2011.

\bibitem{Selig2016}
M.~Selig, G.~Bergh{\"a}user, A.~Raja, P.~Nagler, C.~Sch{\"u}ller, T.~F. Heinz,
  T.~Korn, A.~Chernikov, E.~Malic, and A.~Knorr.
\newblock Excitonic linewidth and coherence lifetime in monolayer transition
  metal dichalcogenides.
\newblock {\em Nat. Comm.}, 7:13279, 2016.

\bibitem{BSE-4-Paleari}
F.~Paleari, H.~P.~C.~Miranda, A.~Molina-S\'anchez, and L.~Wirtz.
\newblock Exciton-Phonon Coupling in the Ultraviolet Absorption and Emission
  Spectra of Bulk Hexagonal Boron Nitride.
\newblock {\em Phys. Rev. Lett.}, 122:187401,  2019.

\bibitem{hamann2013optimized}
D.~Hamann.
\newblock Optimized norm-conserving Vanderbilt pseudopotentials.
\newblock {\em Phys. Rev. B}, 88(8):085117, 2013.

\bibitem{Sangalli2019}
D.~Sangalli, A.~Ferretti, H.~Miranda, C.~Attaccalite, I.~Marri, E.~Cannuccia,
  P.~Melo, M.~Marsili, F.~Paleari, and A.~M. \textit{et al.}
\newblock Many-body perturbation theory calculations using the yambo code.
\newblock {\em J. Phys.: Condens. Matter}, 31:325902, 2019.

\bibitem{Godby1989}
R.~W. Godby and R.~J. Needs.
\newblock Metal-insulator transition in Kohn-Sham theory and quasiparticle
  theory.
\newblock {\em Phys. Rev. Lett.}, 62:1169, 1989.

\bibitem{Cannuccia2011}
E.~Cannuccia.
\newblock {\em Giant polaronic effects in polymers: breakdown of the
  quasiparticle picture}.
\newblock PhD thesis, Rome Tor Vergata University, 2011.

\bibitem{Brinkman-72}
W.~F. Brinkman, T.~M. Rice, P.~W. Anderson, and S.~T. Chui.
\newblock Metallic State of the Electron-Hole Liquid, Particularly in
  Germanium.
\newblock {\em Phys. Rev. Lett.}, 28:961--964,  1972.

\bibitem{landau}
L.~D. LANDAU and E.~M. LIFSHITZ.
\newblock Fluid mechanics, 1993.

\bibitem{Shah-1977}
J.~Shah, R.~F. Leheny, W.~R. Harding, and D.~R. Wight.
\newblock Observation of Electron-Hole Liquid in GaP.
\newblock {\em Phys. Rev. Lett.}, 38:1164--1167,  1977.

\bibitem{Simon-1992}
A.~H. Simon, S.~J. Kirch, and J.~P. Wolfe.
\newblock Excitonic phase diagram in unstressed Ge.
\newblock {\em Phys. Rev. B}, 46:10098--10112,  1992.

\bibitem{Vashishta-PRL-1974}
P.~Vashishta, S.~G. Das, and K.~S. Singwi.
\newblock Thermodynamics of the Electron-Hole Liquid in Ge, Si, and GaAs.
\newblock {\em Phys. Rev. Lett.}, 33:911--914,  1974.

\bibitem{Berciaud2019}
S.~Berciaud.
\newblock Quasi-two-dimensional electron{\textendash}hole droplets.
\newblock {\em Nature Photonics}, 13(4):225--226,  2019.

\bibitem{MoS2-nano-lett}
A.~Rustagi and A.~F. Kemper.
\newblock Theoretical Phase Diagram for the Room-Temperature Electron--Hole
  Liquid in Photoexcited Quasi-Two-Dimensional Monolayer MoS2.
\newblock {\em Nano Letters}, 18(1):455--459,  2018.

\bibitem{Pekh2020}
P.~L. Pekh, P.~V. Ratnikov, and A.~P. Silin.
\newblock Electron-Hole Liquid in Monolayer Transition Metal Dichalcogenide
  Heterostructures.
\newblock {\em JETP Letters}, 111(2):90--95,  2020.

\bibitem{pekh2021phase-1}
P.~L. Pekh, P.~V. Ratnikov, and A.~P. Silin.
\newblock Phase Diagram of Electron--Hole Liquid in Monolayer Heterostructures
  Based on Transition Metal Dichalcogenides.
\newblock {\em Journal of Experimental and Theoretical Physics},
  133(4):494--507,  2021.

\bibitem{saha}
M.~N.~S. D.Sc.
\newblock LIII. Ionization in the solar chromosphere.
\newblock {\em The London, Edinburgh, and Dublin Philosophical Magazine and
  Journal of Science}, 40(238):472--488, 1920.

\bibitem{Valley-pseudospin-PRB}
C.~Yang, Z.~Song, X.~Sun, and J.~Lu.
\newblock Valley pseudospin in monolayer
  $\mathrm{Mo}{\mathrm{Si}}_{2}{\mathrm{N}}_{4}$ and
  $\mathrm{Mo}{\mathrm{Si}}_{2}{\mathrm{As}}_{4}$.
\newblock {\em Phys. Rev. B}, 103:035308,  2021.

\bibitem{nanotech-MoSi2As4}
H.~Yao, C.~Zhang, Q.~Wang, J.~Li, Y.~Yu, F.~Xu, B.~Wang, and Y.~Wei.
\newblock Novel Two-Dimensional Layered MoSi2Z4 (Z = P, As): New Promising
  Optoelectronic Materials.
\newblock {\em Nanomaterials}, 11(3):559, 2021.

\bibitem{Giuliani-Giovanni-Vignale}
G.~F. Giuliani and G.~Vignale.
\newblock {\em Quantum Theory of the Electron Liquid}.
\newblock Cambridge U. Press, New York, 2005.

\bibitem{Bergersen_1975}
B.~Bergersen, P.~Jena, and A.~J. Berlinsky.
\newblock Electron-hole liquid in heavily doped n-type ge and si.
\newblock {\em Journal of Physics C: Solid State Physics}, 8(9):1377,  1975.

\bibitem{Angel-Rubio2011_r0}
P.~Cudazzo, I.~V. Tokatly, and A.~Rubio.
\newblock Dielectric screening in two-dimensional insulators: Implications for
  excitonic and impurity states in graphane.
\newblock {\em Phys. Rev. B}, 84:085406,  2011.

\bibitem{MacDonald_r0}
F.~Wu, F.~Qu, and A.~H. MacDonald.
\newblock Exciton band structure of monolayer ${\mathrm{MoS}}_{2}$.
\newblock {\em Phys. Rev. B}, 91:075310,  2015.

\bibitem{Combescot_1972}
M.~Combescot and P.~Nozieres.
\newblock Condensation of excitons in germanium and silicon.
\newblock {\em Journal of Physics C: Solid State Physics}, 5(17):2369,  1972.

\bibitem{Simon-2002}
R.~Shimano, M.~Nagai, K.~Horiuch, and M.~Kuwata-Gonokami.
\newblock Formation of a High ${T}_{c}$ Electron-Hole Liquid in Diamond.
\newblock {\em Phys. Rev. Lett.}, 88:057404,  2002.

\bibitem{Thomas-73}
G.~A. Thomas, T.~G. Phillips, T.~M. Rice, and J.~C. Hensel.
\newblock Temperature-Dependent Luminescence from the Electron-Hole Liquid in
  Ge.
\newblock {\em Phys. Rev. Lett.}, 31:386--389,  1973.

\bibitem{thomas-PRL-1974}
G.~A. Thomas, T.~M. Rice, and J.~C. Hensel.
\newblock Liquid-Gas Phase Diagram of an Electron-Hole Fluid.
\newblock {\em Phys. Rev. Lett.}, 33:219--222,  1974.

\bibitem{BGR-Kalt-1992}
H.~Kalt and M.~Rinker.
\newblock Band-gap renormalization in semiconductors with multiple inequivalent
  valleys.
\newblock {\em Phys. Rev. B}, 45:1139--1154,  1992.

\bibitem{Tikhodeev_1985}
S.~G. Tikhodeev.
\newblock The electron-hole liquid in a semiconductor.
\newblock {\em Soviet Physics Uspekhi}, 28(1):1,  1985.

\bibitem{Keldysh-86}
L.~V. Keldysh.
\newblock The electron-hole liquid in semiconductors.
\newblock {\em Contemporary Physics}, 27(5):395--428, 1986.

\bibitem{Hensel1978}
J.~Hensel, T.~Phillips, and G.~Thomas.
\newblock The Electron-Hole Liquid in Semiconductors: Experimental Aspects.
\newblock In {\em Solid State Physics}, pages 87--314. Elsevier, 1978.

\bibitem{Yin2023}
Y.~Yin, Q.~Gong, M.~Yi, and W.~Guo.
\newblock Emerging Versatile Two-Dimensional {MoSi}$_2$N$_4$ Family.
\newblock {\em Advanced Functional Materials},  2023.

\bibitem{PhysRevB.96.115409}
V.~Shahnazaryan, I.~Iorsh, I.~A. Shelykh, and O.~Kyriienko.
\newblock Exciton-exciton interaction in transition-metal dichalcogenide
  monolayers.
\newblock {\em Phys. Rev. B}, 96:115409,  2017.

\bibitem{PhysRevB.103.045426}
D.~Erkensten, S.~Brem, and E.~Malic.
\newblock Exciton-exciton interaction in transition metal dichalcogenide
  monolayers and van der Waals heterostructures.
\newblock {\em Phys. Rev. B}, 103:045426,  2021.

\bibitem{Tartakovskii2019}
A.~Tartakovskii.
\newblock Excitons in 2D heterostructures.
\newblock {\em Nature Reviews Physics}, 2(1):8–9,  2019.

\bibitem{Jiang2021}
Y.~Jiang, S.~Chen, W.~Zheng, B.~Zheng, and A.~Pan.
\newblock Interlayer exciton formation, relaxation, and transport in TMD van
  der Waals heterostructures.
\newblock {\em Light: Science and Applications}, 10(1),  2021.

\bibitem{Li2018}
Z.~Li, T.~Wang, Z.~Lu, C.~Jin, Y.~Chen, Y.~Meng, Z.~Lian, T.~Taniguchi,
  K.~Watanabe, S.~Zhang, D.~Smirnov, and S.-F. Shi.
\newblock Revealing the biexciton and trion-exciton complexes in BN
  encapsulated WSe2.
\newblock {\em Nature Communications}, 9(1),  2018.

\bibitem{Plechinger2015}
G.~Plechinger, P.~Nagler, J.~Kraus, N.~Paradiso, C.~Strunk, C.~Sch\"{u}ller,
  and T.~Korn.
\newblock Identification of excitons, trions and biexcitons in single-layer
  WS2: Identification of excitons, trions and biexcitons in single-layer WS2.
\newblock {\em physica status solidi (RRL) - Rapid Research Letters},
  9(8):457–461,  2015.

\bibitem{Cho2021}
Y.~Cho, S.~M. Greene, and T.~C. Berkelbach.
\newblock Simulations of Trions and Biexcitons in Layered Hybrid
  Organic-Inorganic Lead Halide Perovskites.
\newblock {\em Phys. Rev. Lett.}, 126:216402,  2021.

\bibitem{Poonia2023}
A.~K. Poonia, P.~Yadav, B.~Mondal, D.~Mandal, P.~Taank, M.~Shrivastava, A.~Nag,
  A.~Agarwal, and K.~Adarsh.
\newblock Room-Temperature Electron-Hole Condensation in Direct-Band-Gap
  Semiconductor Nanocrystals.
\newblock {\em Phys. Rev. Appl.}, 20:L021002,  2023.

\bibitem{Morita2022}
Y.~Morita, K.~Yoshioka, and M.~Kuwata-Gonokami.
\newblock Observation of Bose-Einstein condensates of excitons in a bulk
  semiconductor.
\newblock {\em Nature Communications}, 13(1),  2022.

\bibitem{Mogi2022}
H.~Mogi, Y.~Arashida, R.~Kikuchi, R.~Mizuno, J.~Wakabayashi, N.~Wada,
  Y.~Miyata, A.~Taninaka, S.~Yoshida, O.~Takeuchi, and H.~Shigekawa.
\newblock Ultrafast nanoscale exciton dynamics via laser-combined scanning
  tunneling microscopy in atomically thin materials.
\newblock {\em npj 2D Materials and Applications}, 6(1),  2022.

\bibitem{Zhang2023}
X.~Zhang and E.~Kioupakis.
\newblock Phonon-assisted optical absorption of SiC polytypes from first
  principles.
\newblock {\em Phys. Rev. B}, 107:115207,  2023.

\bibitem{Hong2020}
J.~Hong, R.~Senga, T.~Pichler, and K.~Suenaga.
\newblock Probing Exciton Dispersions of Freestanding Monolayer
  ${\mathrm{WSe}}_{2}$ by Momentum-Resolved Electron Energy-Loss Spectroscopy.
\newblock {\em Phys. Rev. Lett.}, 124:087401,  2020.

\bibitem{Qiu2021}
D.~Y. Qiu, G.~Cohen, D.~Novichkova, and S.~Refaely-Abramson.
\newblock Signatures of Dimensionality and Symmetry in Exciton Band Structure:
  Consequences for Exciton Dynamics and Transport.
\newblock {\em Nano Letters}, 21(18):7644–7650,  2021.

\end{thebibliography}
